# Combinators: A Centennial View

Stephen Wolfram*

*We give a modern computational introduction to the S,K combinators invented by Moses Schönfinkel in 1920, and present a variety of new results and ideas about combinators. We explore the spectrum of behavior obtained with small combinator expressions, showing a variety of approaches to analysis and visualization. We discuss the implications of evaluation strategies, and of multiway systems representing all possible strategies. We show how causal graphs introduced in recent models of fundamental physics can be applied to combinators, as well as describing how combinators introduce a new form of treelike separation. We give a variety of new results on minimal combinator expressions, as well as showing how empirical computation theory and computational complexity theory can be done with combinators. We also suggest that when viewed in terms of ongoing computation, the S combinator alone may be capable of universal computation.*

## Ultimate Symbolic Abstraction

Before Turing machines, before lambda calculus—even before Gödel's theorem—there were combinators. They were the very first abstract examples ever to be constructed of what we now know as universal computation—and they were first presented on December 7, 1920. In an alternative version of history our whole computing infrastructure might have been built on them. But as it is, for a century, they have remained for the most part a kind of curiosity—and a pinnacle of abstraction, and obscurity.

It's not hard to see why. In their original form from 1920, there were two basic combinators, s and k, which followed the simple replacement rules (now represented very cleanly in terms of patterns in the Wolfram Language):

```
s[x_][y_][z_] → x[z][y[z]]
k[x_][y_] → x
```

The idea was that any symbolic structure could be generated from some combination of s's and k's. As an example, consider a[b[a][c]]. We're not saying what a, b and c are; they're just symbolic objects. But given a, b and c how do we construct a[b[a][c]]? Well, we can do it with the s, k combinators.

A version of this document with immediately executable code is available at https://writings.stephenwolfram.com/2020/12/combinators-a-centennial-view/
*Email: s.wolfram@wolfram.com

Consider the (admittedly obscure) object:

s[s[k[s]][s[k[k]][s[k[s]][k]]]][s[k[s[s[k][k]]]][k]]

(sometimes instead written **S**(**S**(**KS**)(**S**(**KK**)(**S**(**KS**)**K**)))(**S**(**K**(**S**(**SKK**)))**K**)).

Now treat this like a function and apply it to a,b,c s[s[k[s]][s[k[k]][s[k[s]][k]]]][s[k[s[s[k][k]]]][k]][a][b][c]. Then watch what happens when we repeatedly use the s, k combinator replacement rules:

s[s[k[s]][s[k[k]][s[k[s]][k]]]][s[k[s[s[k][k]]]][k]][a][b][c]
s[k[s]][s[k[k]][s[k[s]][k]]][a][s[k[s[s[k][k]]]][k][a]][b][c]
k[s][a][s[k[k]][s[k[s]][k]][a]][s[k[s[s[k][k]]]][k][a]][b][c]
s[s[k[k]][s[k[s]][k]][a]][s[k[s[s[k][k]]]][k][a]][b][c]
s[k[k]][s[k[s]][k]][a][b][s[k[s[s[k][k]]]][k][a][b]][c]
k[k][a][s[k[s]][k][a]][b][s[k[s[s[k][k]]]][k][a][b]][c]
k[s[k[s]][k][a]][b][s[k[s[s[k][k]]]][k][a][b]][c]
s[k[s]][k][a][s[k[s[s[k][k]]]][k][a][b]][c]
k[s][a][k[a]][s[k[s[s[k][k]]]][k][a][b]][c]
s[k[a]][s[k[s[s[k][k]]]][k][a][b]][c]
k[a][c][s[k[s[s[k][k]]]][k][a][b][c]]
a[s[k[s[s[k][k]]]][k][a][b][c]]
a[k[s[s[k][k]]][a][k[a]][b][c]]
a[s[s[k][k]][k[a]][b][c]]
a[s[k][k][b][k[a][b]][c]]
a[k[b][k[b]][k[a][b]][c]]
a[b[k[a][b]][c]]
a[b[a][c]]

Or, a tiny bit less obscurely:

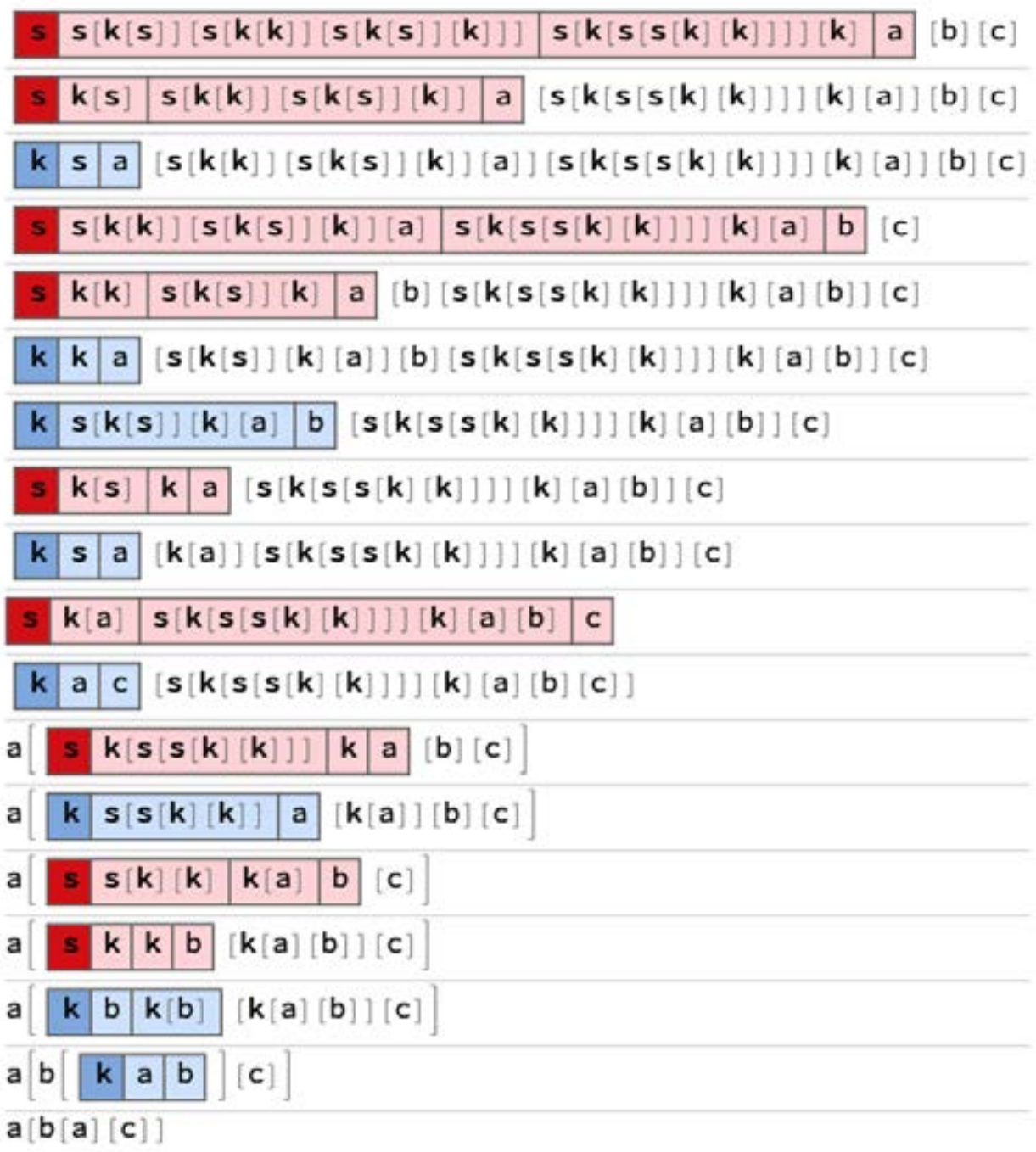

After a number of steps, we get a[b[a][c]]! And the point is that whatever symbolic construction we want, we can always set up some combination of s's and k's that will eventually do it for us—and ultimately be computation universal. They're equivalent to Turing machines, lambda calculus and all those other systems we know are universal. But they were discovered before any of these systems.

By the way, here's the Wolfram Language way to get the result above (//. repeatedly applies the rules until nothing changes anymore):

```
s[s[k[s]][s[k[k]][s[k[s]][k]]]][s[k[s[s[k][k]]]][k]][a][b][c] //. {s[x_][y_][z_] → x[z][y[z]], k[x_][y_] → x}
```

```
a[b[a][c]]
```

And, yes, it's no accident that it's extremely easy and natural to work with combinators in the Wolfram Language—because in fact combinators were part of the deep ancestry of the core design of the Wolfram Language.

For me, though, combinators also have another profound personal resonance. They're examples of very simple computational systems that turn out (as we'll see at length here) to show the same remarkable complexity of behavior that I've spent so many years studying across the computational universe.

A century ago—particularly without actual computers on which to do experiments—the conceptual framework that I've developed for thinking about the computational universe didn't exist. But I've always thought that of all systems, combinators were perhaps the earliest great "near miss" to what I've ended up discovering in the computational universe.

## Computing with Combinators

Let's say we want to use combinators to do a computation on something. The first question is: how should we represent the "something"? Well, the obvious answer is: just use structures built out of combinators!

For example, let's say we want to represent integers. Here's an (at first bizarre-seeming) way to do that. Take s[k] and repeatedly apply s[s[k[s]][k]]. Then we'll get a sequence of combinator expressions:

```
s[k]
s[s[k[s]][k]][s[k]]
s[s[k[s]][k]][s[s[k[s]][k]][s[k]]]
s[s[k[s]][k]][s[s[k[s]][k]][s[s[k[s]][k]][s[k]]]]
s[s[k[s]][k]][s[s[k[s]][k]][s[s[k[s]][k]][s[s[k[s]][k]][s[k]]]]]
s[s[k[s]][k]][s[s[k[s]][k]][s[s[k[s]][k]][s[s[k[s]][k]][s[s[k[s]][k]][s[k]]]]]]
⋮
```

On their own, these expressions are inert under the s and k rules. But take each one (say e) and form e[s][k]. Here's what happens for example to the third case above when you then apply the s and k rules:

```
s[s[k[s]][k]][s[s[k[s]][k]][s[k]]][s][k]
s[k[s]][k][s][s[s[k[s]][k]][s[k]][s]][k]
k[s][s][k[s]][s[s[k[s]][k]][s[k]][s]][k]
s[k[s]][s[s[k[s]][k]][s[k]][s]][k]
k[s][k][s[s[k[s]][k]][s[k]][s][k]]
s[s[s[k[s]][k]][s[k]][s][k]]
s[s[k[s]][k][s][s[k][s]][k]]
s[k[s][s][k[s]][s[k][s]][k]]
s[s[k[s]][s[k][s]][k]]
s[k[s][k][s[k][s][k]]]
s[s[s[k][s][k]]]
s[s[k[k][s[k]]]]
s[s[k]]
```

To get this in the Wolfram Language, we can use Nest, which nestedly applies functions:

```
Nest[f, x, 4]
```

```
f[f[f[f[x]]]]
```

Then the final result above is obtained as:

```
Nest[s[s[k[s]][k]], s[k], 2][s][k] //. {s[x_][y_][z_]→x[z][y[z]], k[x_][y_]→x}
```

```
s[s[k]]
```

Here’s an example involving nesting 7 times:

```
Nest[s[s[k[s]][k]], s[k], 7][s][k] //. {s[x_][y_][z_]→x[z][y[z]], k[x_][y_]→x}
```

```
s[s[s[s[s[s[s[k]]]]]]]
```

So this gives us a (perhaps seemingly obscure) way to represent an integer n. Just form:

Nest[s[s[k[s]][k]], s[k], *n*]

This is a combinator representation of n, that we can “decode” by applying to [s][k]. OK, so given two integers represented this way, how would we add them together? Well, there’s a combinator for that! And here it is:

```
s[k[s]][s[k[s[k[s]]]][s[k[k]]]]
```

If we call this plus, then let's compute plus[1][2][s][k], where 1 and 2 are represented by combinators:

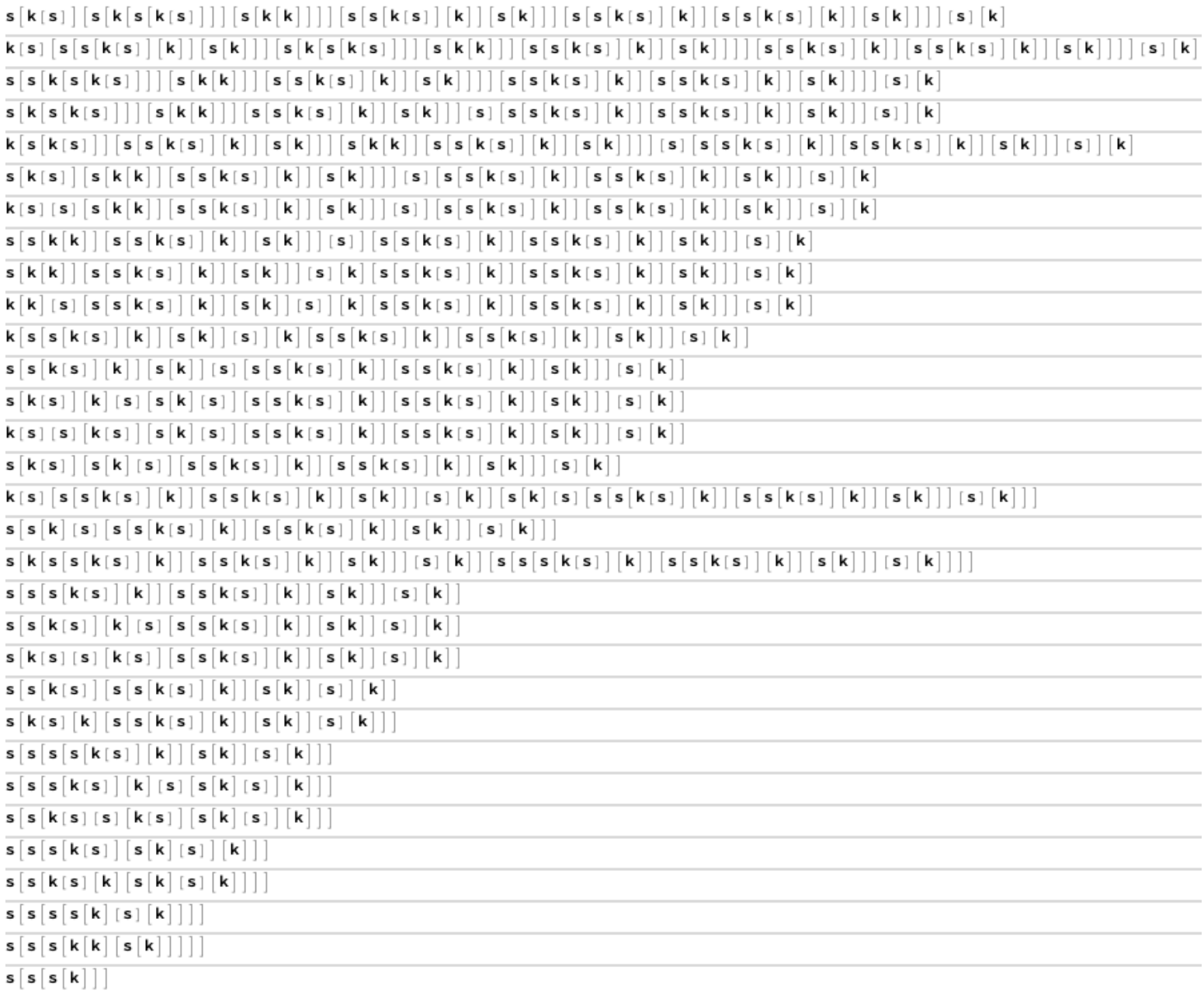

It takes a while, but there's the result: 1 + 2 = 3.

Here's 4 + 3, giving the result s[s[s[s[s[s[s[k]]]]]]] (i.e. 7), albeit after 51 steps:

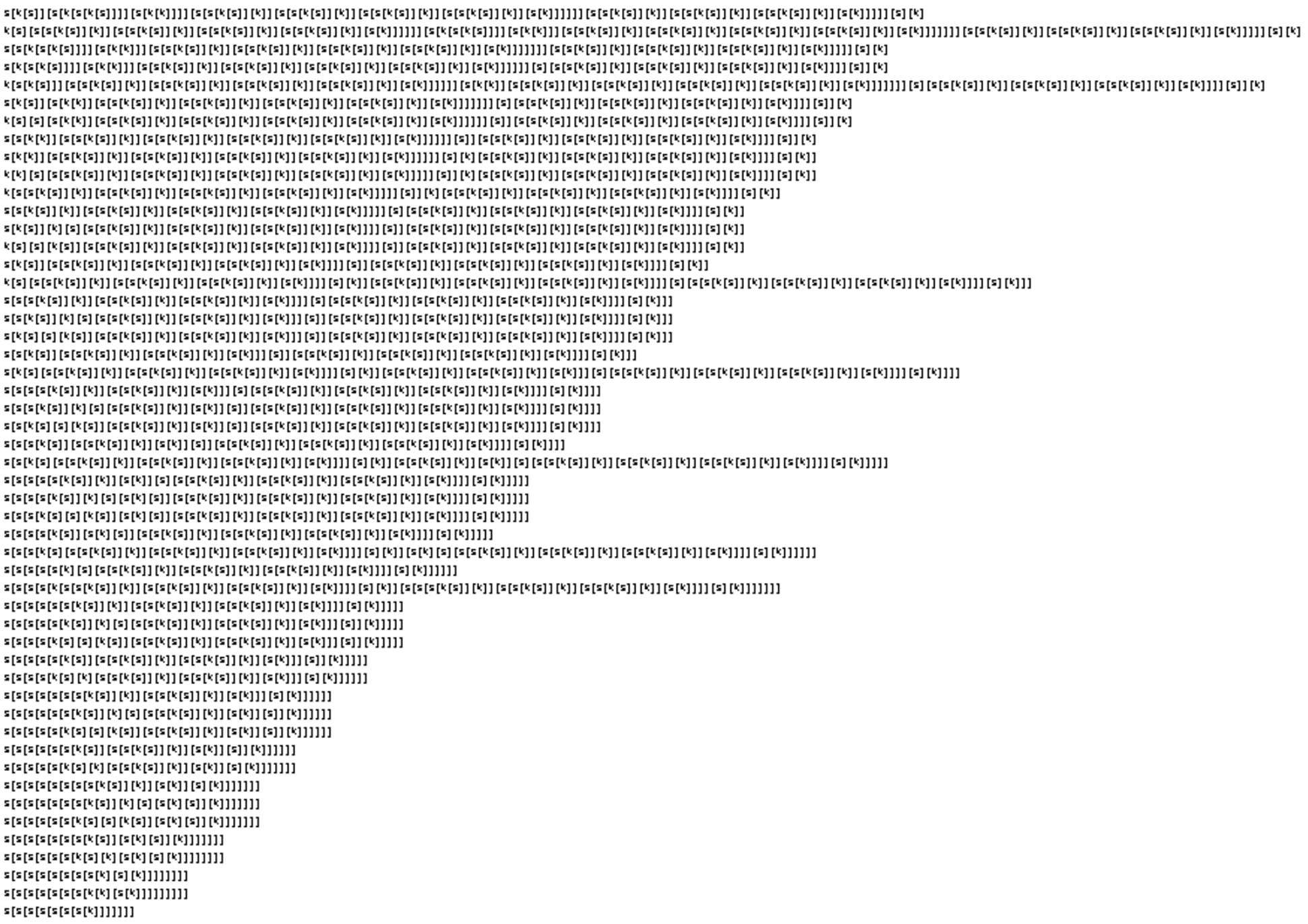

What about doing multiplication? There's a combinator for that too, and it's actually rather simple:

s[k[s]][k]

Here's the computation for 3 × 2—giving 6 after 58 steps:

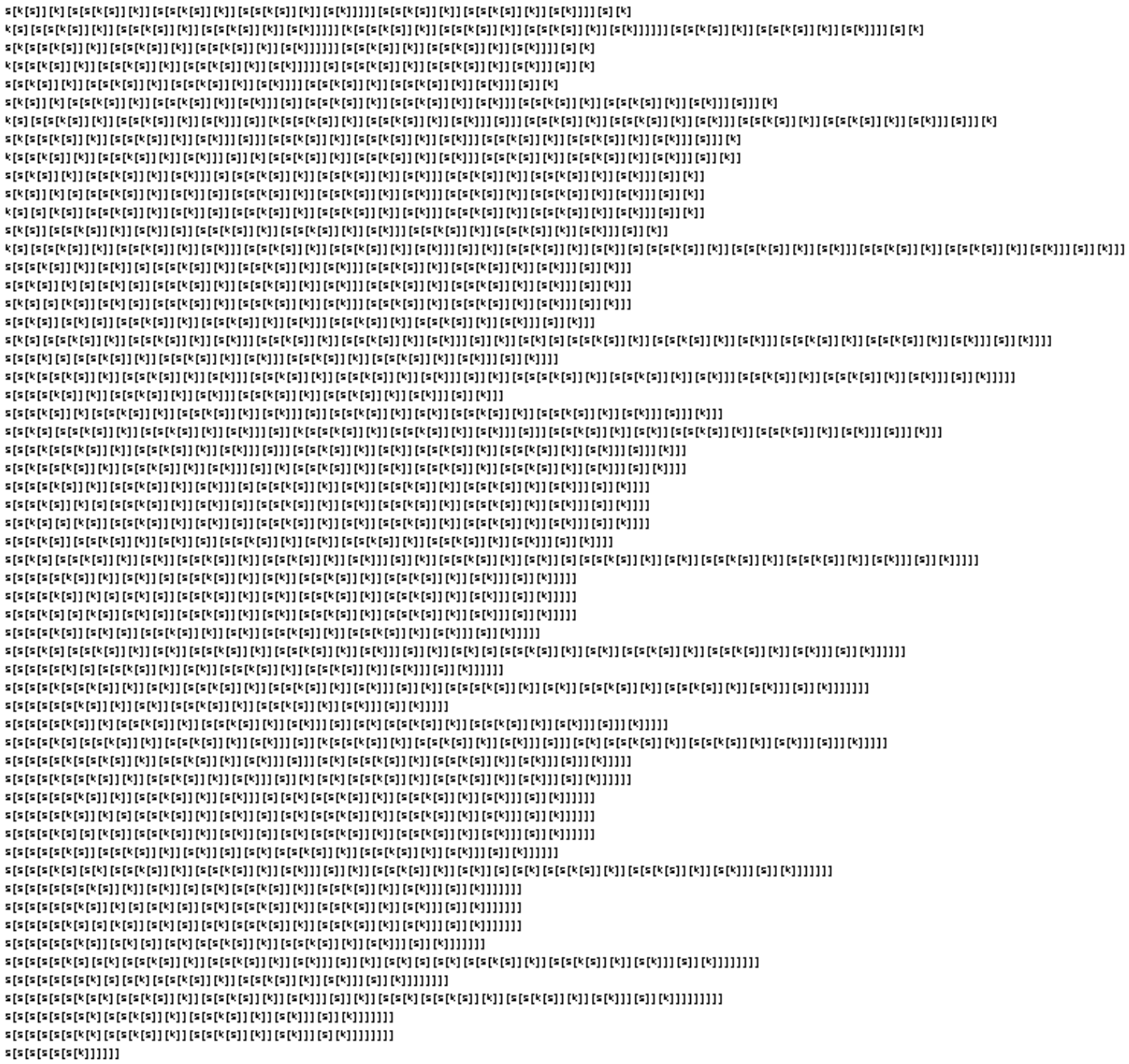

Here's a combinator for power:

s[k[s[s[k][k]]]][k]

And here's the computation of $3^2$ using it (which takes 116 steps):

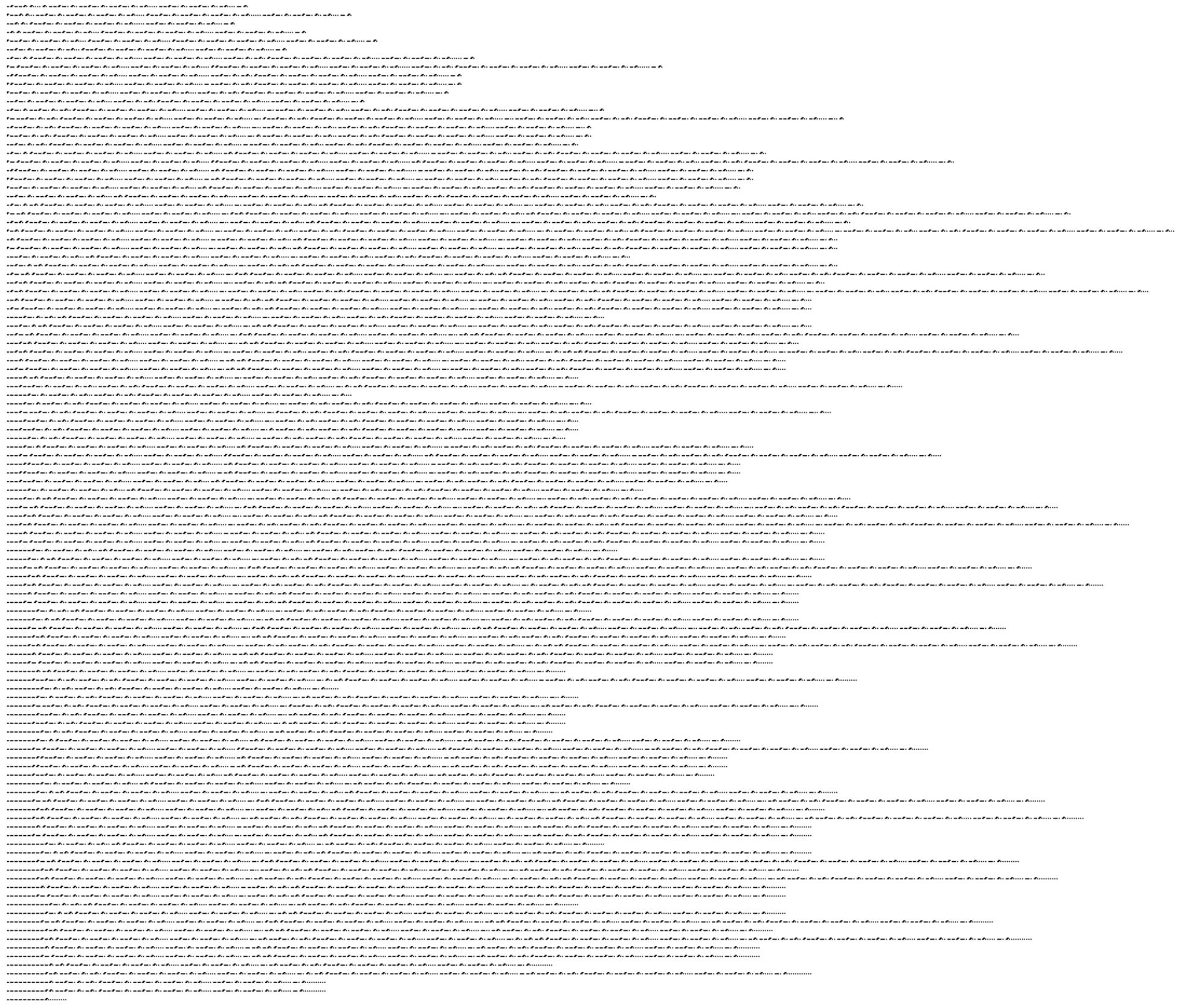

One might think this is a crazy way to compute things. But what's important is that it works, and, by the way, the basic idea for it was invented in 1920.

And while it might seem complicated, it's very elegant. All you need is s and k. Then you can construct everything from them: functions, data, whatever.

So far we're using what's essentially a unary representation of numbers. But we can set up combinators to handle binary numbers instead. Or, for example, we can set up combinators to do logic operations.

Imagine having k stand for true, and s[k] stand for false (so, like If[p,x,y], k[x][y] gives x while s[k][x][y] gives y). Then the minimal combinator for And is just

**s[s[s]][s][s[k]]**

and we can check this works by computing a truth table (FF, FT, TF, TT):

```
s[s][k][s[k]][s[k]]
s[s[k]][k[s[k]]][s[k]]
s[k][s[k]][k[s[k]][s[k]]]
k[k[s[k]][s[k]]][
  s[k][k[s[k]][s[k]]]]
k[s[k]][s[k]]
s[k]
```

```
s[s][k][s[k]][k]
s[s[k]][k[s[k]]][k]
s[k][k][k[s[k]][k]]
k[k[s[k]][k]][k[k[s[k]][k]]]
k[s[k]][k]
s[k]
```

```
s[s][k][k][s[k]]
s[k][k[k]][s[k]]
k[s[k]][k[k][s[k]]]
s[k]
```

```
s[s][k][k][k]
s[k][k[k]][k]
k[k][k[k][k]]
k
```

A search gives the minimal combinator expressions for the 16 possible 2-input Boolean functions:

| | | | | |
|---|---|---|---|---|
| 0 | □□□□ | True | k[k[k]] | K(KK) |
| 1 | □□□■ | Nand | s[s[k[s[s[s][k[k[k]]]]]]][s] | S(S(K(S(SS(K(KK))))))S |
| 2 | □□■□ | Implies | s[s][k[k[k]]] | SS(K(KK)) |
| 3 | □□■■ | Not | s[s][s[s[s[s[k]]][s]]][k[k]] | SS(S(S(S(SK))S))(KK) |
| 4 | □■□□ | | s[k[s[s[s][k[k[k]]]]]][k] | S(K(S(SS(K(KK)))))K |
| 5 | □■□■ | Not | k[s[s[s][k[k[k]]]][s]] | K(S(SS(K(KK)))S) |
| 6 | □■■□ | Equal | s[s][k[s[s[s][k[k[k]]]][s]]] | SS(K(S(SS(K(KK)))S)) |
| 7 | □■■■ | Nor | s[s[s[s[s][k[k[k[k]]]]]][k[s]]] | S(S(S(SS(K(K(KK)))))(KS)) |
| 8 | ■□□□ | Or | s[s[s]][s][s[k]] | S(SS)S(SK) |
| 9 | ■□□■ | Xor | s[s[s[s[s]][s[s[s[k]]]][s]]][k] | S(S(S(SS)(S(S(SK)))S))K |
| 10 | ■□■□ | Last | s[k] | SK |
| 11 | ■□■■ | | s[s[s[k]]][s] | S(S(SK))S |
| 12 | ■■□□ | First | k[k][s] | KKS |
| 13 | ■■□■ | | s[k[s[s[s[s[k]]][s]]]][k] | S(K(S(S(S(SK))S)))K |
| 14 | ■■■□ | And | s[s][k] | SSK |
| 15 | ■■■■ | False | k[k[s[k]]] | K(K(SK)) |

And by combining these (or even just copies of the one for **Nand**) one can make combinators that compute any possible Boolean function. And in fact in general one can—at least in principle—represent any computation by "compiling" it into combinators.

Here's a more elaborate example, from my book A New Kind of Science. This is a combinator that represents one step in the evolution of the rule 110 cellular automaton:

```
s[s[k[s]][s[k[s[s[k][k]]]]][
s[k[k]][s[s[s[s[s[k][k]][k[s[k]]]][k[s[s[k[s]][s[k[s[s[k][k]]]][s[k[k]][
s[s[k[s]][s[k[s[s[k][k]]]][s[k[k]][s[s[k][k]][k[k]]]]]][s[k[k]][s[
s[s[k][k]][k[s[k]]]][k[s[k]]]]]]]]][s[k[k]][s[s[s[k][k]][k[s[k]]]][k[k]]]]]][
s[k[s[s[s[k][k]][k[s[s[s[k][k]][k[s[s[s[k][k]][k[k]]][k[k]]]]][k[k]]]]]]][
s[k[k]][s[s[s[s[s[k][k]][k[s[k]]]]][k[
s[s[k[s]][s[k[s[s[k][k]]]][s[k[k]][s[s[s[k][k]][k[k]]][k[k]]]]]][
s[k[k]][s[s[k[s]][s[k[s[s[k][k]]]][s[k[k]][s[s[k][k]][k[s[k]]]]]]][s[k[k]][
s[s[s[s[s[k][k]][k[k]]][k[s[k]]]][s[s[s[s[k][k]][k[k]]][k[k]]][k[s[k]]]]][
s[s[s[s[s[s[k][k]][k[k]]][k[k]]][k[s[k]]]][k[s[k]]]][
s[s[s[s[s[s[s[k][k]][k[k]]][k[k]]][k[k]]][k[s[k]]]][k[s[k]]]][k[k]]]]]]]]]][
s[s[k[s]][s[k[s[s[k][k]]]][s[k[k]][s[s[k[s]][s[k[s[s[k][k]]]][s[k[k]][
s[s[k][k]][k[k]]]]]][k[k[k]]]]]]][k[k[k]]]]][k[s[k]]]]]]][k[k]]]]]][
s[k[k]][s[k[s[s[k[s]][k]]]][s[s[k][k]][k[s[k]]]]]
```

And, here from the book, are representations of repeatedly applying this combinator to compute—with great effort—three steps in the evolution of rule 110:

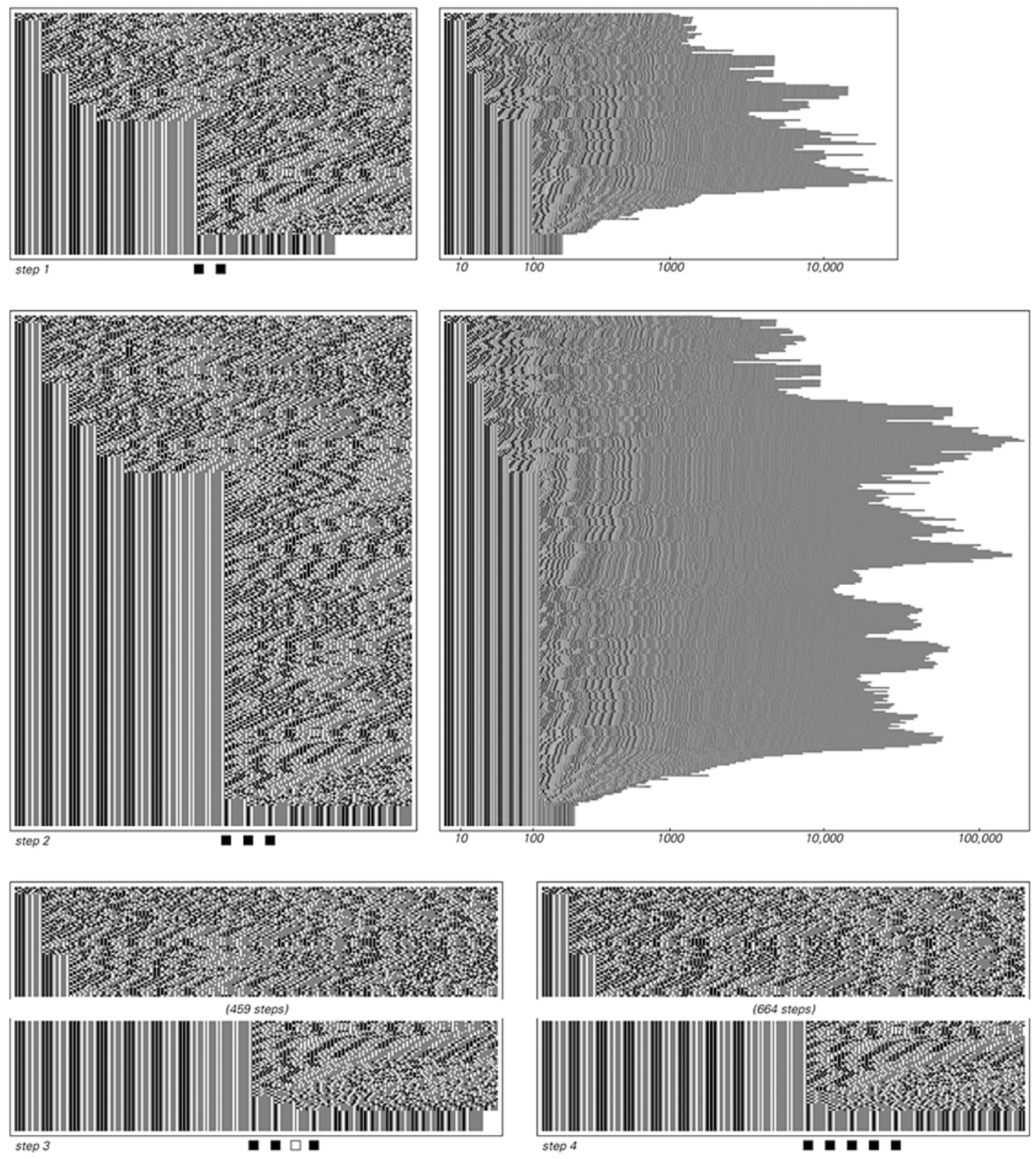

Emulating the rule 110 cellular automaton using combinators. The rule 110 combinator from the previous page is applied once for each step of rule 110 evolution. The initial state is taken to consist of a single black cell.

There's a little further to go, involving fixed-point combinators, etc. But basically, since we know that rule 110 is computation universal, this shows that combinators also are.

## A Hundred Years Later...

Now that a century has passed, what should we think of combinators? In some sense, they still might be the purest way to represent computation that we know. But they're also very hard for us humans to understand.

Still, as computation and the computational paradigm advance, and become more familiar, it seems like on many fronts we're moving ever closer to core ideas of combinators. And indeed the foundational symbolic structure of the Wolfram Language—and much of what I've personally built over the past 40 years—can ultimately be seen as deeply informed by ideas that first arose in combinators.

Computation may be the single most powerful unifying intellectual concept ever. But the actual engineering development of computers and computing has tended to keep different aspects of it apart. There's data. There are data types. There's code. There are functions. There are variables. There's flow of control. And, yes, it may be convenient to keep these things apart in the traditional approach to the engineering of computer systems. But it doesn't need to be that way. And combinators show us that actually there's no need to have any of these distinctions: everything can be together, and can made of the same, dynamic "computational stuff".

It's a very powerful idea. But in its raw form, it's also very disorienting for us humans. Because to understand things, we tend to rely on having "fixed anchors" to which we can attach meaning. And in pure, ever-changing seas of s, k combinators—like the ones we saw above—we just don't have these.

Still, there's a compromise—and in a sense that's exactly what's made it possible for me to build the full-scale computational language that the Wolfram Language now is. The point is that if we're going to be able to represent everything in the world computationally we need the kind of unity and flexibility that combinator-like constructs provide. But we don't just want raw, simple combinators. We need to in effect pre-define lots of combinator-like constructs that have particular meanings related to what we're representing in the world.

At a practical level, the crucial idea is to represent everything as a symbolic expression, and then to say that evaluating these expressions consists in repeatedly applying transformations to them. And, yes, symbolic expressions in the Wolfram Language are just like the expressions we've made out of combinators—except that instead of involving only s's and k's, they involve thousands of different symbolic constructs that we define to represent molecules, or cities or polynomials. But the key point is that—like with combinators—the things we're dealing with are always structurally just nested applications of pure symbolic objects.

Something we immediately learn from combinators is that "data" is really no different from "code"; they can both be represented as symbolic expressions. And both can be the raw material for computation. We also learn that "data" doesn't have to maintain any particular type or structure; not only its content, but also the way it is built up as a symbolic expression can be the dynamic output of a computation.

One might imagine that things like this would just be esoteric matters of principle. But what I've learned in building the Wolfram Language is that actually they're natural and crucially important in having convenient ways to capture computationally how we humans think about things, and the way the world is.

From the early days of practical computing, there was an immediate instinct to imagine that programs should be set up as sequences of instructions saying for example "take a thing, then do this to it, then do that" and so on. The result would be a "procedural" program like:

```
x = f[x]; x = g[x]; x = h[x]; x
```

But as the combinator approach suggests, there's a conceptually much simpler way to write this in which one's just successively applying functions, to make a "functional" program:

```
h[g[f[x]]]
```

(In the Wolfram Language, this can also be written h@g@f@x or x//f//g//h.)

Given the notion that everything is a symbolic expression, one's immediately led to have functions to operate on other functions, like

```
Nest[f, x, 6]
```

```
f[f[f[f[f[f[x]]]]]]
```

or:

```
ReverseApplied[f][a, b]
```

```
f[b, a]
```

This idea of such "higher-order functions" is quintessentially combinator informed—and very elegant and powerful. And as the years go by we're gradually managing to see how to make more and more aspects of it understandable and accessible in the Wolfram Language (think: **Fold**, **MapThread**, **SubsetMap**, **FoldPair**, ...).

OK, but there's one more thing combinators do—and it's their most famous: they allow one to set things up so that one never needs to define variables or name things. In typical programming one might write things like:

```
With[{x = 3}, 1 + x ^ 2]
```

```
f[x_] := 1 + x ^ 2
```

```
Function[x, 1 + x ^ 2]
```

```
x ⟼ 1 + x ^ 2
```

But in none of these cases does it matter what the actual name x is. The x is just a place-holder that's standing for something one's "passing around" in one's code.

But why can't one just "do the plumbing" of specifying how something should be passed around, without explicitly naming anything? In a sense a nested sequence of functions like f[g[x]] is doing a simple case of this; we're not giving a name to the result of g[x]; we're just feeding it as input to f in a "single pipe". And by setting up something like Function[x, 1+x^2] we're constructing a function that doesn't have a name, but which we can still apply to things:

```
Function[x, 1+x^2][4]
```

17

The Wolfram Language gives us an easy way to get rid of the x here too:

```
(1+#^2) &[4]
```

17

In a sense the # ("slot") here acts a like a pronoun in a natural language: we're saying that whatever we're dealing with (which we're not going to name), we want to find "one plus the square of it".

OK, but so what about the general case? Well, that's what combinators provide a way to do.

Consider an expression like:

```
f[g[x][y]][y]
```

Imagine this was called q, and that we wanted q[x][y] to give f[g[x][y]][y]. Is there a way to define q without ever mentioning names of variables? Yes, here's how to do it with s, k combinators:

```
s[s[k[s]][s[k[s[k[f]]]][g]]][k[s[k][k]]]
```

There's no mention of x and y here; the combinator structure is just defining—without naming anything—how to "flow in" whatever one provides as "arguments". Let's watch it happen:

```
s[s[k[s]][s[k[s[k[f]]]][g]]][k[s[k][k]]][x][y]
s[k[s]][s[k[s[k[f]]]][g]][x][k[s[k][k]][x]][y]
k[s][x][s[k[s[k[f]]]][g][x]][k[s[k][k]][x]][y]
s[s[k[s[k[f]]]][g][x]][k[s[k][k]][x]][y]
s[k[s[k[f]]]][g][x][y][k[s[k][k]][x][y]]
k[s[k[f]]][x][g[x]][y][k[s[k][k]][x][y]]
s[k[f]][g[x]][y][k[s[k][k]][x][y]]
k[f][y][g[x][y]][k[s[k][k]][x][y]]
f[g[x][y]][k[s[k][k]][x][y]]
f[g[x][y]][s[k][k][y]]
f[g[x][y]][k[y][k[y]]]
f[g[x][y]][y]
```

Yes, it seems utterly obscure. And try as I might over the years to find a usefully human-understandable "packaging" of this that we could build into the Wolfram Language, I have so far failed.

But it's very interesting—and inspirational—that there's even in principle a way to avoid all named variables. Yes, it's often not a problem to use named variables in writing programs, and the names may even communicate useful information. But there are all sorts of tangles they can get one into.

It’s particularly bad when a name is somehow global, and assigning a value to it affects (potentially insidiously) everything one’s doing. But even if one keeps the scope of a name localized, there are still plenty of problems that can occur.

Consider for example:

```
Function[x, Function[y, 2 x + y]]
```

It’s two nested anonymous functions (AKA lambdas)—and here the x “gets” a, and y “gets” b:

```
Function[x, Function[y, 2 x + y]][a][b]
```

```
2 a + b
```

But what about this:

```
Function[x, Function[x, 2 x + x]]
```

The Wolfram Language conveniently colors things red to indicate that something bad is going on. We’ve got a clash of names, and we don’t know “which x” is supposed to refer to what.

It’s a pretty general problem; it happens even in natural language. If we write “Jane chased Bob. Jane ran fast.” it’s pretty clear what we’re saying. But “Jane chased Jane. Jane ran fast.” is already confused. In natural language, we avoid names with pronouns (which are basically the analog of # in the Wolfram Language). And because of the (traditional) gender setup in English “Jane chased Bob. She ran fast.” happens to work. But “The cat chased the mouse. It ran fast.” again doesn’t.

But combinators solve all this, by in effect giving a symbolic procedure to describe what reference goes where. And, yes, by now computers can easily follow this (at least if they deal with symbolic expressions, like in the Wolfram Language). But the passage of a century—and even our experience with computation—doesn’t seem to have made it much easier for us humans to follow it.

By the way, it’s worth mentioning one more “famous” feature of combinators—that actually had been independently invented before combinators—and that these days, rather ahistorically, usually goes by the name “currying”. It’s pretty common—say in the Wolfram Language—to have functions that naturally take multiple arguments. GeoDistance[a, b] or Plus[a, b, c] (or a+b+c) are examples. But in trying to uniformize as much as possible, combinators just make all “functions” nominally have only one argument.

To set up things that “really” have multiple arguments, one uses structures like f[x][y][z]. From the point of standard mathematics, this is very weird: one expects “functions” to just “take an argument and return a result”, and “map one space to another” (say real numbers to complex numbers).

But if one's thinking "sufficiently symbolically" it's fine. And in the Wolfram Language—with its fundamentally symbolic character (and distant ancestry in combinator concepts)—one can just as well make a definition like

```
f[x_][y_] := x + y
```

as:

```
f[x_, y_] := x + y
```

Back in 1980—even though I don't think I knew about combinators yet at that time—I actually tried in my SMP system that was a predecessor to the Wolfram Language the idea of having f[x][y] be able to be equivalent to f[x,y]. But it was a bit like forcing every verb to be intransitive—and there were many situations in which it was quite unnatural, and hard to understand.

## Combinators in the Wild: Some Zoology

So far we've been talking about combinators that are set up to compute specific things that we want to compute. But what if we just pick possible combinators "from the wild", say at random? What will they do?

In the past, that might not have seemed like a question that was worth asking. But I've now spent decades studying the abstract computational universe of simple programs—and building a whole "new kind of science" around the discoveries I've made about how they behave. And with that conceptual framework it now becomes very interesting to look at combinators "in the wild" and see how they behave.

So let's begin at the beginning. The simplest s, k combinator expressions that won't just remain unchanged under the combinator rules have to have size 3. There are a total of 16 such expressions:

```
{s[s][s], s[s[s]], s[s][k], s[s[k]], s[k][s], s[k[s]], s[k][k],
  s[k[k]], k[s][s], k[s[s]], k[s][k], k[s[k]], k[k][s], k[k[s]], k[k][k], k[k[k]]}
```

And none of them do anything interesting: they either don't change at all, or, as in for example k[s][s], they immediately give a single symbol (here s).

But what about larger combinator expressions? The total number of possible combinator expressions of size n grows like

{2, 4, 16, 80, 448, 2688, 16 896, 109 824, 732 160, 4 978 688}

or in general

$$2^n \,\textbf{CatalanNumber}[n-1] = \frac{2^n \,\textbf{Binomial}[2\,n-2,\, n-1]}{n}$$

or asymptotically:

$$\frac{8^n}{4\sqrt{\pi}\, n^{3/2}}$$

At size 4, again nothing too interesting happens. With all the 80 possible expressions, the longest it takes to reach a fixed point is 3 steps, and that happens in 4 cases:

```
s[k][s][s]
k[s][s[s]]
s

s[k][s][k]
k[k][s[k]]
k

s[k][k][s]
k[s][k[s]]
s

s[k][k][k]
k[k][k[k]]
k
```

At size 5, the longest it takes to reach a fixed point is 4 steps, and that happens in 10 cases out of 448:

```
s[s][s][k][s]
s[k][s[k]][s]
k[s][s[k][s]]
s

s[s][s][k][k]
s[k][s[k]][k]
k[k][s[k][k]]
k

s[s][k][s][s]
s[s][k[s]][s]
s[s][k[s][s]]
s[s][s]

s[s[k]][s][s]
s[k][s][s[s]]
k[s[s]][s[s[s]]]
s[s]

s[s][k][s][k]
s[s][k[s]][k]
s[k][k[s][k]]
s[k][s]

s[s[k]][s][k]
s[k][k][s[k]]
k[s[k]][k[s[k]]]
s[k]

s[s][k][k][s]
s[k][k[k]][s]
k[s][k[k][s]]
s

s[s[k]][k][s]
s[k][s][k[s]]
k[k[s]][s[k[s]]]
k[s]

s[s][k][k][k]
s[k][k[k]][k]
k[k][k[k][k]]
k

s[s[k]][k][k]
s[k][k][k[k]]
k[k[k]][k[k[k]]]
k[k]
```

At size 6, there is a slightly broader distribution of “halting times”:

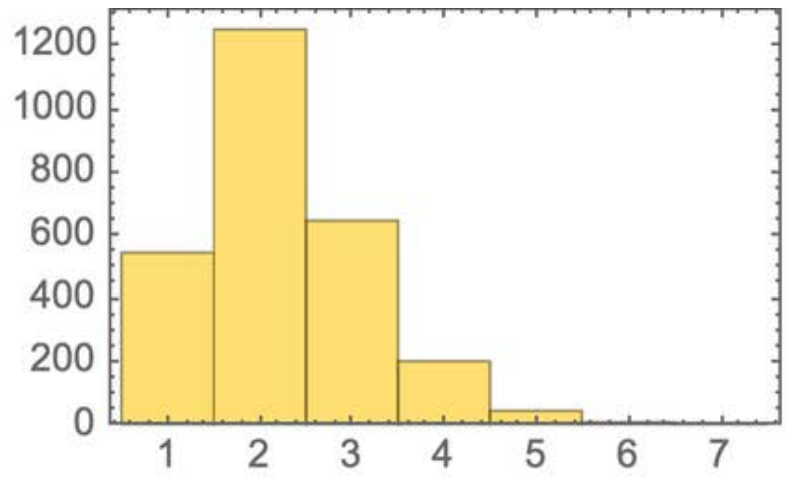

The longest halting time is 7, achieved by:

```
s[s[s]][s][k][s]
s[s][k][s[k]][s]
s[s[k]][k[s[k]]][s]
s[k][s][k[s[k]][s]]
k[k[s[k]][s]][s[k[s[k]][s]]]
k[s[k]][s]
s[k]

s[s[s]][s][k][k]
s[s][k][s[k]][k]
s[s[k]][k[s[k]]][k]
s[k][k][k[s[k]][k]]
k[k[s[k]][k]][k[k[s[k]][k]]]
k[s[k]][k]
s[k]
```

Meanwhile, the largest expressions created are of size 10 (in the sense that they contain a total of 10 s’s or k’s):

```
s[s[s]][s][s][s]
s[s][s][s[s]][s]
s[s[s]][s[s[s]]][s]
s[s][s][s[s[s]][s]]
s[s[s[s]][s]][s[s[s[s]][s]]]

s[s][s][s[s]][s]
s[s[s]][s[s[s]]][s]
s[s][s][s[s[s]][s]]
s[s[s[s]][s]][s[s[s[s]][s]]]

s[s[s]][s][s][k]
s[s][s][s[s]][k]
s[s[s]][s[s[s]]][k]
s[s][k][s[s[s]][k]]
s[s[s[s]][k]][k[s[s[s]][k]]]

s[s][s][s[s]][k]
s[s[s]][s[s[s]]][k]
s[s][k][s[s[s]][k]]
s[s[s[s]][k]][k[s[s[s]][k]]]
```

The distribution of final sizes is a little odd:

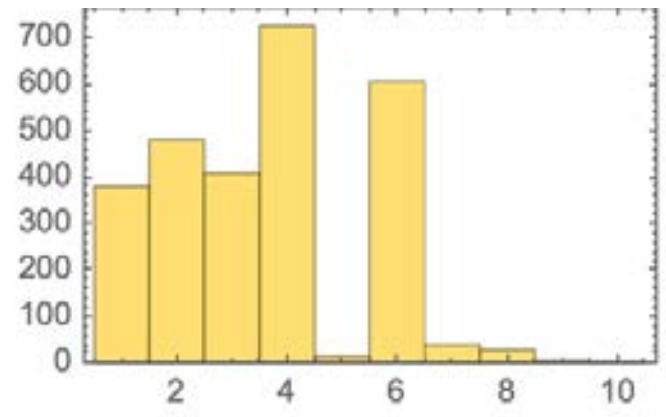

For size n ≤ 5, there's actually a gap with no final states of size n – 1 generated. But at size 6, out of 2688 expressions, there are just 12 that give size 5 (about 0.4%).

OK, so what's going to happen if we go to size 7? Now there are 16,896 possible expressions. And there's something new: two never stabilize (**S**(**SS**)**SSSS**, **SSS**(**SS**)**SS**):

{s[s[s]][s][s][s][s], s[s][s][s[s]][s][s]}

After one step, the first one of these evolves to the second, but then this is what happens over the next few steps (we'll see other visualizations of this later):

```
s[s][s][s[s]][s][s]
s[s[s]][s[s[s]]][s][s]
s[s][s][s[s[s]][s]][s]
s[s[s[s]][s]][s[s[s[s]][s]]][s]
s[s[s]][s][s][s[s[s[s]][s]][s]]
s[s][s][s[s]][s[s[s[s]][s]][s]]
s[s[s]][s[s[s]]][s[s[s[s]][s]][s]]
s[s][s[s[s[s]][s]][s]][s[s[s]][s[s[s[s]][s]][s]]]
s[s[s[s]][s[s[s[s]][s]][s]]][s[s[s[s]][s]][s][s[s[s]][s[s[s[s]][s]][s]]]]
```

The total size (i.e. LeafCount, or "number of s's") grows like:

{7, 8, 8, 11, 11, 11, 12, 17, 25, 33, 41, 50, 59, 87, 115, 149,
187, 215, 243, 272, 301, 389, 398, 413, 422, 431, 440, 450, 460, 491, 533}

A log plot shows that after an initial transient the size grows roughly exponentially:

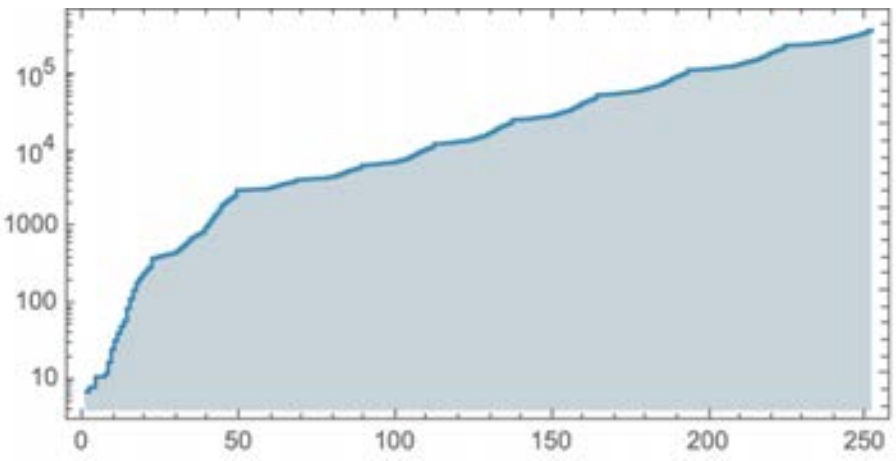

And looking at successive ratios one sees some elaborate fine structure:

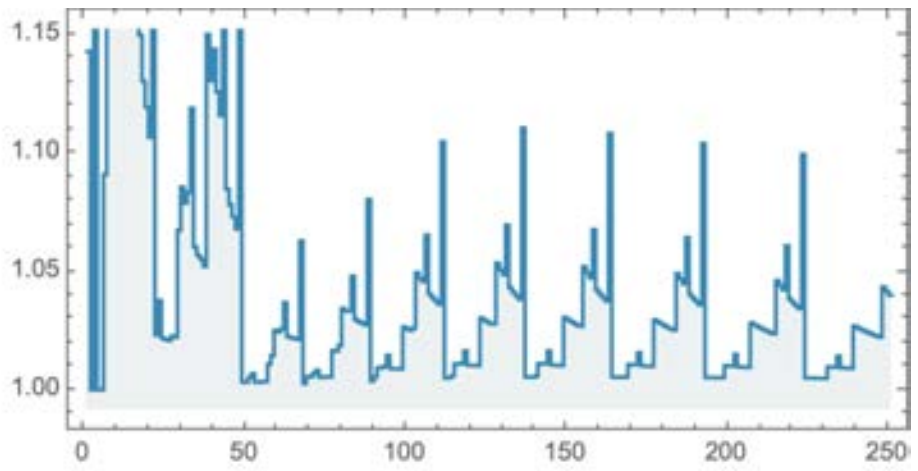

What is this ultimately doing? With a little effort, one finds that the sizes have a length-83 transient, followed by sequences of values of length 23 + 2n, in which the second differences of successive sizes are given by:

```
Join[38 {0, 0, 0, 12, −17} 2^n + {0, 1, 0, −135, 189}, Table[0, n],
  38 {0, 1, 0, 0, 1, −1, 0, 0, 0, 4} 2^n + {12, −13, 0, 6, −7, 1, 0, 1, 0, −27},
  Table[0, n + 2], 228 {0, 1, 0, 0, 1, −1} 2^n + 2 {6, −20, 0, 3, −17, 14}]
```

The final sequence of sizes is obtained by concatenating these blocks and computing Accumulate[Accumulate[*list*]]—giving an the asymptotic size that appears to be of the form $a\sqrt{t}\ 2^{b+\sqrt{t}}$. So, yes, we can ultimately “figure out what’s going on” with this little size-7 combinator (and we’ll see some more details later). But it’s remarkable how complicated it is.

OK, but let’s go back and look at the other size-7 expressions. The halting time distribution (ignoring the 2 cases that don’t halt) basically falls off exponentially, but shows a couple of outliers:

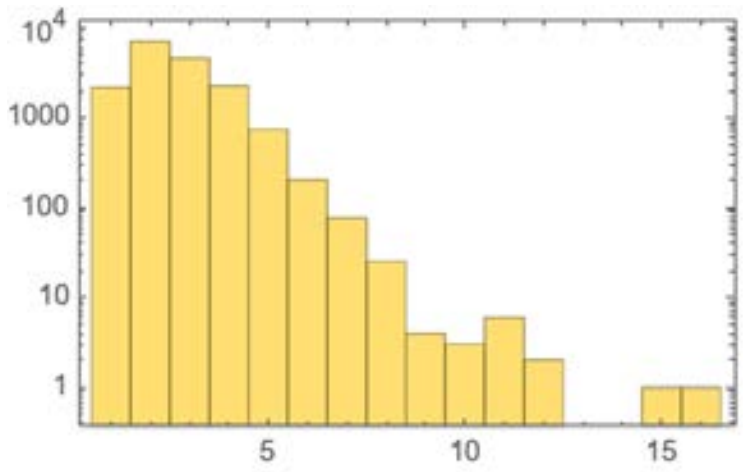

The maximum finite halting time is 16 steps, achieved by s[s[s[s]]][s][s][s] (**S**(**S**(**SS**))**SSS**):

```
s[s[s[s]]][s][s][s]
s[s[s]][s][s[s]][s]
s[s][s[s]][s[s[s]]][s]
s[s[s[s]]][s[s][s[s[s]]]][s]
s[s[s]][s][s[s][s[s[s]]][s]]
s[s][s[s][s[s[s]]][s]][s[s[s][s[s[s]]][s]]]
s[s[s[s][s[s[s]]][s]]][s[s][s[s[s]]][s][s[s[s][s[s[s]]][s]]]]
s[s[s[s][s[s[s]][s]]]][s[s][s[s[s]]][s][s[s[s][s[s[s]]][s]]]]
s[s[s[s][s[s[s]][s]]]][s[s][s[s[s]][s]][s[s[s][s[s[s]]][s]]]]
s[s[s[s][s[s[s]][s]]]][s[s[s[s][s[s[s]]][s]]][s[s[s]][s][s[s[s][s[s[s]]][s]]]]]
s[s[s[s][s[s[s]][s]]]][s[s[s[s][s[s[s]][s]]]][s[s[s]][s][s[s[s][s[s[s]]][s]]]]]
s[s[s[s][s[s[s]][s]]]][s[s[s[s][s[s[s]][s]]]][s[s][s[s[s][s[s[s]]][s]]][s[s[s[s][s[s[s]]][s]]]]]]
s[s[s[s][s[s[s]][s]]]][s[s[s[s][s[s[s]][s]]]][s[s[s[s][s[s[s]]][s]]]][s[s[s][s[s[s]]][s]][s[s[s[s][s[s[s]]][s]]]]]]
s[s[s[s][s[s[s]][s]]]][s[s[s[s][s[s[s]][s]]]][s[s[s[s][s[s[s]][s]]]]][s[s[s][s[s[s]]][s]][s[s[s[s][s[s[s]]][s]]]]]]
s[s[s[s][s[s[s]][s]]]][s[s[s[s][s[s[s]][s]]]][s[s[s[s][s[s[s]][s]]]]][s[s[s][s[s[s]][s]]][s[s[s[s][s[s[s]]][s]]]]]]
s[s[s[s][s[s[s]][s]]]][s[s[s[s][s[s[s]][s]]]][s[s[s[s][s[s[s]][s]]]]][s[s[s][s[s[s]][s]]][s[s[s[s][s[s[s]][s]]]]]]]
```

And the distribution of final sizes is (with the maximum of 41 being achieved by the maximum-halting-time expression we’ve just seen):

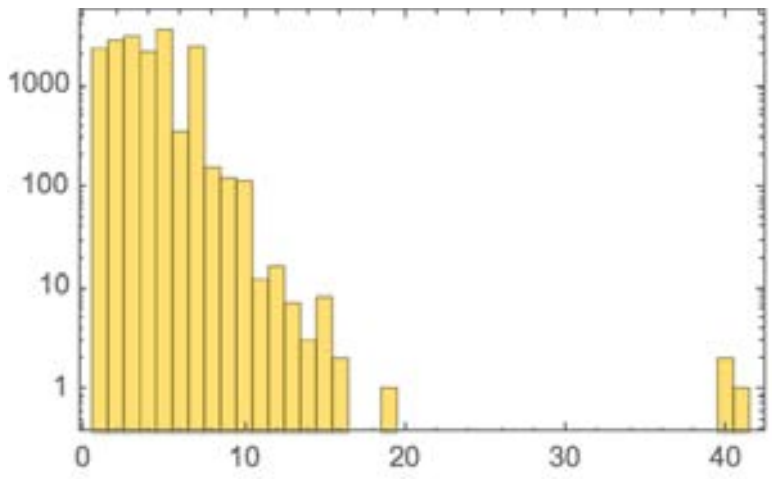

OK, so what happens at size 8? There are 109,824 possible combinator expressions. And it's fairly easy to find out that all but 76 of these go to fixed points within at most 50 steps (the longest survivor is s[s][s][s[s[s]]][k][k] (**SSS**(**S**(**SS**))**KK**), which halts after 44 steps):

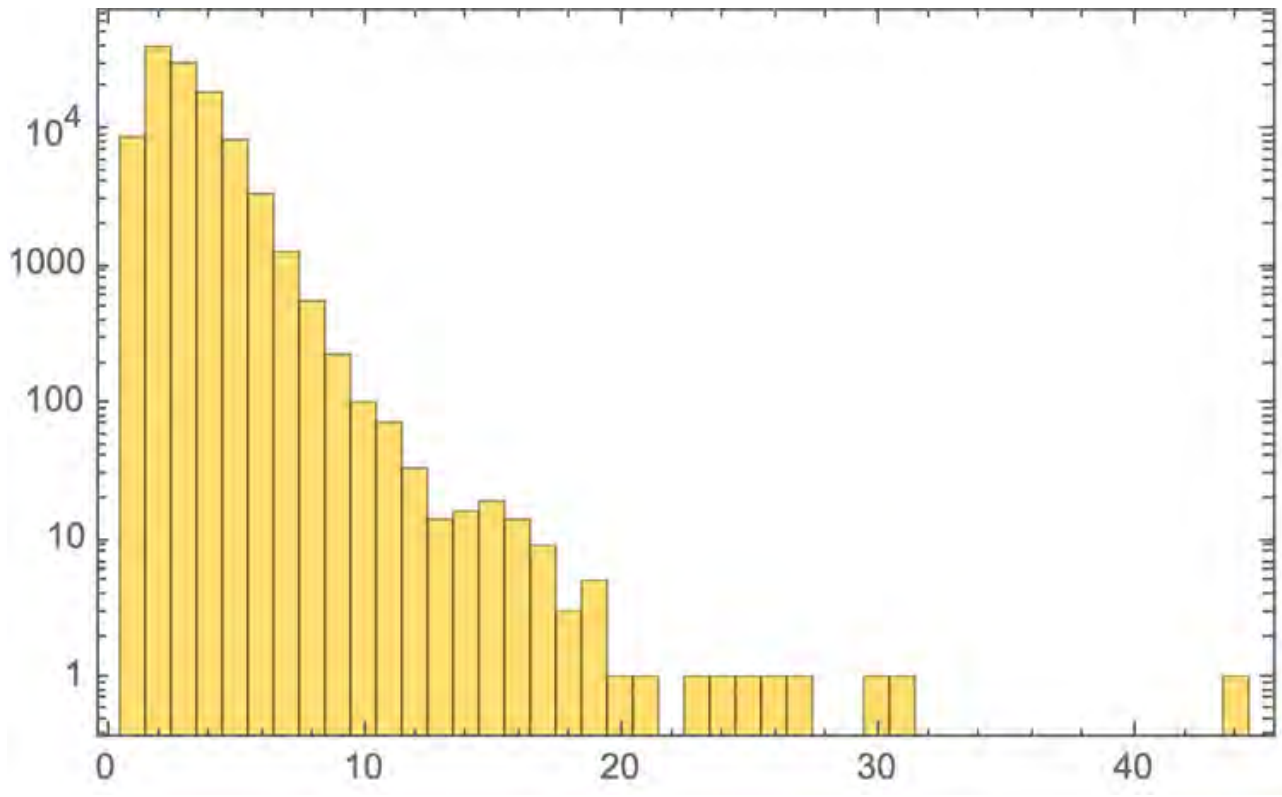

The final fixed points in these cases are mostly quite small; this is the distribution of their sizes:

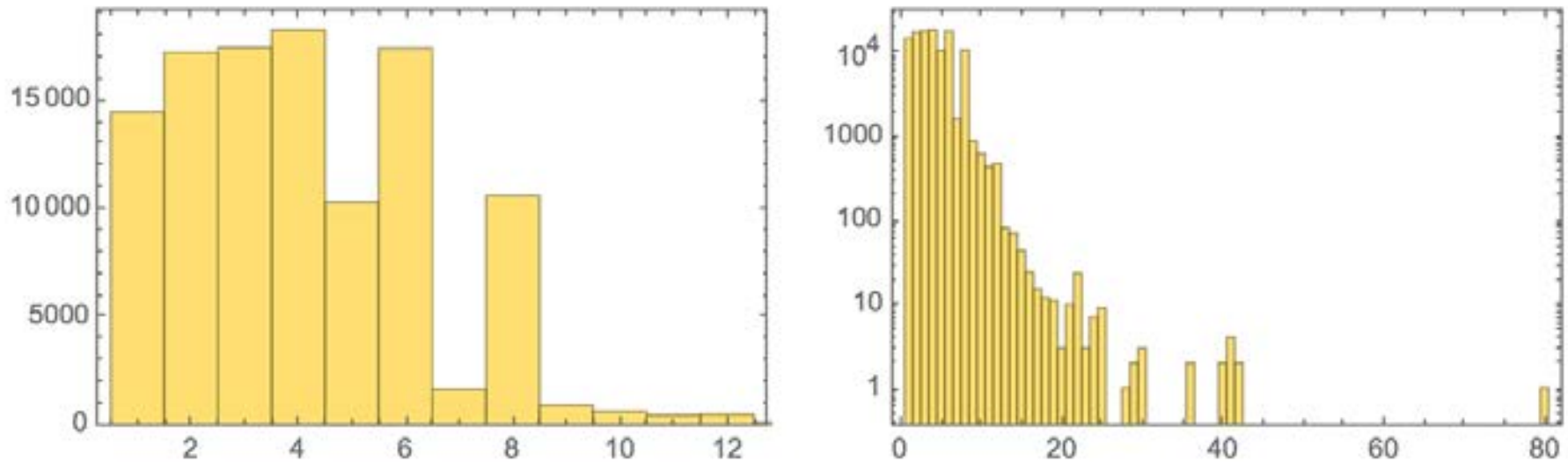

And here is a comparison between halting times and final sizes:

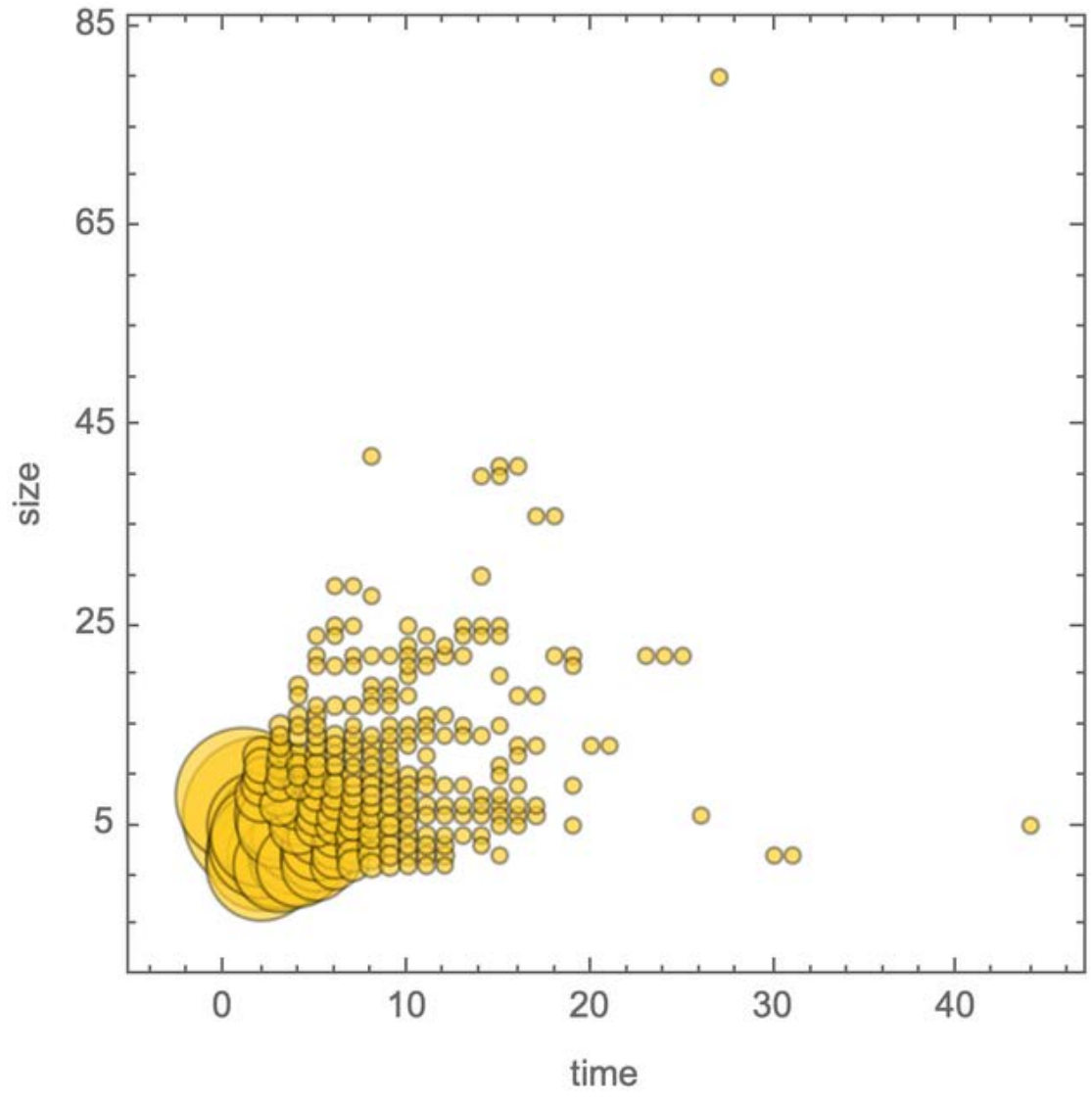

The outlier for size is s[s][k][s[s[s]][s]][s] (**SSK**(**S**(**SS**)**S**)**S**), which evolves in 27 steps to a fixed expression of size 80 (along the way reaching an intermediate expression of size 86):

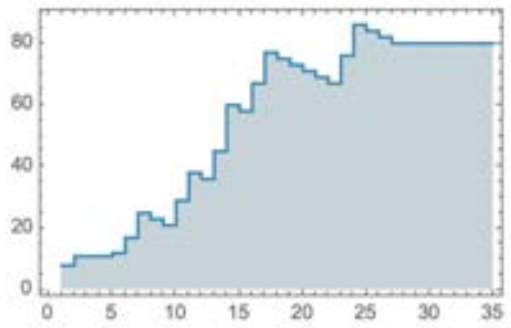

Among combinator expressions that halt in less than 50 steps, the maximum intermediate expression size of 275 is achieved for s[s][s][s[s[s][k]]][k] (**SSS**(**S**(**SSK**))**K**) (which ultimately evolves to s[s[s[s][k]]][k] after 26 steps):

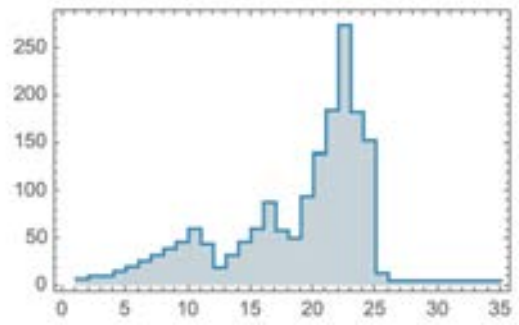

So what about size-8 expressions that don't halt after 50 steps? There are altogether 76—with 46 of these being inequivalent (in the sense that they don't quickly evolve to others in the set).

Here's how these 46 expressions grow (at least until they reach size 10,000):

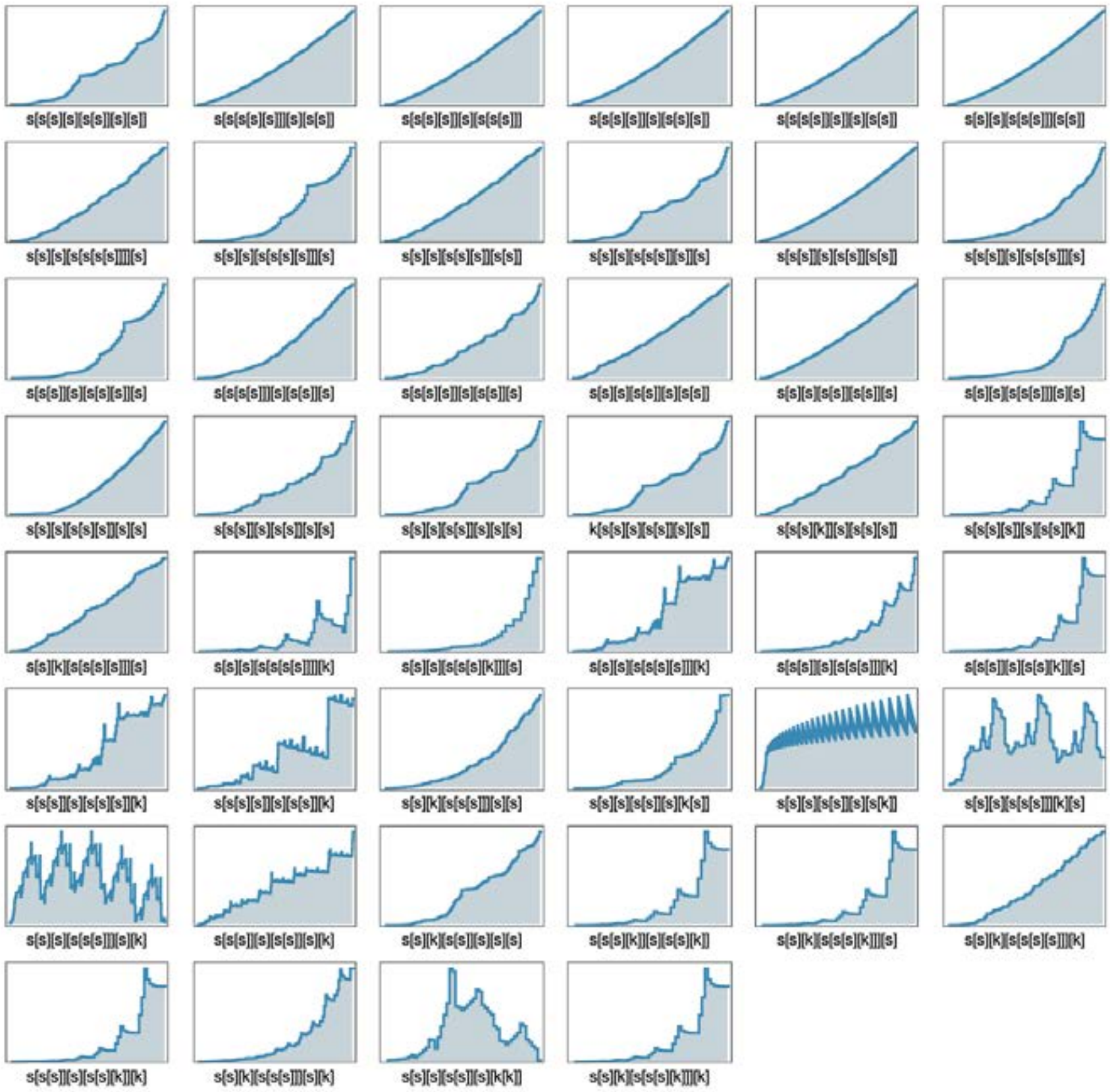

Some of these actually end up halting. In fact, s[s][s][s[s]][s][k[k]] (**SSS**(**SS**)**S**(**KK**)) halts after just 52 steps, with final result k[s[k][k[s[k][k]]]] (**K**(**SK**(**K**(**SKK**)))), having achieved a maximum expression size of 433:

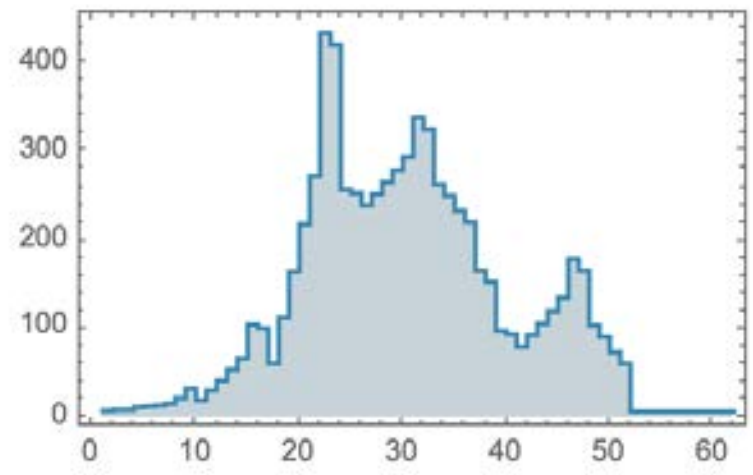

The next shortest halting time occurs for s[s][s][s[s[s]]][k][s] (**SSS**(**S**(**SS**))**KS**), which takes 89 steps to produce an expression of size 65:

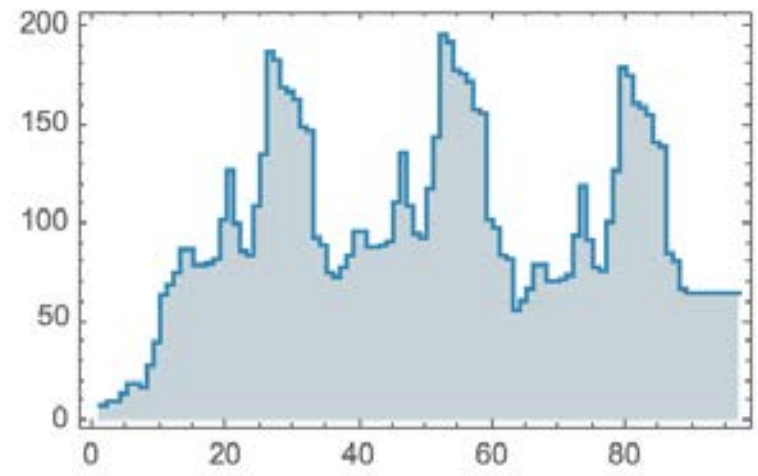

Then we have s[s][s][s[s[s]]][s][k] (**SSS**(**S**(**SS**))**SK**), which halts (giving the size-10 s[k][s[s[s[s[s[s]]][s]]][k]]), but only after 325 steps:

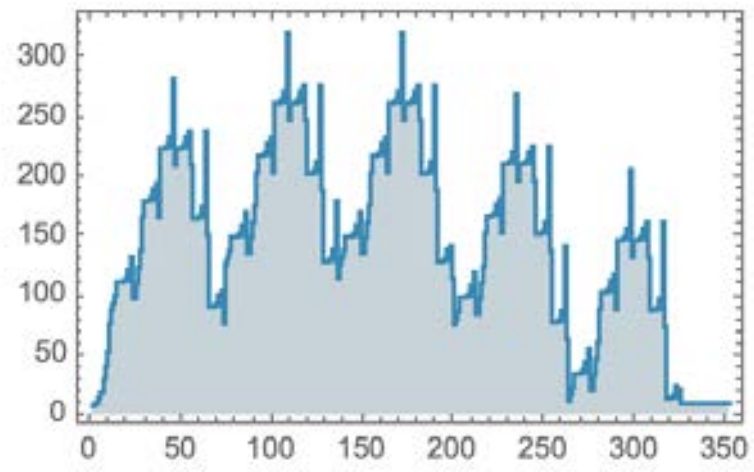

There's also a still-larger case to be seen: s[s[s][s]][s][s[s]][k] (**S**(**SSS**)**S**(**SS**)**K**), which exhibits an interesting "IntegerExponent-like" nested pattern of growth, but finally halts after 1958 steps, having achieved a maximum intermediate expression size of 21,720 along the way:

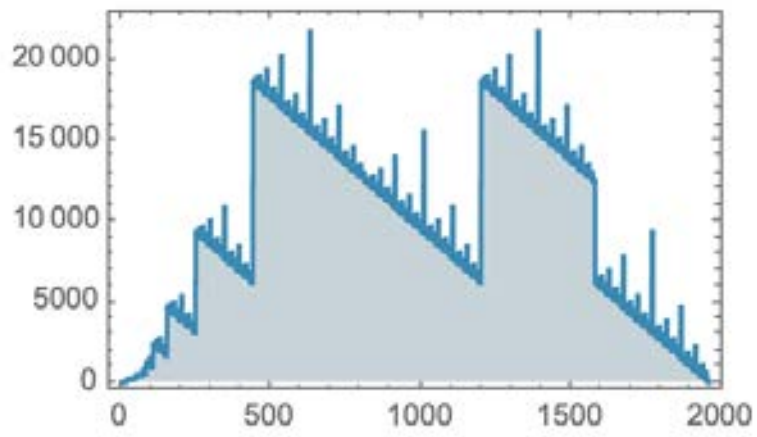

What about the other expressions? s[s][s][s[s]][s][s[k]] (**SSS**(**SS**)**S**(**SK**)) shows very regular $\sqrt{t}$ growth in size:

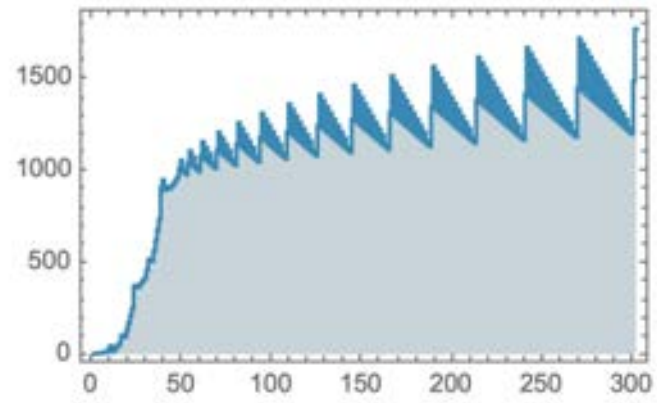

In the other cases, there's no such obvious regularity. But one can start to get a sense of what happens by plotting differences between sizes on successive steps:

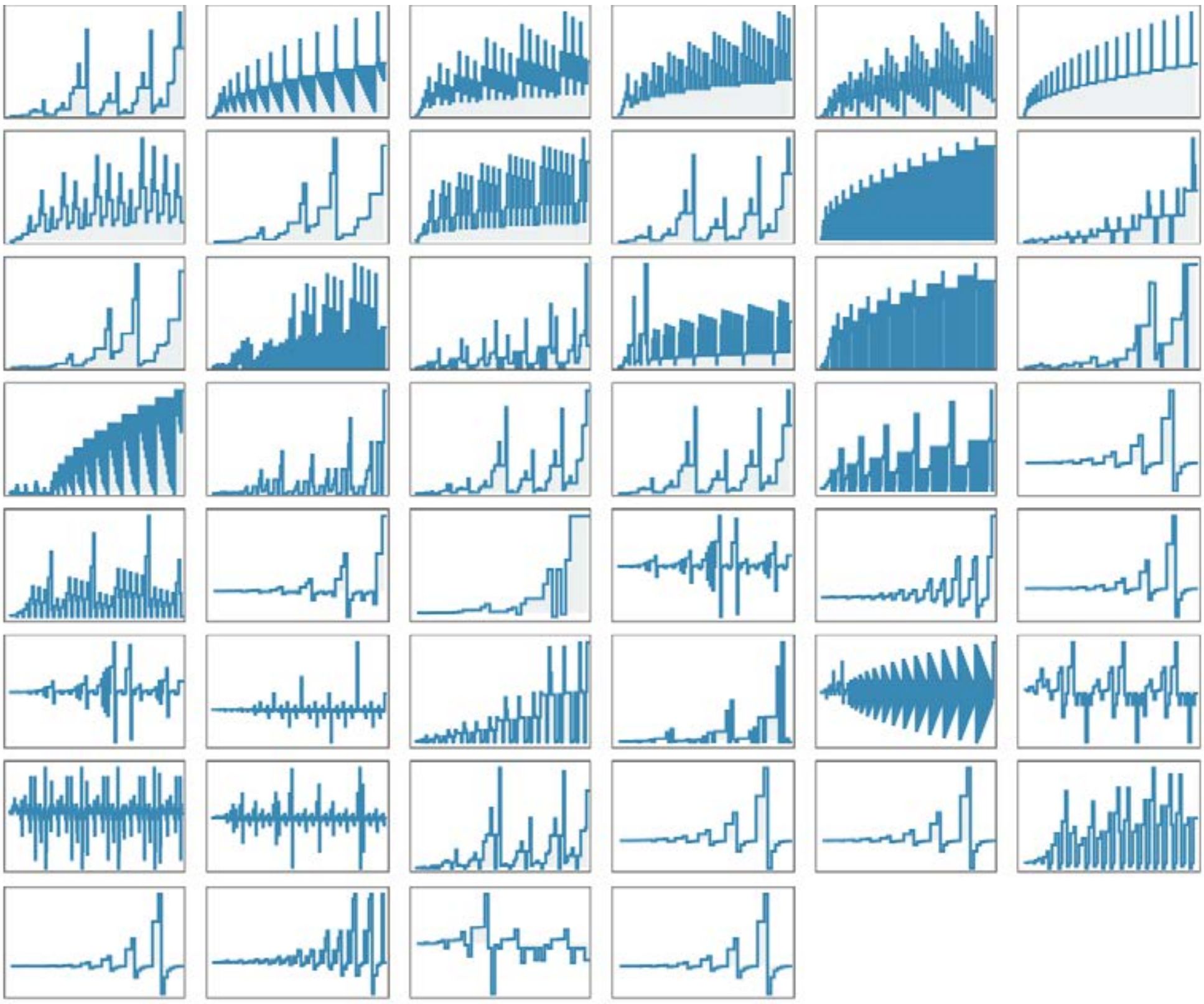

There are some obvious cases of regularity here. Several show a regular pattern of linearly increasing differences, implying overall $t^2$ growth in size:

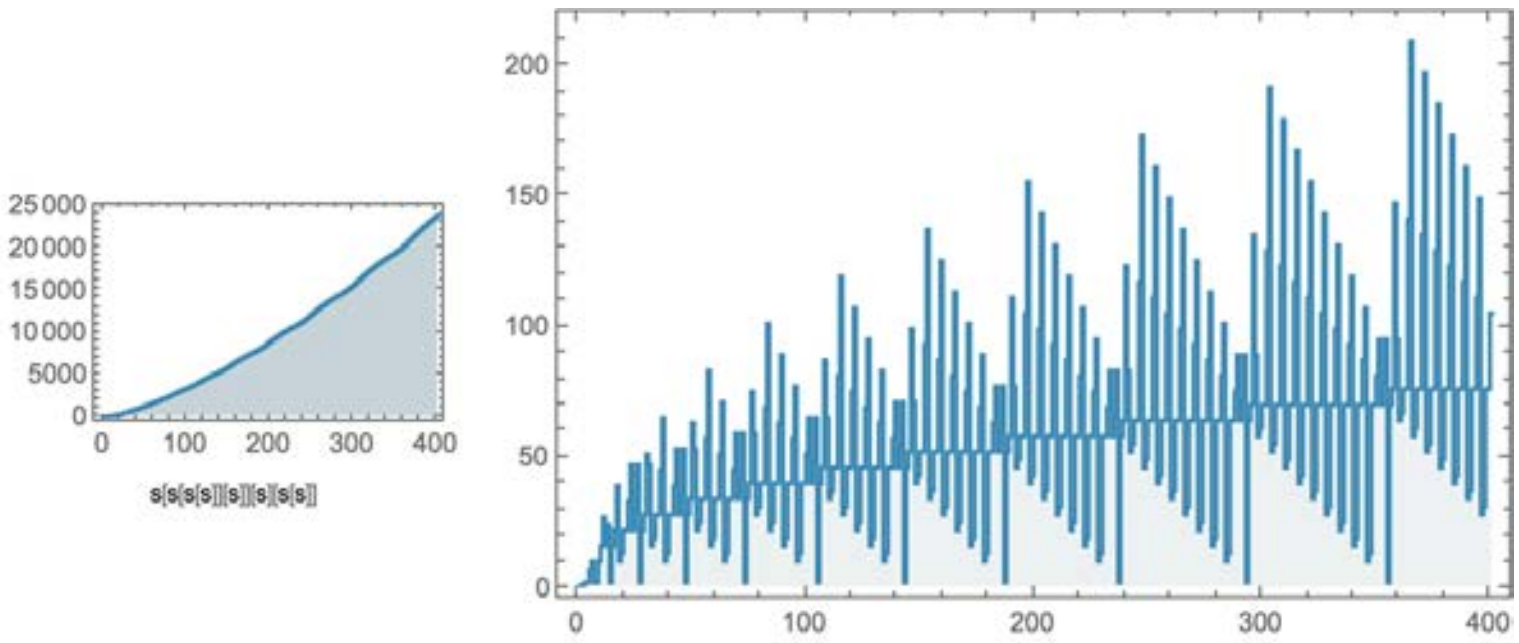

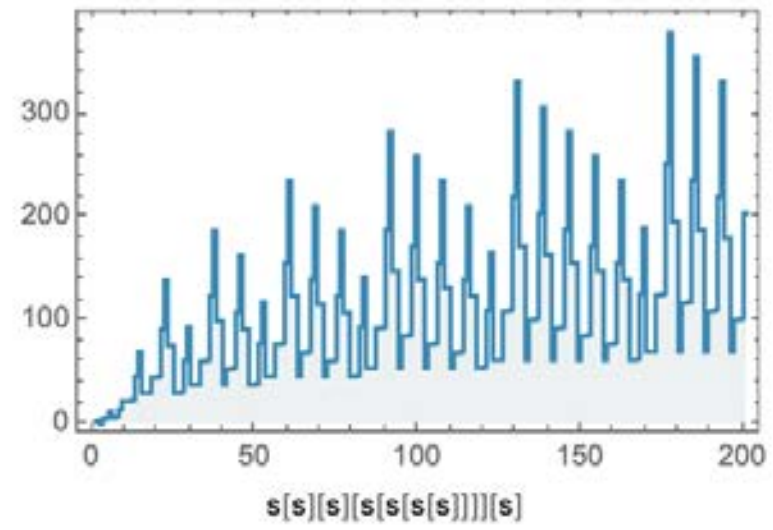

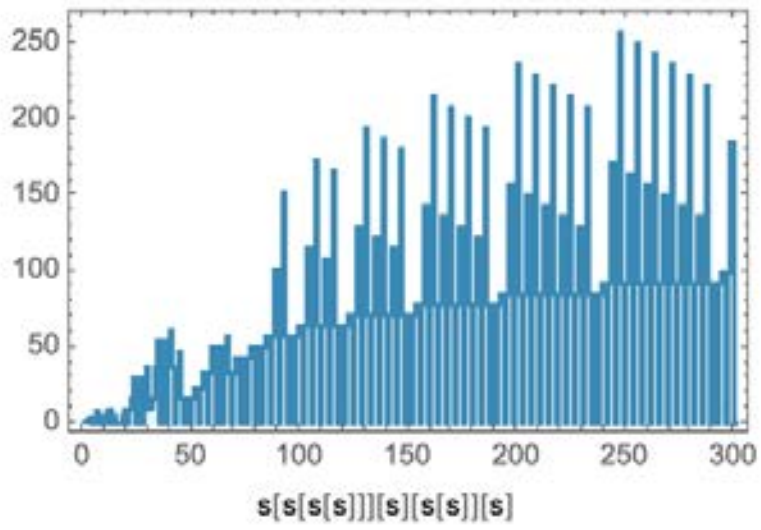

Others show regular $\sqrt{t}$ growth in differences, leading to $t^{3/2}$ growth in size:

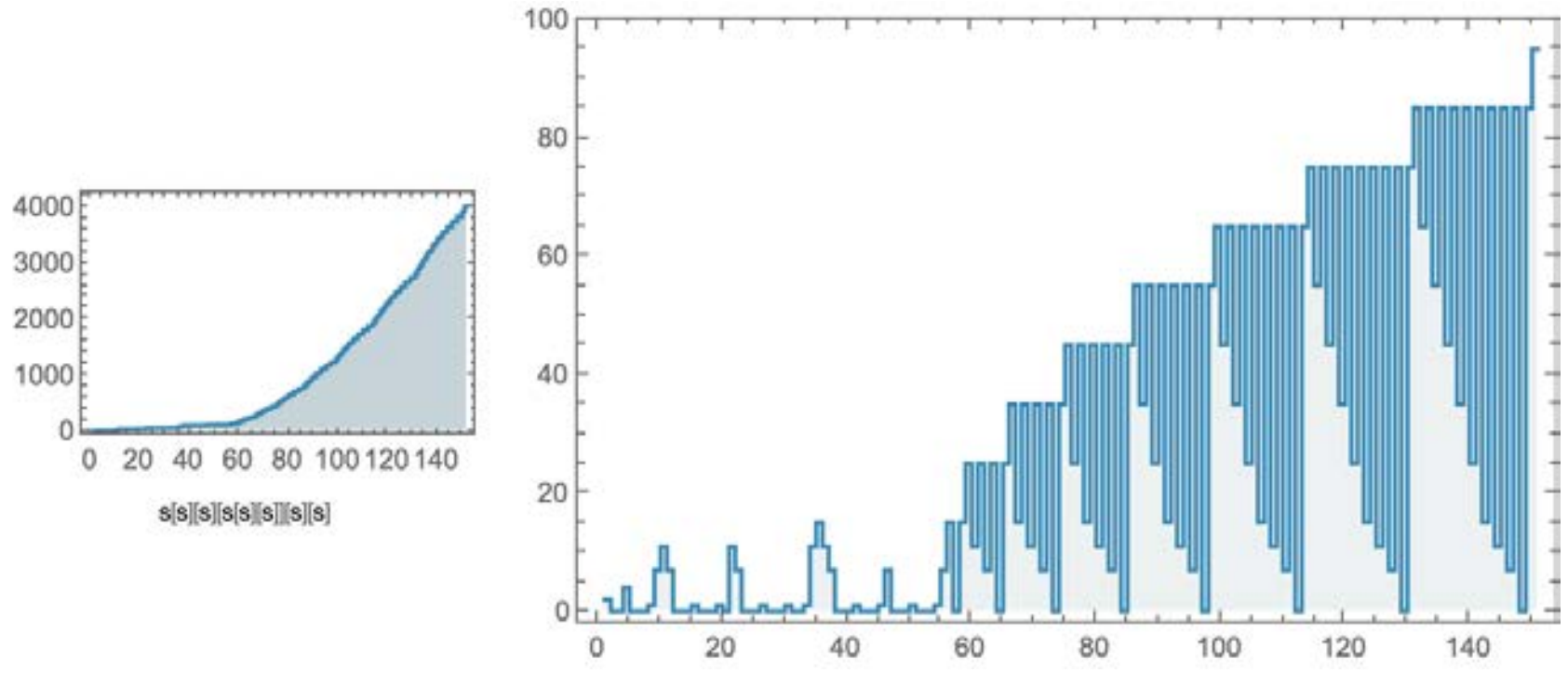

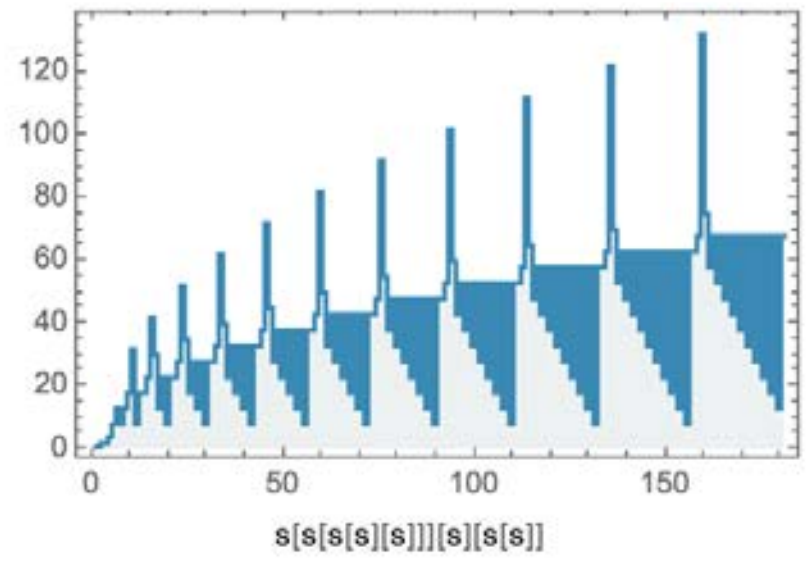

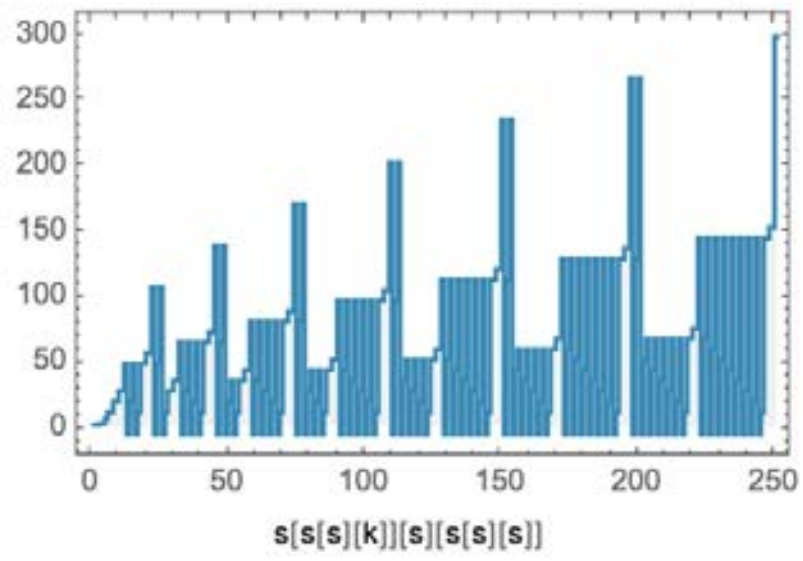

Others have pure exponential growth:

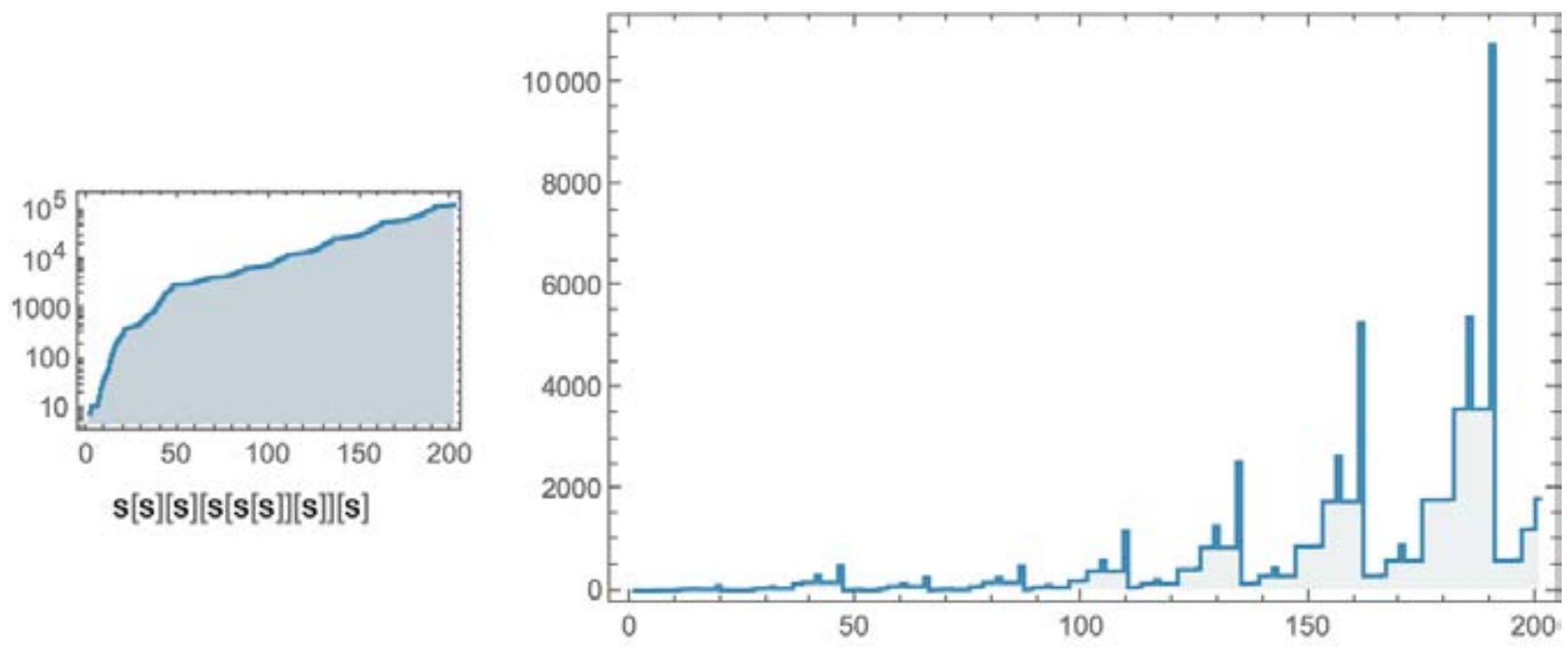

There are quite a few that have regular but below-exponential growth, much like the size-7 case s[s][s][s[s]][s][s] (**SSS**(**SS**)**SS**) with $\sim 2^{\sqrt{t}}$ growth:

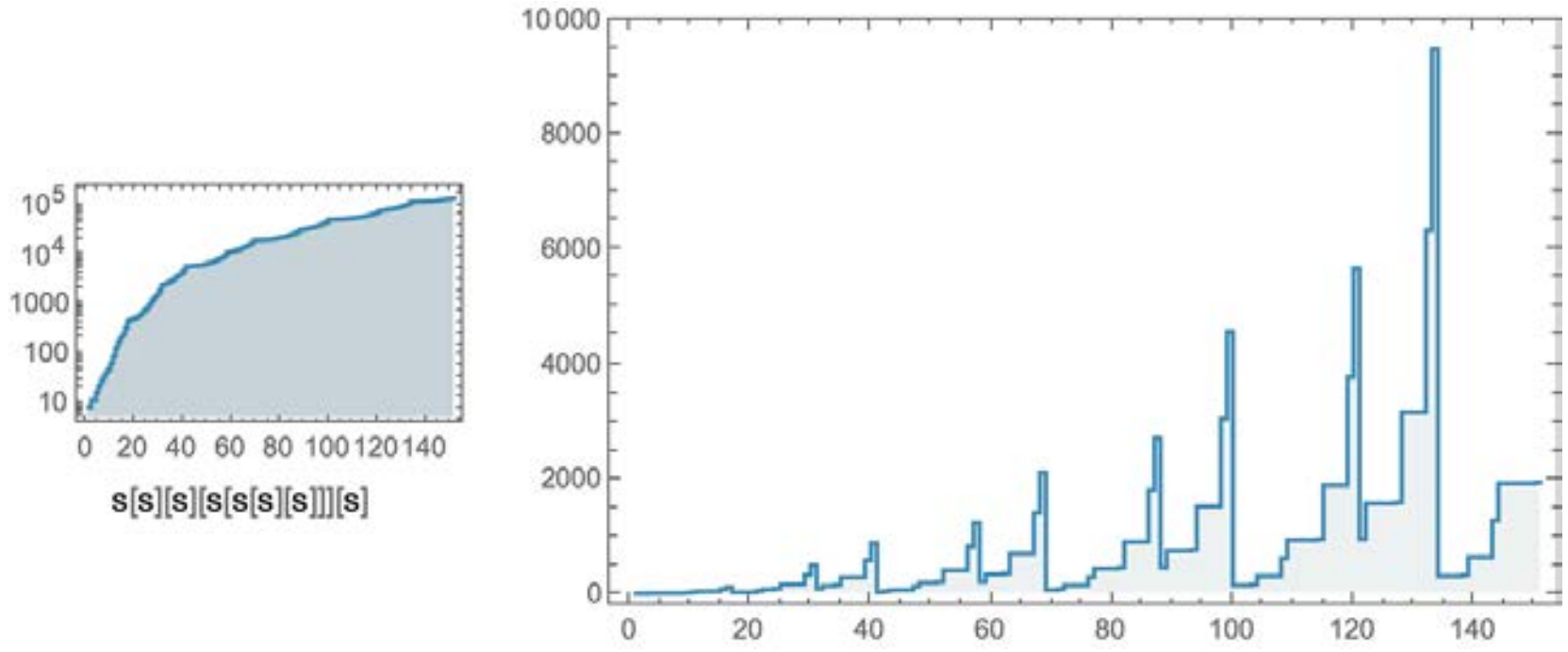

All the cases we've just looked at only involve s. When we allow k as well, there's for example s[s][s][s[s[s][s]]][k] (**SSS**(**S**(**SSS**))**K**)—which shows regular, essentially "stair-like" growth:

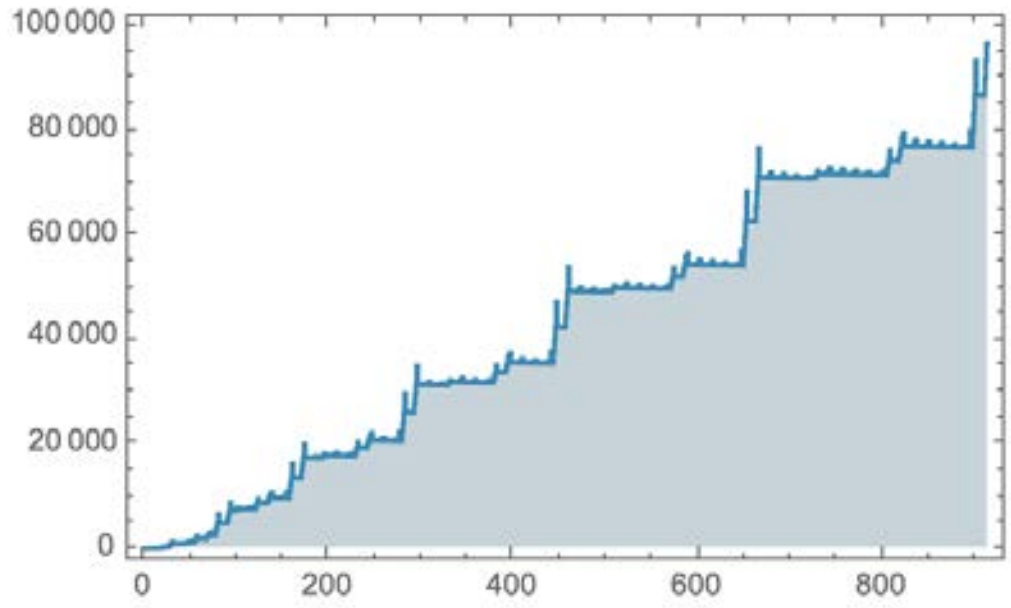

There's also a case like s[s[s]][s][s[s]][s][k] (**S**(**SS**)**S**(**SS**)**SK**):

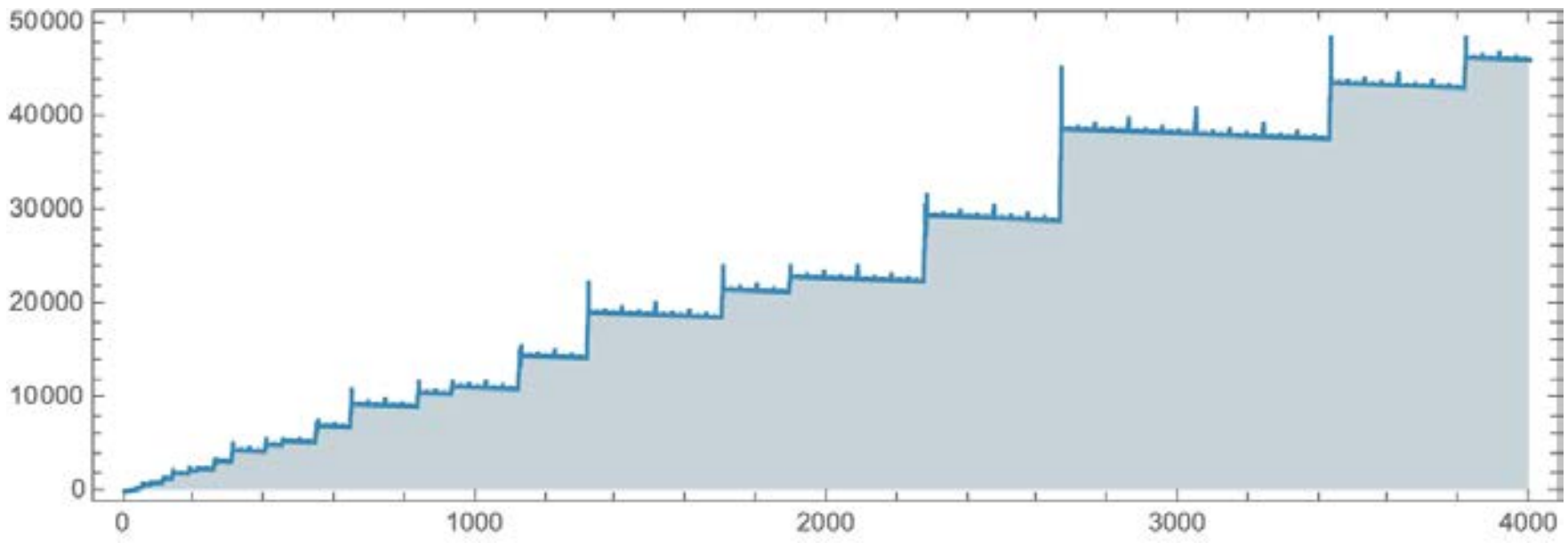

On a small scale, this appears somewhat regular, but the larger-scale structure, as revealed by taking differences, it doesn't seem so regular (though it does have a certain "**IntegerExponent**-like" look):

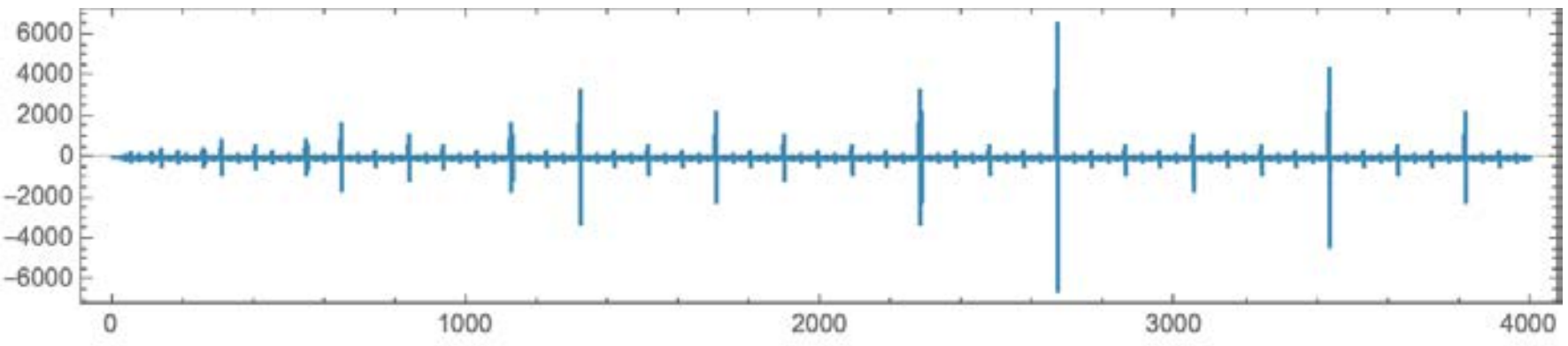

It's not clear what will happen in this case. The overall form of the behavior looks a bit similar to examples above that eventually terminate. Continuing for 50,000 steps, though, here's what's happened:

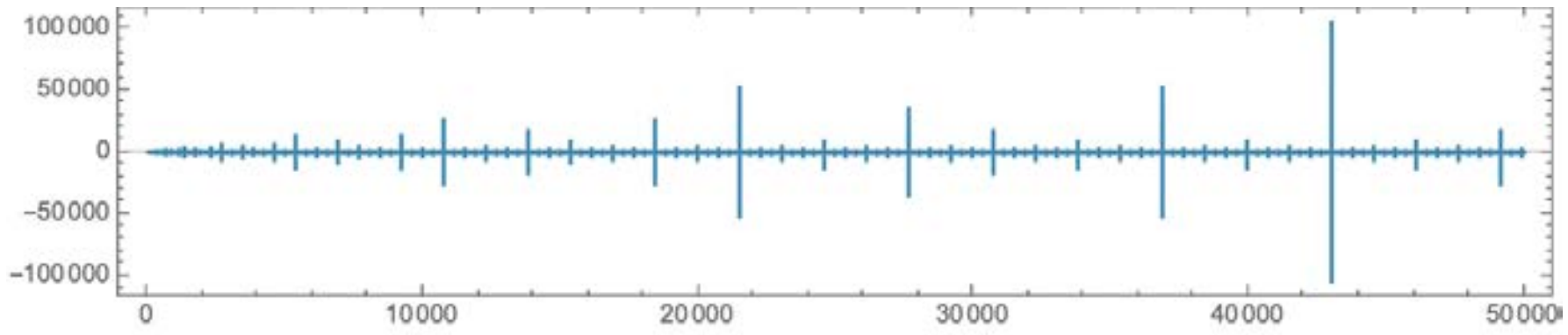

And in fact it turns out that the size-difference peaks continue to get higher—having values of the form 6 $(17\times 2^n+1)$ and occurring at positions of the form 2 $(9\times 2^{n+2}+n-18)$.

Here's another example: s[s][s][s[s]][s][k[s]] (**SSS**(**SS**)**S**(**KS**)). The overall growth in this case—at least for 200 steps—looks somewhat irregular:

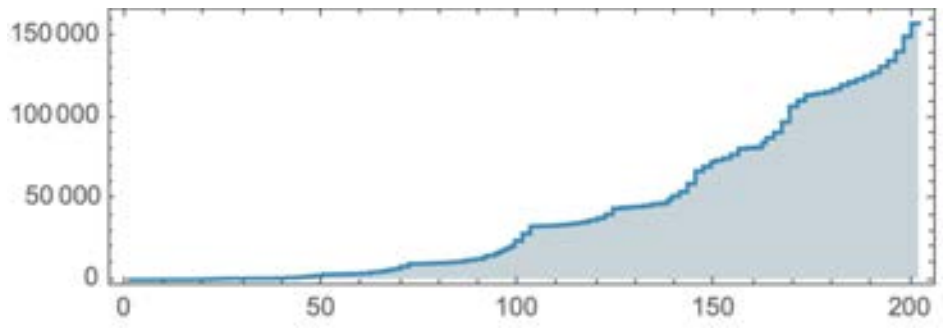

And taking differences reveals a fairly complex pattern of behavior:

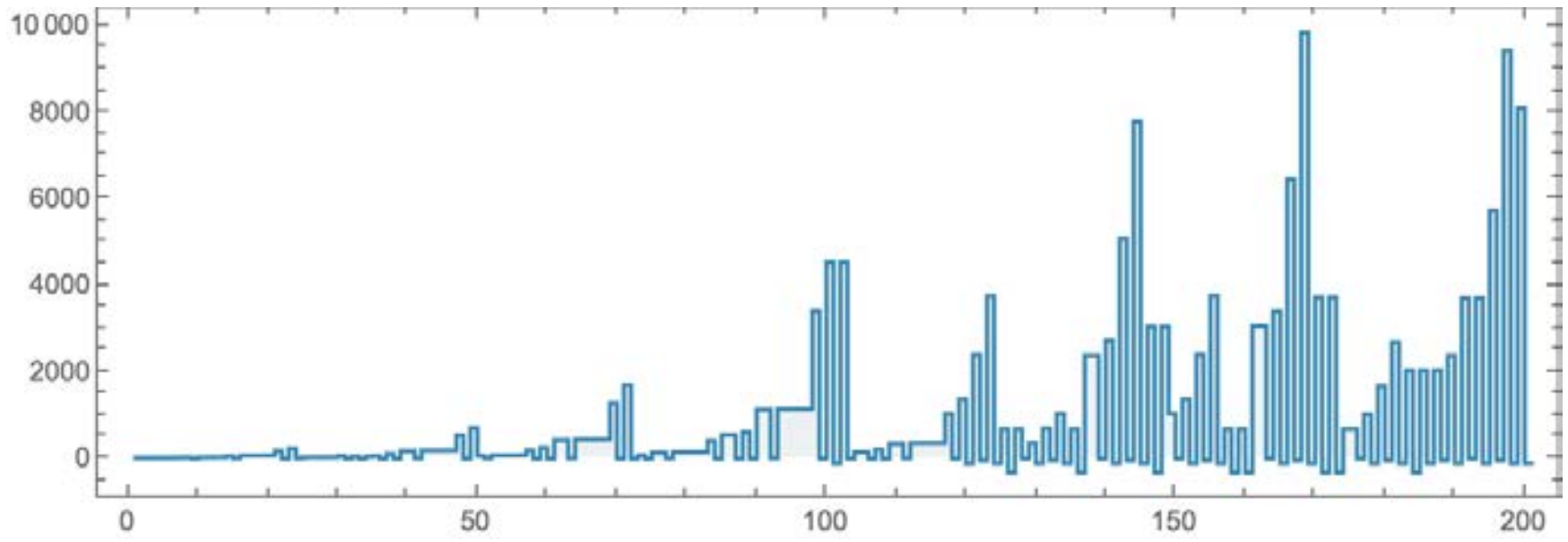

But after 1000 steps there appears to be some regularity to be seen:

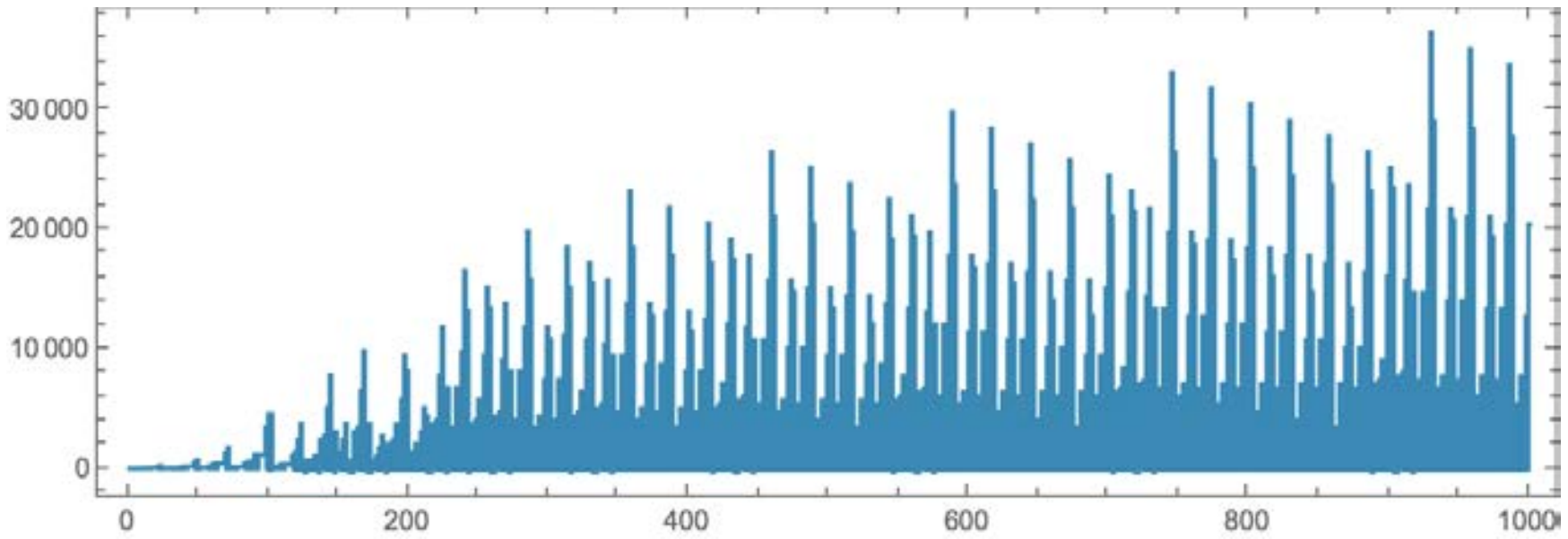

And even after 2000 steps the regularity is more obvious:

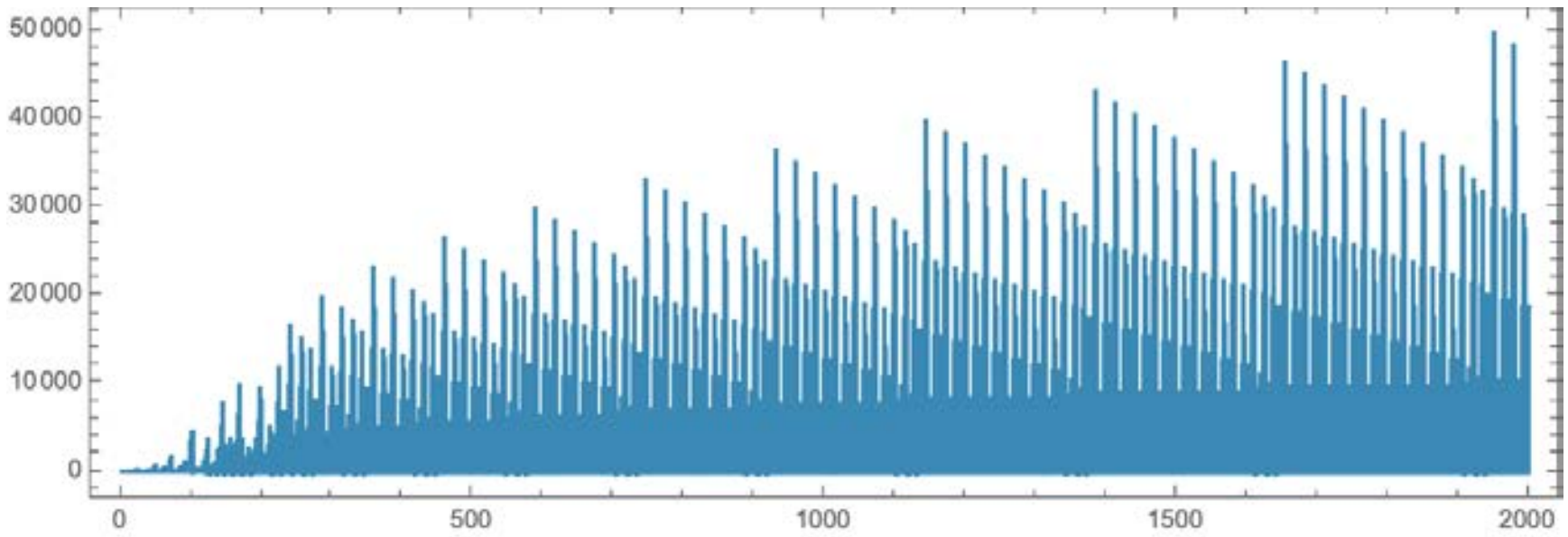

There's a long transient, but after that there are systematic peaks in the size difference, with the $n^{th}$ peak having height $16487 + 3320\,n$ and occurring at step $14\,n^2+59\,n+284$. (And, yes, it's pretty weird to see all these strange numbers cropping up.)

What happens if we look at size-10 combinator expressions? There's a lot of repeating of behavior that we've seen with smaller expressions. But some new things do seem to happen.

After 1000 steps s[s][k][s[s][k][s[s]][s]][k] (**SSK**(**SSK**(**SS**)**S**)**K**) seems to be doing something quite complicated when one looks at its size differences:

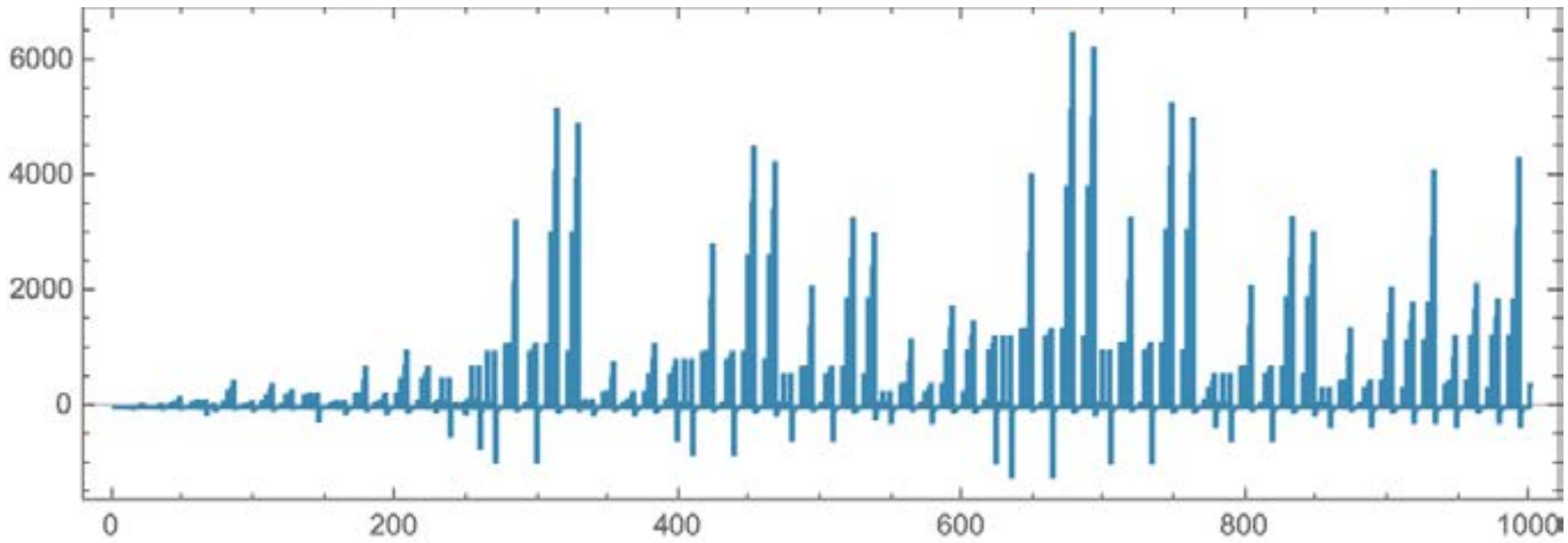

But it turns out that this is just a transient, and after 1000 steps or so, the system settles into a pattern of continual growth similar to ones we've seen before:

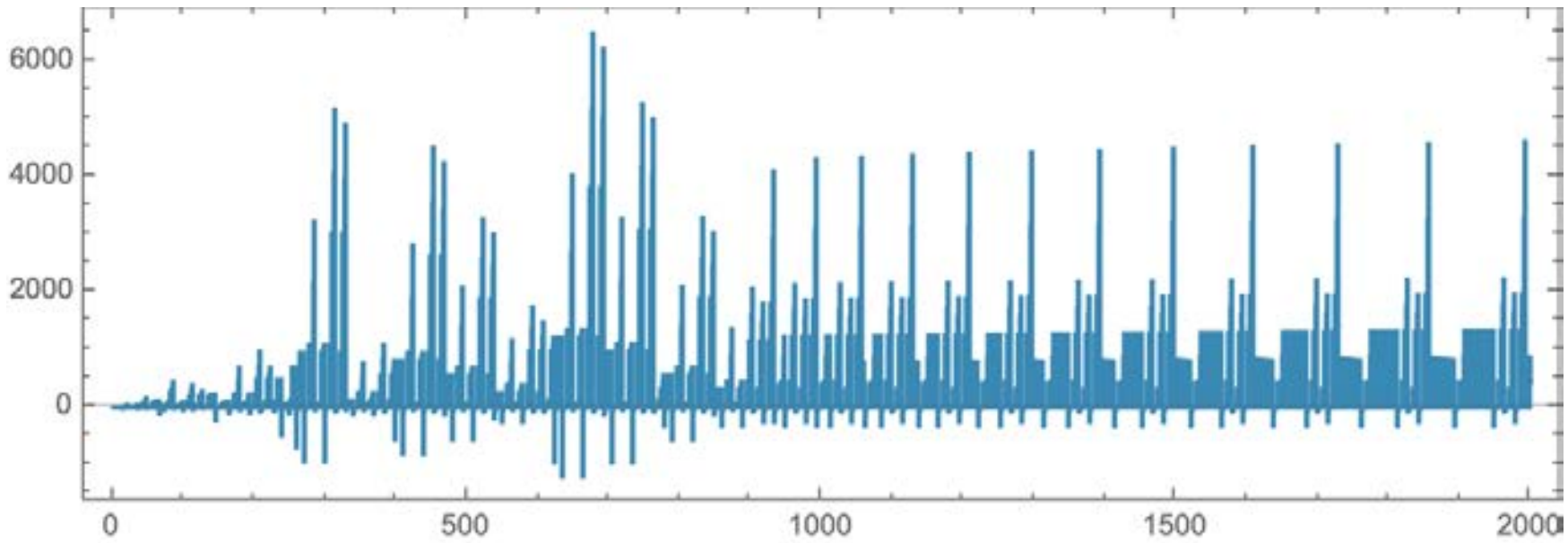

Another example is s[s][k][s[s][k][s[s]][s]][s] (**SSK**(**SSK**(**SS**)**S**)**S**). After 2000 steps there seems to be some regularity, and some irregularity:

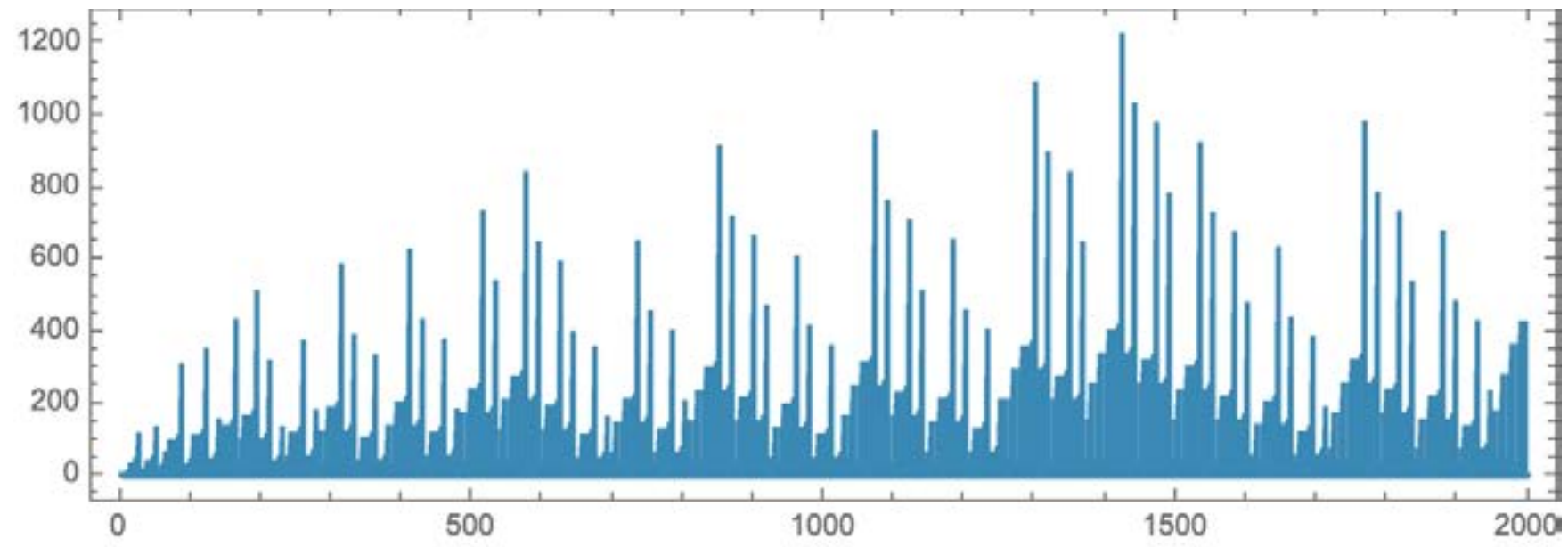

And basically this continues:

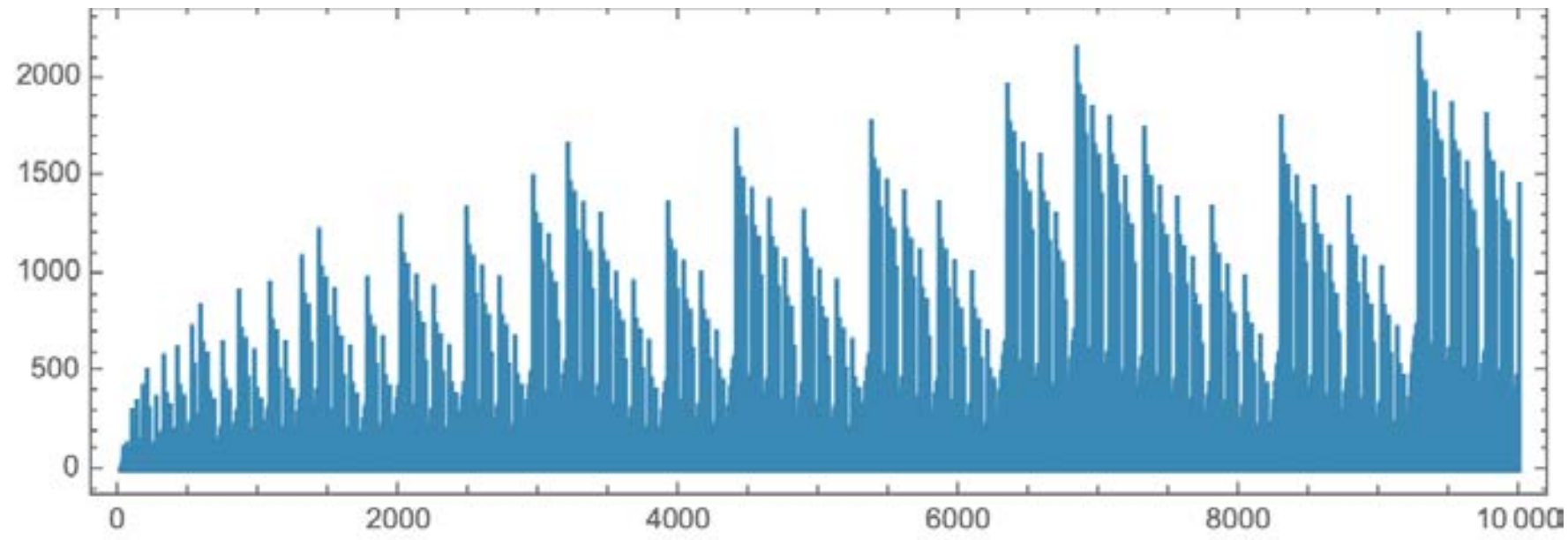

s[s][s][s[s[s[k]]]][s][s[k]] (**SSS**(**S**(**S**(**SK**)))**S**(**SK**)) is a fairly rare example of "nested-like" growth that continues forever (after a million steps, the size obtained is 597,871,806):

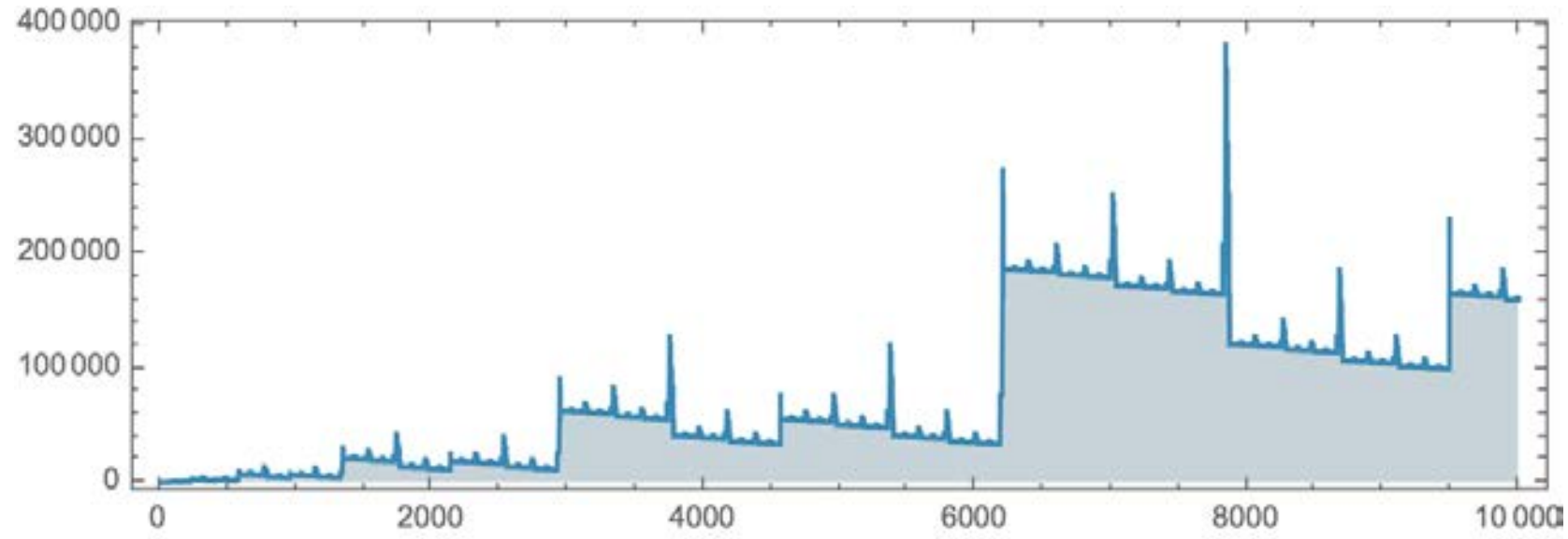

As a final example, consider s[s[s]][s][s][s][s[s][k[k]]] (**S**(**SS**)**SSS**(**SS**(**KK**))). Here's what this does for the first 1000 steps:

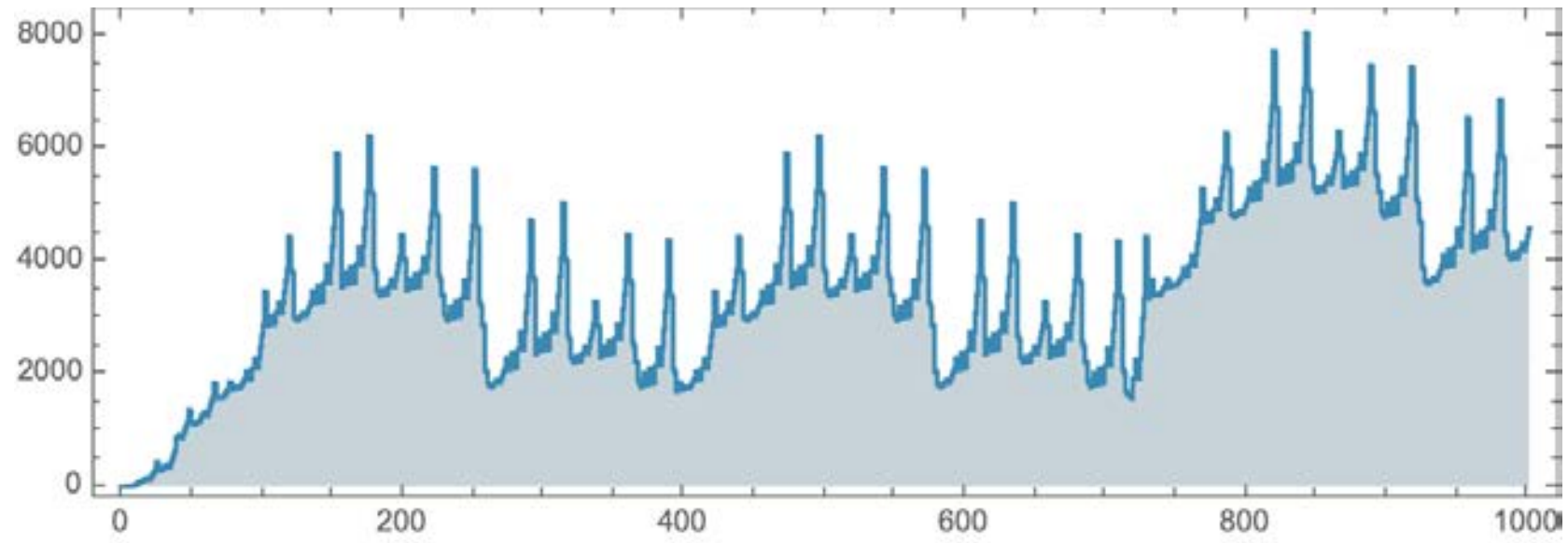

It looks somewhat complicated, but seems to be growing slowly. But then around step 4750 it suddenly jumps up, quickly reaching size 51,462:

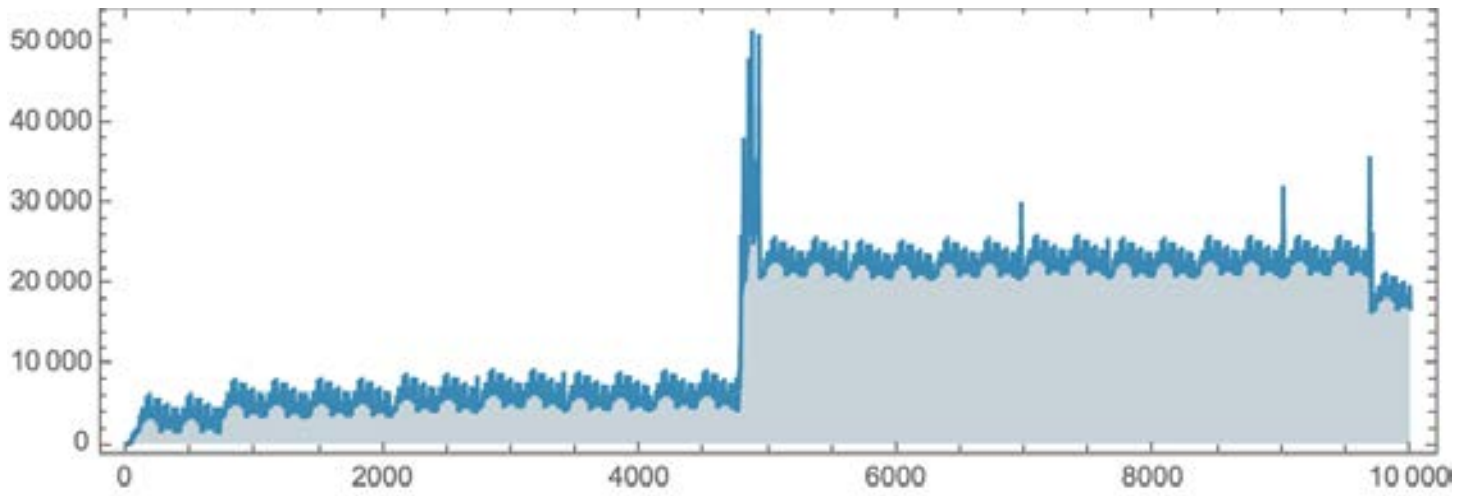

Keep going further, and there are more jumps:

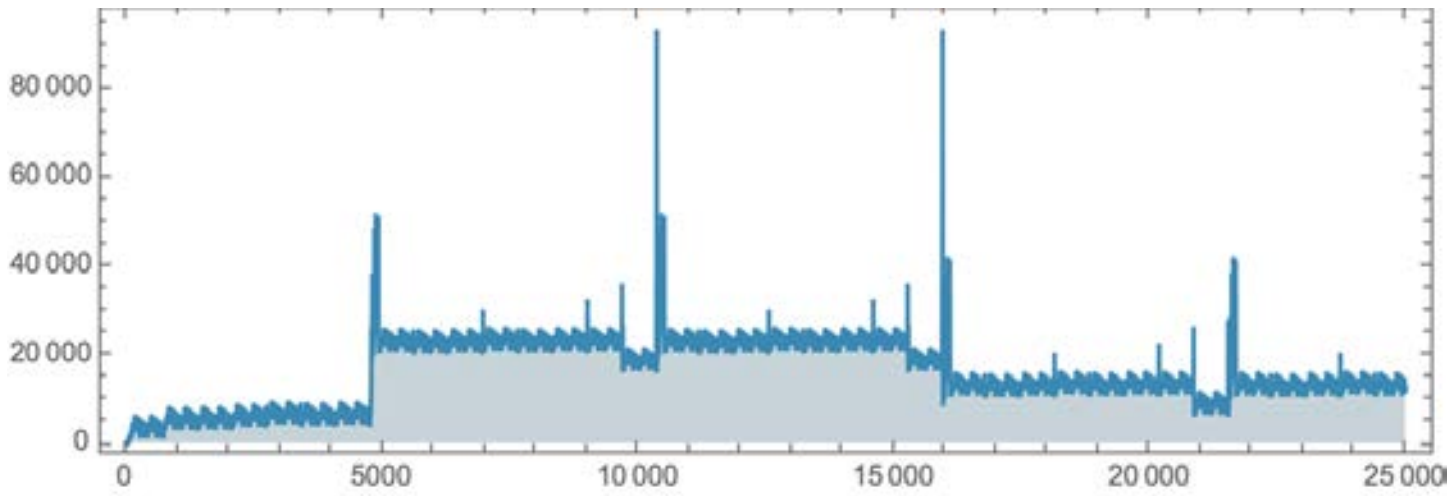

After 100,000 steps there's a definite pattern of jumps—but it's not quite regular:

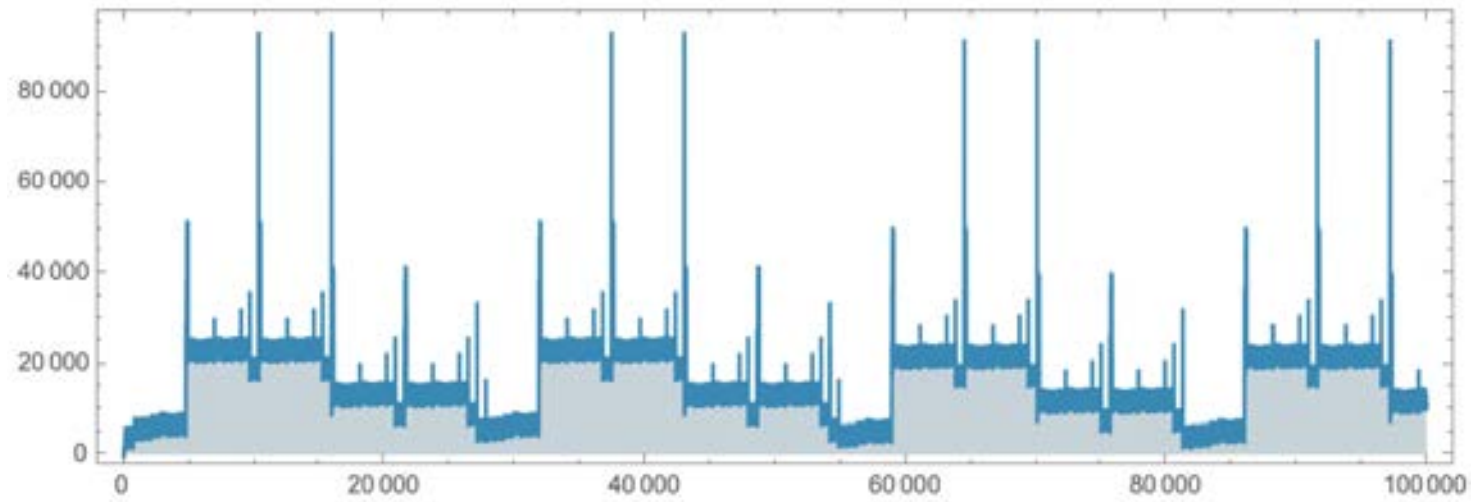

So what's going to happen? Mostly it seems to be maintaining a size of a few thousand or more. But then, after 218,703 steps, it dips down, to size 319. So, one might think, perhaps it's going to "die out". Keep going longer, and at step 34,339,093 it gets down to size 27, even though by step 36,536,622 it's at size 105,723.

Keep going even longer, and one sees it dipping down in size again (here shown in a down-sampled log plot):

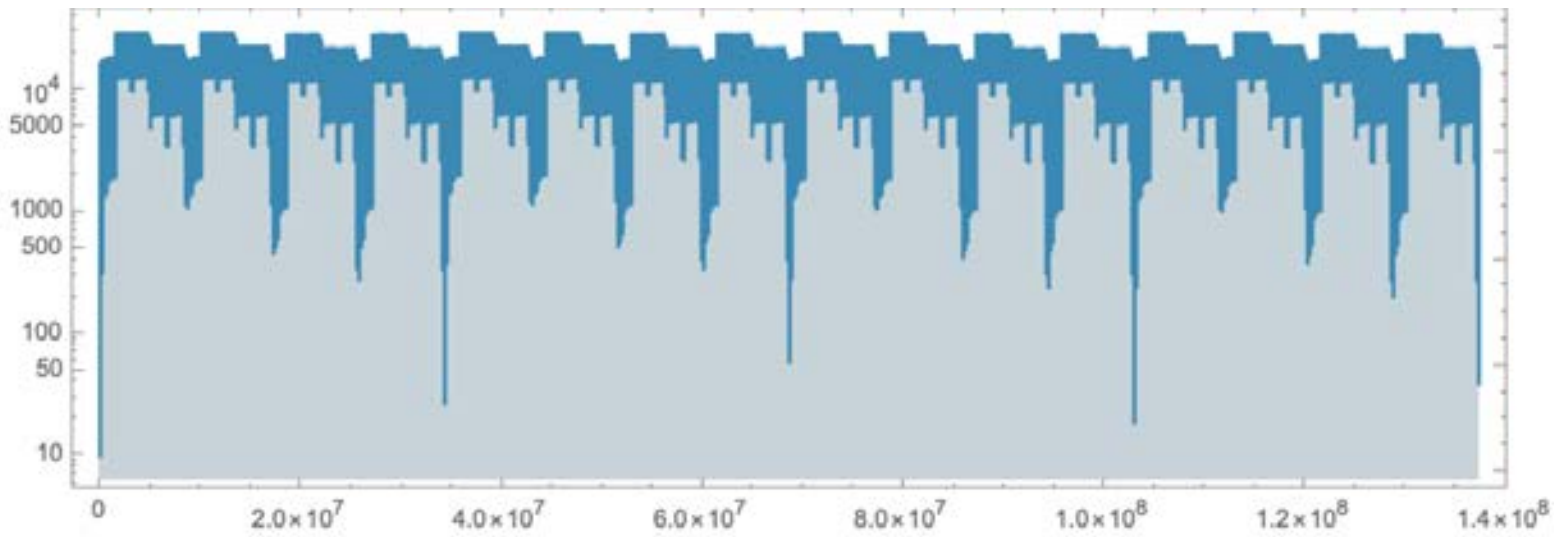

But, then, suddenly, boom. At step 137,356,329 it stops, reaching a fixed point of size 39. And, yes, it's totally remarkable that a tiny combinator expression like s[s[s]][s][s][s][s[s][k[k]]] (**S**(**SS**)**SSS**(**SS**(**KK**))) can do all this.

If one hasn't seen it before, this kind of complexity would be quite shocking. But after spending so long exploring the computational universe, I've become used to it. And now I just view each new case I see as yet more evidence for my Principle of Computational Equivalence.

A central fact about s,k combinators is that they're computation universal. And this tells us that whatever computation we want to do, it'll always be possible to "write a combinator program"—i.e. to create a combinator expression—that'll do it. And from this it follows that—just like with the halting problem for Turing machines—the problem of whether a combinator will halt is in general undecidable.

But the new thing we're seeing here is that it's difficult to figure out what will happen not just "in general" for complicated expressions set up to do particular computations but also for simple combinator expressions that one might "find in the wild". But the Principle of Computational Equivalence tells us why this happens.

Because it says that even simple programs—and simple combinator expressions—can lead to computations that are as sophisticated as anything. And this means that their behavior can be computationally irreducible, so that the only way to find out what will happen is essentially just to run each step and see what happens. So then if one wants to know what will happen in an infinite time, one may have to do an effectively infinite amount of computation to find out.

Might there be another way to formulate our questions about the behavior of combinators? Ultimately we could use any computation universal system to represent what combinators do. But some formulations may connect more immediately with existing ideas—say mathematical ones. And for example I think it's conceivable that the sequences of combinator sizes we've seen above could be obtained in a more "direct numerical way", perhaps from something like nestedly recursive functions (I discovered this particular example in 2003):

```
f[n_] := 3 f[n – f[n – 1]]
f[n_ /; n < 1] = 1
```

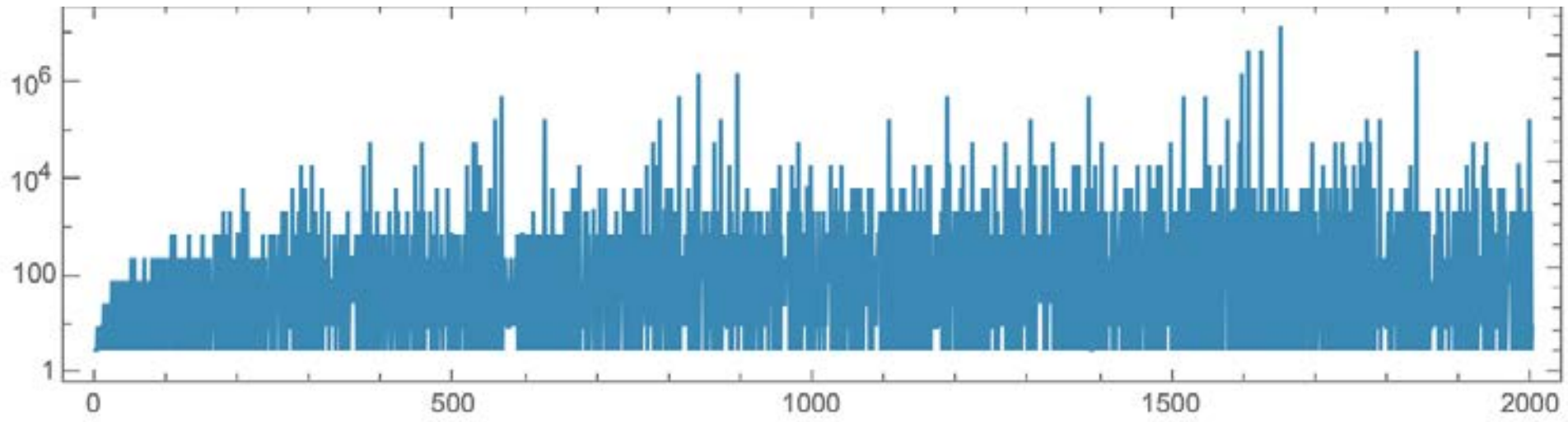

# Visualizing Combinators

One of the issues in studying combinators is that it's so hard to visualize what they're doing. It's not like with cellular automata where one can make arrays of black and white cells and readily use our visual system to get an impression of what's going on. Consider for example the combinator evolution:

```
s[s][k][s[s[s]][s]][s]
s[s[s[s]][s]][k[s[s[s]][s]]][s]
s[s[s]][s][s][k[s[s[s]][s]][s]]
s[s][s][s[s]][k[s[s[s]][s]][s]]
s[s[s]][s[s[s]]][k[s[s[s]][s]][s]]
s[s][k[s[s[s]][s]][s]][s[s[s]][k[s[s[s]][s]][s]]]
s[s[s[s]][k[s[s[s]][s]][s]]][k[s[s[s]][s]][s][s[s[s]][k[s[s[s]][s]][s]]]]
s[s[s[s]][s[s[s]][s]]][k[s[s[s]][s]][s][s[s[s]][k[s[s[s]][s]][s]]]]
```

In a cellular automaton the rule would be operating on neighboring elements, and so there'd be locality to everything that's happening. But here the combinator rules are effectively moving whole chunks around at a time, so it's really hard to visually trace what's happening.

But even before we get to this issue, can we make the mass of brackets and letters in something like

s[s[s[s]][k[s[s[s]][s]][s]]][k[s[s[s]][s]][s][s[s[s]][k[s[s[s]][s]][s]]]]

easier to read? For example, do we really need all those brackets? In the Wolfram Language, for example, instead of writing

a[b[c[d[e]]]]

we can equivalently write

a@b@c@d@e

thereby avoiding brackets.

But using @ doesn't avoid all grouping indications. For example, to represent

a[b][c][d][e]

with @ we'd have to write:

(((a@b)@c)@d)@e

In our combinator expression above, we had 24 pairs of brackets. By using @, we can reduce this to 10:

(s@(s@s@s)@(k@(s@s@s)@s)@s)@((k@(s@s@s)@s)@s)@(s@s@s)@(k@(s@s@s)@s)@s

And we don't really need to show the @, so we can make this smaller:

(s(sss)(k(sss)s)s)((k(sss)s)s)(sss)(k(sss)s)s

When combinators were first introduced a century ago, the focus was on "multi-argument-function-like" expressions such as a[b][c] (as appear in the rules for s and k), rather than on "nested-function-like" expressions such as a[b[c]]. So instead of thinking of function application as "right associative"—so that a[b[c]] can be written without parentheses as a@b@c—people instead thought of function application as left associative—so that a[b][c] could be written without parentheses. (Confusingly, people often used @ as the symbol for this left-associative function application.)

As it’s turned out, the f[g[x]] form is much more common in practice than f[g][x], and in 30+ years there hasn’t been much of a call for a notation for left-associative function application in the Wolfram Language. But in celebration of the centenary of combinators, we’ve decided to introduce Application (indicated by •) to represent left-associative function application.

So this means that a[b][c][d][e] can now be written

a•b•c•d•e

without parentheses. Of course, now a[b[c[d[e]]]] needs parentheses:

a•(b•(c•(d•e)))

In this notation the rules for s and k can be written without brackets as:

{s•x_•y_•z_ → x•z•(y•z), k•x_•y_ → x}

Our combinator expression above becomes

s•(s•(s•s)•(k•(s•(s•s)•s)•s))•(k•(s•(s•s)•s)•s•(s•(s•s)•(k•(s•(s•s)•s)•s)))

or without the function application character

s(s(ss)(k(s(ss)s)s))(k(s(ss)s)s(s(ss)(k(s(ss)s)s)))

which now involves 13 pairs of parentheses.

Needless to say, if you consider all possible combinator expressions, left and right associativity on average do just the same in terms of parenthesis counts: for size-n combinator expressions, both on average need (n-2)/2 pairs; the number of cases needing k pairs is

Binomial[n – 1, k – 1] Binomial[n, k – 1]/k

(the "Catalan triangle"). (Without associativity, we're dealing with our standard representation of combinator expressions, which always requires n – 1 pairs of brackets.)

By the way, the number of “right-associative” parenthesis pairs is just the number of subparts of the combinator expression that match _[_][_], while for left-associative parenthesis pairs it’s the number that match _[_[_]]. (The number of brackets in the no-associativity case is the number of matches of _[_].)

If we look at the parenthesis/bracket count in the evolution of the smallest non-terminating combinator expression from above s[s][s][s[s]][s][s] (otherwise known as s•s•s•(s•s)•s•s) we find:

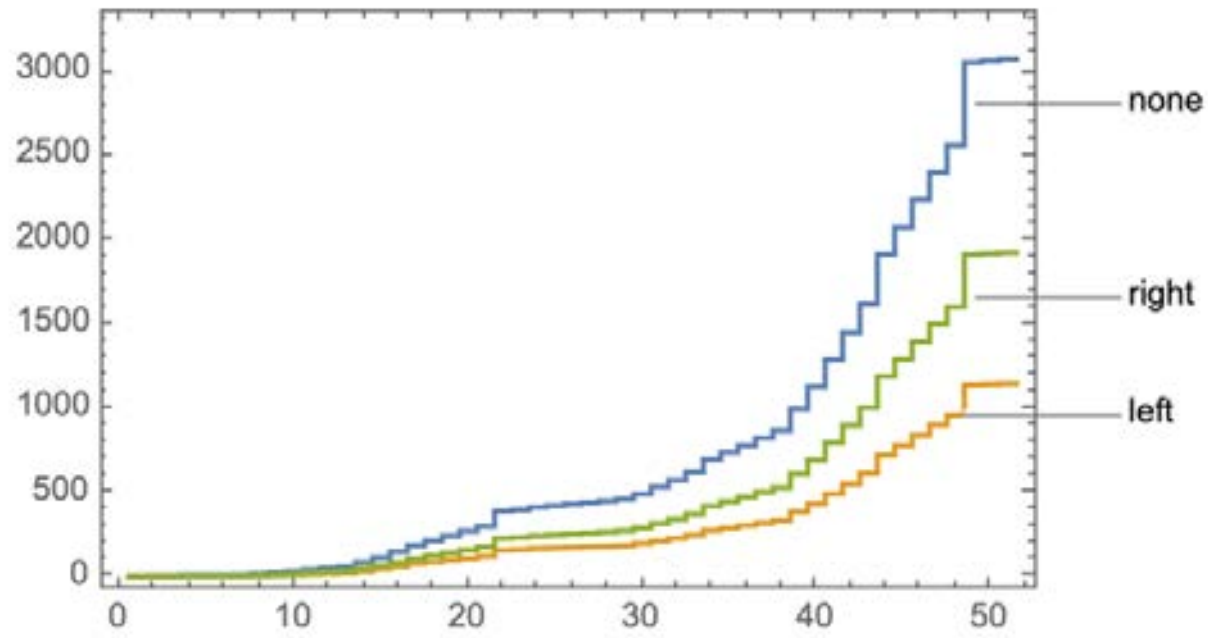

Or in other words, in this case, left associativity leads on average to about 62% of the number of parentheses of right associativity. We'll look at this in more detail later, but for growing combinator expressions, it'll almost always turn out to be the case that left associativity is the "parenthesis-avoidance winner".

But even with our "best parenthesis avoidance" it's still very hard to see what's going on from the textual form:

| left | right | none |
|---|---|---|
| sss(ss)ss | ((((ss)s)ss)s)s | s[s][s][s[s]][s][s] |
| s(ss)(s(ss))ss | (((sss)sss)s)s | s[s[s]][s[s[s]]][s][s] |
| sss(s(ss)s)s | (((ss)s)(sss)s)s | s[s][s][s[s[s]][s]][s] |
| s(s(ss)s)(s(s(ss)s))s | ((s(sss)s)s(sss)s)s | s[s[s[s]][s]][s[s[s[s]][s]]][s] |
| s(ss)ss(s(s(ss)s)s) | (((sss)s)s)(s(sss)s)s | s[s[s]][s][s][s[s[s[s]][s]][s]] |
| sss(ss)(s(s(ss)s)s) | (((ss)s)ss)(s(sss)s)s | s[s][s][s[s]][s[s[s[s]][s]][s]] |
| s(ss)(s(ss))(s(s(ss)s)s) | ((sss)sss)(s(sss)s)s | s[s[s]][s[s[s]]][s[s[s[s]][s]][s]] |
| ss(s(s(ss)s)s)(s(ss)(s(s(ss)s)s)) | ((ss)(s(sss)s)s)(sss)(s(sss)s)s | s[s][s[s[s[s]][s]][s]][s[s[s]][s[s[s[s]][s]][s]]] |

So what about getting rid of parentheses altogether? Well, we can always use so-called Polish (or Łukasiewicz) "prefix" notation—in which we write f[x] as •fx and f[g[x]] as •f•gx. And in this case our combinator expression from above becomes:

••s••s•ss••k••s•ssss•••k••s•ssss••s•ss••k••s•ssss

Alternatively—like a traditional HP calculator—we can use reverse Polish "postfix" notation, in which f[x] is fx• and f[g[x]] is fgx•• (and • is like HP ENTER):

ssss••ksss••s••s•••ksss••s••s•sss••ksss••s••s••••

The total number of • symbols is always equal to the number of pairs of brackets in our standard "non-associative" functional form:

| Polish | reverse Polish |
|---|---|
| •••••sss•ssss | ss•s•ss••s•s• |
| ••••s•ss•s•ssss | sss••sss•••s•s• |
| ••••sss••s•ssss | ss•s•sss••s••s• |
| •••s••s•sss•s••s•ssss | ssss••s••ssss••s•••s• |
| ••••s•ssss••s••s•ssss | sss••s•s•ssss••s••s•• |
| ••••sss•ss••s••s•ssss | ss•s•ss••ssss••s••s•• |
| •••s•ss•s•ss••s••s•ssss | sss••sss•••ssss••s••s•• |
| •••ss••s••s•ssss••s•ss••s••s•ssss | ss•ssss••s••s••sss••ssss••s••s••• |

What if we look at this on a larger scale, "cellular automaton style", with s being and • being ? Here's the not-very-enlightening result:

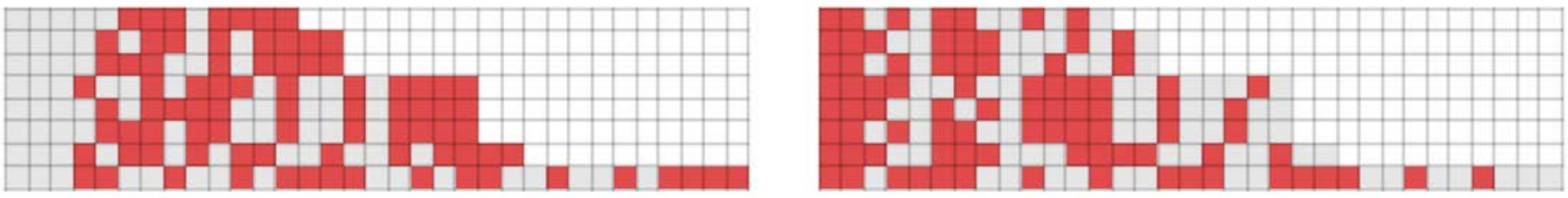

Polish

reverse Polish

Running for 50 steps, and fixing the aspect ratio, we get (for the Polish case):

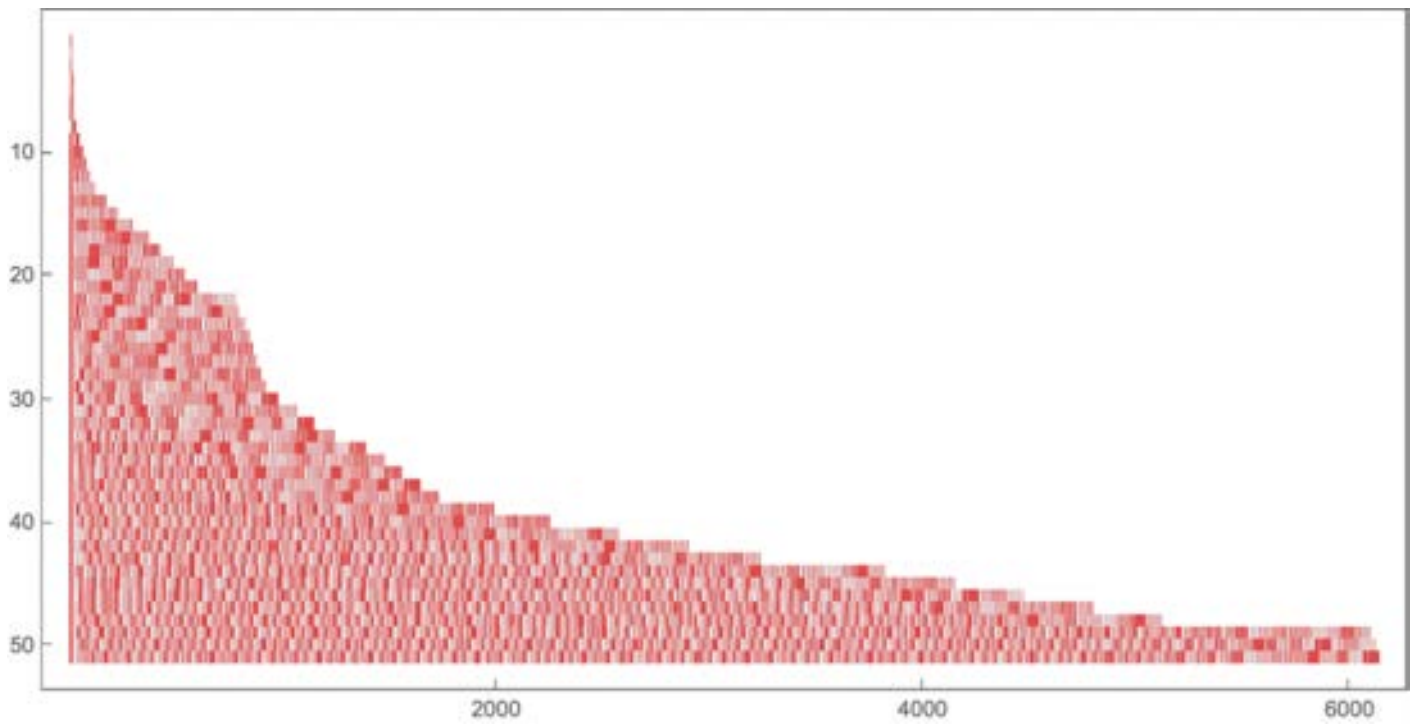

We can make the same kinds of pictures from our bracket representation too. We just take a string like s[s][s][s[s]][s][s] and render each successive character as a cell of some color. (It's particularly easy if we've only got one basic combinator—say s—because then we only need colors for the opening and closing brackets.) We can also make "cellular automaton style" pictures from parenthesis representations like sss(ss)ss . Again, all we do is to render each successive character as a cell of some color.

The results essentially always tend to look much like the reverse Polish case above. Occasionally, though, they reveal at least something about the "innards of the computation". Like here's the terminating combinator expression s[s][s][s[s[s]]][k][s]] from above, rendered in right-associative form:

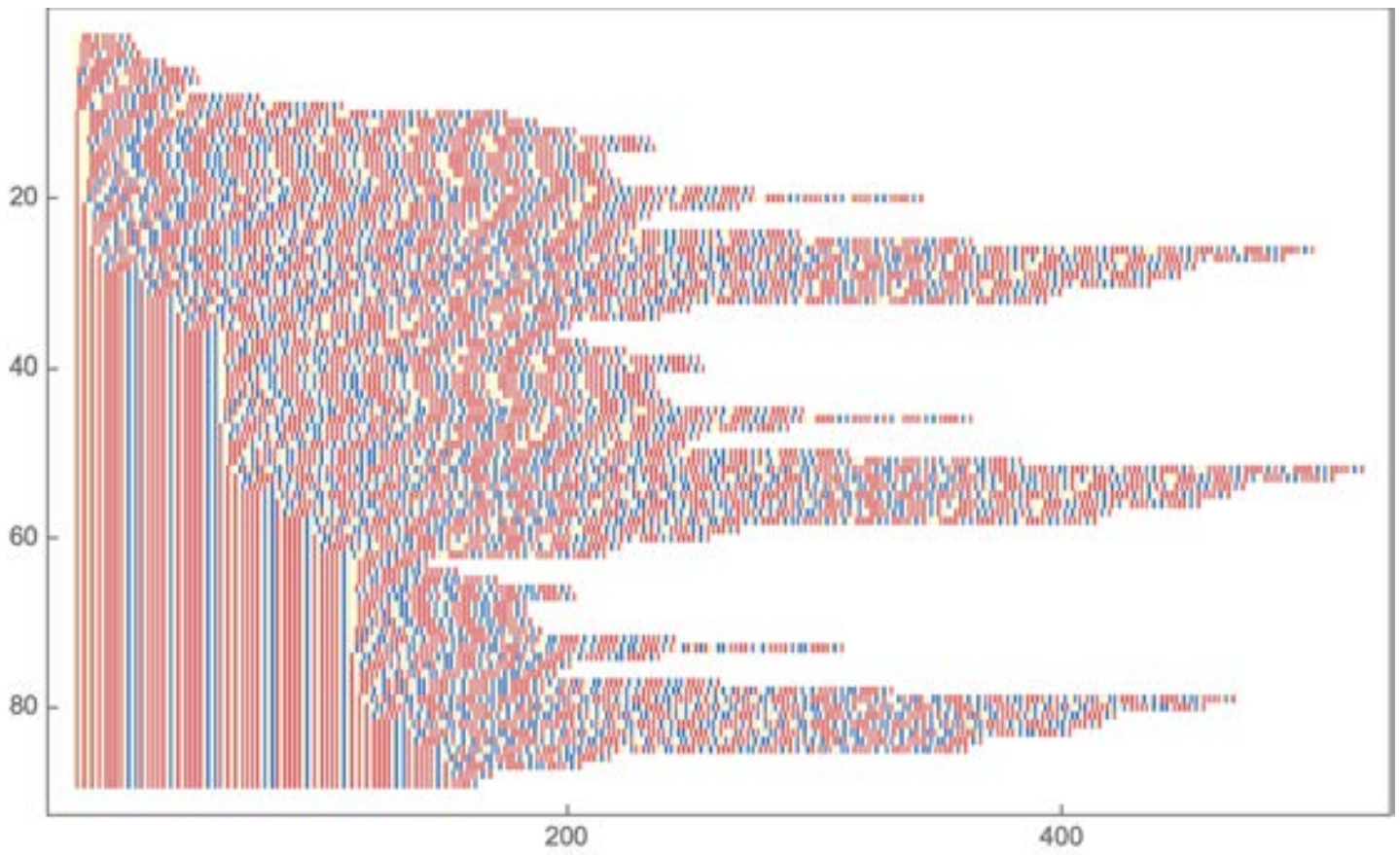

Pictures like this in a sense convert all combinator expressions to sequences. But combinator expressions are in fact hierarchical structures, formed by nested invocations of symbolic "functions". One way to represent the hierarchical structure of

**s[s][s][s[s]][s][s]**

is through a hierarchy of nested boxes:

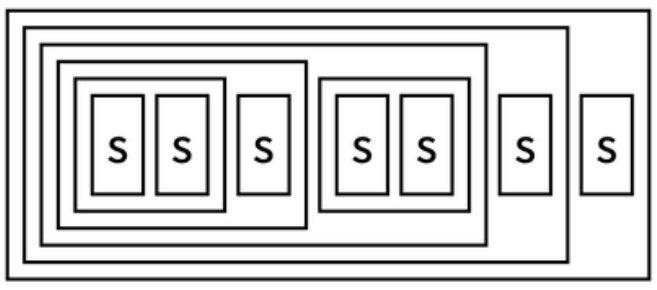

We can color each box by its depth in the expression:

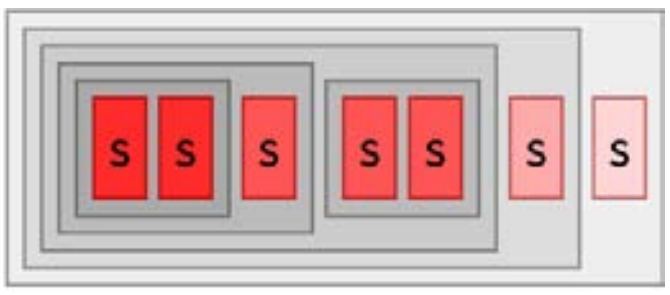

But now to represent the expression all we really need do is show the basic combinators in a color representing its depth. And doing this, we can visualize the terminating combinator evolution above as:

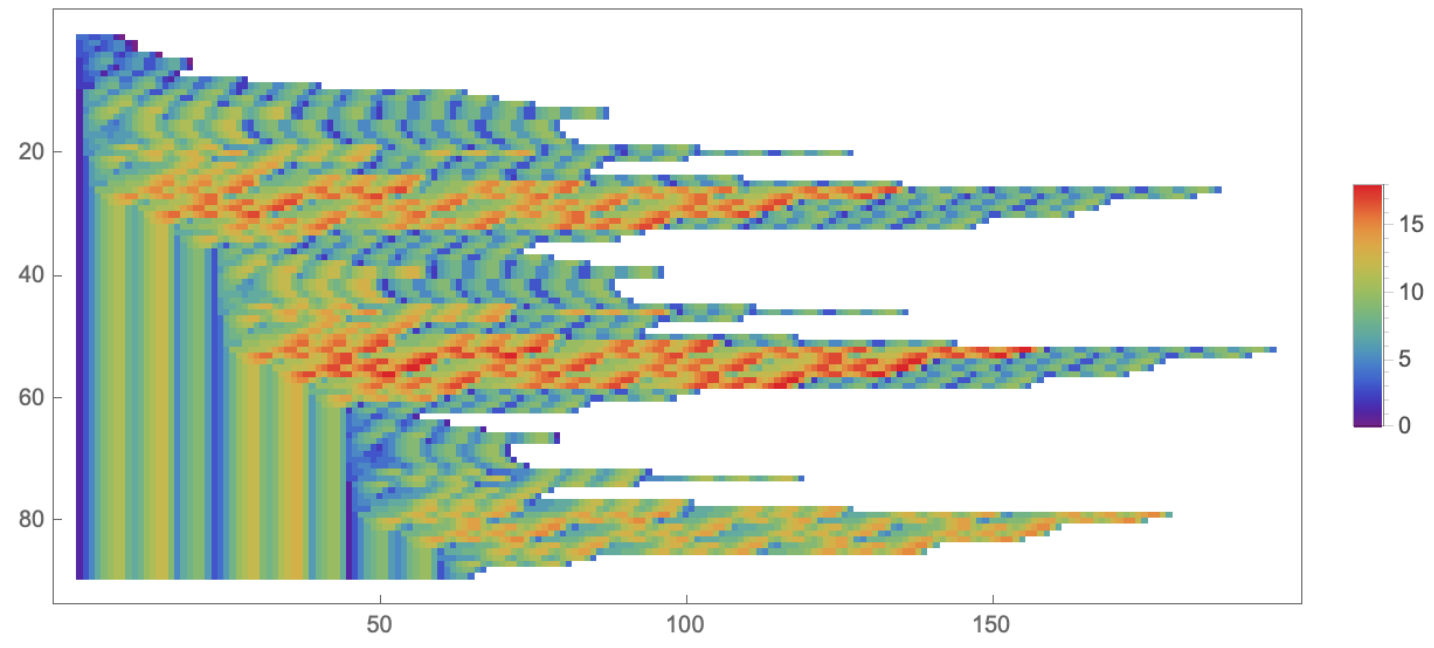

We can also render this in 3D (with the height being the "depth" in the expression):

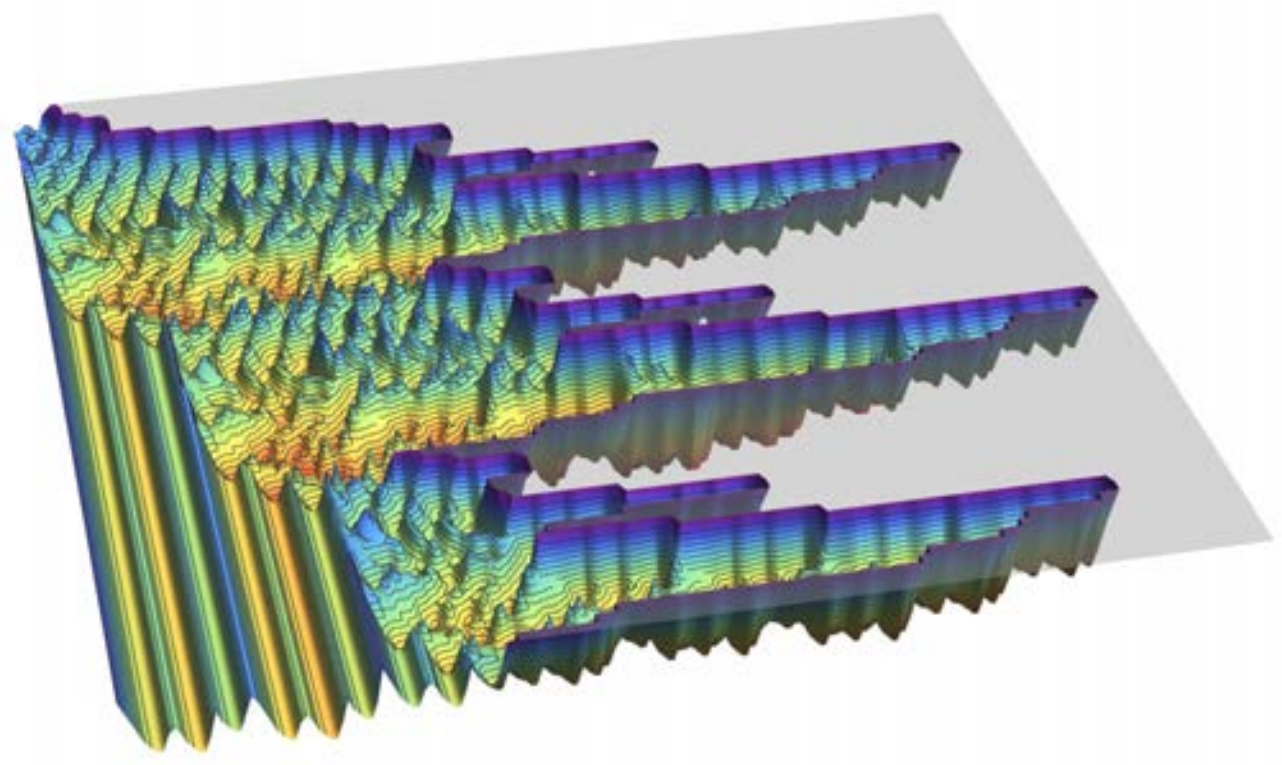

To test out visualizations like these, let's look (as above) at all the size-8 combinator expressions with distinct evolutions that don't terminate within 50 steps. Here's the "depth map" for each case:

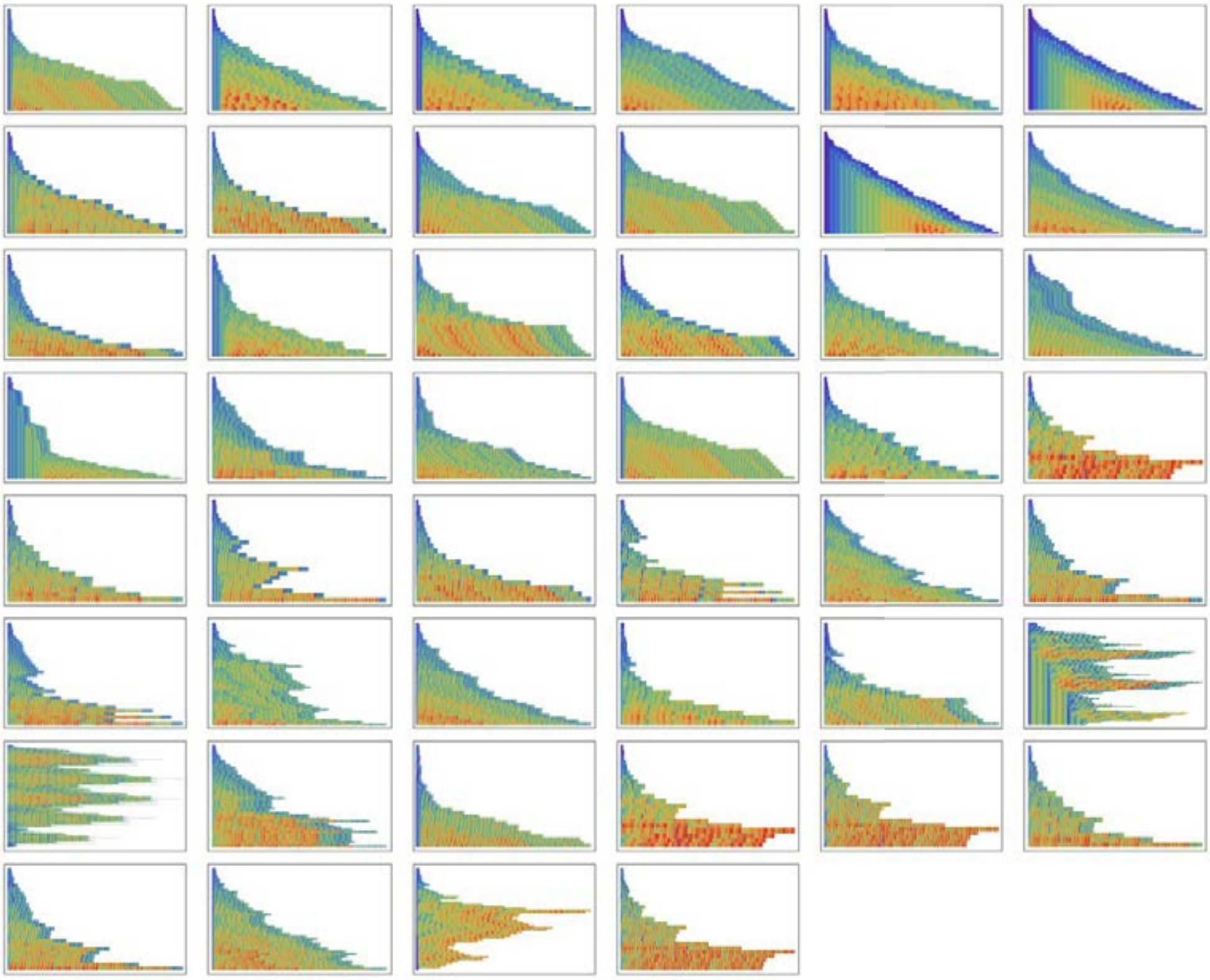

In these pictures we're drawing a cell for each element in the "stringified version" of the combinator expression at each step, then coloring it by depth. But given a particular combinator expression, one can consider other ways to indicate the depth of each element. Here are a few possibilities, shown for step 8 in the evolution of s[s][s][s[s]][s][s] (**SSS**(**SS**)**SS**) (note that the first of these is essentially the "indentation level" that might be used if each s, k were "pretty printed" on a separate line):

s[s[s[s]][s[s[s[s]][s]][s]]][s[s[s[s]][s]][s][s[s[s]][s[s[s[s]][s]][s]]]]

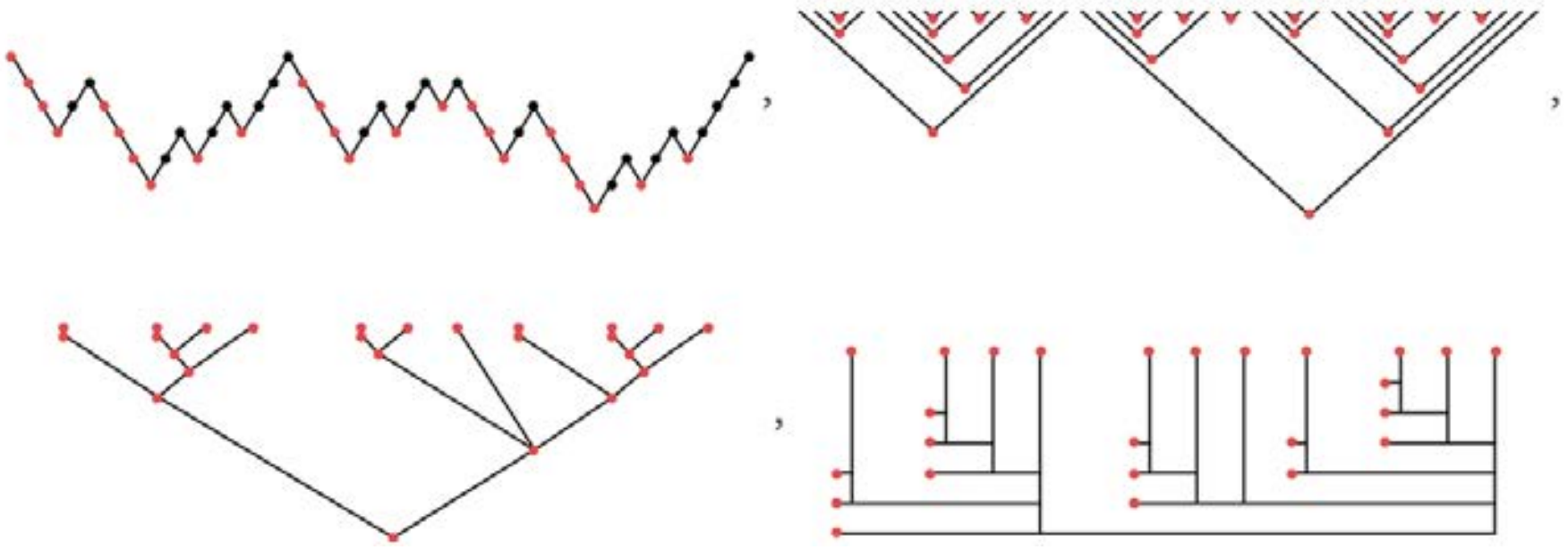

And this is what one gets on a series of steps:

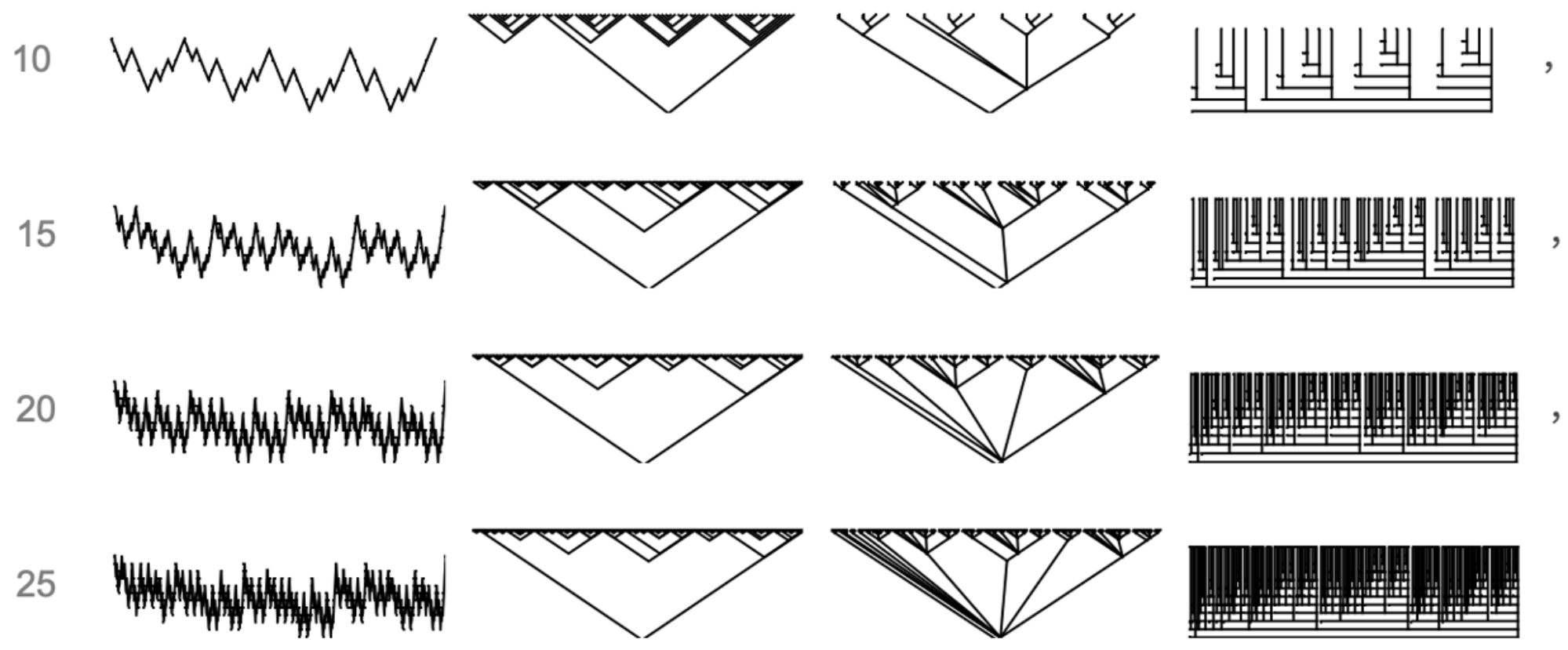

But in a sense a more direct visualization of combinator expressions is as trees, as for example in:

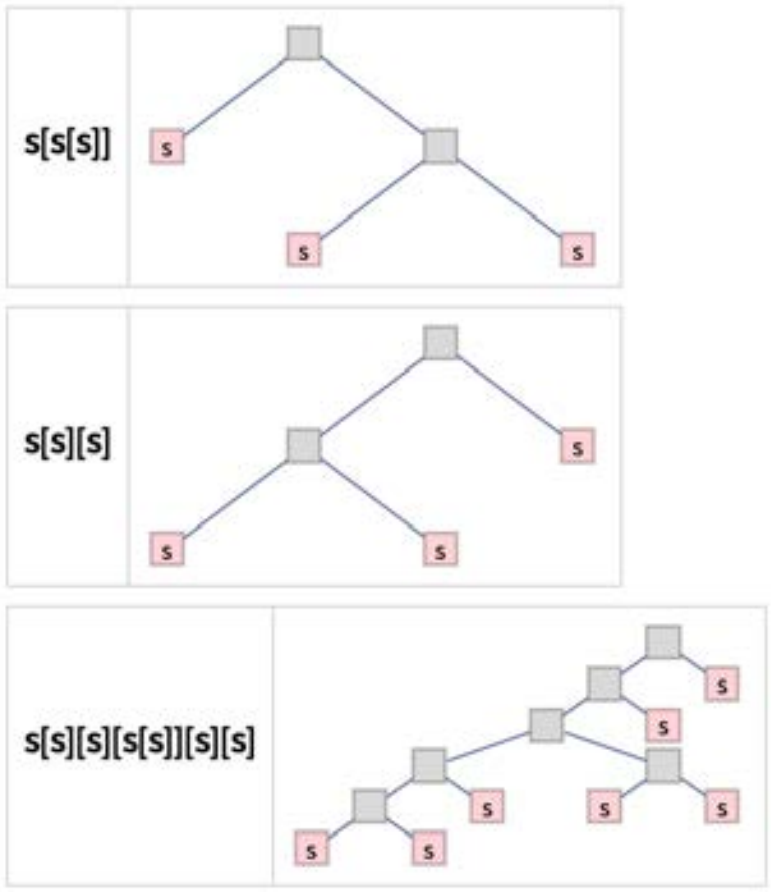

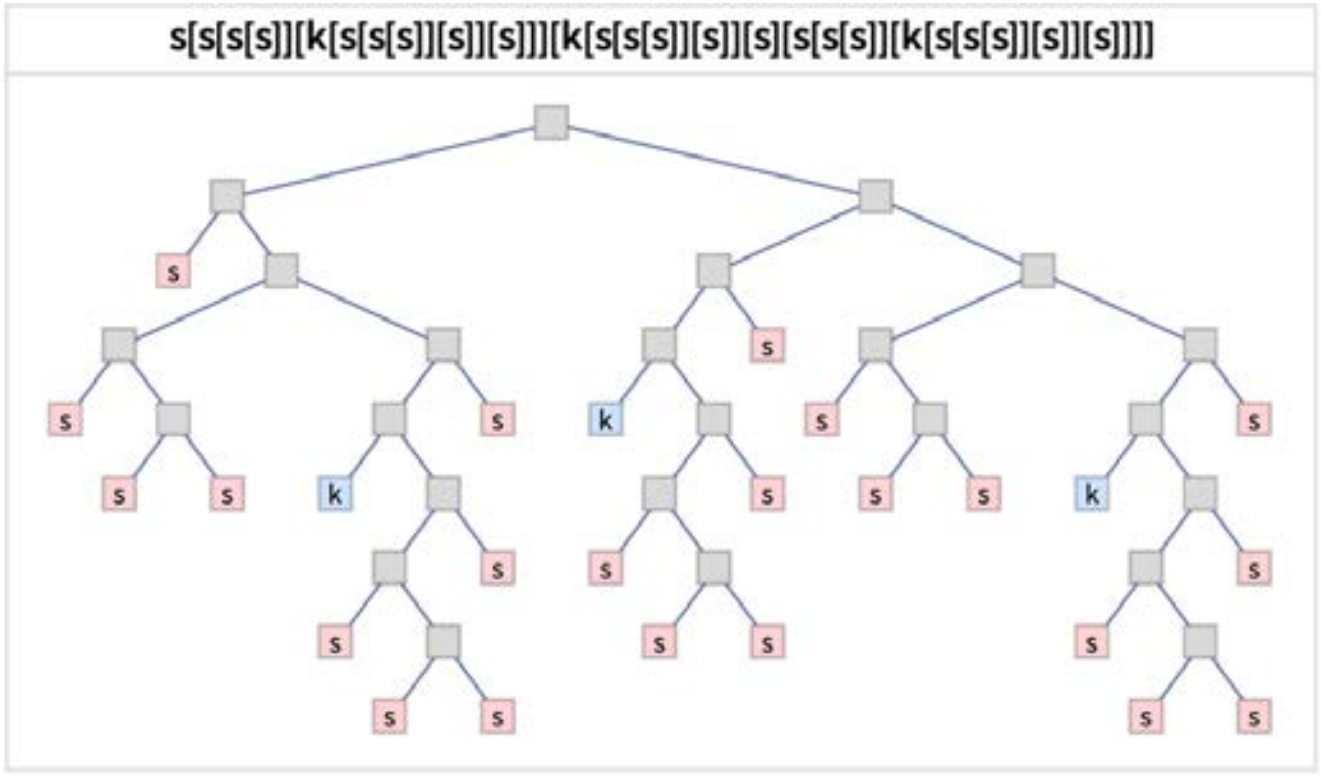

Note that these trees can be somewhat simplified by treating them as left or right "associative", and essentially pulling left or right leaves into the "branch nodes".

But using the original trees, we can ask for example what the trees for the expressions produced by the evolution of s[s][s][s[s]][s][s] (**SSS**(**SS**)**SS**) are. Here are the results for the first 15 steps:

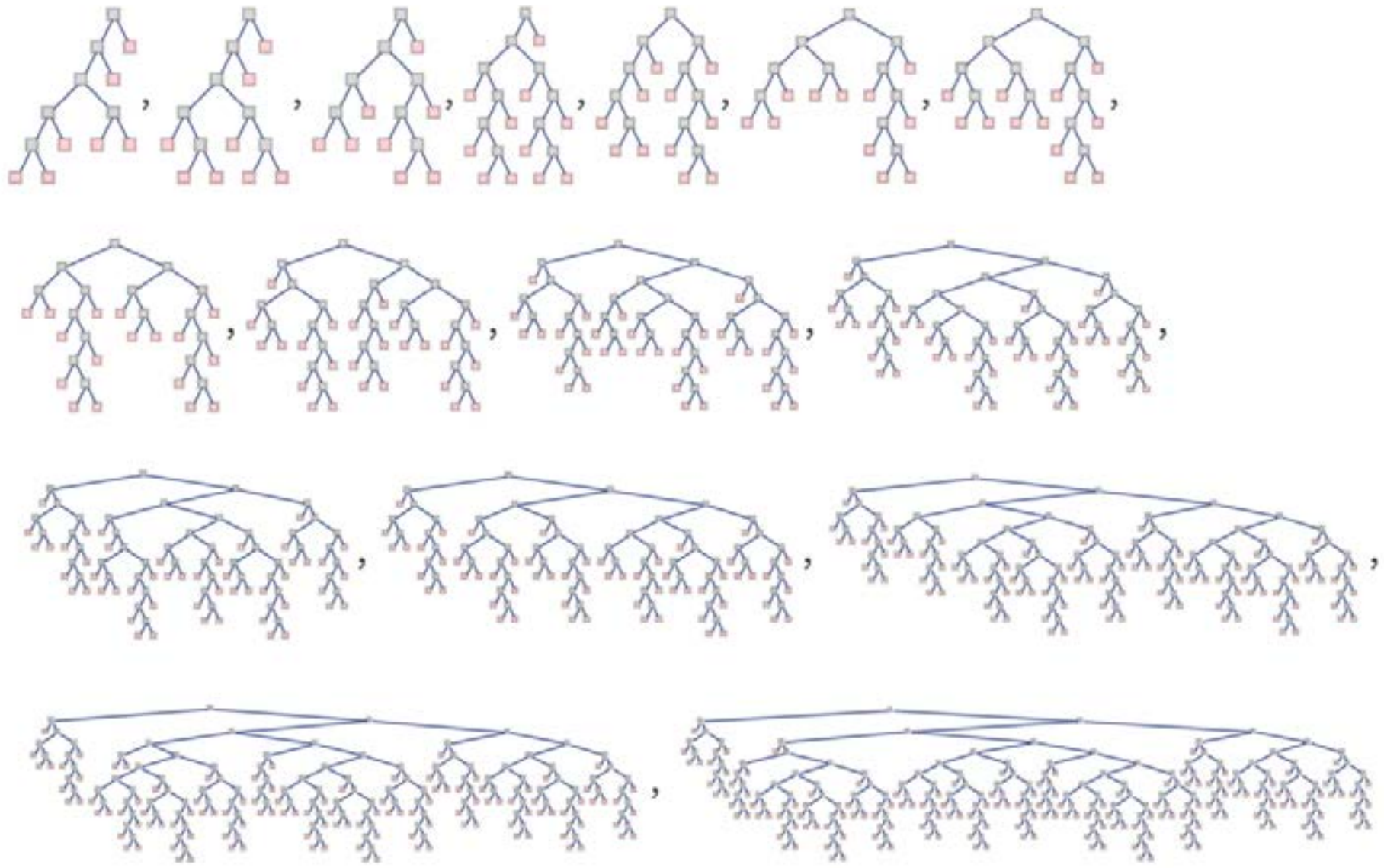

In a different rendering, these become:

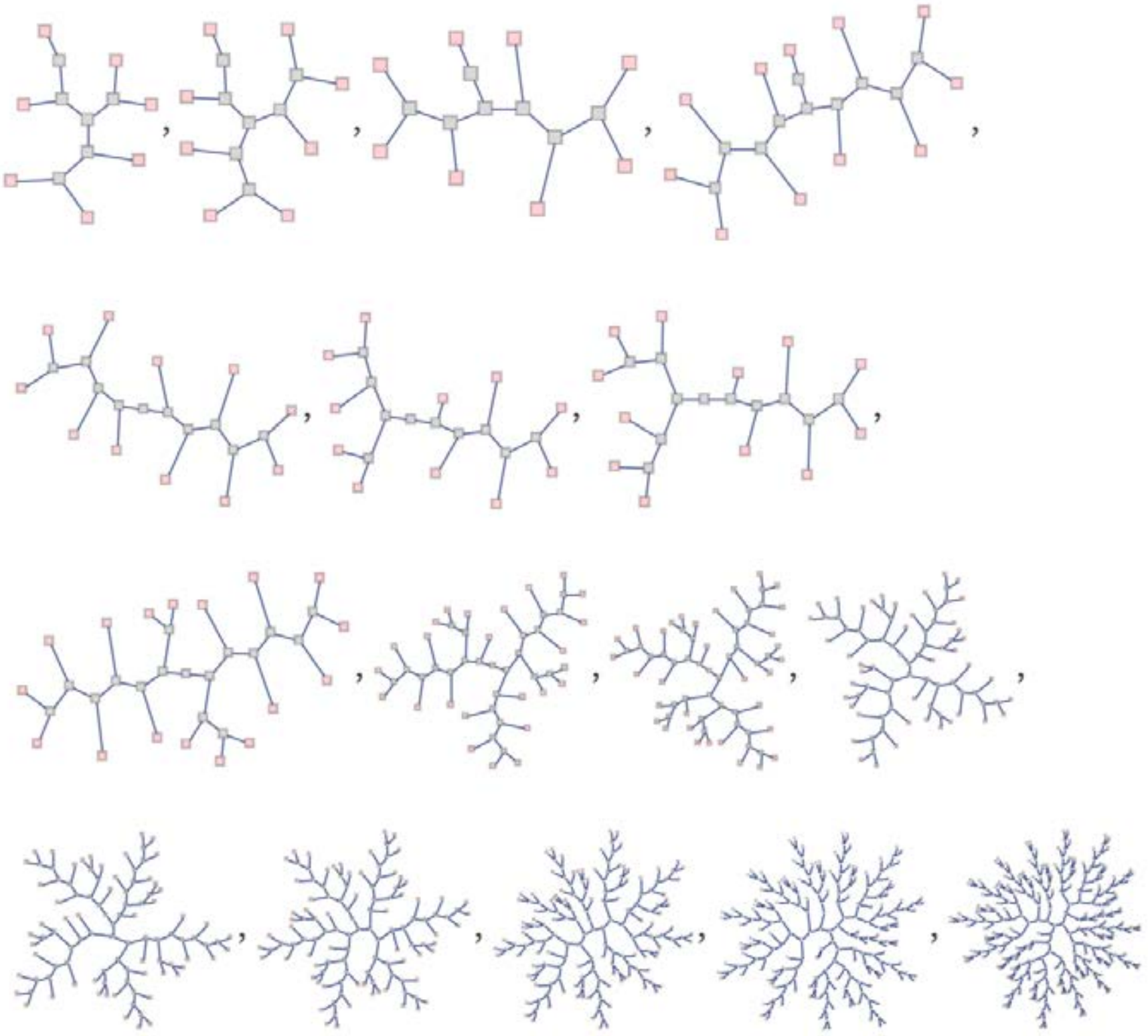

OK, so these are representations of the combinator expressions on successive steps. But where are the rules being applied at each step? As we'll discuss in much more detail in the next section, in the way we've done things so far we're always doing just one update at each step. Here's an example of where the updates are happening in a particular case:

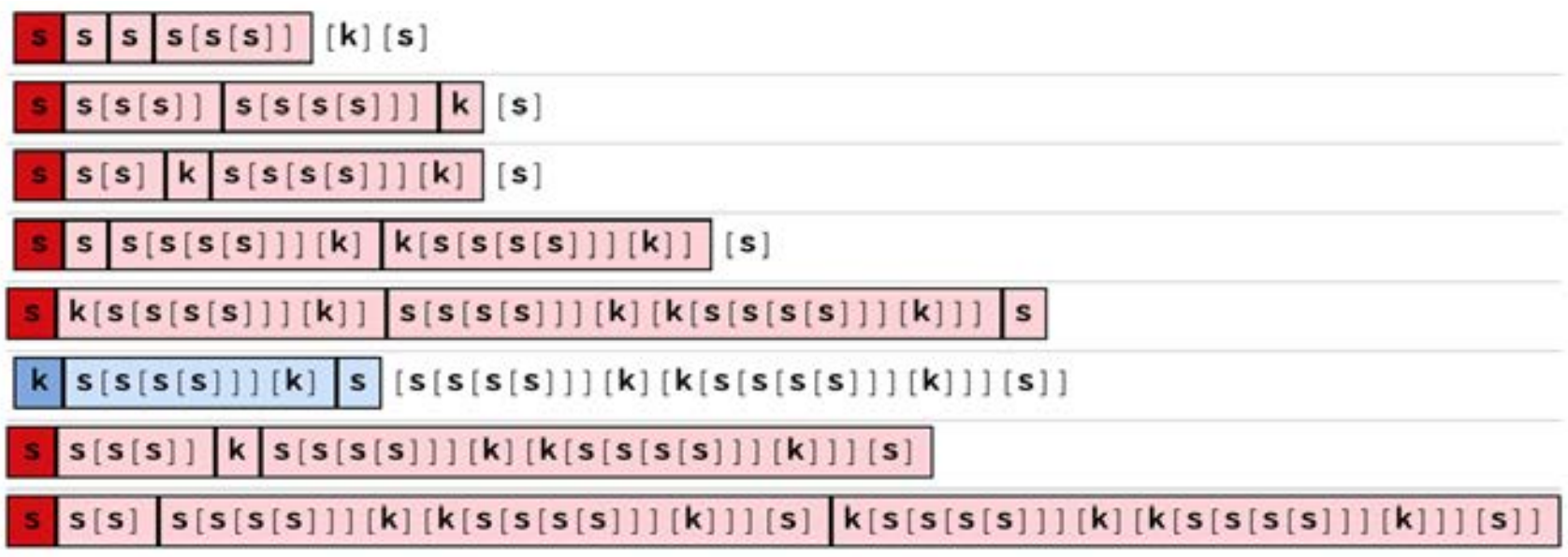

Continuing longer we get (note that some lines have wrapped in this display):

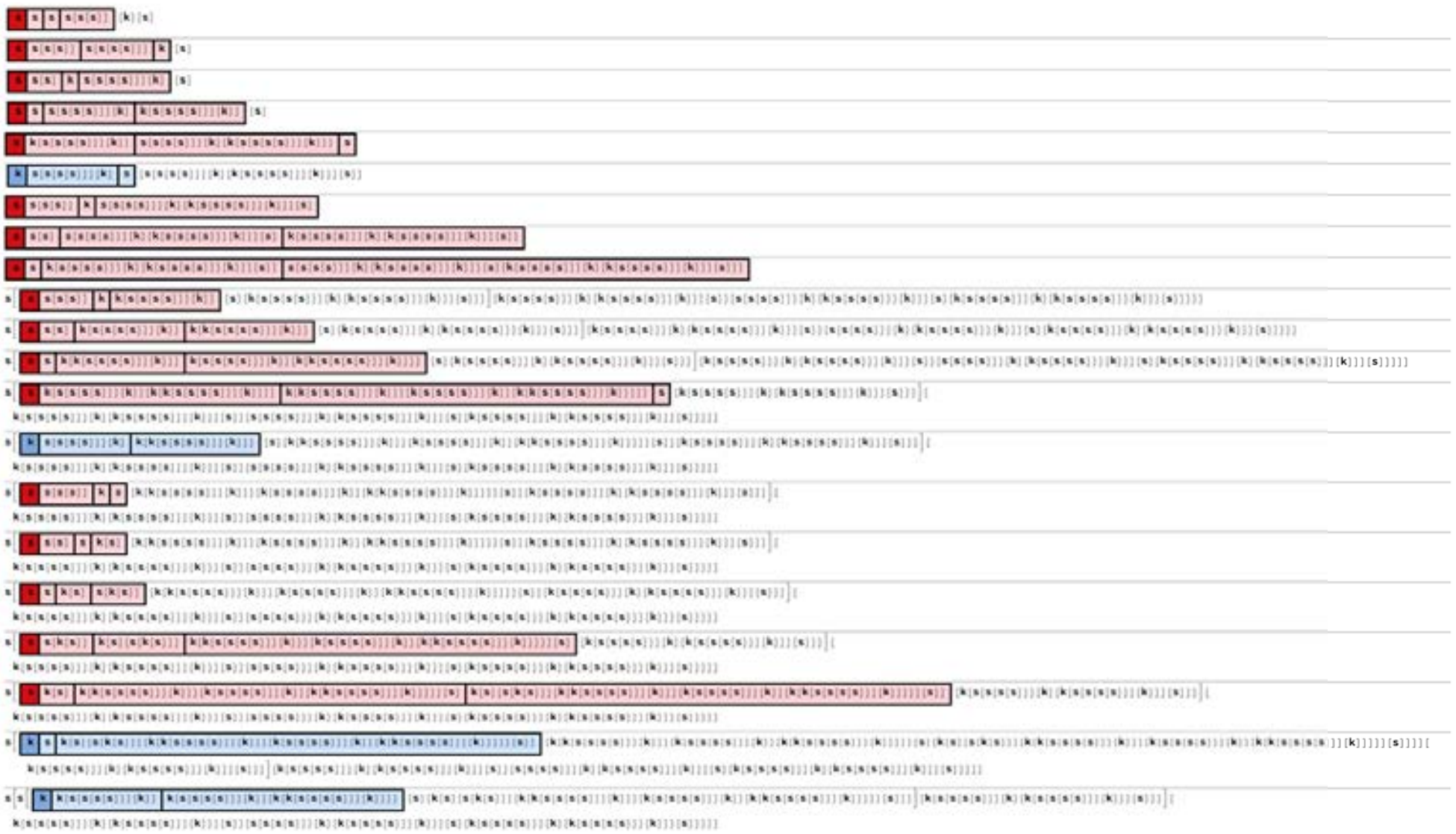

A feature of the way we're writing out combinator expressions is that the "input" to any combinator rule always corresponds to a contiguous span within the expression as we display it. So when we show the total size of combinator expressions on each step in an evolution, we can display which part is getting rewritten:

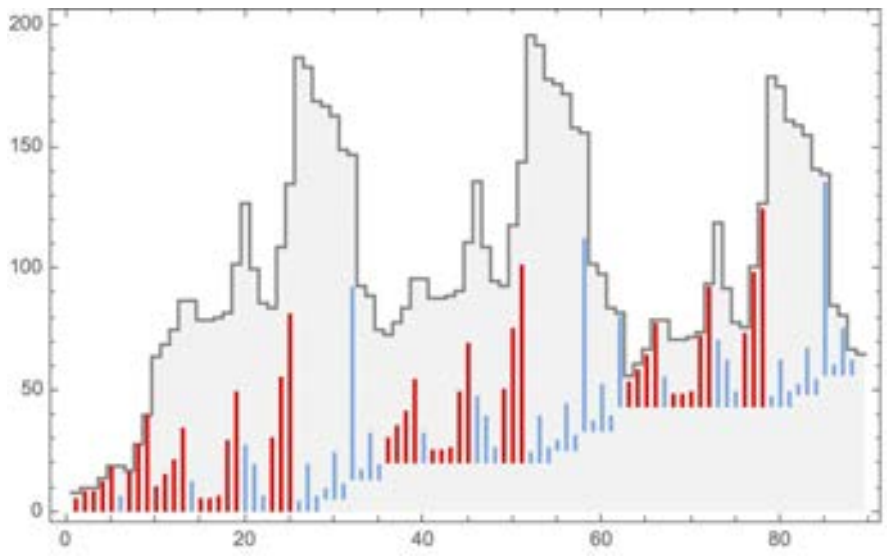

Notice that, as expected, application of the S rule tends to increase size, while the K rule decreases it.

Here is the distribution of rule applications for all the examples we showed above:

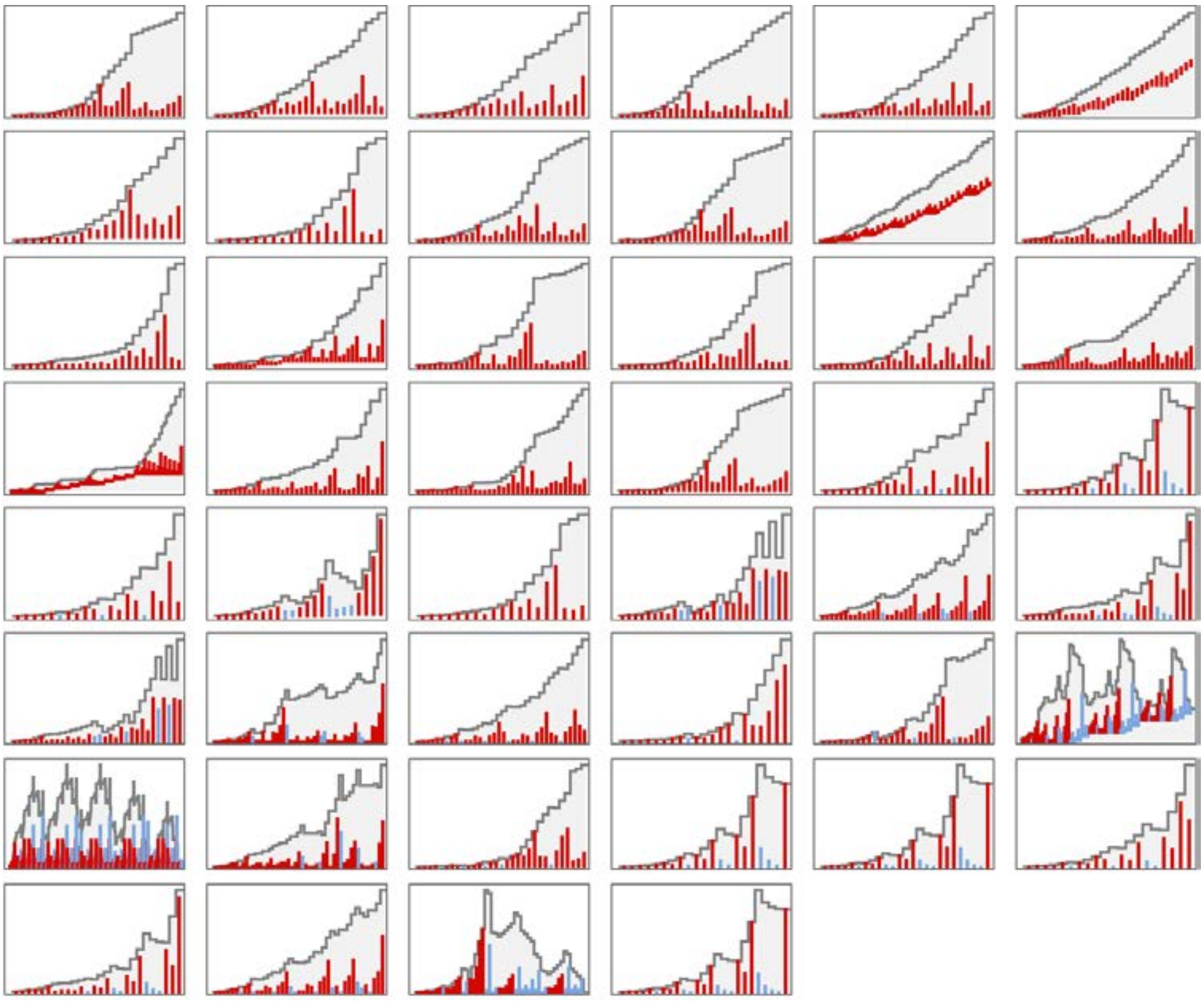

We can combine multiple forms of visualization by including depths:

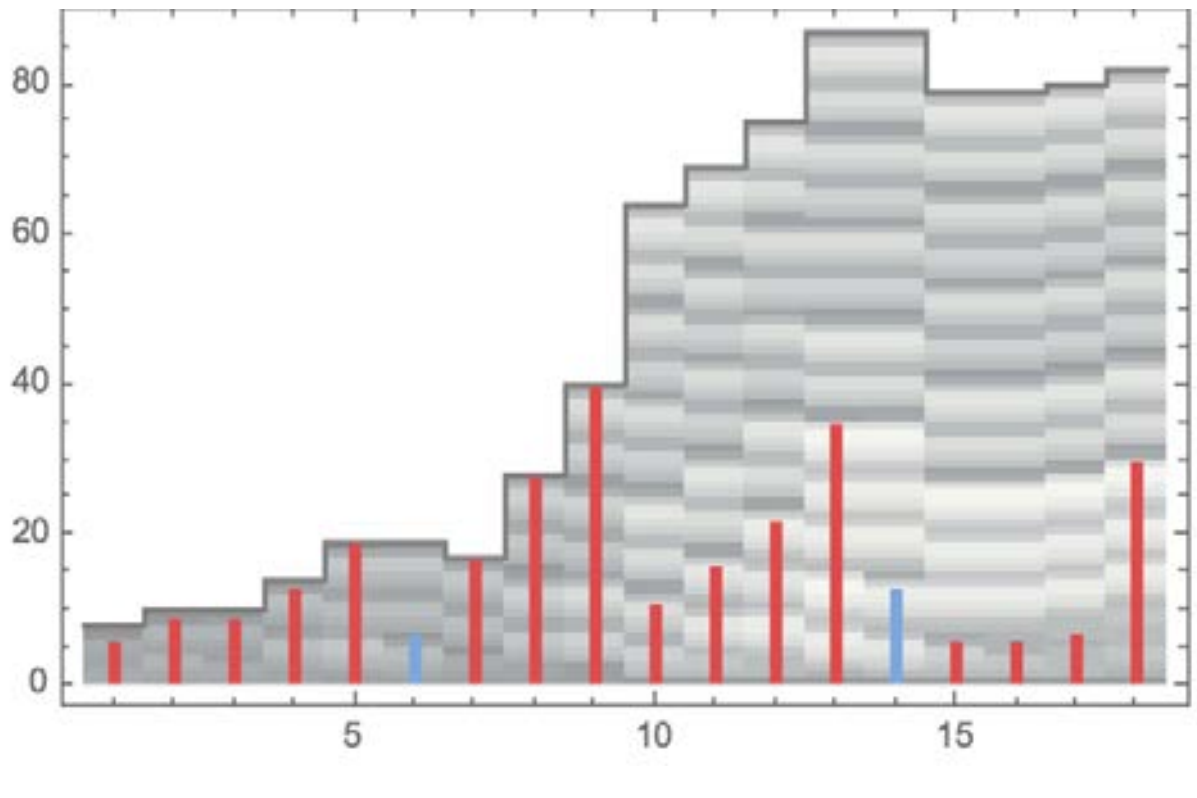

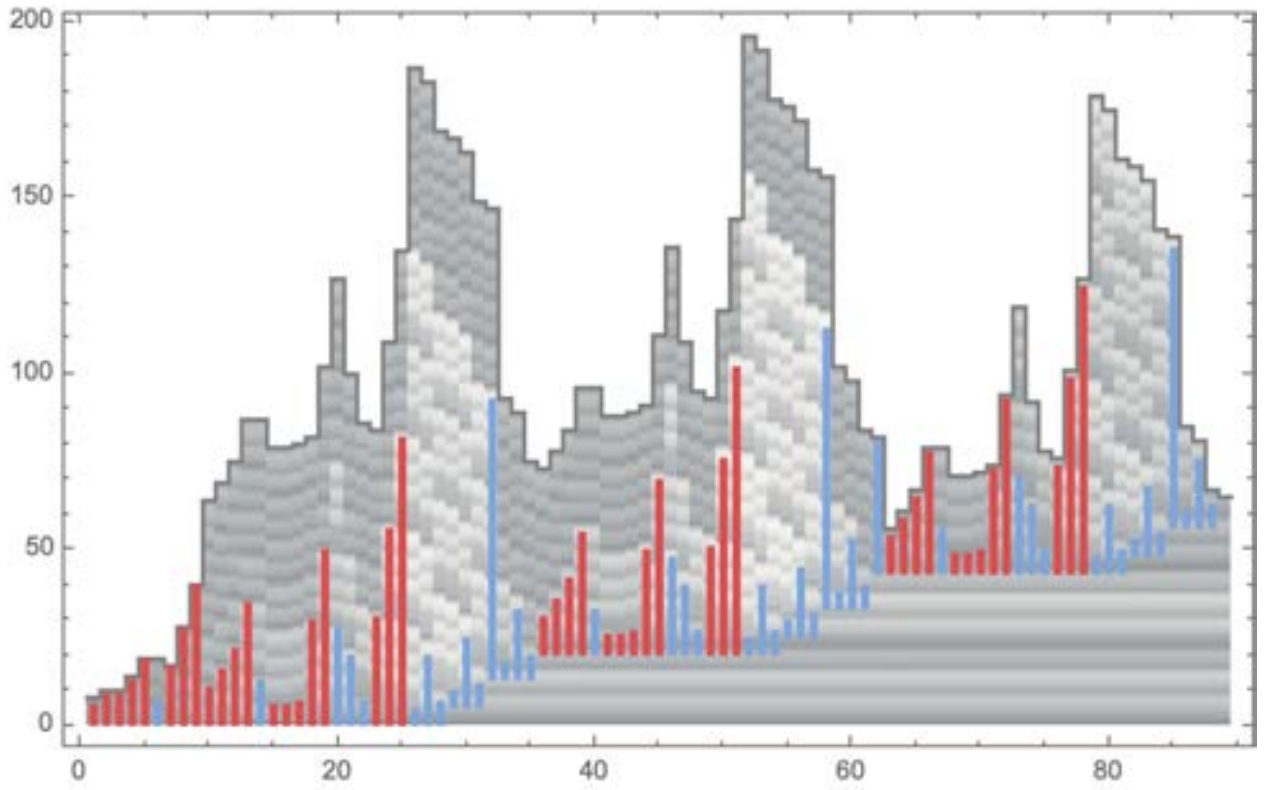

We can also do the same in 3D:

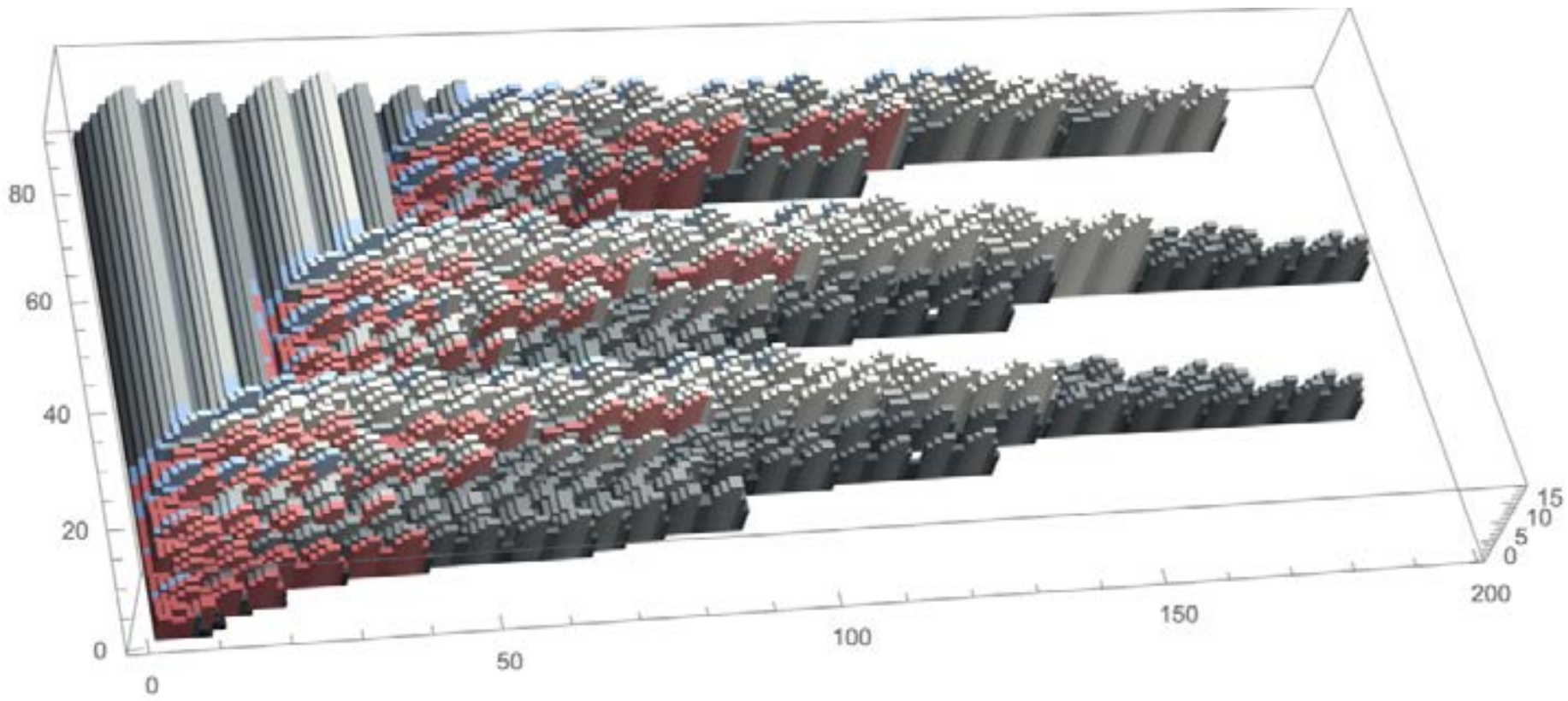

So what about the underlying trees? Here are the S, K combinator rules in terms of trees:

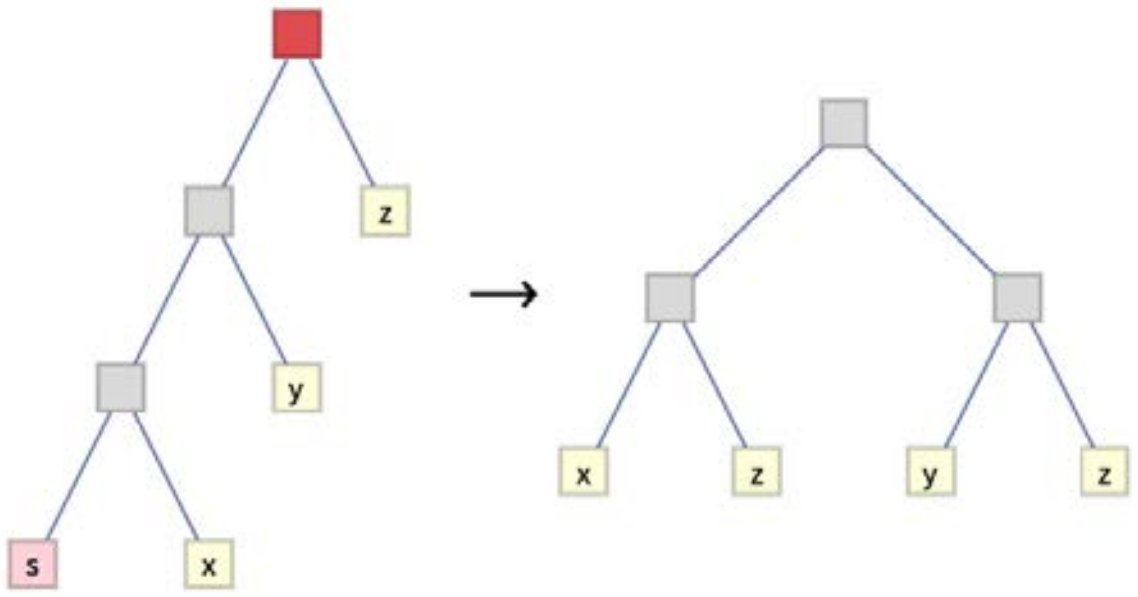

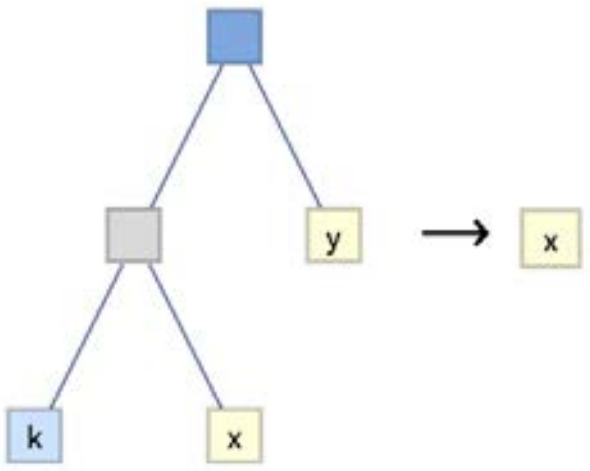

And here are the updates for the first few steps of the evolution of s[s][s][s[s[s]]][k][s] (**SSS**(**S**(**SS**))**KS**):

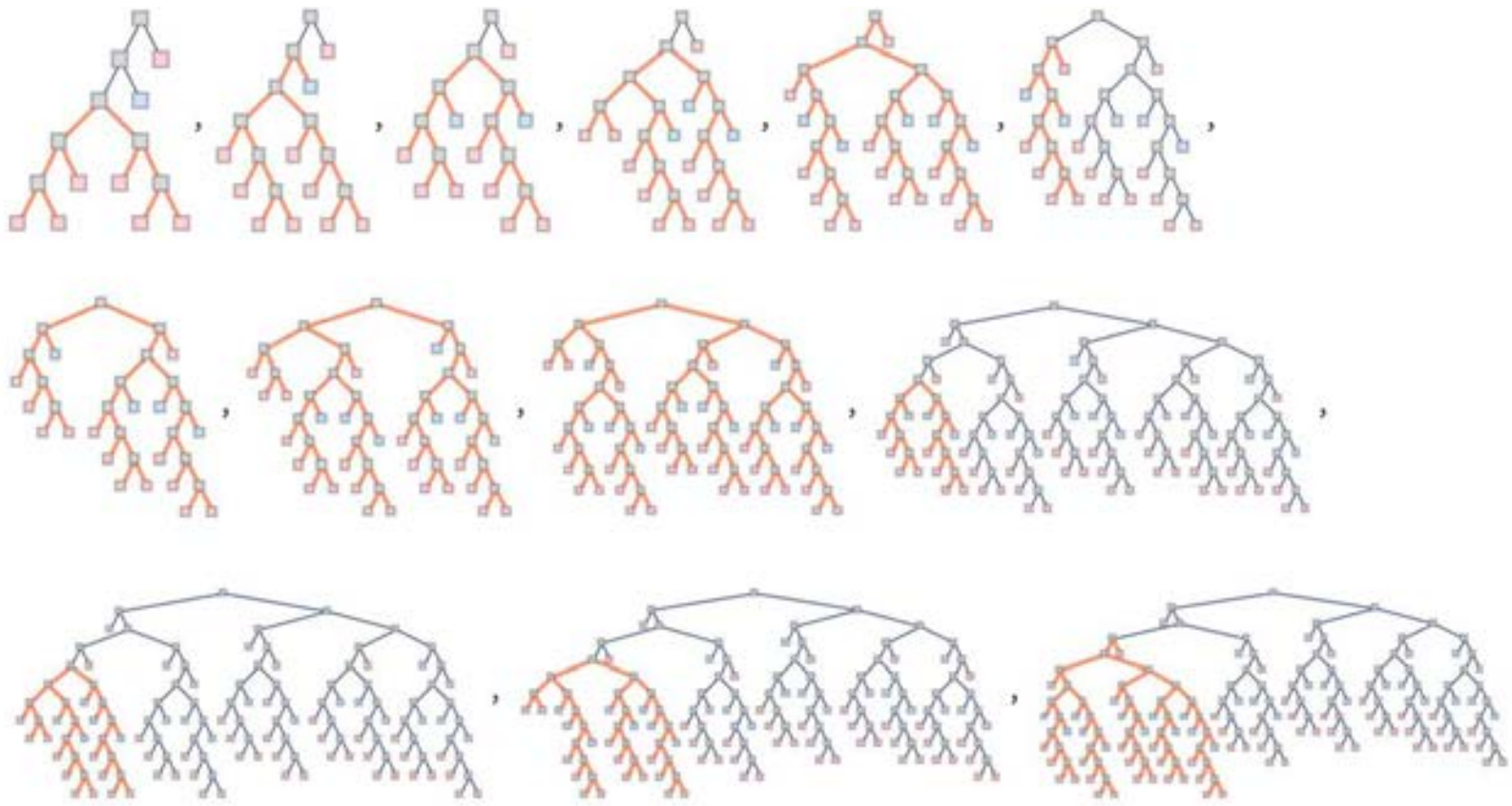

In these pictures we are effectively at each step highlighting the “first” subtree matching s[_][_][_] or k[_][_]. To get a sense of the whole evolution, we can also simply count the number of subtrees with a given general structure (like _[_][_] or _[_[_]]) that occur at a given step (see also below):

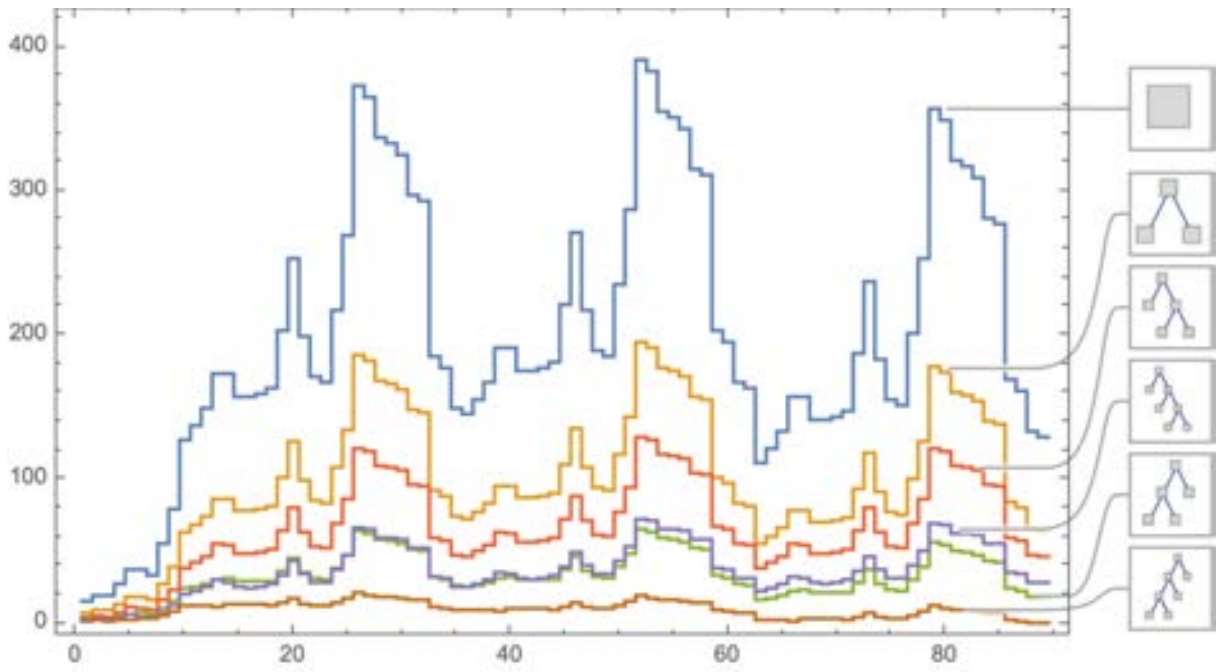

One more indication of the behavior of combinators comes from looking at tree depths. In addition to the total depth (i.e. Wolfram Language `Depth`) of the combinator tree, one can also look at the depth at which update events happen (here with the total size shown underneath):

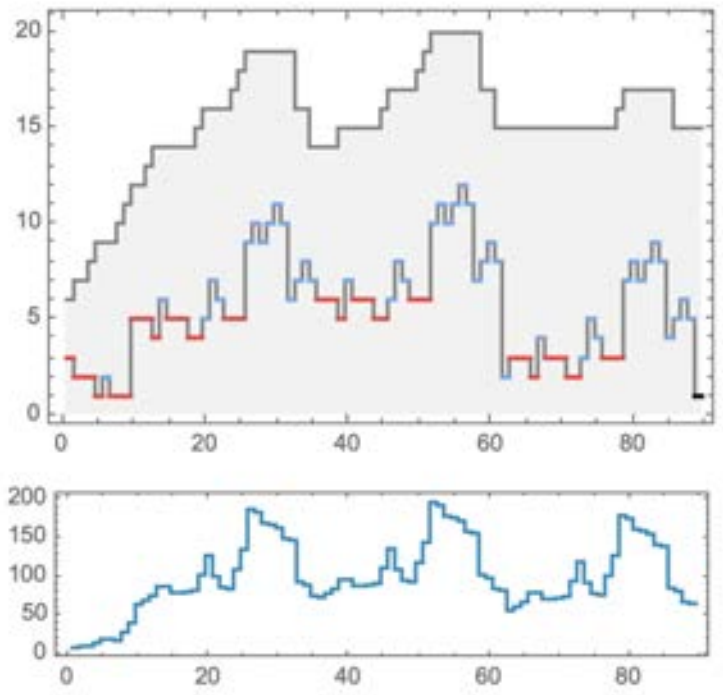

Here are the depth profiles for the rules shown above:

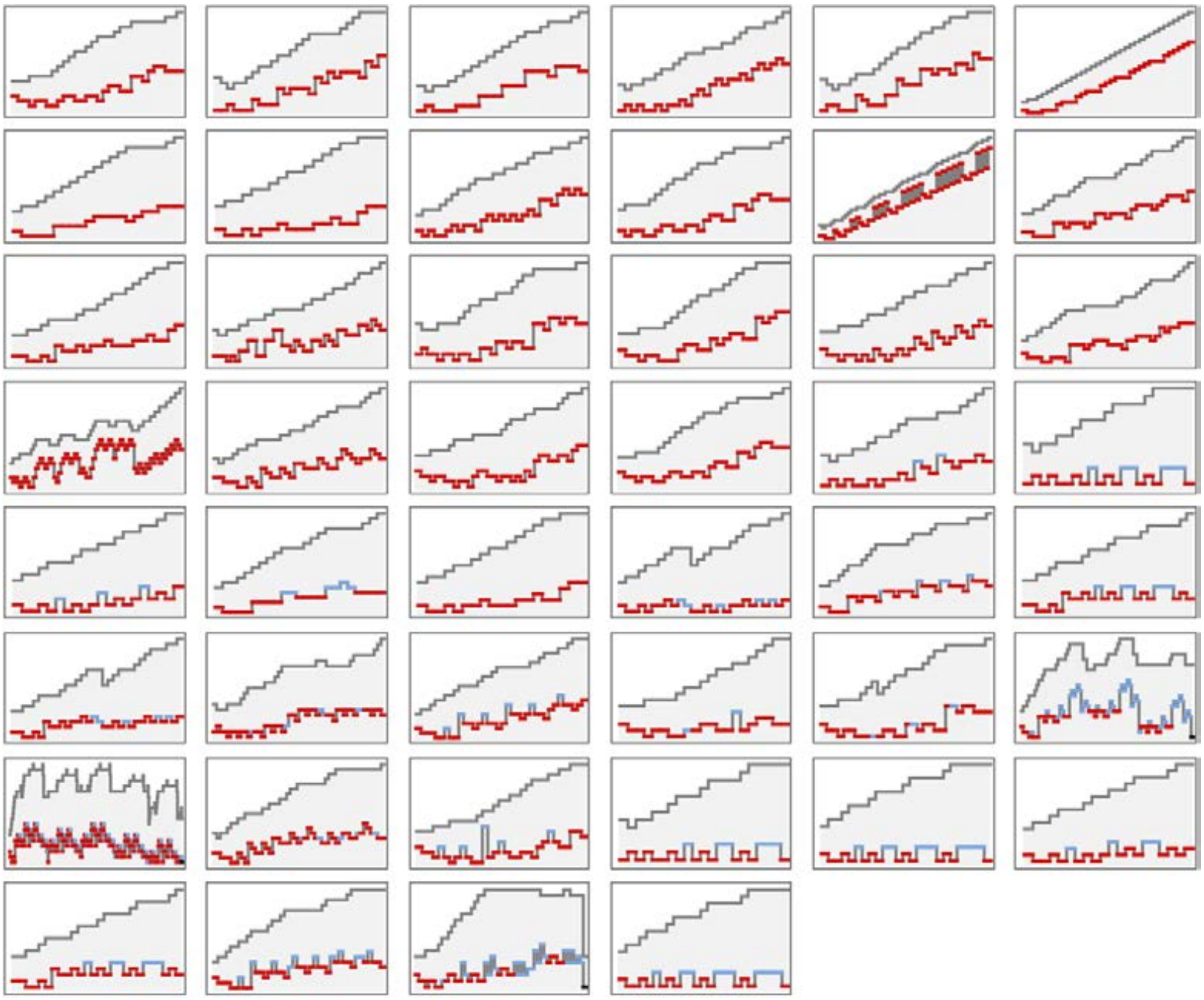

Not surprisingly, total depth tends to increase when growth continues. But it is notable that—except when termination is near at hand—it seems like (at least with our current updating scheme) updates tend to be made to “high-level” (i.e. low-depth) parts of the expression tree.

When we write out a combinator expression like the size-33

**s[s[s[s]][s[s[s[s]][s]][s]]][s[s[s]][s][s[s[s]][s[s[s[s]][s]][s]]][s[s[s[s]][s[s[s[s]][s]][s]]]]]**

or show it as a tree

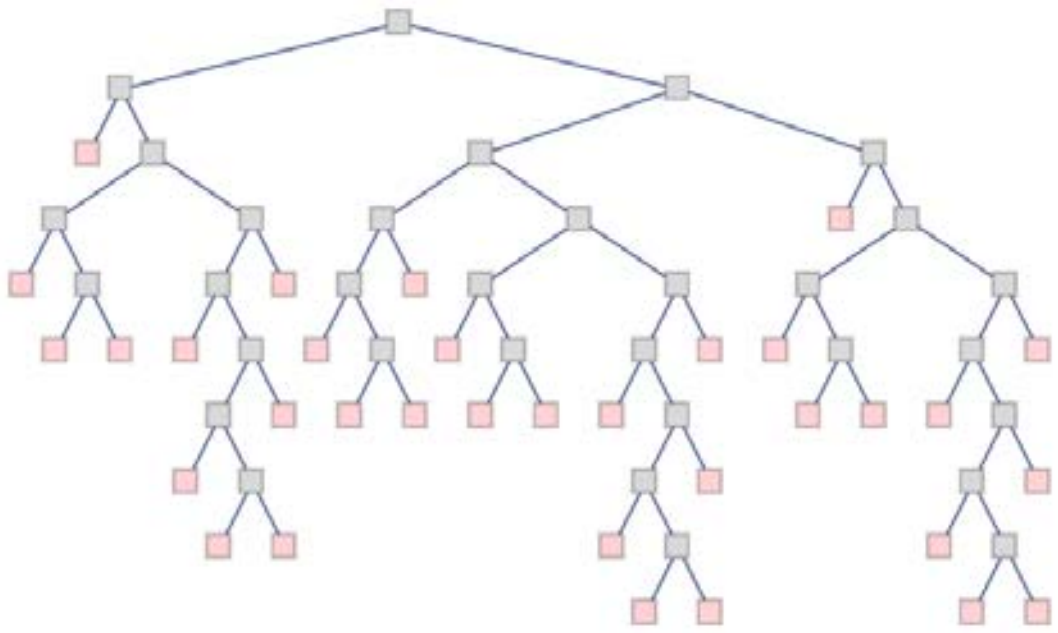

we're in a sense being very wasteful, because we're repeating the same subexpressions many times. In fact, in this particular expression, there are 65 subexpressions altogether—but only 11 distinct ones.

So how can we represent a combinator expression making as much use as possible of the commonality of these subexpressions? Well, instead of using a tree for the combinator expression, we can use a directed acyclic graph (DAG) in which we start from a node representing the whole expression, and then show how it breaks down into shared subexpressions, with each shared subexpression represented by a node.

To see how this works, let's consider first the trivial case of f[x]. We can represent this as a tree—in which the root represents the whole expression f[x], and has one connection to the head f, and another to the argument x:

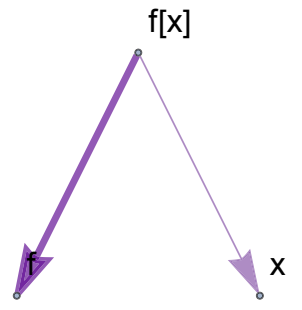

The expression f[g[x]] is still a tree:

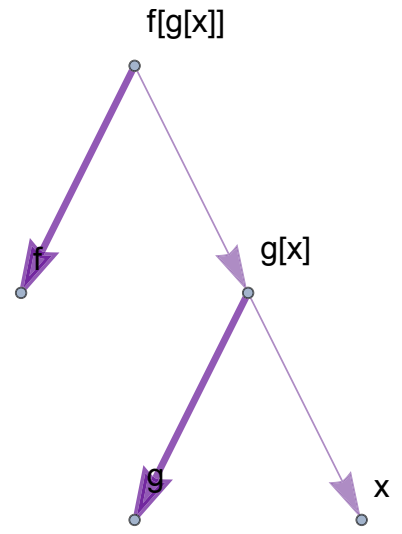

But in f[f[x]] there is a "shared subexpression" (which in this case is just f), and the graph is no longer a tree:

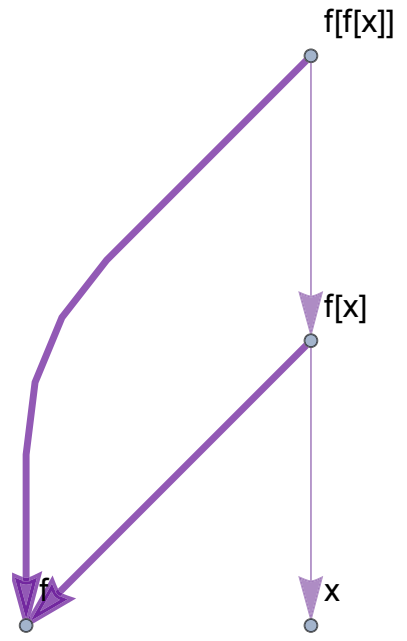

For f[x][f[x][f[x]]], f[x] is a shared subexpression:

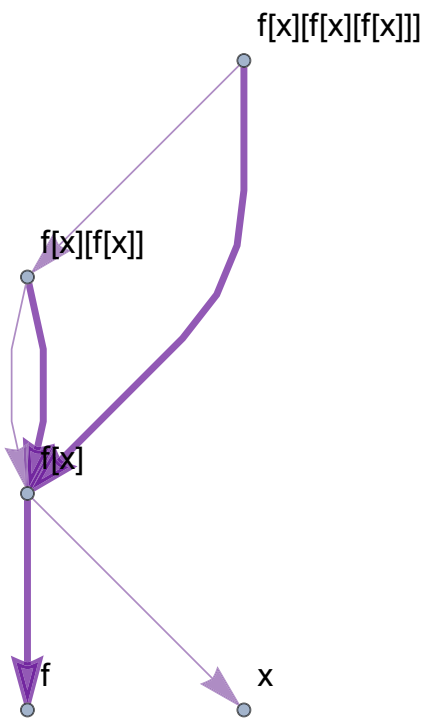

For s[s][s][s[s]][s][s] things get a bit more complicated:

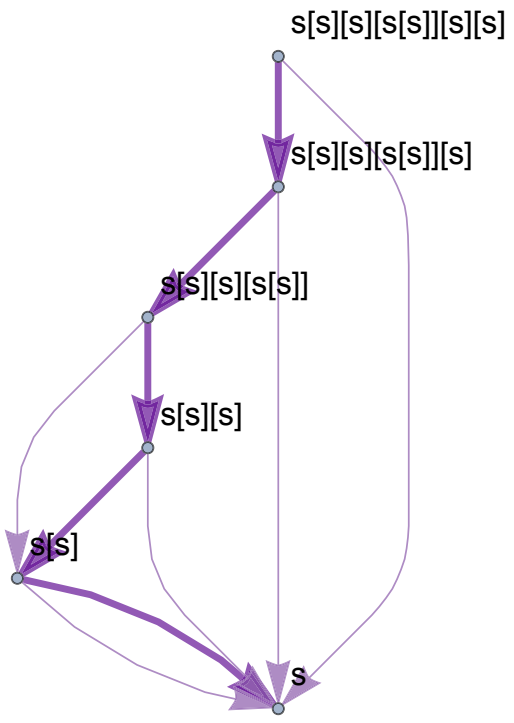

For the size-33 expression above, the DAG representation is

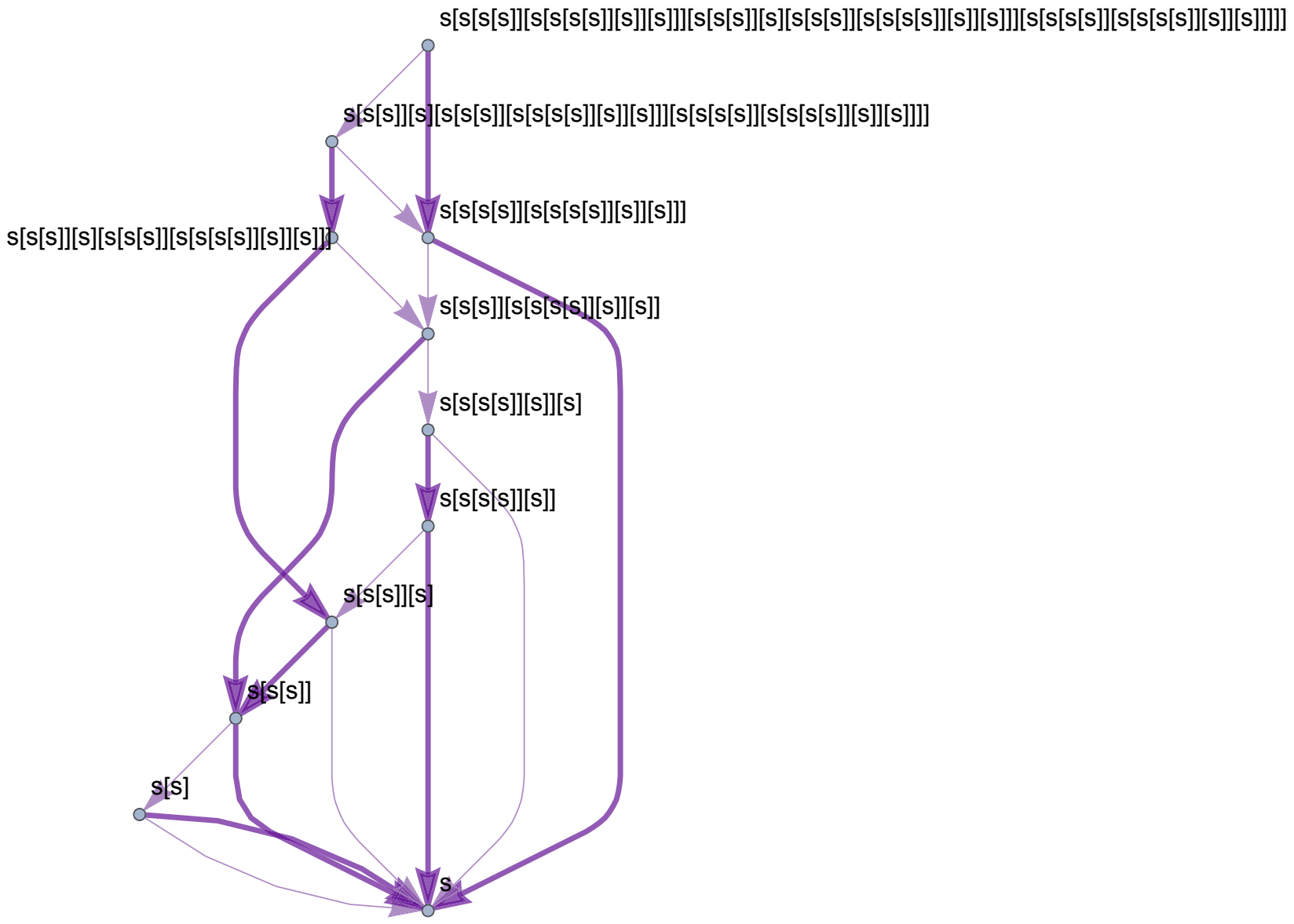

where the nodes correspond to the 11 distinct subexpression of the whole expression that appears at the root.

So what does combinator evolution look like in terms of DAGs? Here are the first 15 steps in the evolution of s[s][s][s[s]][s][s]:

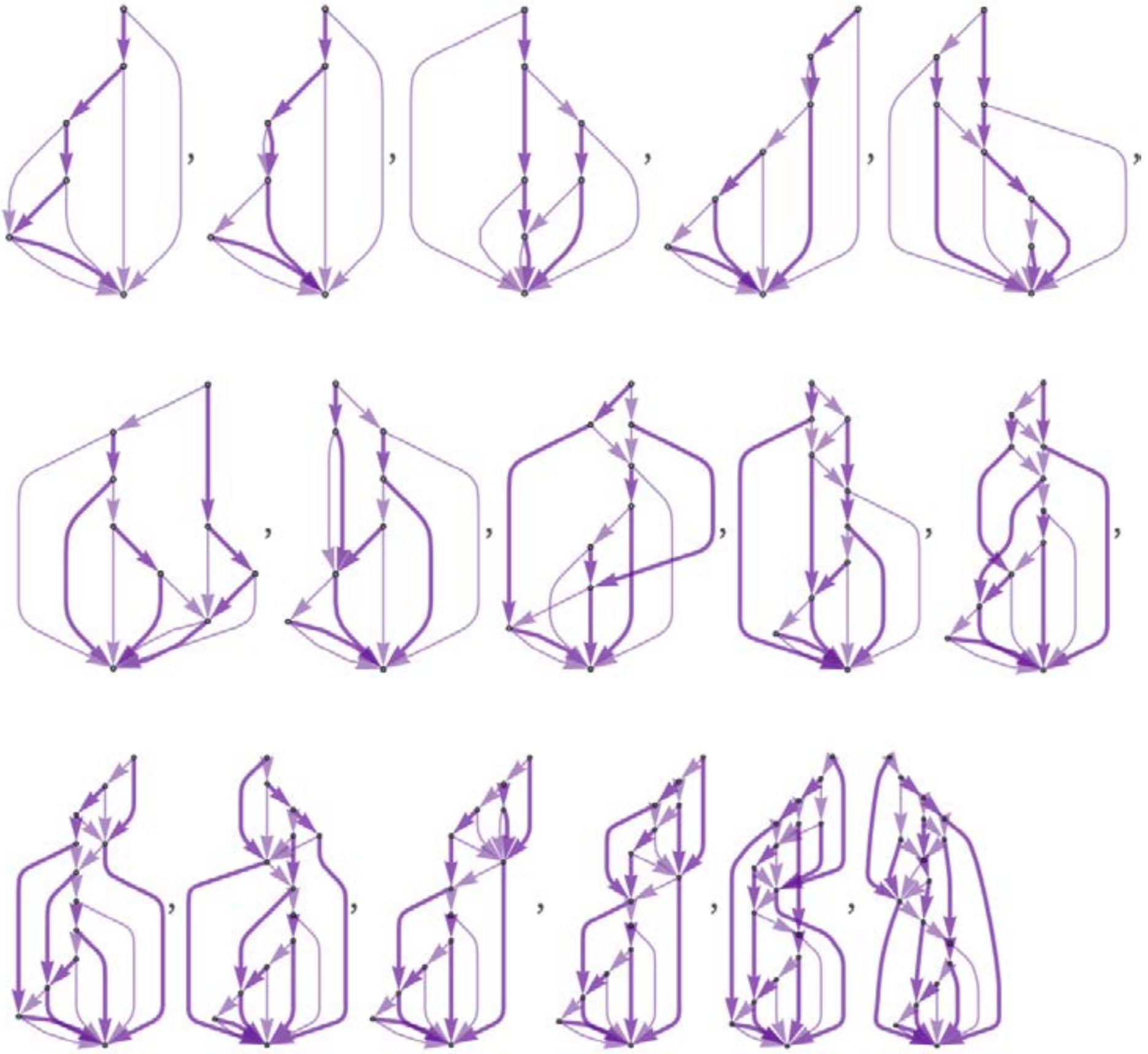

And here are some later steps:

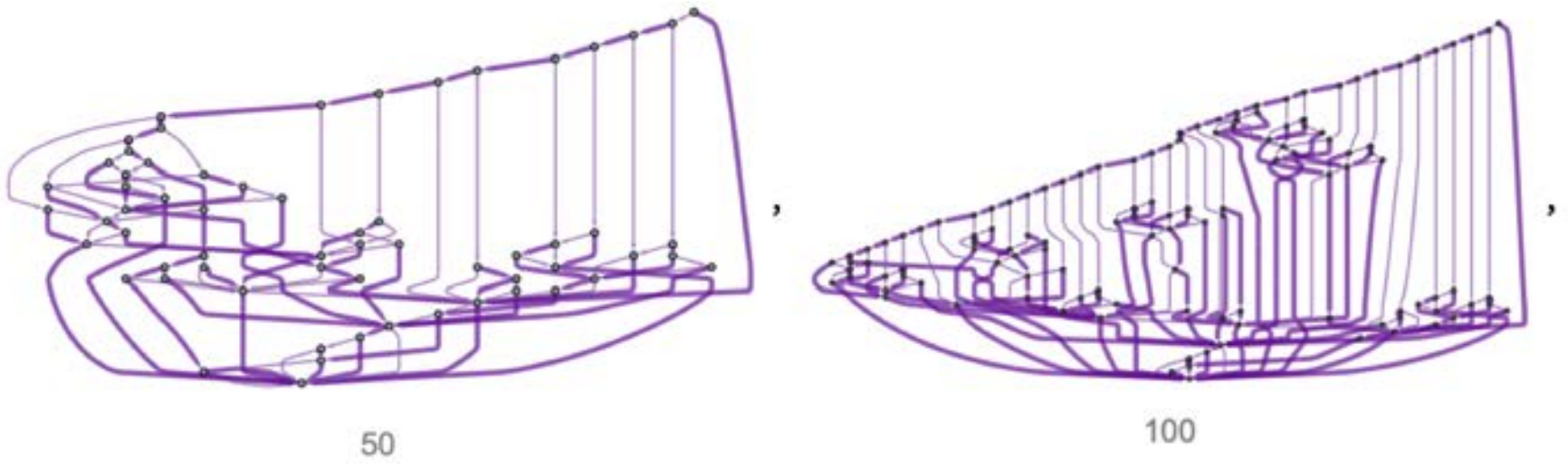

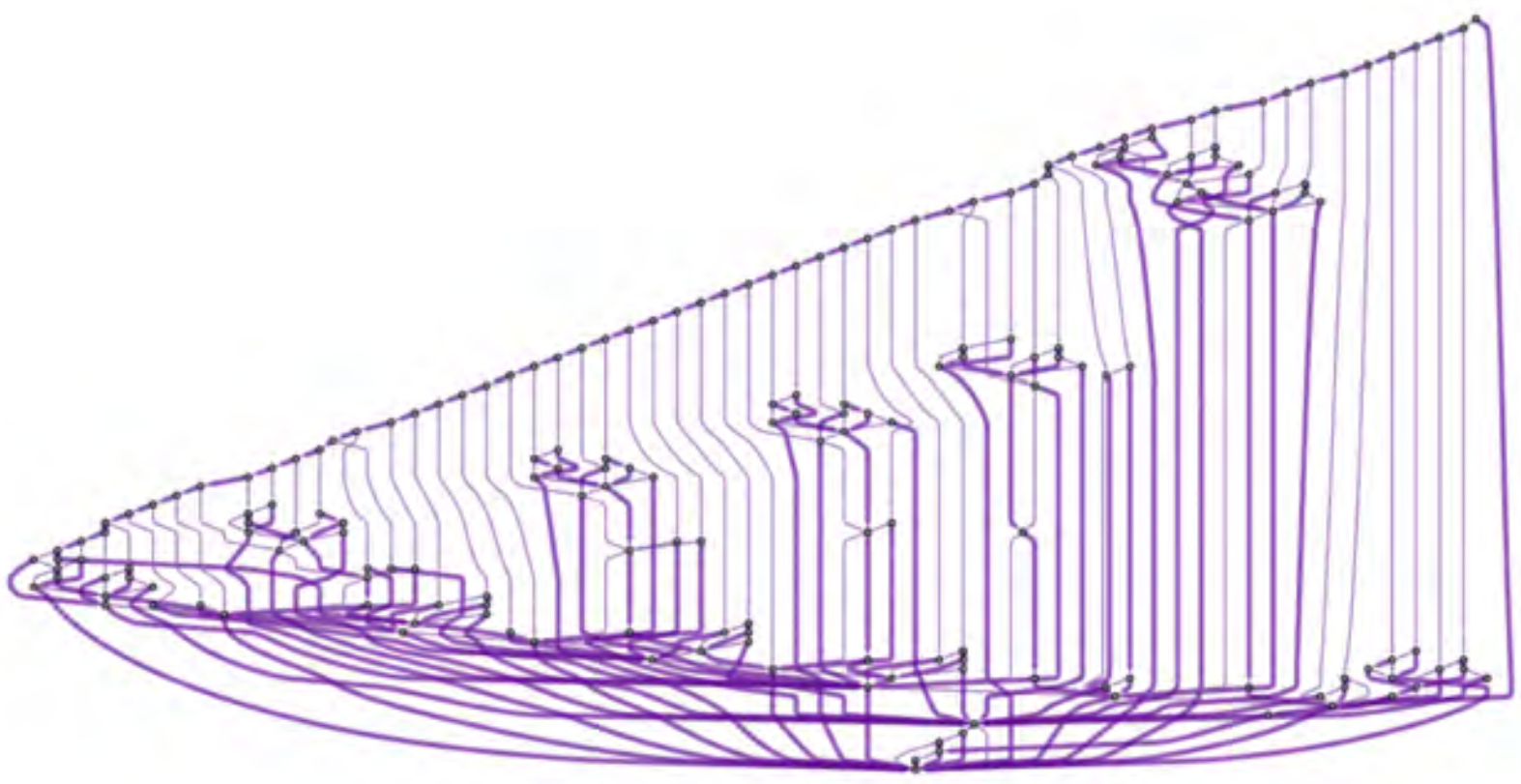

150

Sharing all common subexpressions is in a sense a maximally reduced way to specify a combinator expression. And even when the total size of the expressions is growing roughly exponentially, the number of distinct subexpressions may grow only linearly—here roughly like 1.24 t:

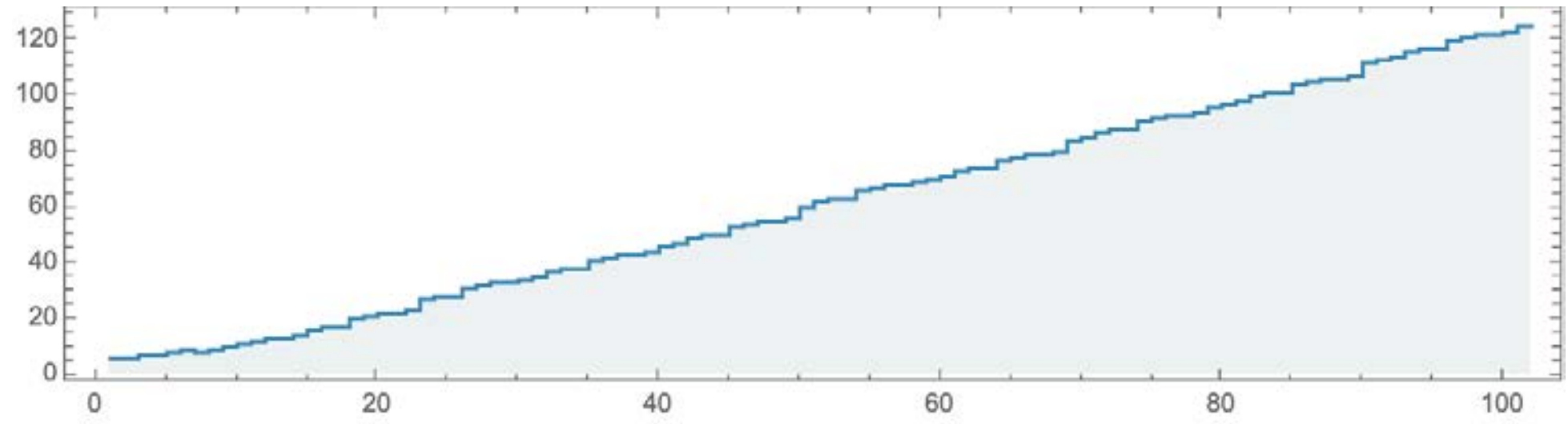

Looking at successive differences suggests a fairly simple pattern:

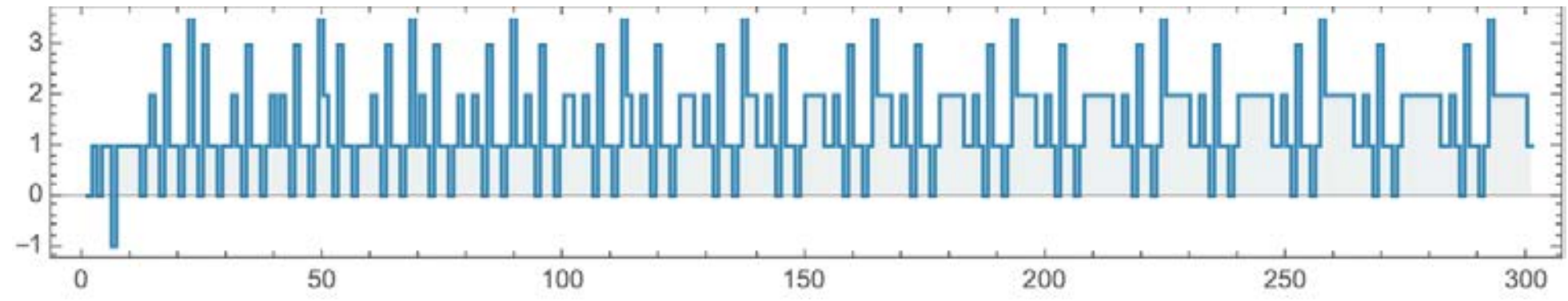

Here are the DAG representations of the result of 50 steps in the evolution of the 46 “growing size-7” combinator expressions above:

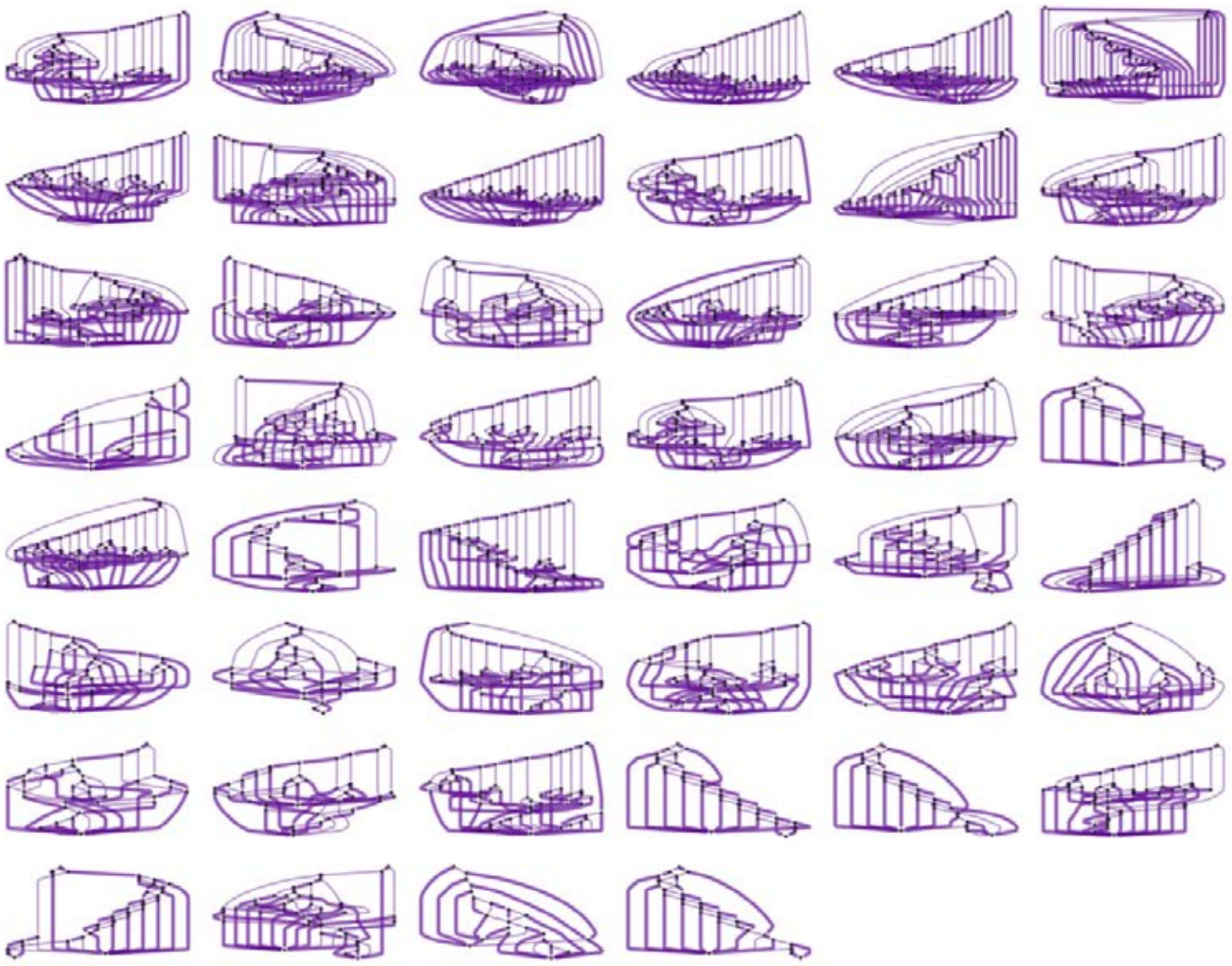

It's notable that some of these show considerable complexity, while others have a rather simple structure.

## Updating Schemes and Multiway Systems

The world of combinators as we've discussed it so far may seem complicated. But we've actually so far been consistently making a big simplification. And it has to do with how the combinator rules are applied.

Consider the combinator expression:

s[s[s][s][s[s][k[k][s]][s]]][s][s][s[k[s][k]][k][s]]

There are 6 places (some overlapping) at which s[_][_][_] or k[_][_] matches some subpart of this expression:

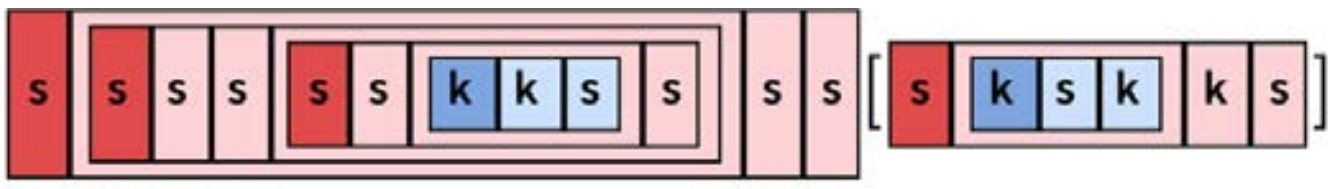

One can see the same thing in the tree form of the expression (the matches are indicated at the roots of their subtrees):

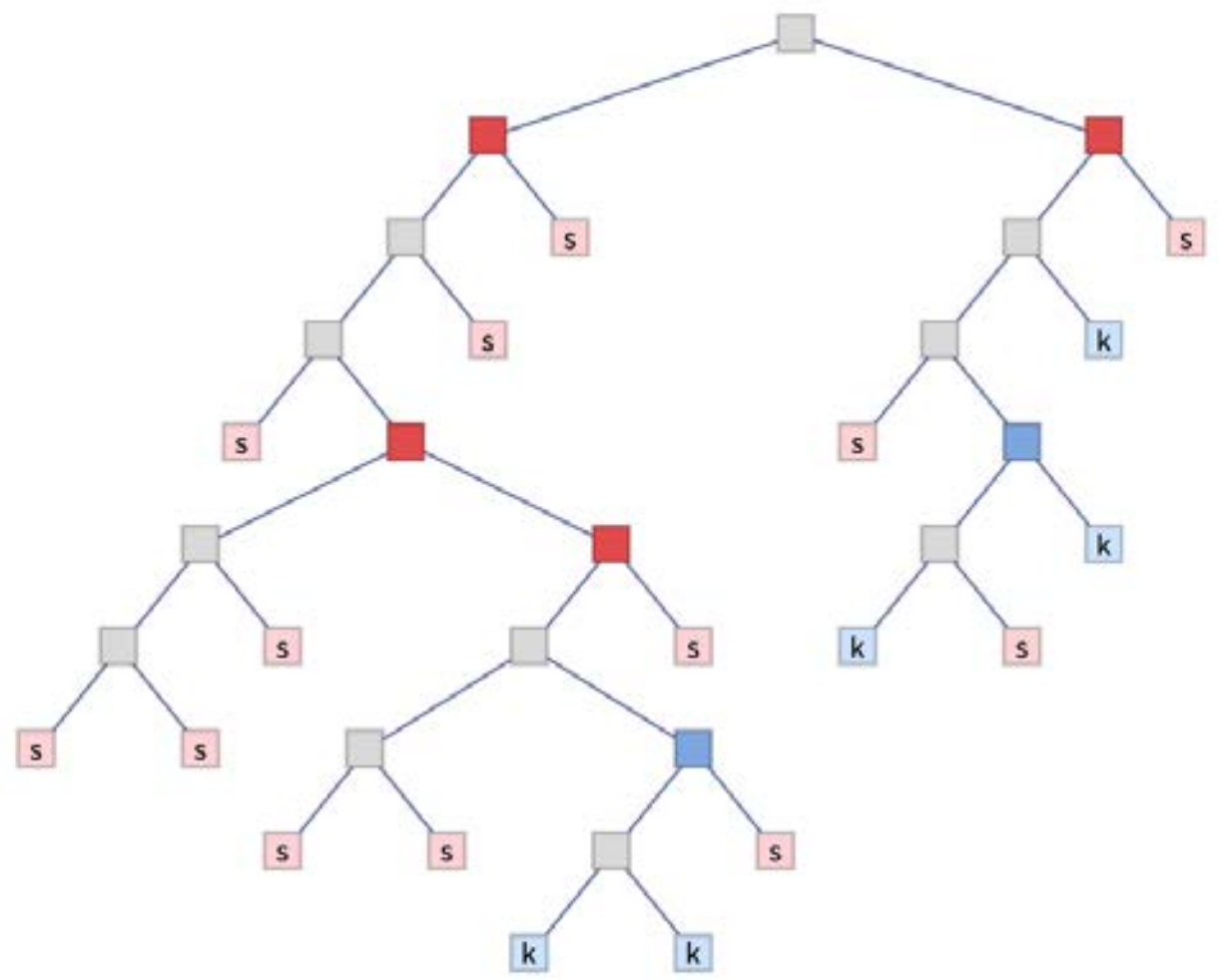

But now the question is: if one's applying combinator rules, which of these matches should one use?

What we've done so far is to follow a particular strategy—usually called leftmost outermost—which can be thought of as looking at the combinator expression as we normally write it out with brackets etc. and applying the first match we encounter in a left-to-right scan, or in this case:

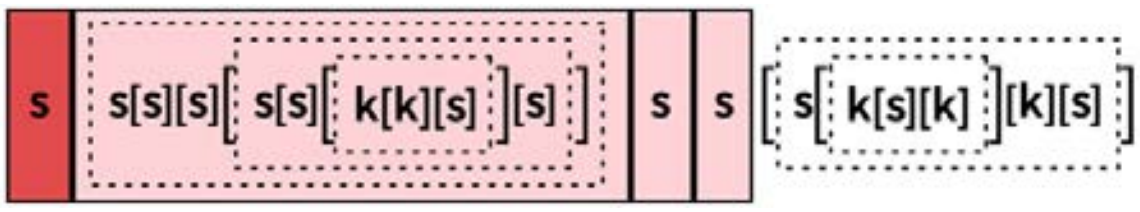

In the Wolfram Language we can find the positions of the matches just using:

```
expr = s[s[s][s][s[s][k[k][s]][s]]][s][s][s[k[s][k]][k][s]]
```

```
pos = Position[expr, s[_][_][_] | k[_][_]]
```

{{0, 0, 0, 1, 1, 0, 1}, {0, 0, 0, 1, 1}, {0, 0, 0, 1}, {0}, {1, 0, 0, 1}, {1}}

This shows—as above—where these matches are in the expression:

```
expr = s[s[s][s][s[s][k[k][s]][s]]][s][s][s[k[s][k]][k][s]];
```

```
pos = Position[expr, s[_][_][_] | k[_][_]];

MapAt[Framed, expr, pos]
```

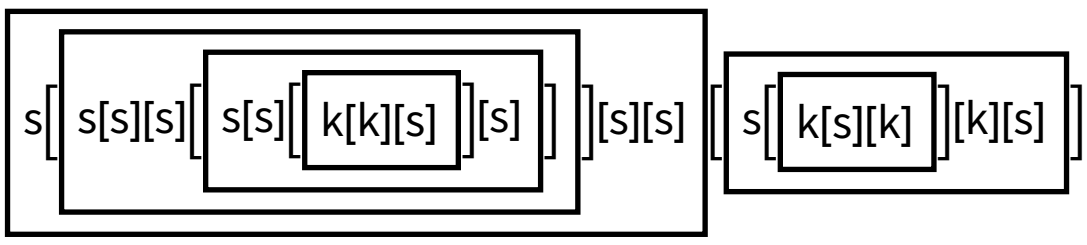

Here are the matches, in the order provided by Position:

| | |
|---|---|
| {0, 0, 0, 1, 1, 0, 1} | s[s[s][s][s[s][k[k][s]]][s]]][s][s][s[k[s][k]][k][s]] |
| {0, 0, 0, 1, 1} | s[s[s][s][s[s][k[k][s]]][s]]][s][s][s[k[s][k]][k][s]] |
| {0, 0, 0, 1} | s[s[s][s][s[s][k[k][s]]][s]]][s][s][s[k[s][k]][k][s]] |
| {0} | s[s[s][s][s[s][k[k][s]]][s]]][s][s][s[k[s][k]][k][s]] |
| {1, 0, 0, 1} | s[s[s][s][s[s][k[k][s]]][s]]][s][s][s[k[s][k]][k][s]] |
| {1} | s[s[s][s][s[s][k[k][s]]][s]]][s][s][s[k[s][k]][k][s]] |

The leftmost-outermost match here is the one with position {0}.

In general the series of indices that specify the position of a subexpression say whether to reach the subexpression one should go left or right at each level as one descends the expression tree. An index 0 says to go to "head", i.e. the f in f[x], or the f[a][b] in f[a][b][c]; an index 1 says to the "first argument", i.e. the x in f[x], or the c in f[a][b][c]. The length of the list of indices gives the depth of the corresponding subexpression.

We'll talk in the next section about how leftmost outermost—and other schemes—are defined in terms of indices. But here the thing to notice is that in our example here Position doesn't give us part {0} first; instead it gives us {0,0,0,1,1,0,1}:

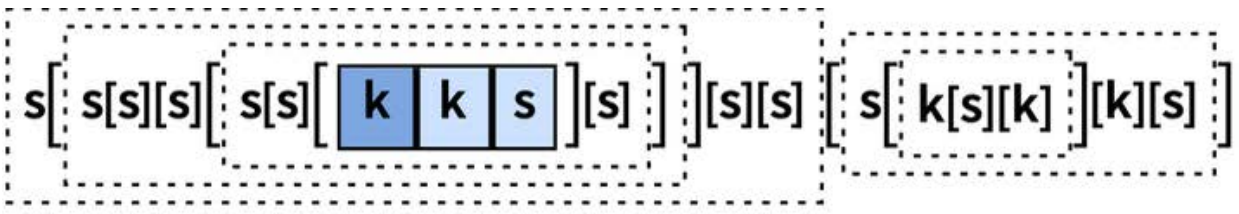

And what's happening is that Position is doing a depth-first traversal of the expression tree to look for matches, so it first descends all the way down the left-hand tree branches—and since it finds a match there, that's what it returns. In the taxonomy we'll discuss in the next section, this corresponds to a leftmost-innermost scheme, though here we'll refer to it as "depth first".

Now consider the example of s[s][s][k[s][s]]. Here is what it does first with the leftmost-outermost strategy we've been using so far, and second with the new strategy:

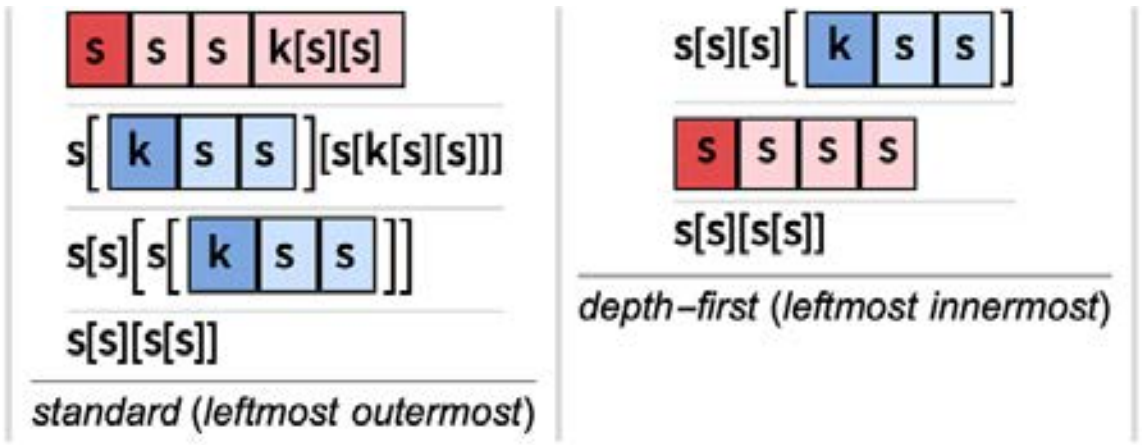

There are two important things to notice. First, that in both cases the final result is the same. And second, that the steps taken—and the total number required to get to the final result—is different in the two cases.

Let's consider a larger example: s[s][s][s[s[s]]][k][s]] (**SSS**(**S**(**SS**))**KS**). With our standard strategy we saw above that the evolution of this expression terminates after 89 steps, giving an expression of size 65. With the depth-first strategy the evolution still terminates with the same expression of size 65, but now it takes only 29 steps:

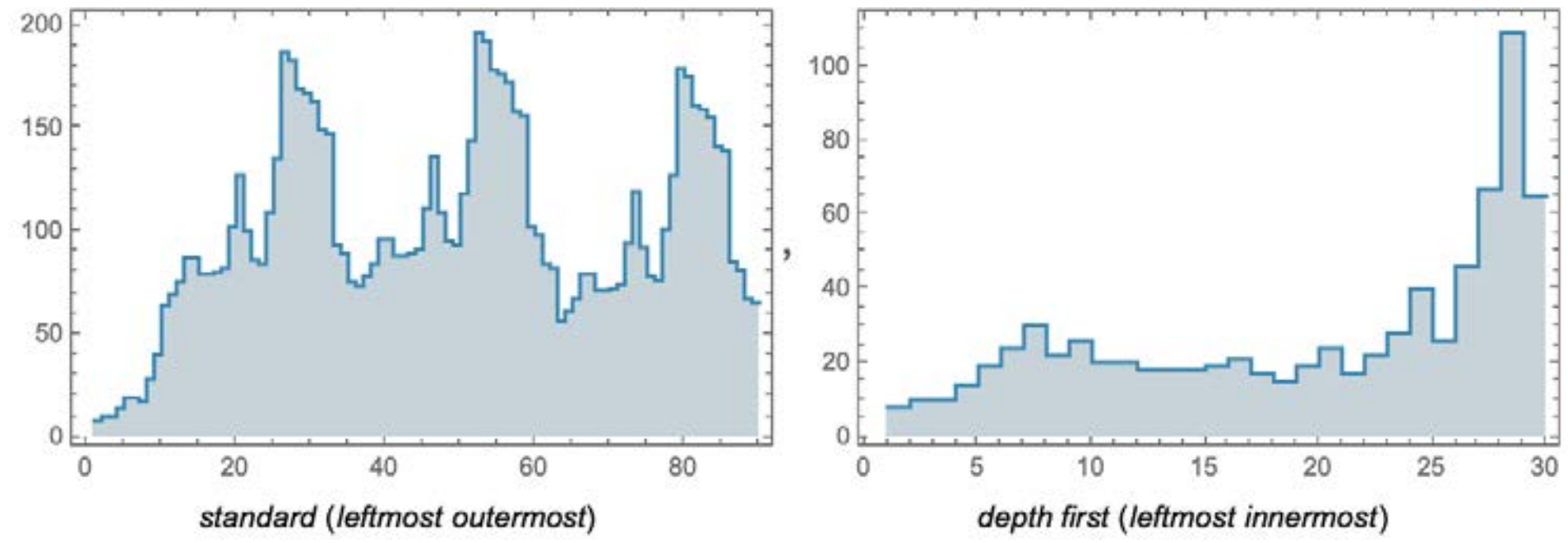

It's an important feature of combinator expression evolution that when it terminates—whatever strategy one's used—the result must always be the same. (This "confluence" property—that we'll discuss more later—is closely related to the concept of causal invariance in our models of physics.)

What happens when the evolution doesn't terminate? Let's consider the simplest non-terminating case we found above: s[s][s][s[s]][s][s] (**SSS**(**SS**)**SS**). Here's how the sizes increase with the two strategies we've discussed:

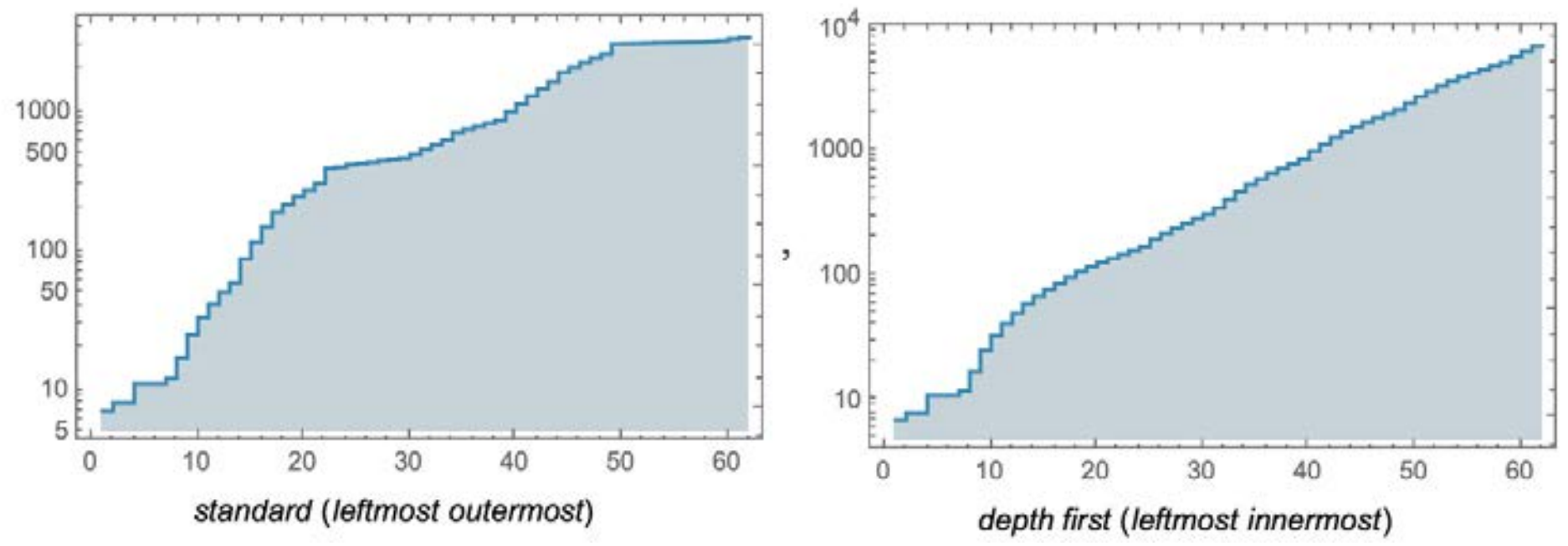

The difference is more obvious if we plot the ratios of sizes on successive steps:

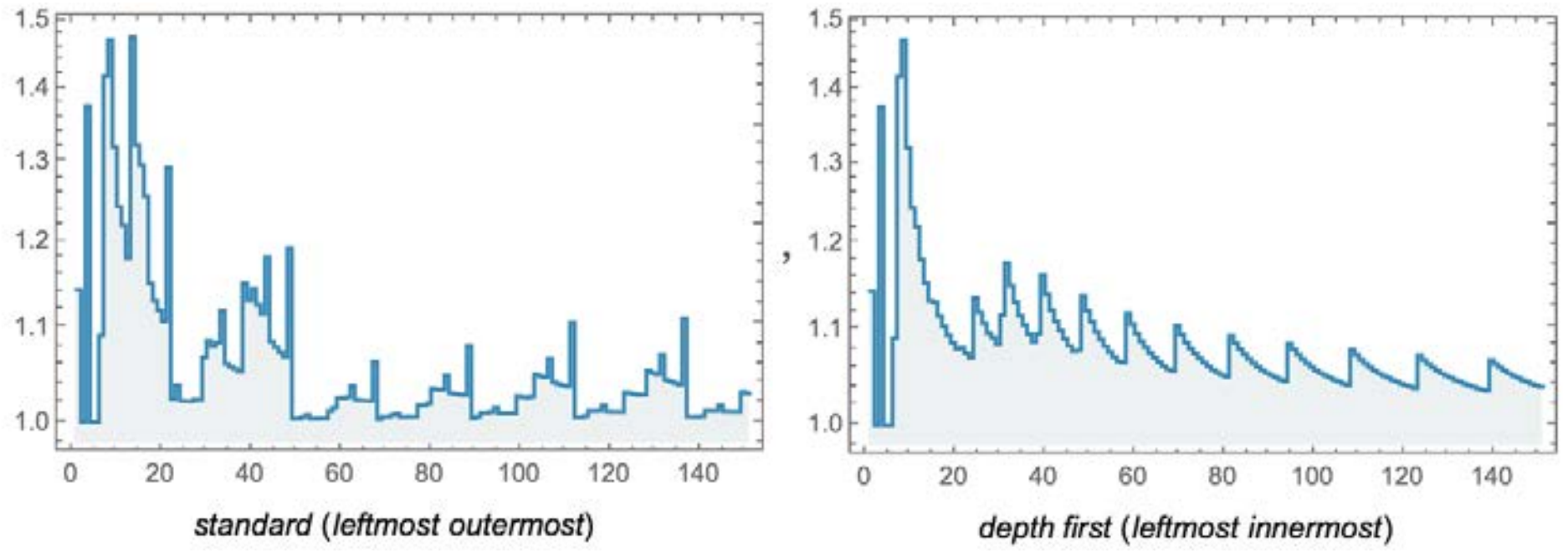

In both these pairs of pictures, we can see that the two strategies start off producing the same results, but soon diverge.

OK, so we've looked at two particular strategies for picking which updates to do. But is there a general way to explore all possibilities? It turns out that there is—and it's to use multiway systems, of exactly the kind that are also important in our Physics Project.

The idea is to make a multiway graph in which there's an edge to represent each possible update that can be performed from each possible "state" (i.e. combinator expression). Here's what this looks like for the example of s[s][s][k[s][s]] (**SSS**(**KSS**)) above:

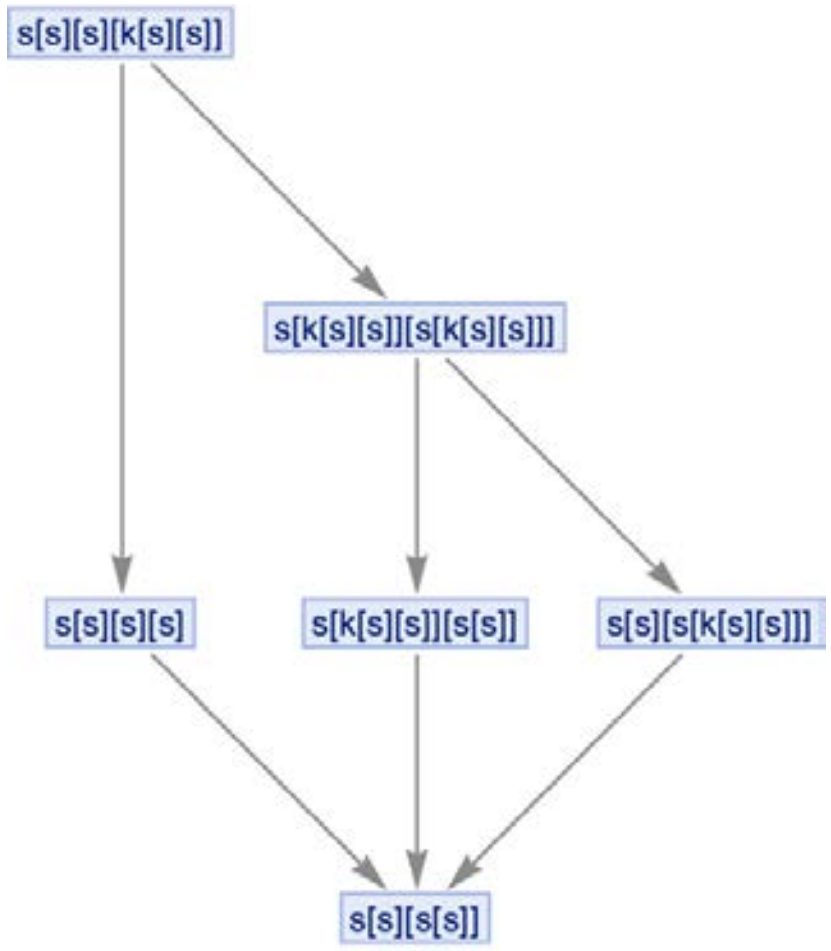

Here's what we get if we include all the "updating events":

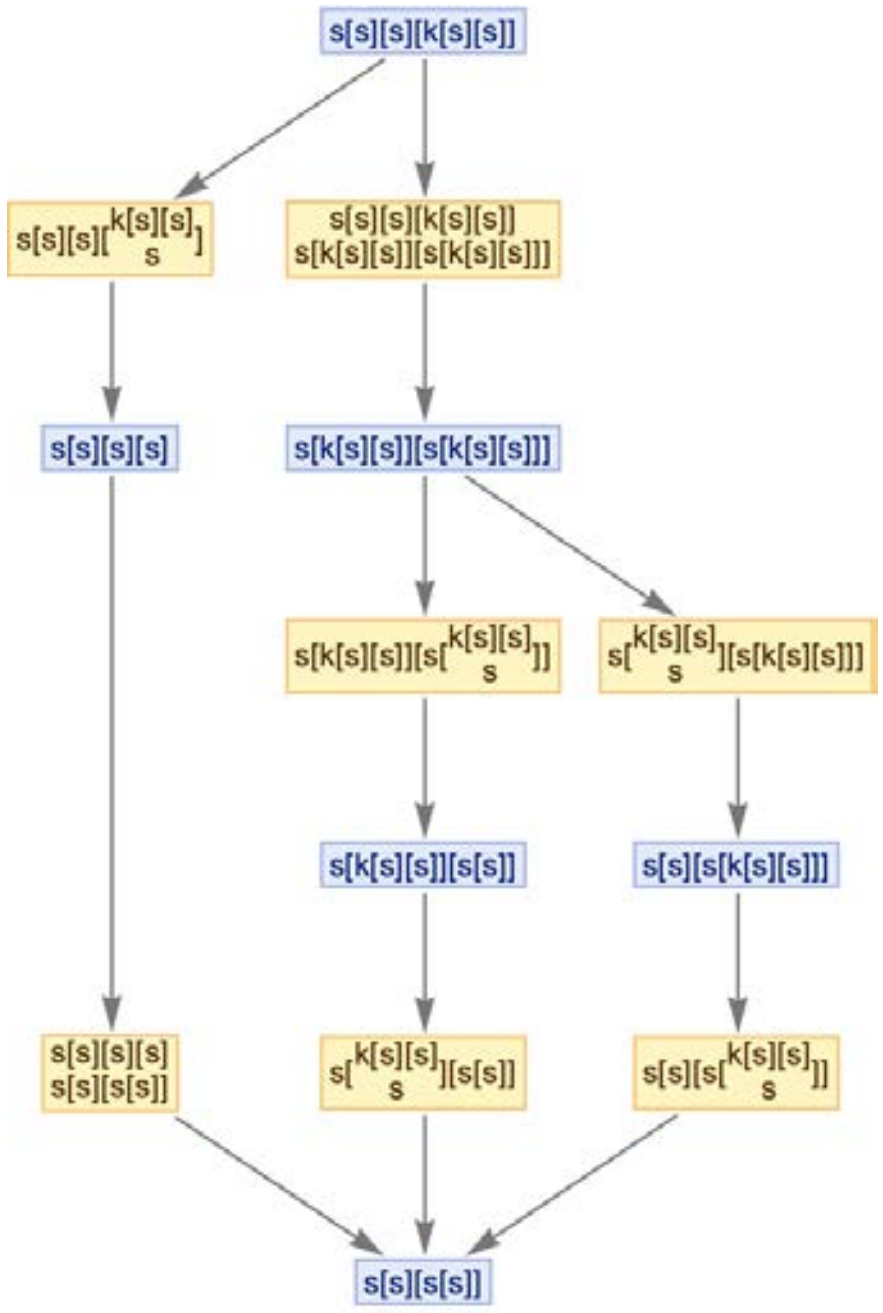

Now each possible sequence of updating events corresponds to a path in the multiway graph. The two particular strategies we used above correspond to these paths:

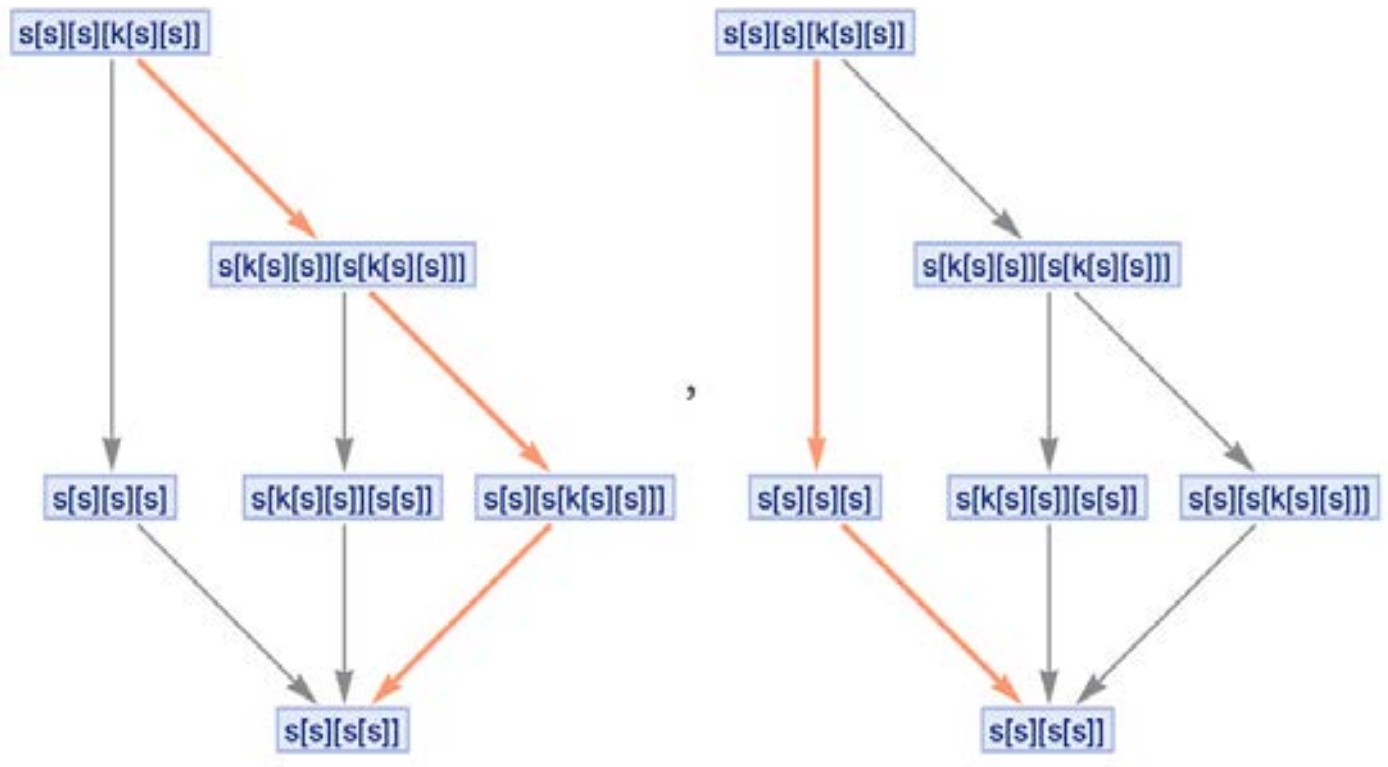

*standard (leftmost outermost)* *depth first (leftmost innermost)*

We see that even at the first step here, there are two possible ways to go. But in addition to branching, there is also merging, and indeed whichever branch one takes, it's inevitable that one will end up at the same final state—in effect the unique "result" of applying the combinator rules.

Here's a slightly more complicated case, where there starts out being a unique path, but then after 4 steps, there's a branch, but after a few more steps, everything converges again to a unique final result:

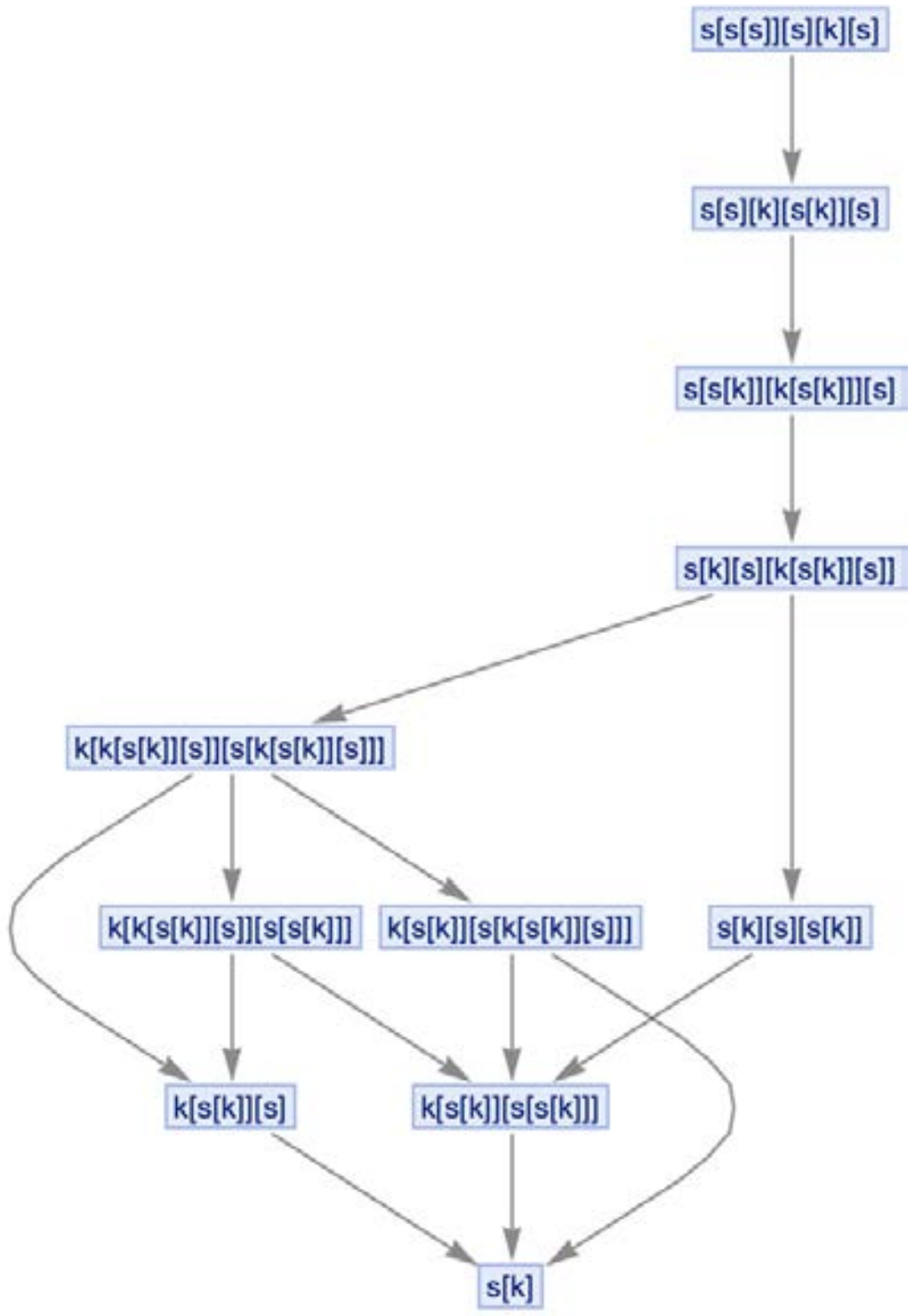

For combinator expressions of size 4, there's never any branching in the multiway graph. At size 5 the multiway graphs that occur are:

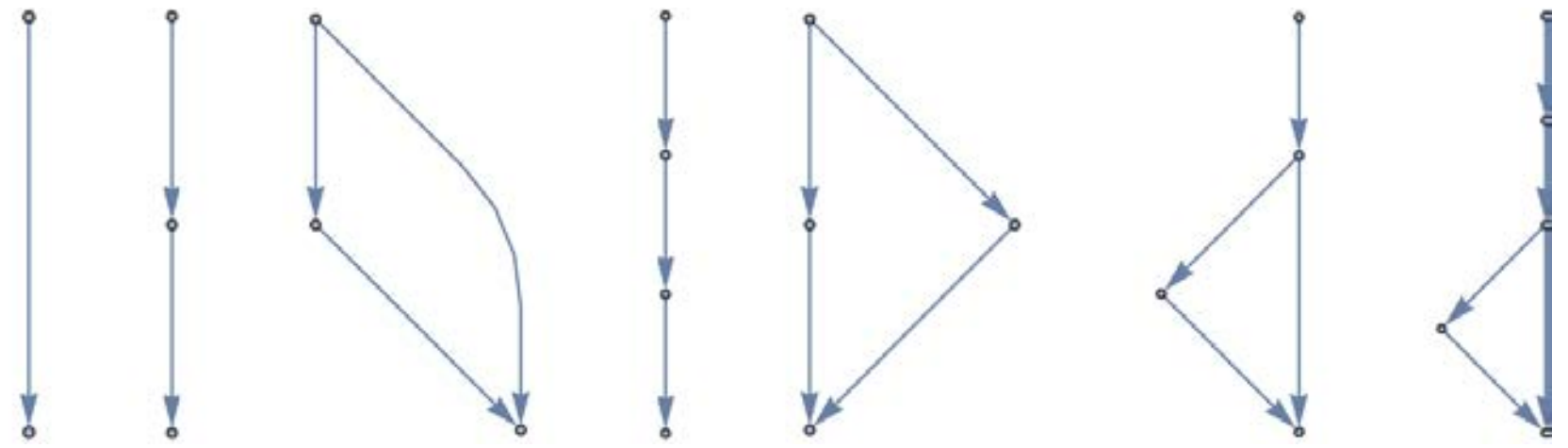

At size 6 the 2688 possible combinator expressions yield the following multiway graphs, with the one shown above being basically as complicated as it gets:

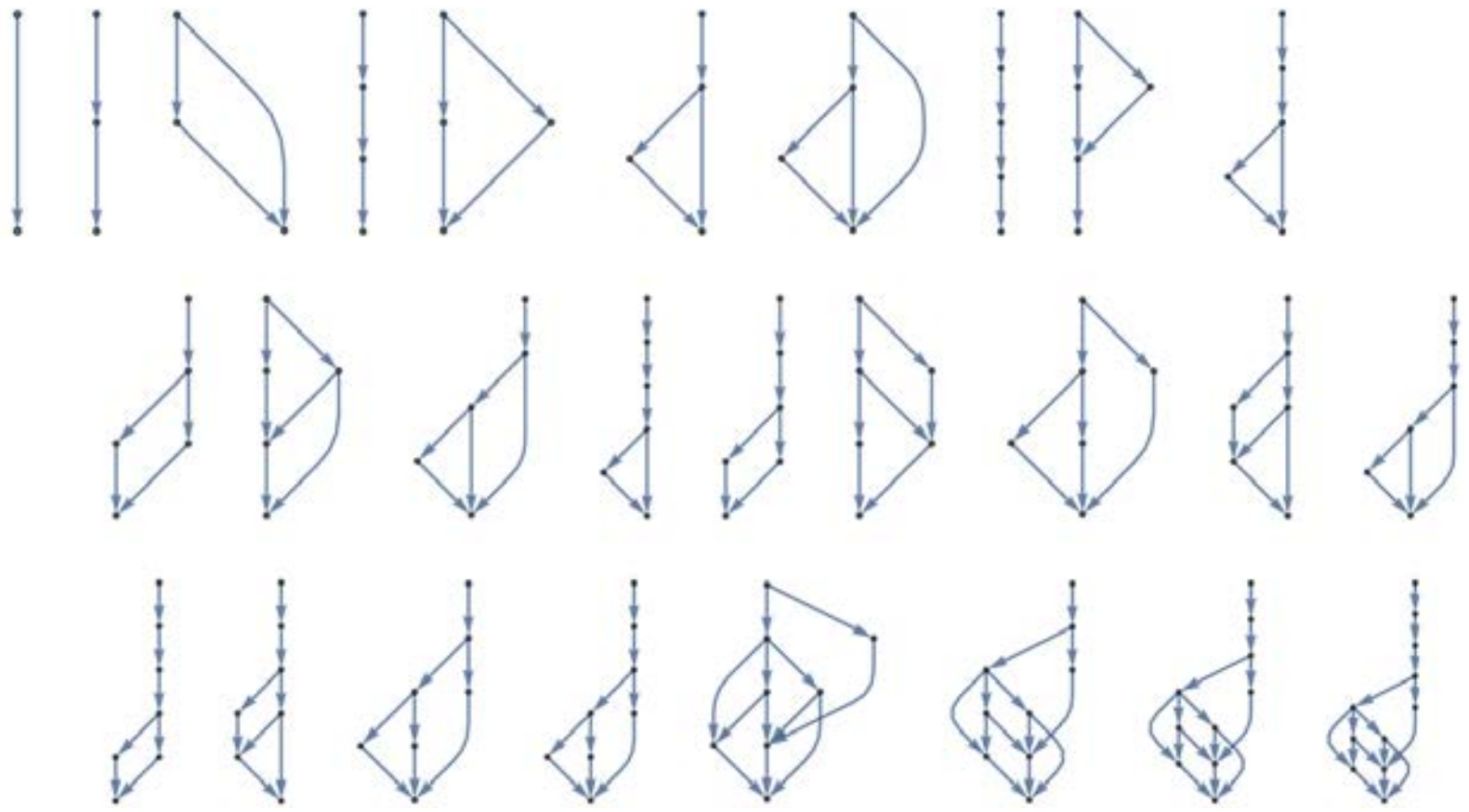

At size 7, much more starts being able to happen. There are rather regular structures like:

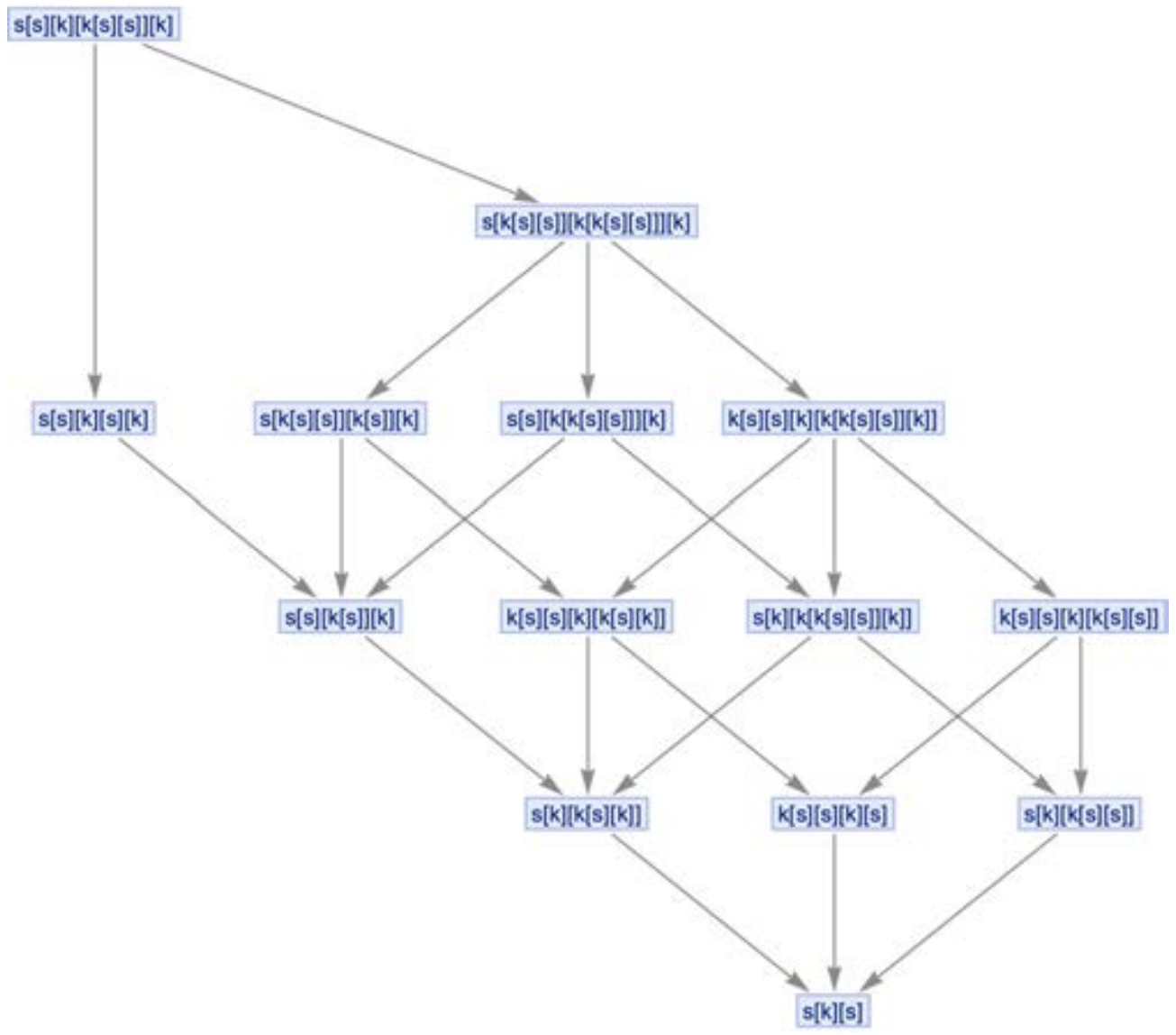

As well as cases like:

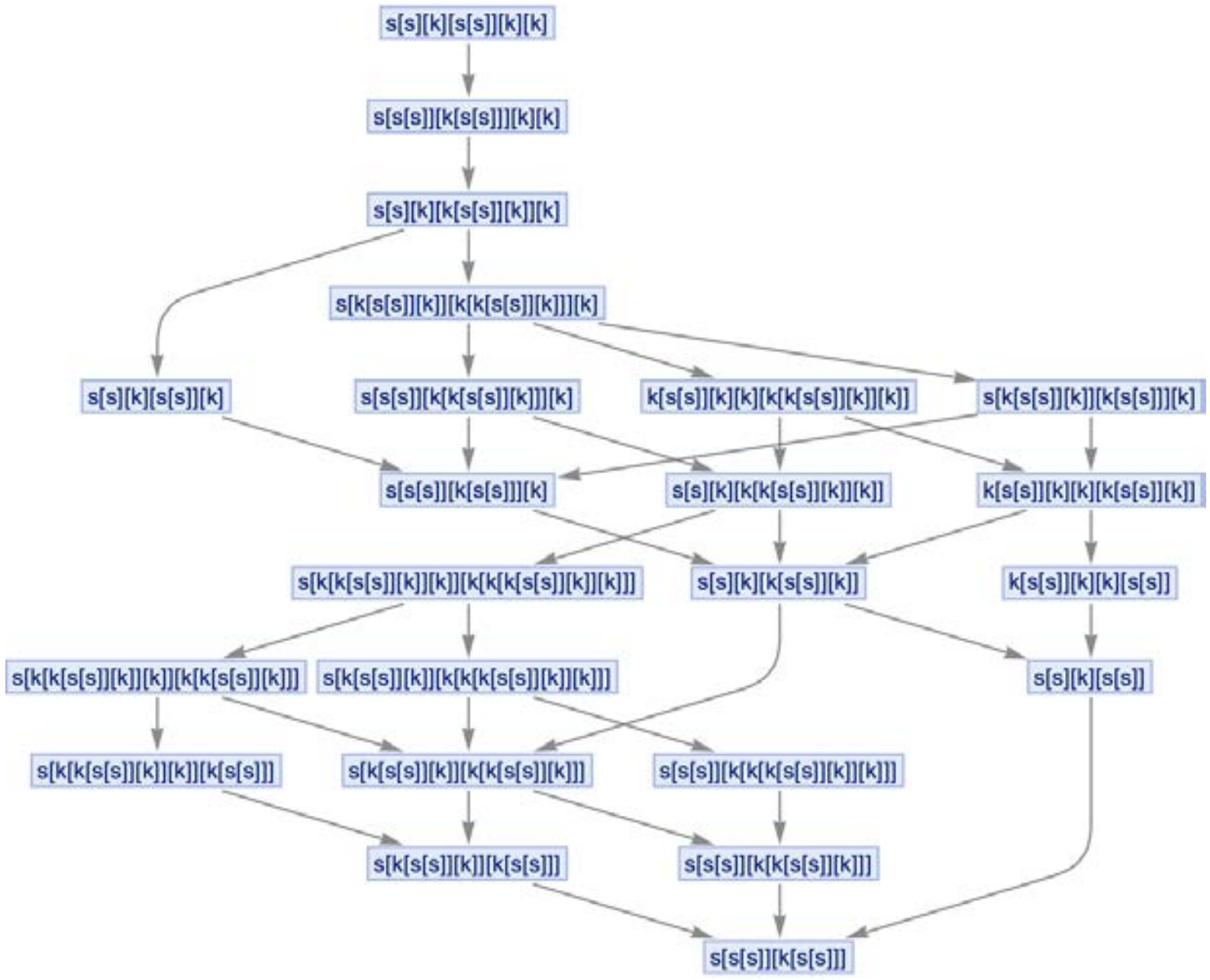

This can be summarized by giving just the size of each intermediate expression, here showing the path defined by our standard leftmost-outermost updating strategy:

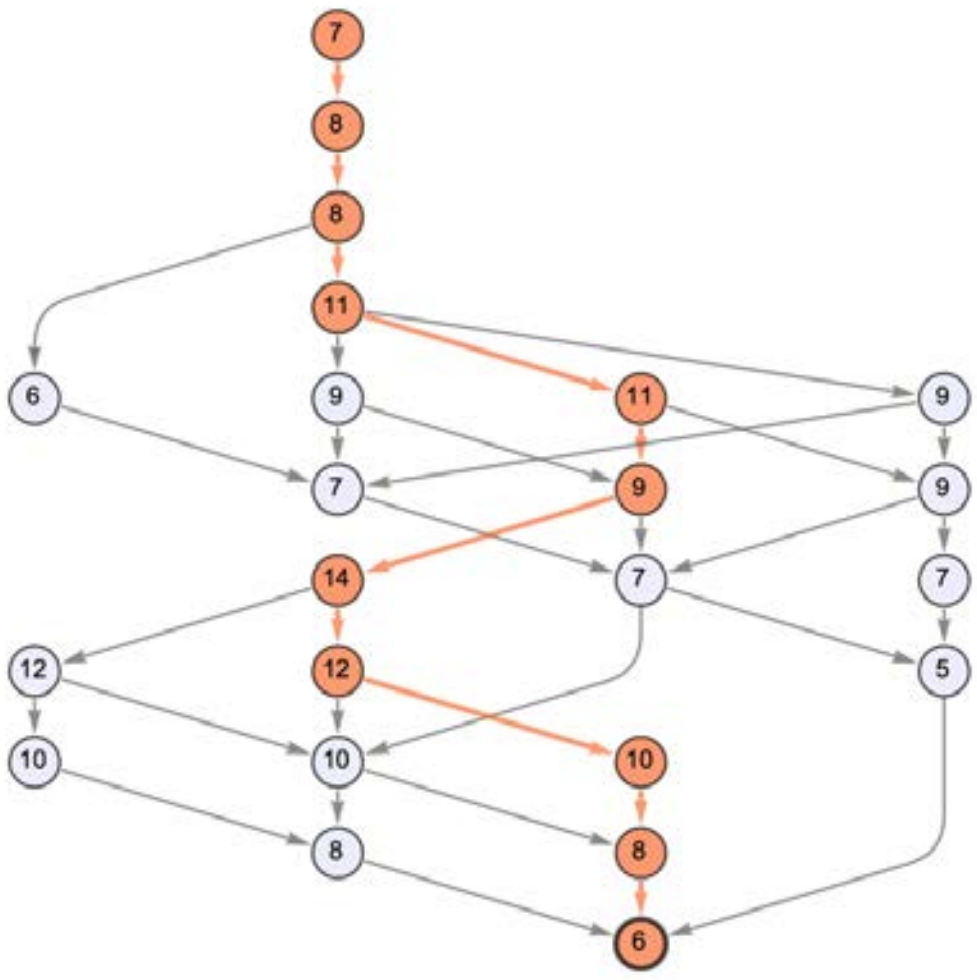

By comparison, here is the path defined by the depth-first strategy above:

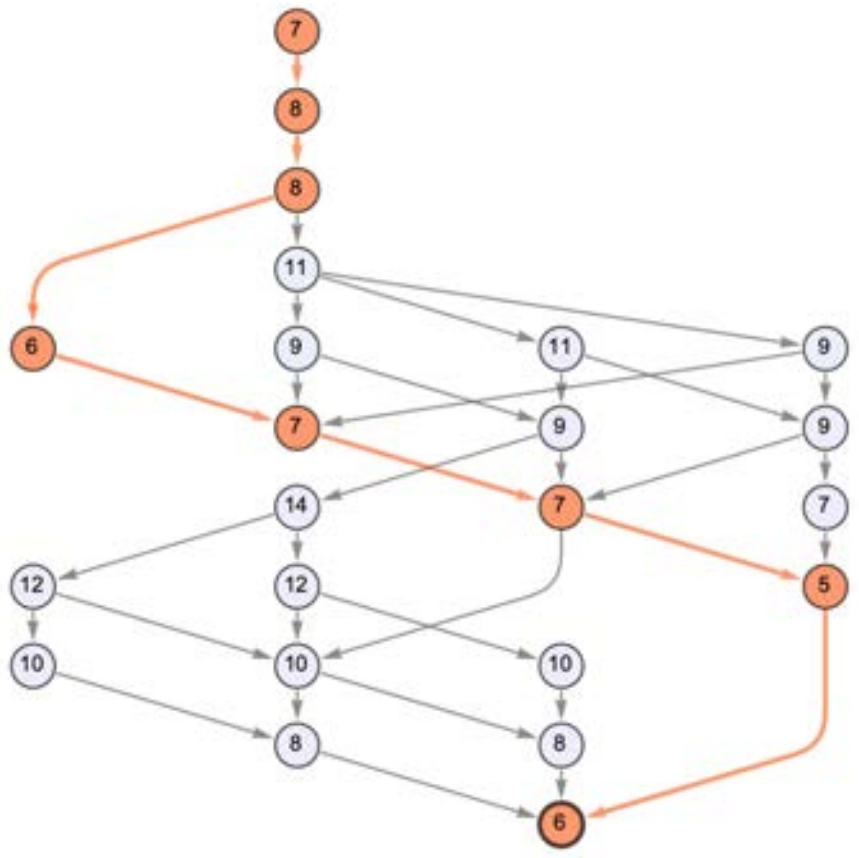

s[s][s][s[s[k]]][k] (**SSS**(**S**(**SK**))**K**) is a case where leftmost outermost-evaluation avoids longer paths and larger intermediate expressions

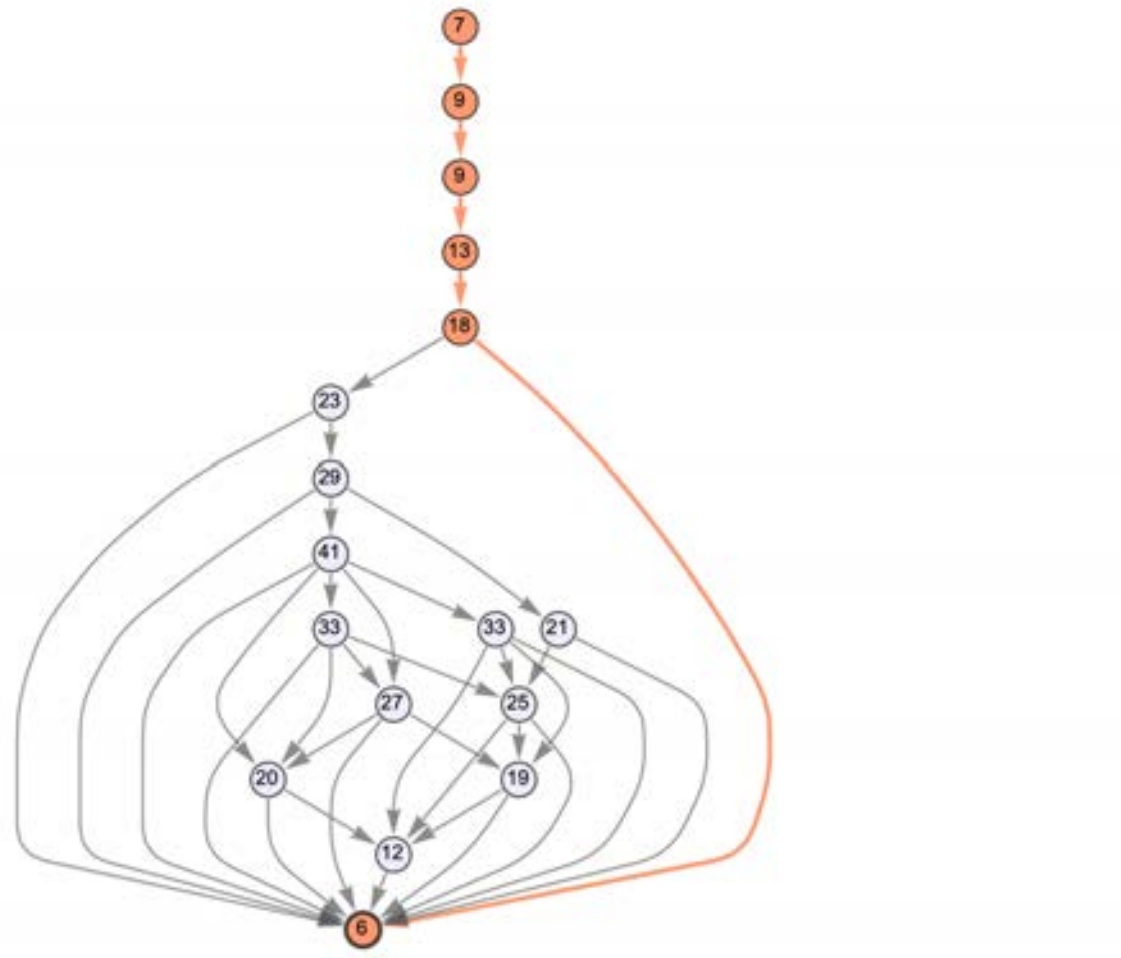

while depth-first evaluation takes more steps:

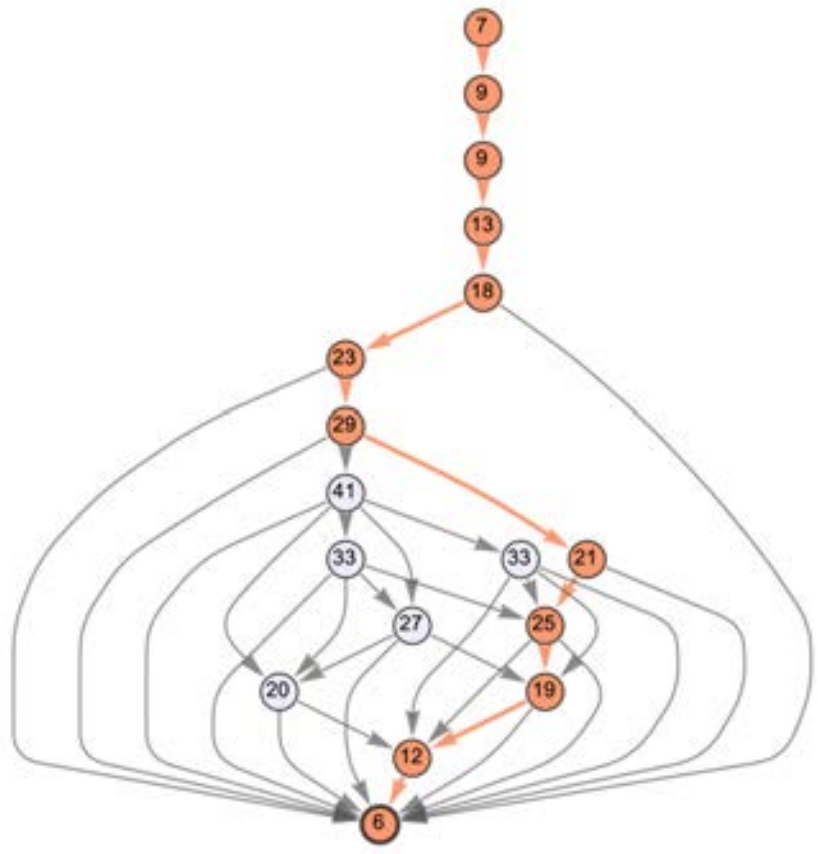

s[s[s]][s][s[s]][s] (**S**(**SS**)**S**(**SS**)**S**) gives a larger but more uniform multiway graph (s[s[s[s]]][s][s][s] evolves directly to s[s[s]][s][s[s]][s]):

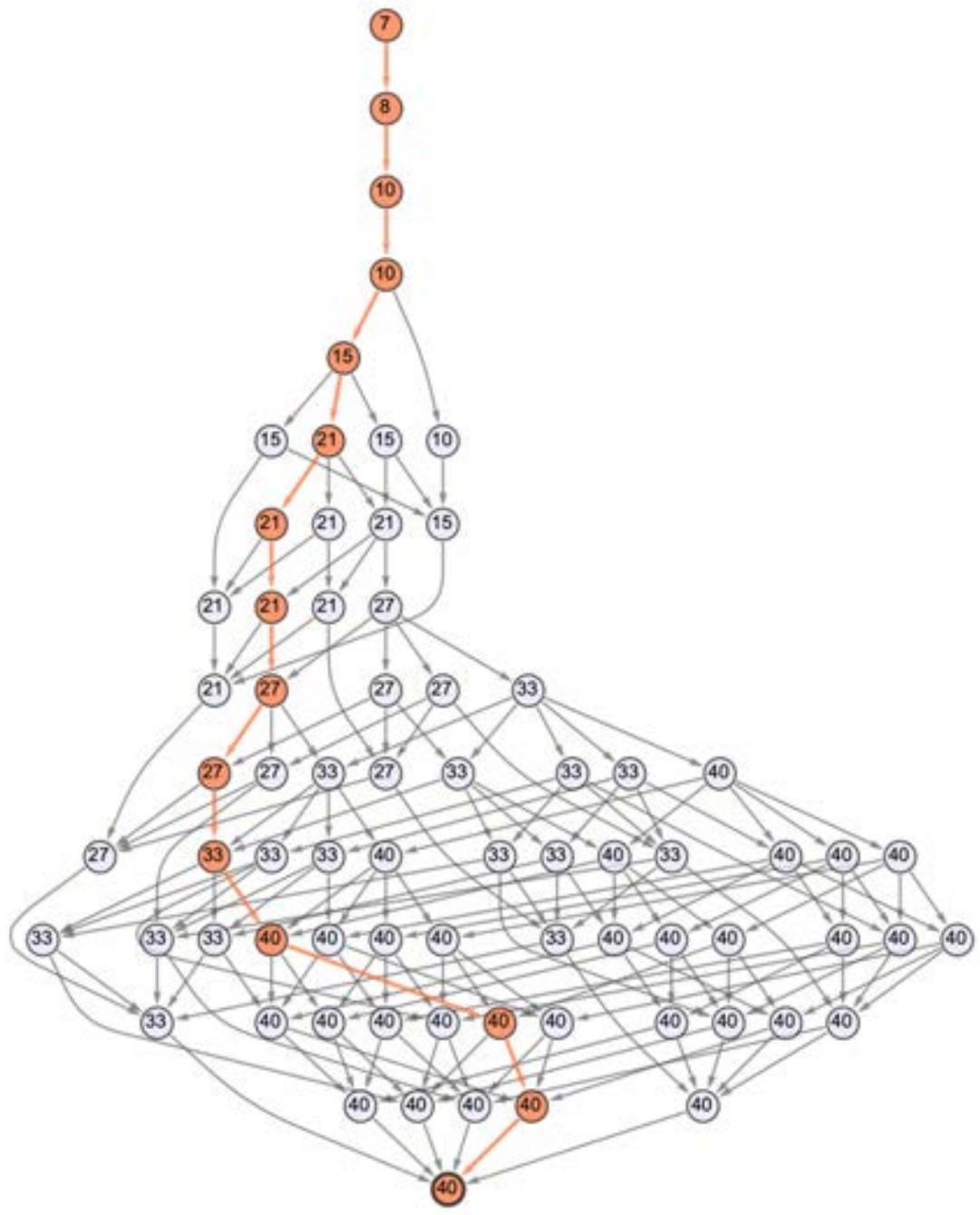

Depth-first evaluation gives a slightly shorter path:

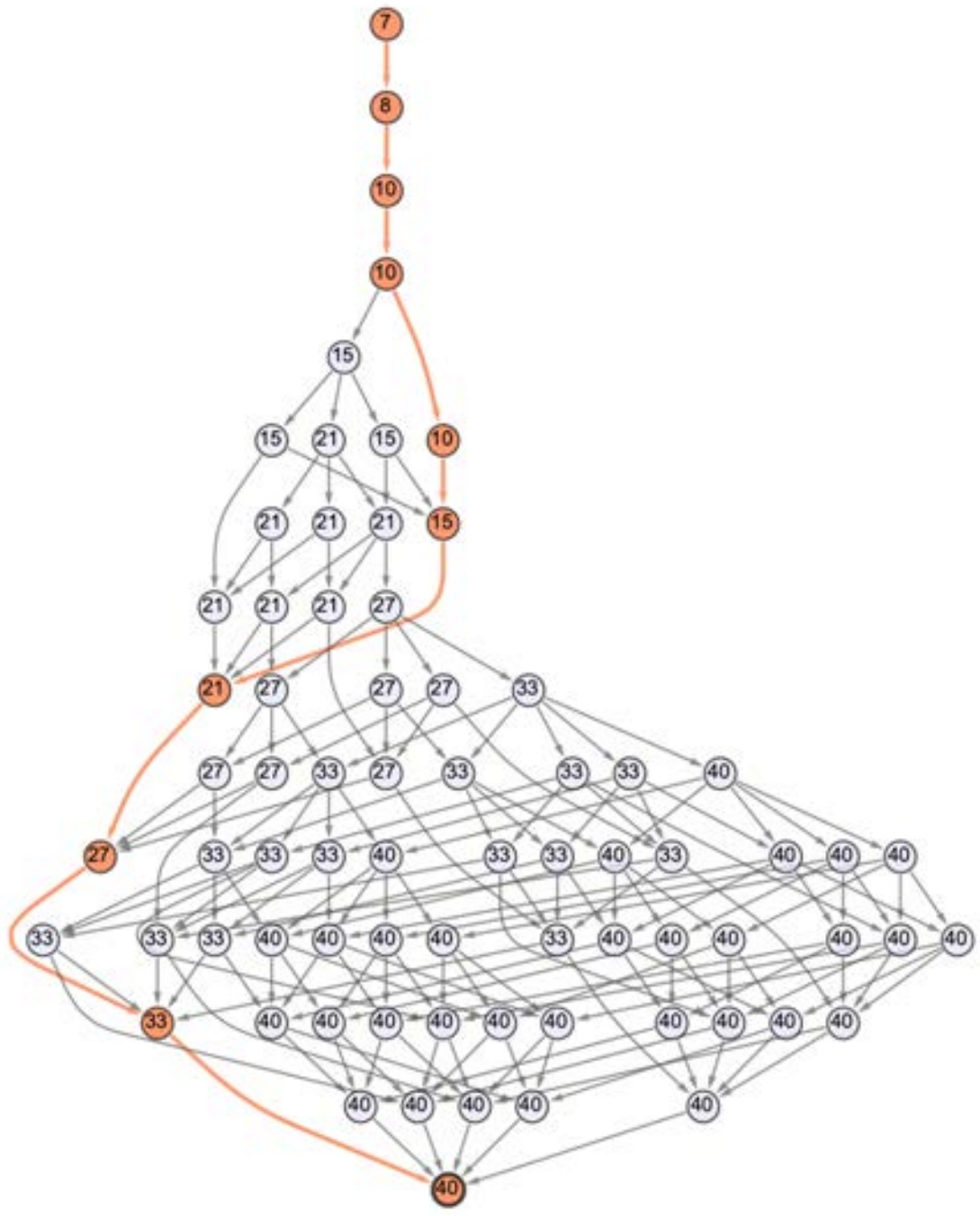

Among size-7 expressions, the largest finite multiway graph (with 94 nodes) is for s[s[s[s]]][s][s][k] (**S**(**S**(**SS**))**SSK**):

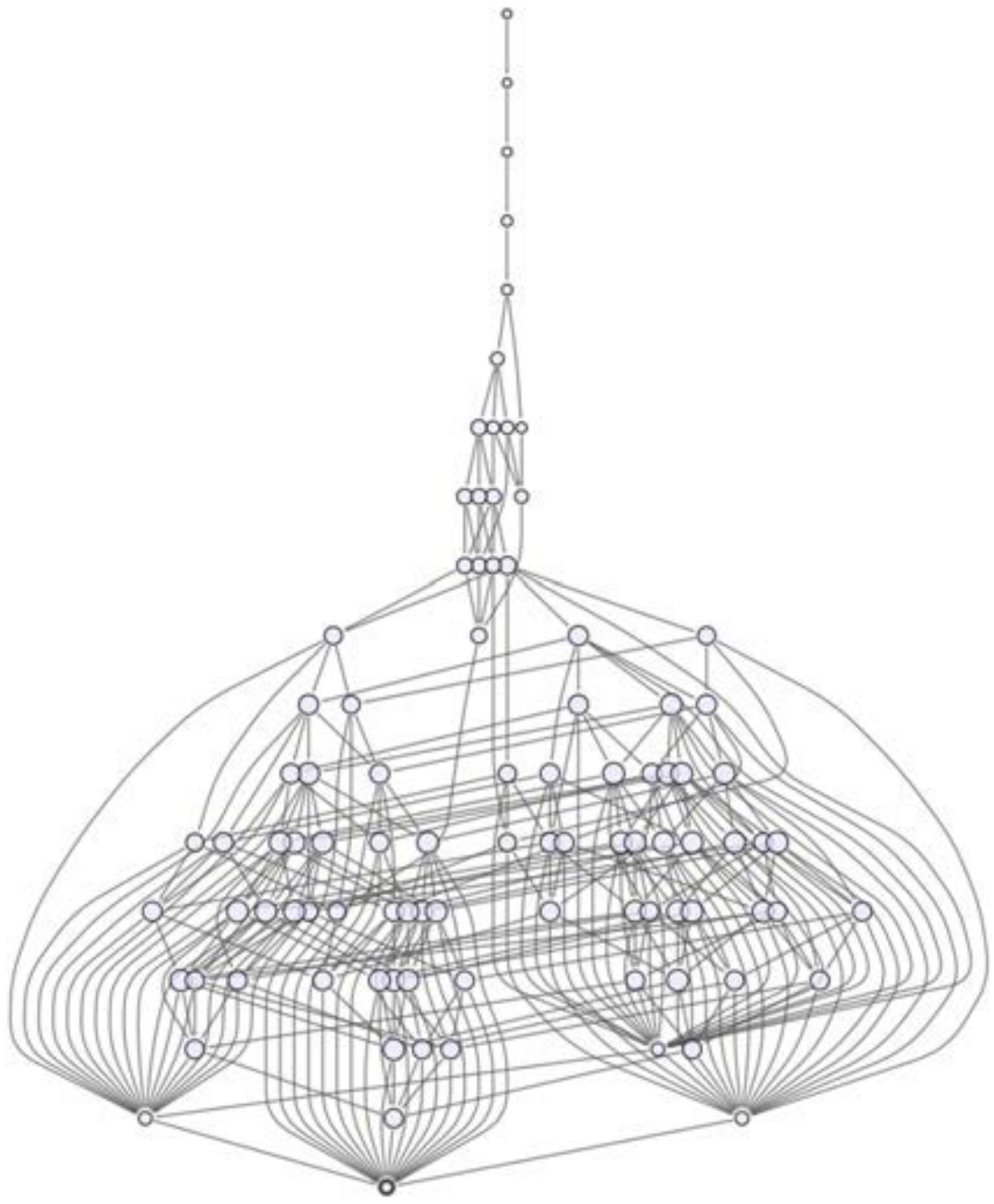

Depending on the path, this can take between 10 and 18 steps to reach its final state:

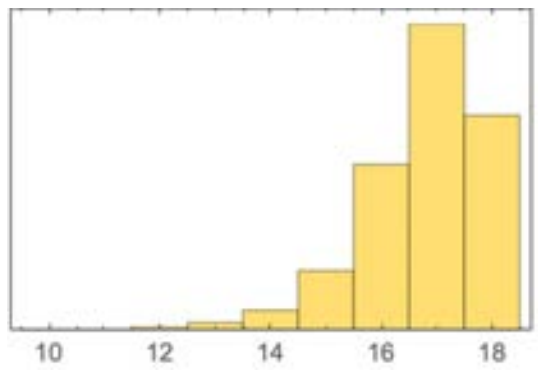

Our standard leftmost outermost strategy takes 12 steps; the depth-first takes 13 steps:

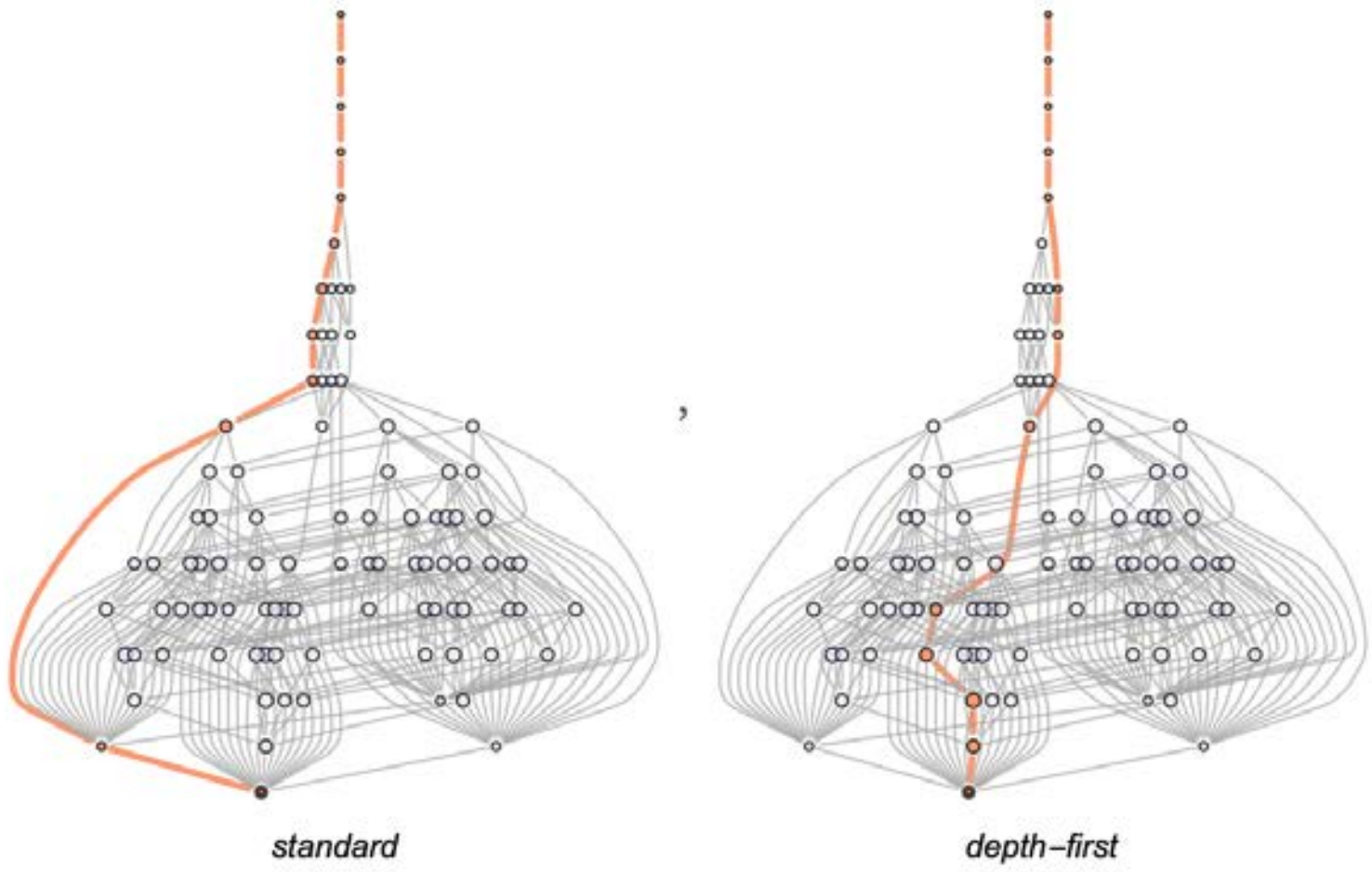

But among size-7 combinator expressions there are basically two that do not lead to finite multiway systems: s[s[s]][s][s][s][k] (**S**(**SS**)**SSSK**) (which evolves immediately to s[s][s][s[s]][s][k]) and s[s[s]][s][s][s][s] (**S**(**SS**)**SSSS**) (which evolves immediately to s[s][s][s[s]][s][s]).

Let's consider s[s[s]][s][s][s][k]. For 8 steps there's a unique path of evolution. But at step 9, the evolution branches

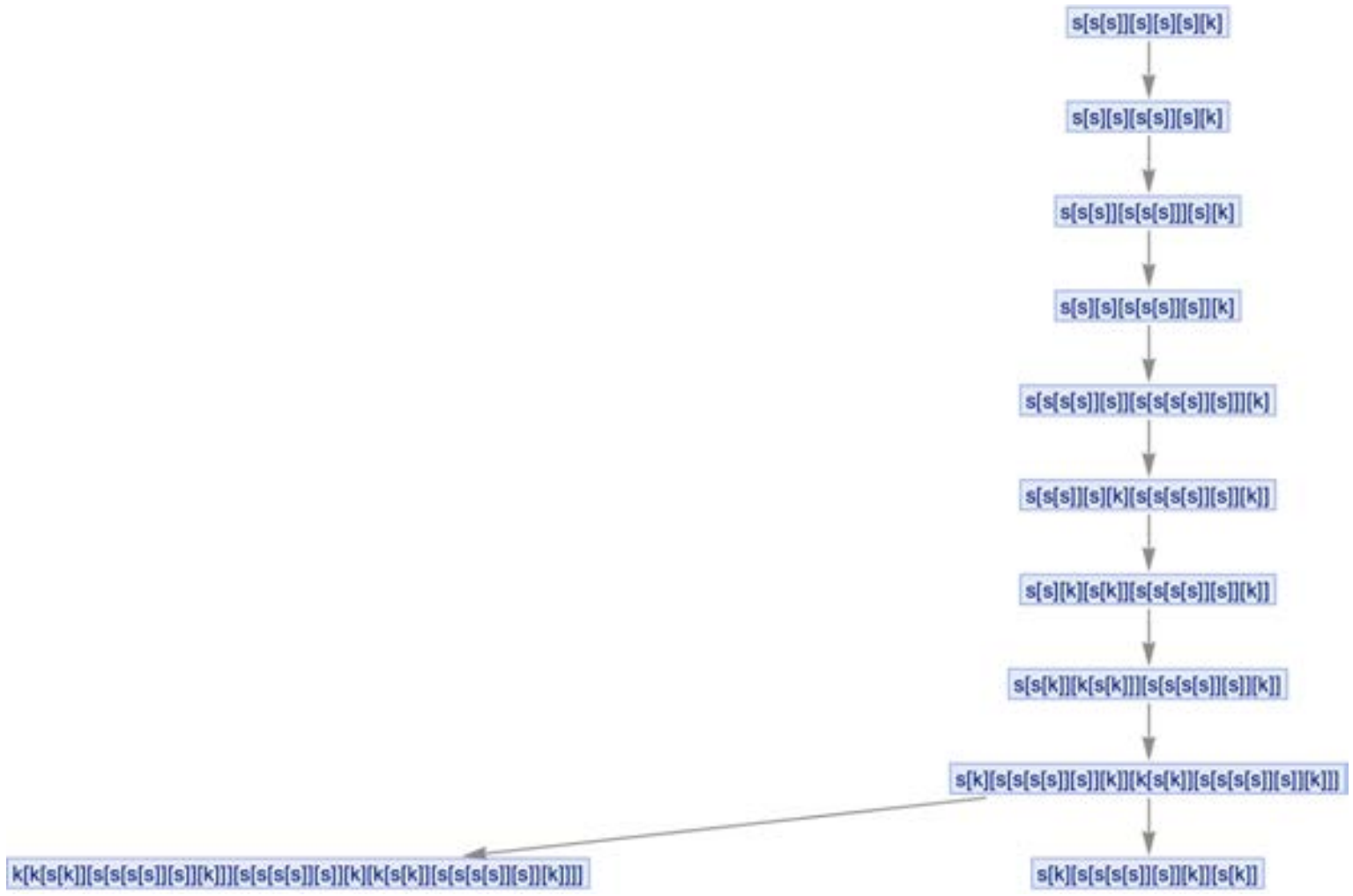

as a result of there being two distinct possible updating events:

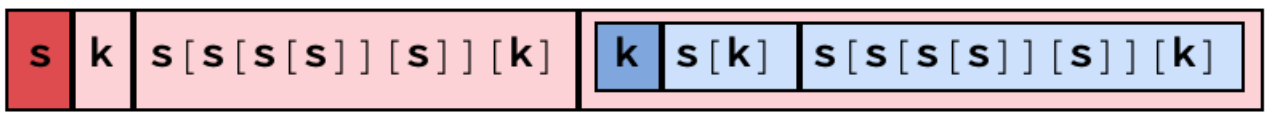

Continuing for 14 steps we get a fairly complex multiway system:

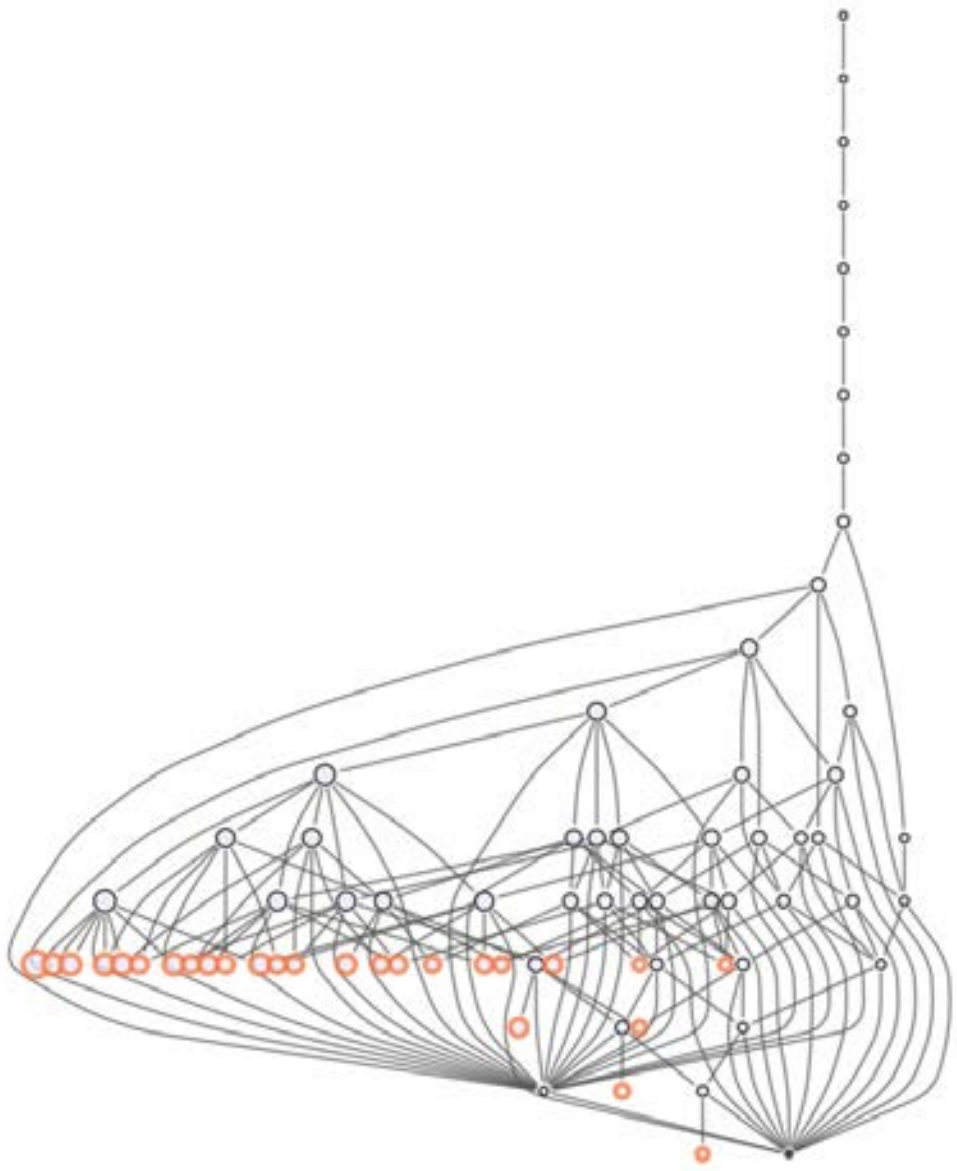

But this isn't "finished"; the nodes circled in red correspond to expressions that are not fixed points, and will evolve further. So what happens with particular evaluation orders?

Here are the results for our two updating schemes:

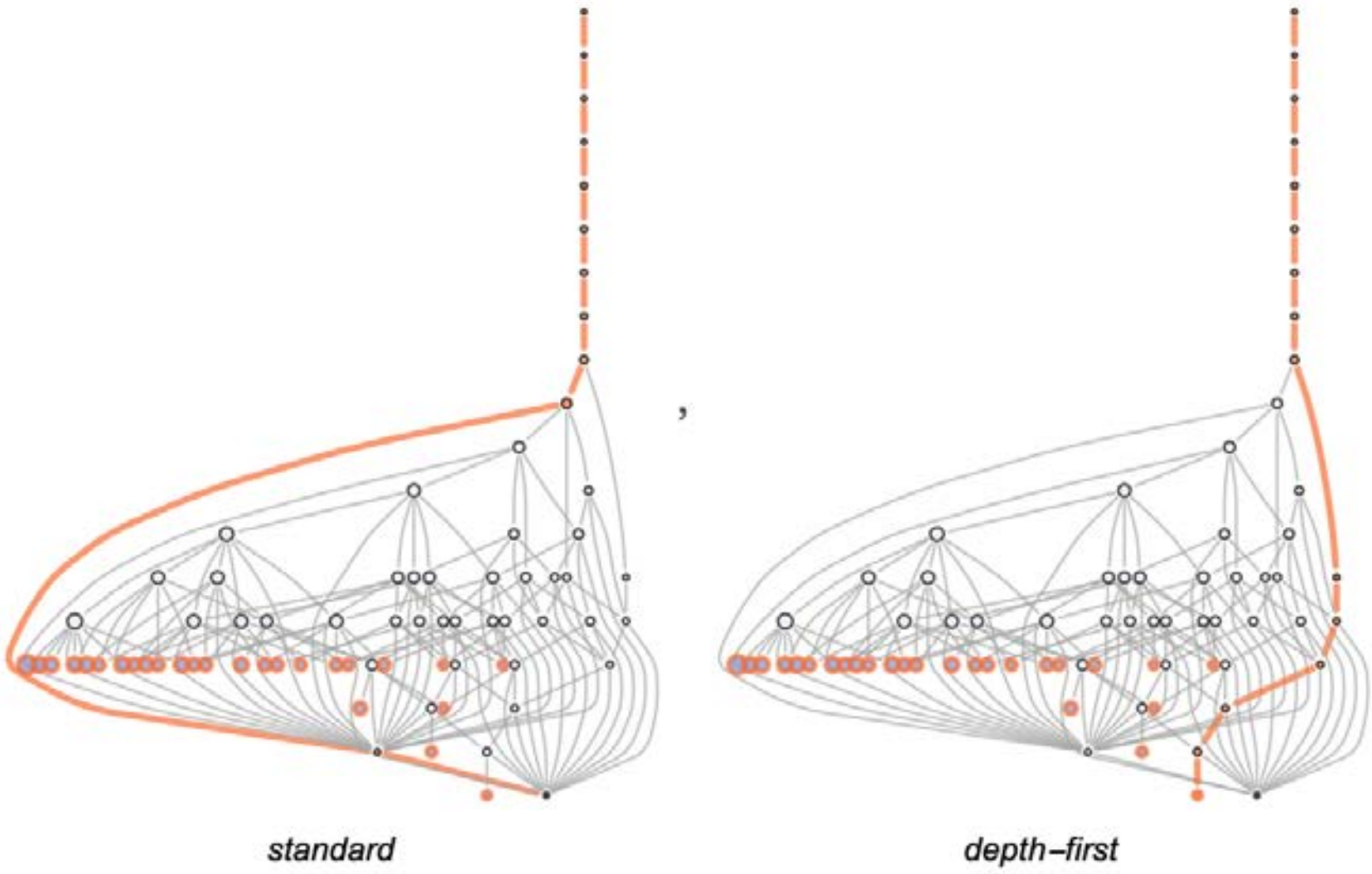

Something important is visible here: the leftmost-outermost path leads (in 12 steps) to a fixed-point node, while the depth-first path goes to a node that will evolve further. In other words, at least as far as we can see in this multiway graph, leftmost-outermost evaluation terminates while depth first does not.

There is just a single fixed point visible (s[k]), but there are many "unfinished paths". What will happen with these? Let's look at depth-first evaluation. Even though it hasn't terminated after 14 steps, it does so after 29 steps—yielding the same final result s[k]:

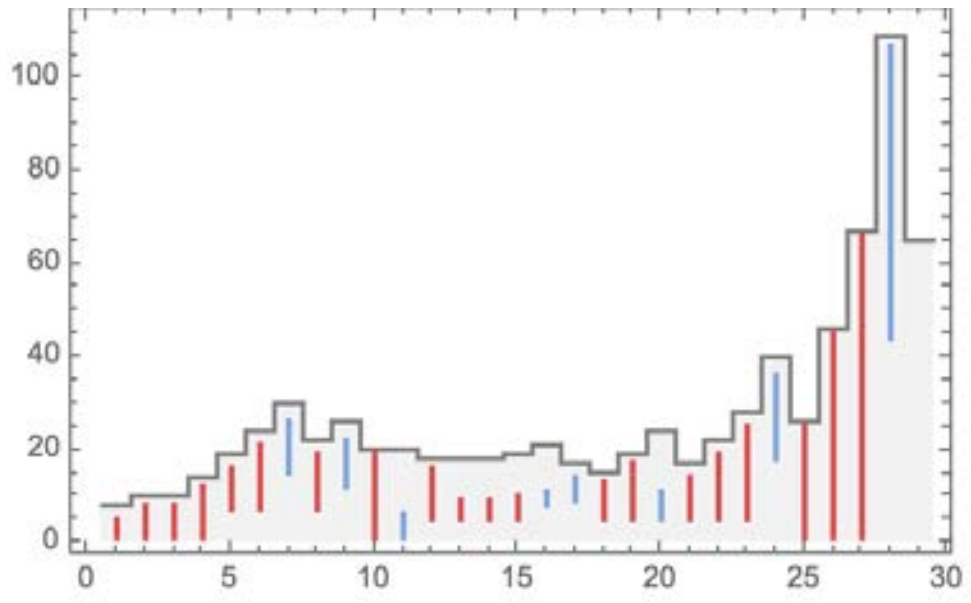

And indeed it turns out to be a general result (known since the 1940s) that if a combinator evolution path is going to terminate, it must terminate in a unique fixed point, but it's also possible that the path won't terminate at all.

Here's what happens after 17 steps. We see more and more paths leading to the fixed point, but we also see an increasing number of "unfinished paths" being generated:

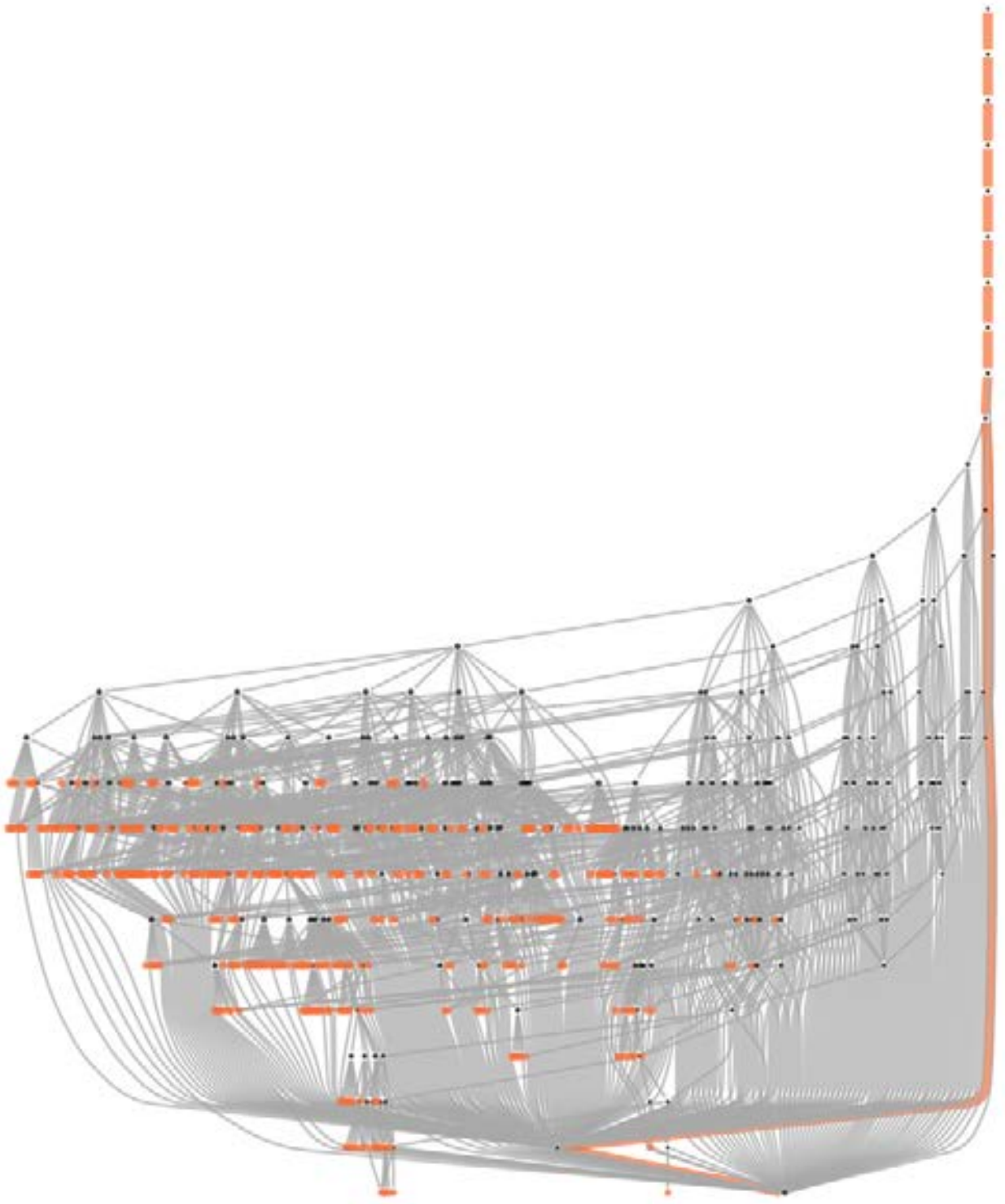

Let's now come back to the other case we mentioned above: s[s[s]][s][s][s][s] (**S**(**SS**)**SSSS**). For 12 steps the evolution is unique:

```
s[s[s]][s][s][s][s]
s[s][s][s[s]][s][s]
s[s[s]][s[s[s]]][s][s]
s[s][s][s[s[s]][s]][s]
s[s[s[s]][s]][s[s[s[s]][s]]][s]
s[s[s]][s][s][s[s[s[s]][s]][s]]
s[s][s][s[s]][s[s[s[s]][s]][s]]
s[s[s]][s[s[s]]][s[s[s[s]][s]][s]]
s[s][s[s[s[s]][s]][s]][s[s[s]][s[s[s[s]][s]][s]]]
s[s[s[s]][s[s[s[s]][s]][s]]][s[s[s[s]][s]][s][s[s[s]][s[s[s[s]][s]][s]]]]
s[s[s[s]][s[s[s[s]][s]][s]]][s[s[s]][s][s[s[s]][s[s[s[s]][s]][s]]][s[s[s[s]][s[s[s[s]][s]][s]]]]]
s[s[s[s]][s[s[s[s]][s]][s]]][s[s][s[s[s]][s[s[s[s]][s]][s]]][s[s[s[s]][s[s[s[s]][s]][s]]]][s[s[s[s]][s[s[s[s]][s]][s]]]]]
s[s[s[s]][s[s[s[s]][s]][s]]][
  s[s[s[s[s]][s[s[s[s]][s]][s]]]][s[s[s]][s[s[s[s]][s]][s]][s[s[s[s]][s[s[s[s]][s]][s]]]]][s[s[s[s]][s[s[s[s]][s]][s]]]]]
```

But at that step there are two possible updating events:

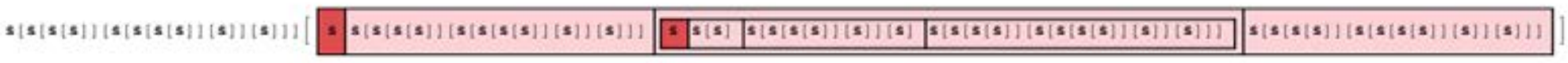

And from there on out, there's rapid growth in the multiway graph:

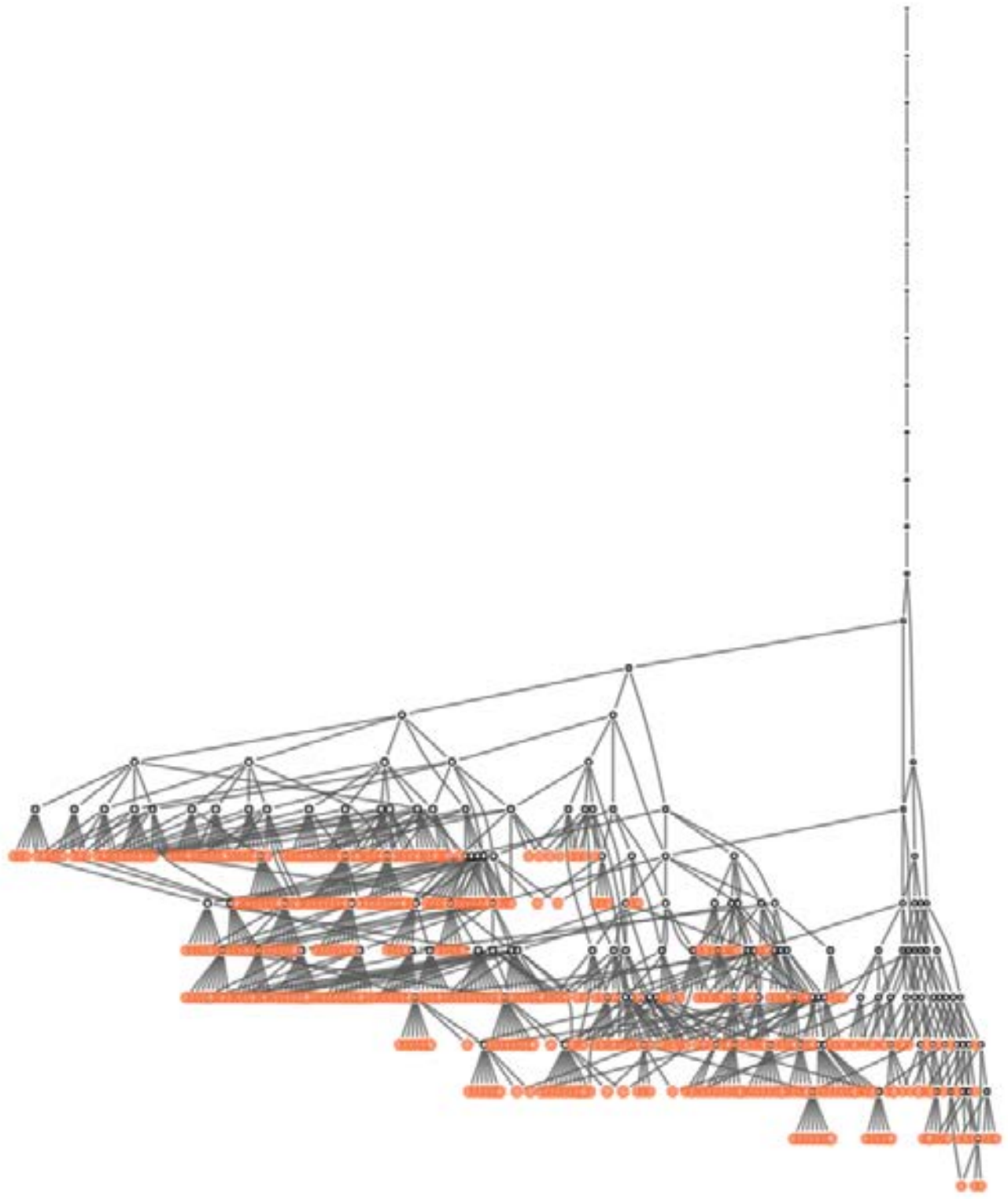

And what's important here is that there are no fixed points: there is no possible evaluation strategy that leads to a fixed point. And what we're seeing here is an example of a general result: if there is a fixed point in a combinator evolution, then leftmost-outermost evaluation will always find it.

In a sense, leftmost-outermost evaluation is the "most conservative" evaluation strategy, with the least propensity for ending up with "runaway evolution". Its "conservatism" is on display if one compares growth from it and from depth-first evaluation in this case:

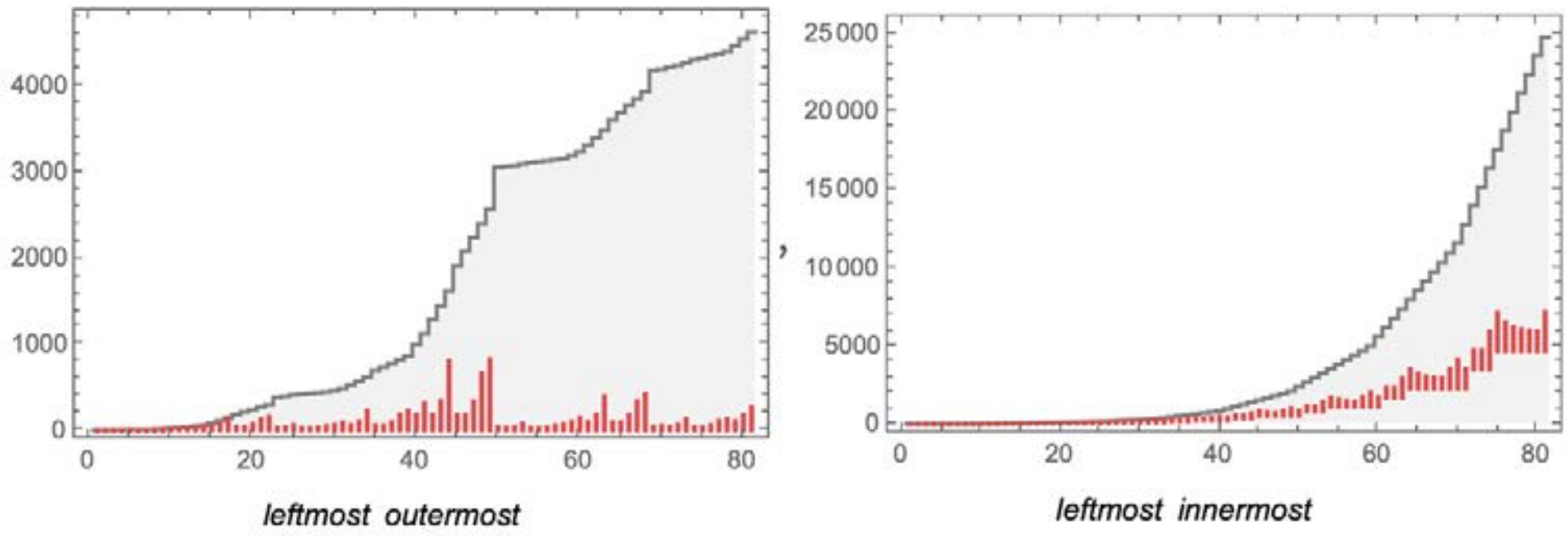

Looking at the multiway graph—as well as others—a notable feature is the presence of "long necks": for many steps every evaluation strategy leads to the same sequence of expressions, and there is just one possible match at each step.

But how long can this go on? For size 8 and below it's always limited (the longest "neck" at size 7 is for s[s[s]][s][s][s][s] and is of length 13; for size 8 it is no longer, but is of length 13 for s[s[s[s]][s][s][s][s]] and k[s[s[s]][s][s][s][s]]). But at size 9 there are four cases (3 distinct) for which growth continues forever, but is always unique:

{s[s[s[s]]][s[s[s][s]]][s], s[s[s[s]]][s[s[s]]][s[s]], s[s[s]][s][s[s[s][s]][s]], s[s[s]][k][s[s[s][s]][s]]}

And as one might expect, all these show rather regular patterns of growth:

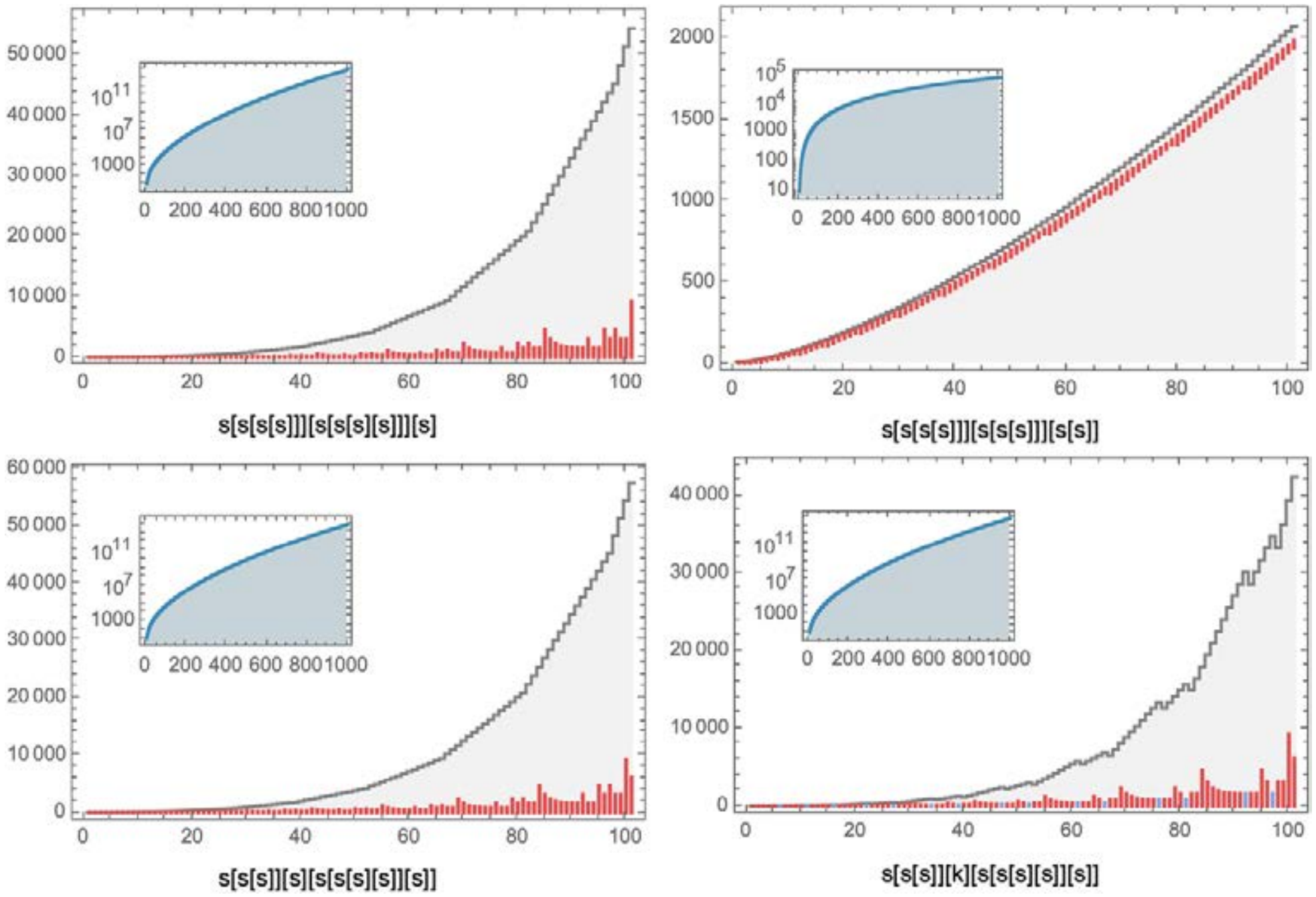

The second differences are given in the first and third cases by repeats of (for successive n):

Join[{0, 0, 1}, Table[0, n], {7, 0, 0, 1, 0, $3\,(2^{n+2}-3)$}]

In the second they are given by repeats of

Join[Table[0, n], {2}]

and in the final case by repeats of

Join[{0, 1}, Table[0, n], {$-3\times 2^{n+3}+18$, $3\times 2^{n+3}-11$, 0, 1, 0, $-3\times 2^{n+3}+2$, $9\times 2^{n+2}-11$}]

## The Question of Evaluation Order

As a computational language designer, it's an issue I've been chasing for 40 years: what's the best way to define the order in which one evaluates (i.e. computes) things? The good news is that in a well-designed language (like the Wolfram Language!) it fundamentally doesn't matter, at least much of the time. But in thinking about combinators—and the way they

evolve—evaluation order suddenly becomes a central issue. And in fact it's also a central issue in our new model of physics—where it corresponds to the choice of reference frame, for relativity, quantum mechanics and beyond.

Let's talk first about evaluation order as it shows up in the symbolic structure of the Wolfram Language. Imagine you're doing this computation:

```
Length[Join[{a, b}, {c, d, e}]]
```

```
5
```

The result is unsurprising. But what's actually going on here? Well, first you're computing **Join[...]**:

```
Join[{a, b}, {c, d, e}]
```

```
{a, b, c, d, e}
```

Then you're taking the result, and providing it as argument to **Length**, which then does its job, and gives the result 5. And in general in the Wolfram Language, if you're computing f[g[x]] what'll happen is that x will be evaluated first, followed by g[x], and finally f[g[x]]. (Actually, the head f in f[x] is the very first thing evaluated, and in f[x, y] one evaluates f, then x, then y and then f[x, y].)

And usually this is exactly what one wants, and what people implicitly expect. But there are cases where it isn't. For example, let's say you've defined x = 1 (i.e. **Set**[x, 1]). Now you want to say x = 2 (**Set**[x,2]). If the x evaluated first, you'd get **Set**[1,2], which doesn't make any sense. Instead, you want **Set** to "hold its first argument", and "consume it" without first evaluating it. And in the Wolfram Language this happens automatically because **Set** has attribute **HoldFirst**.

How is this relevant to combinators? Well, basically, the standard evaluation order used by the Wolfram Language is like the depth-first (leftmost-innermost) scheme we described above, while what happens when functions have **Hold** attributes is like the leftmost-outermost scheme.

But, OK, so if we have something like f[a[x],y] we usually first evaluate a[x], then use the result to compute f[a[x],y]. And that's pretty easy to understand if a[x], say, immediately, evaluates to something like 4 that doesn't itself need to be evaluated. But what happens when in f[a[x],y], a[x] evaluates to b[x] which then evaluates to c[x] and so on? Do you do the complete chain of "subevaluations" before you "come back up" to evaluate y, and f[...]?

What's the analog of this for combinators? Basically it's whether when you do an update based on a particular match in a combinator expression, you then just keep on "updating the update", or whether instead you go on and find the next match in the expression before doing anything with the result of the update. The "updating the update" scheme is basically what we've called our depth-first scheme, and it's essentially what the Wolfram Language does in its automatic evaluation process.

Imagine we give the combinator rules as Wolfram Language assignments:

```
s[x_][y_][z_] := x[z][y[z]]
```

```
k[x_][y_] := x
```

Then—by virtue of the standard evaluation process in the Wolfram Language—every time we enter a combinator expression these rules will automatically be repeatedly applied, until a fixed point is reached:

```
s[s][s][s[s[s]]][k][s]
```

```
s[s[s[s[s[s[s[s]]][k]][k[s[s[s[s[s]]][k]]]]][s[s[s[s[s]]][k]]]][k[s[s[s[s[s[s[s]]][k]][k[s[s[s[s[s]]][k]]]]][s[s[s[s[s]]][k]]]]]][
  s[s[s[s[s[s[s]]][k]][k[s[s[s[s[s]]][k]]]]][s[s[s[s[s]]][k]]]]]
```

What exactly is happening "inside" here? If we trace it in a simpler case, we can see that there is repeated evaluation, with a depth-first (AKA leftmost-innermost) scheme for deciding what to evaluate:

```
Dataset[Trace[s[k[k][k]][s][s]]]
```

| | | | | | |
|---|---|---|---|---|---|
| k[k][k] | s[k] | s[k][s] | s[k][s][s] | k[s][s[s]] | s |
| k | | | | | |

Of course, given the assignment above for s, if one enters a combinator expression—like s[s][s][s[s]][s][s]]—whose evaluation doesn't terminate, there'll be trouble, much as if we define x=x+1 (or x={x}) and ask for x. Back when I was first doing language design people often told me that issues like this meant that a language that used automatic infinite evaluation "just couldn't work". But 40+ years later I think I can say with confidence that "programming with infinite evaluation, assuming fixed points" works just great in practice—and in rare cases where there isn't going to be a fixed point one has to do something more careful anyway.

In the Wolfram Language, that's all about specifically applying rules, rather than just having it happen automatically. Let's say we clear our assignments for s and k:

```
Clear[s, k]
```

Now no transformations associated with s and k will automatically be made:

```
s[s][s][s[s[s]]][k][s]
```

```
s[s][s][s[s[s]]][k][s]
```

But by using /. (ReplaceAll) we can ask that the s, k transformation rules be applied once:

```
s[s][s][s[s[s]]][k][s] /. {s[x_][y_][z_] → x[z][y[z]], k[x_][y_] → x}
```

```
s[s[s[s]]][s[s[s[s]]]][k][s]
```

With FixedPointList we can go on applying the rule until we reach a fixed point:

```
In[•]:= FixedPointList[# /. {s[x_][y_][z_] → x[z][y[z]], k[x_][y_] → x} &, s[s][s][s[s[s]]][k][s]]

Out[•]= {s[s][s][s[s[s]]][k][s],
  s[s[s[s]]][s[s[s[s]]]][k][s], s[s[s]][k][s[s[s[s]]][k]][s],
  s[s][s[s[s[s]]][k]][k[s[s[s[s]]][k]]][s],
  s[k[s[s[s[s]]][k]]][s[s[s[s]]][k][k[s[s[s[s]]][k]]]][s],
  ⋮
   s[s[s[s[s[s[s]]][k]][k[s[s[s[s[s]]][k]]]]][s[s[s[s[s]]][k]]]]],
  s[s[s[s[s[s[s]]][k]][k[s[s[s[s[s]]][k]]]]][s[s[s[s[s]]][k]]]][k[s[
       s[s[s[s[s[s]]][k]][k[s[s[s[s[s]]][k]]]]][s[s[s[s[s]]][k]]]]]]][
   s[s[s[s[s[s[s]]][k]][k[s[s[s[s[s]]][k]]]]][s[s[s[s[s]]][k]]]]]}
```

It takes 26 steps—which is different from the 89 steps for our leftmost-outermost evaluation, or the 29 steps for leftmost-innermost (depth-first) evaluation. And, yes, the difference is the result of /. in effect applying rules on the basis of a different scheme than the ones we've considered so far.

But, OK, so how can we parametrize possible schemes? Let's go back to the combinator expression from the beginning of the previous section:

```
s[s[s][s][s[s][k[k][s]][s]]][s][s][s[k[s][k]][k][s]]
```

Here are the positions of possible matches in this expression:

```
Position[s[s[s][s][s[s][k[k][s]][s]]][s][s][s[k[s][k]][k][s]], s[_][_][_] | k[_][_]]

{{0, 0, 0, 1, 1, 0, 1}, {0, 0, 0, 1, 1}, {0, 0, 0, 1}, {0}, {1, 0, 0, 1}, {1}}
```

An evaluation scheme must define a way to say which of these matches to actually do at each step. In general we can apply pretty much any algorithm to determine this. But a convenient approach is to think about sorting the list of positions by particular criteria, and then for example using the first k positions in the result.

Given a list of positions, there are two obvious potential types of sorting criteria to use: ones based on the lengths of the position specifications, and ones based on their contents. For example, we might choose (as Sort by default does) to sort shorter position specifications first:

```
Sort[{{0, 0, 0, 1, 1, 0, 1}, {0, 0, 0, 1, 1}, {0, 0, 0, 1}, {0}, {1, 0, 0, 1}, {1}}]

{{0}, {1}, {0, 0, 0, 1}, {1, 0, 0, 1}, {0, 0, 0, 1, 1}, {0, 0, 0, 1, 1, 0, 1}}
```

But what do the shorter position specifications correspond to? They're the more "outer" parts of the combinator expression, higher on the tree. And when we say we're using an "outermost" evaluation scheme, what we mean is that we're considering matches higher on the tree first.

Given two position specifications of the same length, we then need a way to compare these. An obvious one is lexicographic—with 0 sorted before 1. And this corresponds to taking f before x in f[x], or taking the leftmost object first.

We have to decide whether to sort first by length and then by content, or the other way around. But if we enumerate all choices, here's what we get:

| | | | | | | |
|---|---|---|---|---|---|---|
| *leftmost outermost* | 0 | 0001 | 00011 | 0001101 | 1 | 1001 |
| *leftmost innermost* | 0001101 | 00011 | 0001 | 0 | 1001 | 1 |
| *rightmost outermost* | 1 | 1001 | 0 | 0001 | 00011 | 0001101 |
| *rightmost innermost* | 1001 | 1 | 0001101 | 00011 | 0001 | 0 |
| *outermost leftmost* | 0 | 1 | 0001 | 1001 | 00011 | 0001101 |
| *outermost rightmost* | 1 | 0 | 1001 | 0001 | 00011 | 0001101 |
| *innermost leftmost* | 0001101 | 00011 | 0001 | 1001 | 0 | 1 |
| *innermost rightmost* | 0001101 | 00011 | 1001 | 0001 | 1 | 0 |

And here's where the first match with each scheme occurs in the expression tree:

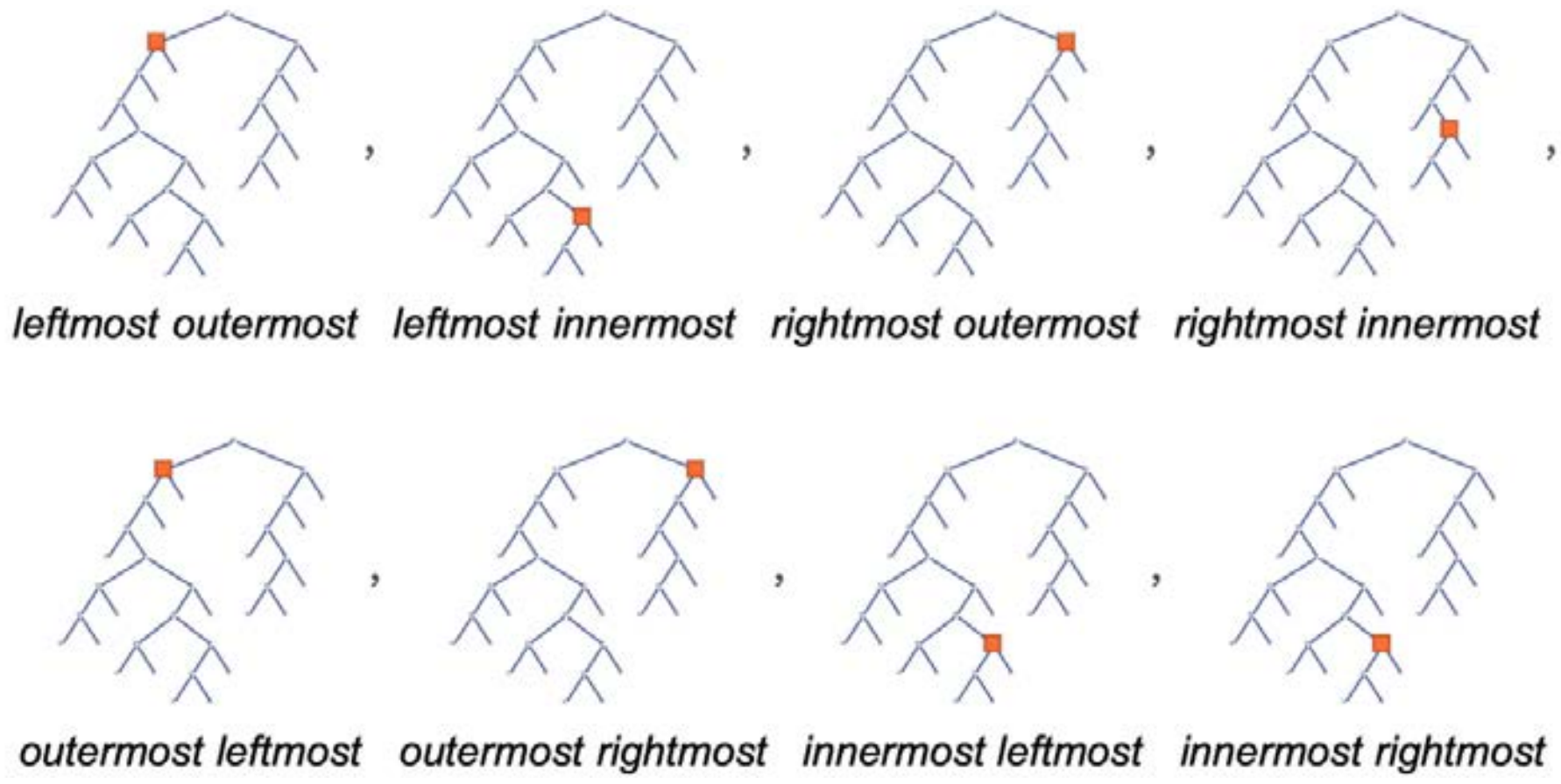

So what happens if we use these schemes in our combinator evolution? Here's the result for the terminating example s[s][s][s[s[s]]][k][s] above, always keeping only the first match with a given sorting criterion, and at each step showing where the matches were applied:

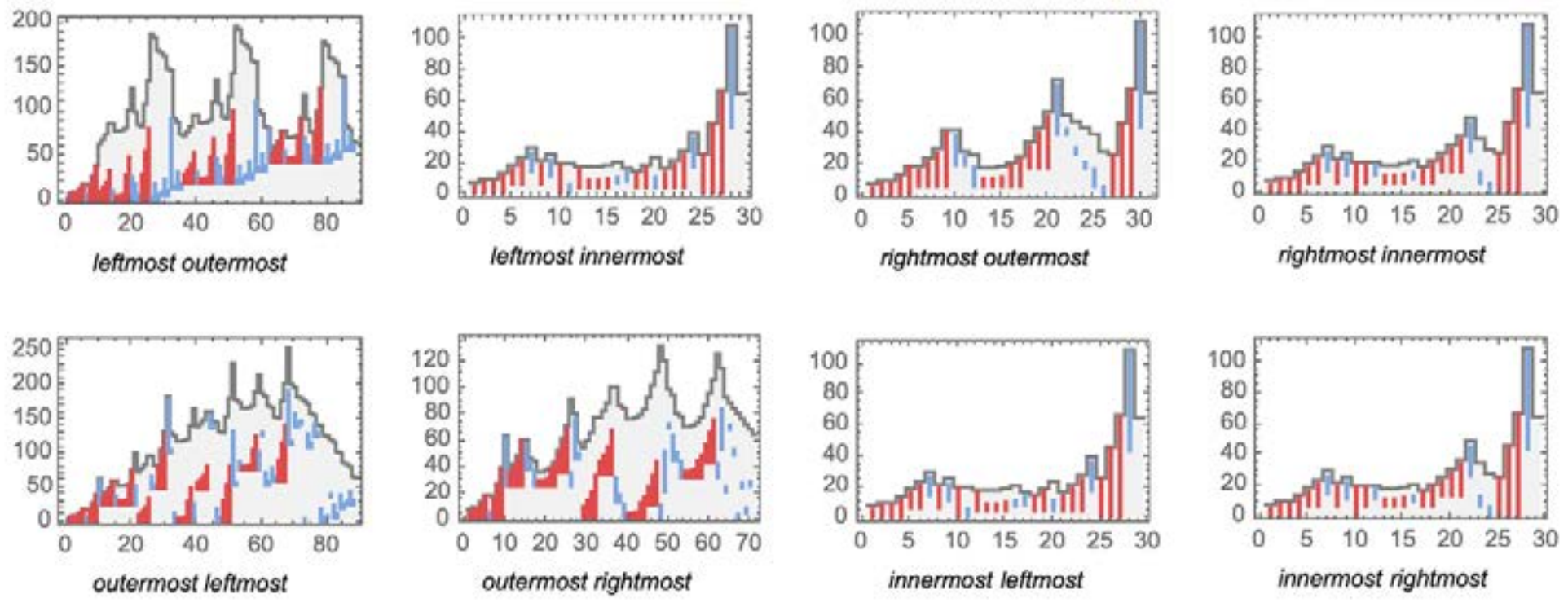

Here now are the results if we allow the first up to 2 matches from each sorted list to be applied:

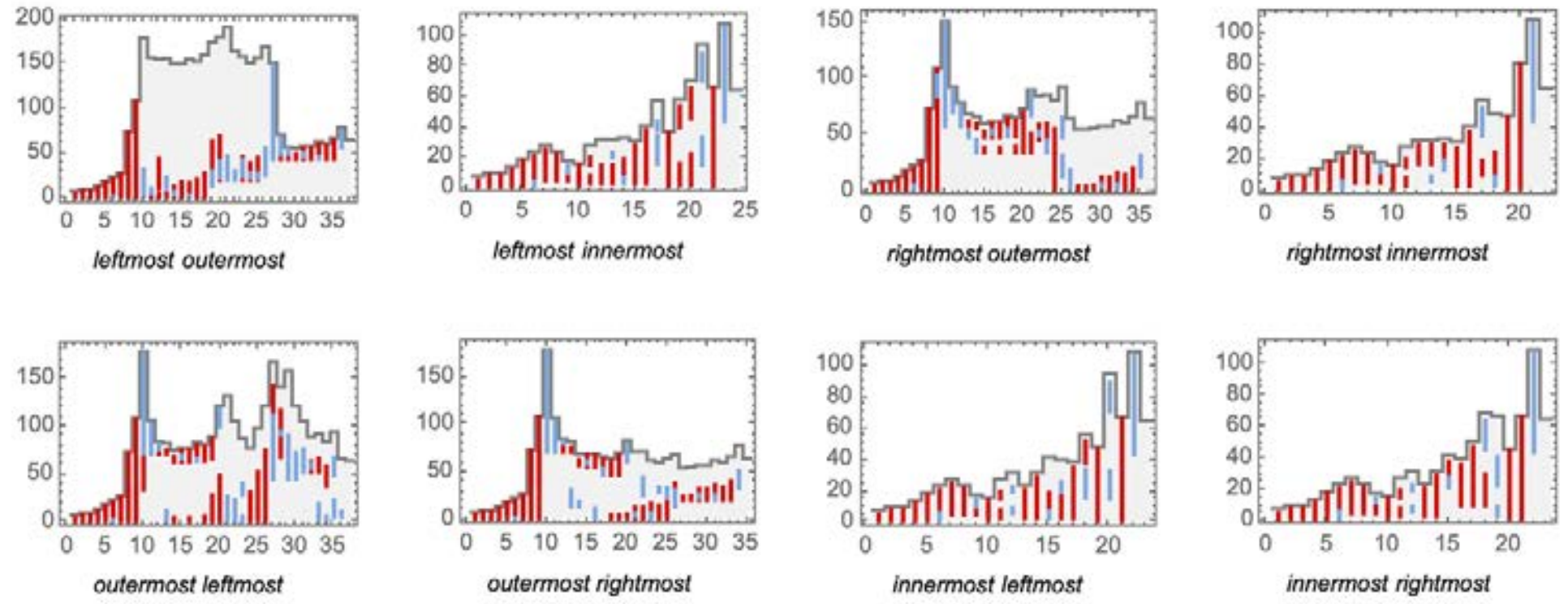

Here are the results for leftmost outermost, allowing up to between 1 and 8 updates at each step:

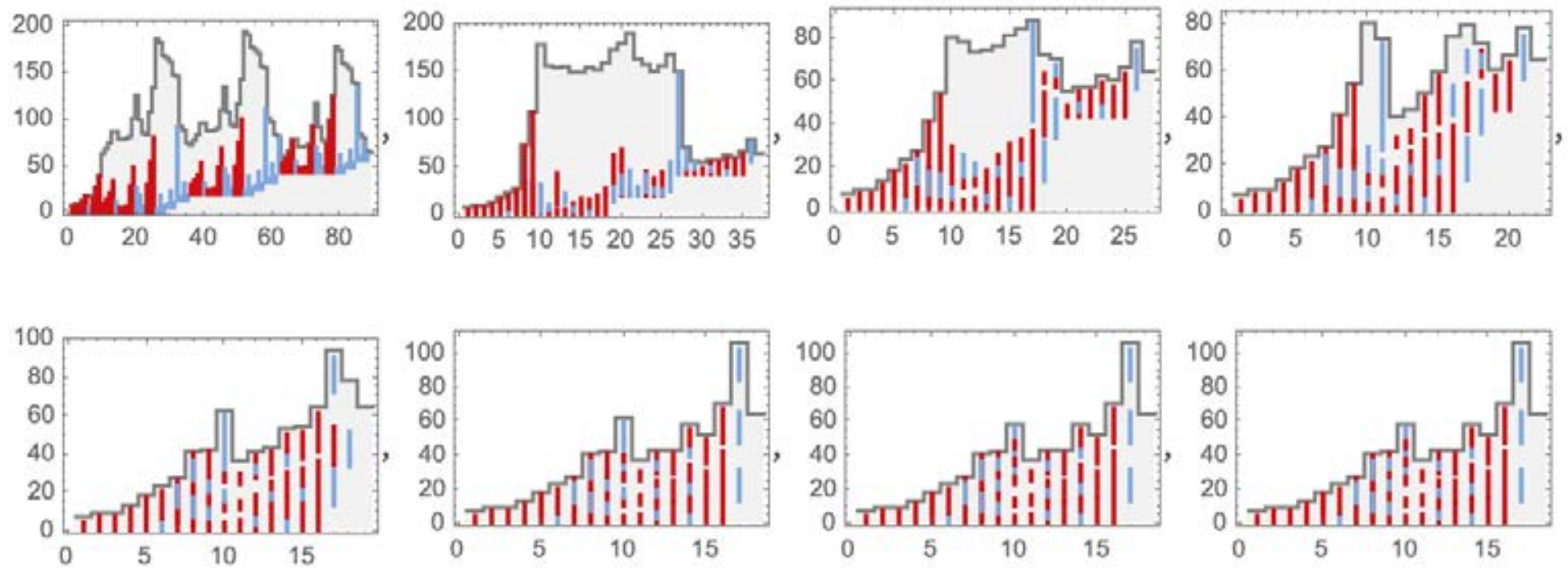

And here's a table of the "time to reach the fixed point" with different evaluation schemes, allowing different numbers of updates at each step:

| | 1 | 2 | 3 | 4 | 5 | 6 | 7 | 8 |
|---|---|---|---|---|---|---|---|---|
| *leftmost outermost* | 89 | 37 | 27 | 22 | 19 | 18 | 18 | 18 |
| *leftmost innermost* | 29 | 24 | 24 | 21 | 19 | 18 | 18 | 18 |
| *rightmost outermost* | 31 | 36 | 28 | 21 | 19 | 19 | 18 | 18 |
| *rightmost innermost* | 29 | 22 | 23 | 23 | 19 | 18 | 18 | 18 |
| *outermost leftmost* | 89 | 37 | 27 | 22 | 19 | 18 | 18 | 18 |
| *outermost rightmost* | 71 | 35 | 26 | 20 | 19 | 19 | 18 | 18 |
| *innermost leftmost* | 29 | 23 | 23 | 23 | 20 | 18 | 18 | 18 |
| *innermost rightmost* | 29 | 23 | 22 | 25 | 22 | 18 | 18 | 18 |

Not too surprisingly, the time to reach the fixed point always decreases when the number of updates that can be done at each step increases.

For the somewhat simpler terminating example s[s[s[s]]][s][s][s] (**S**(**S**(**SS**))**SSS**) we can explicitly look at the updates on the trees at each step for each of the different schemes:

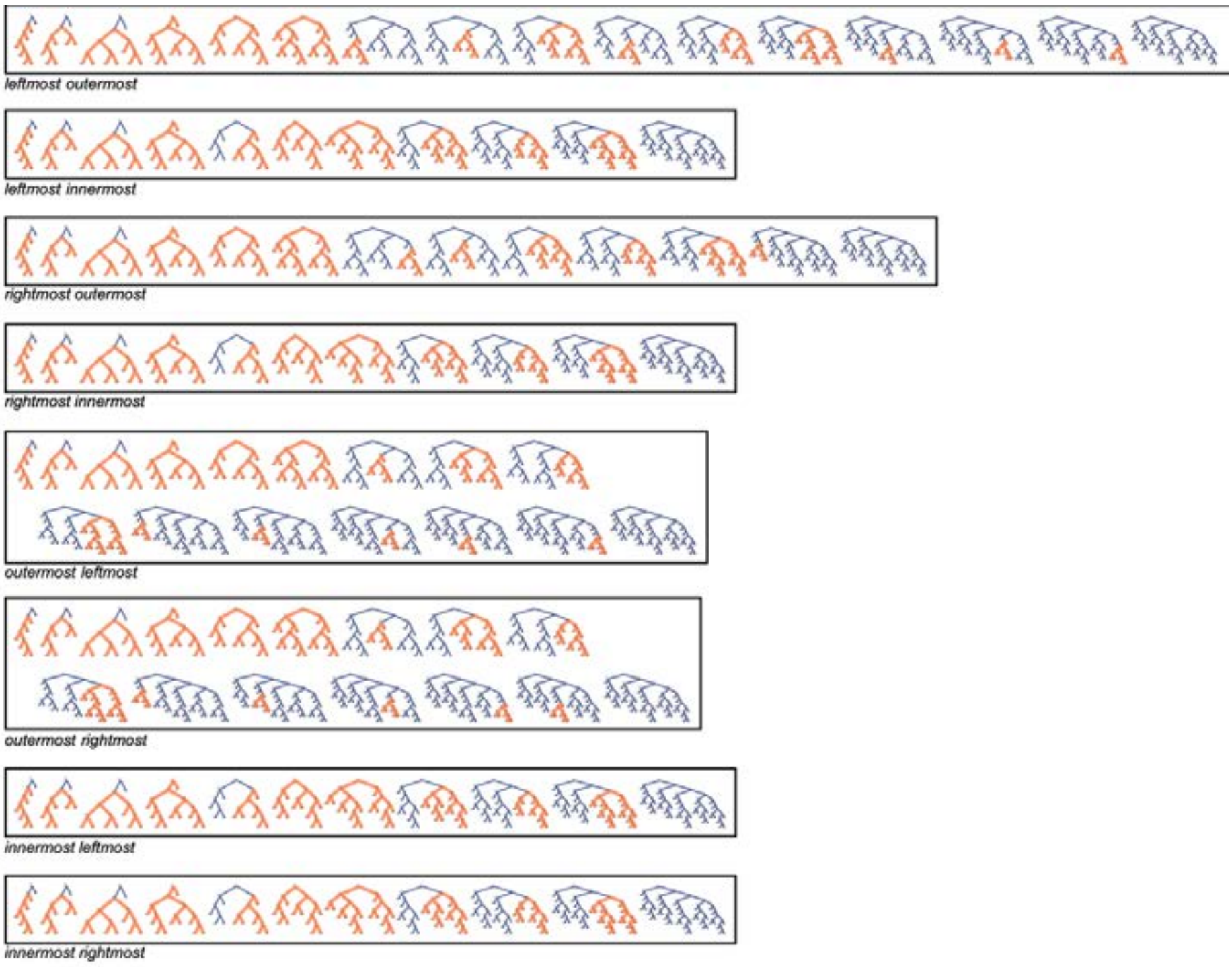

OK, so what about a combinator expression that does not terminate? What will these different evaluation schemes do? Here are the results for s[s[s]][s][s][s][s] (**S**(**SS**)**SSSS**) over the course of 50 steps, in each case using only one match at each step:

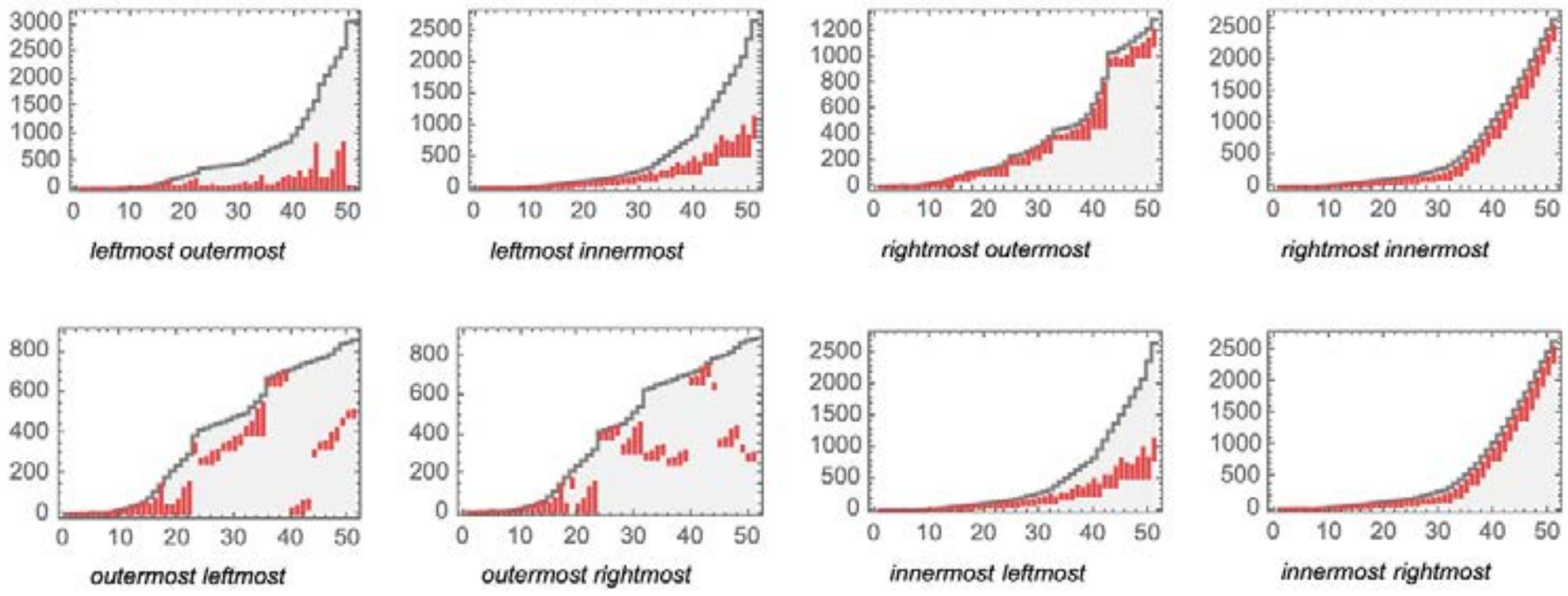

And here is what happens if we allow successively more matches (selected in leftmost outermost order) to be used at each step:

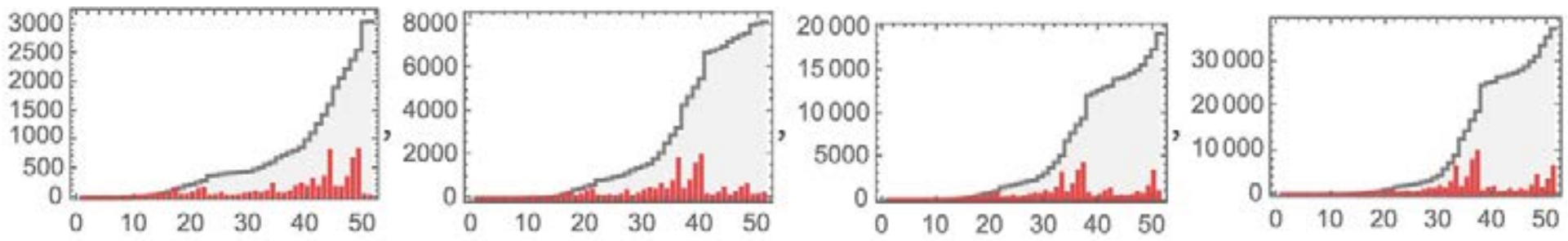

Not surprisingly, the more matches allowed, the faster the growth in size (and, yes, looking at pictures like this suggests studying a kind of "continuum limit" or "mean field theory" for combinator evolution):

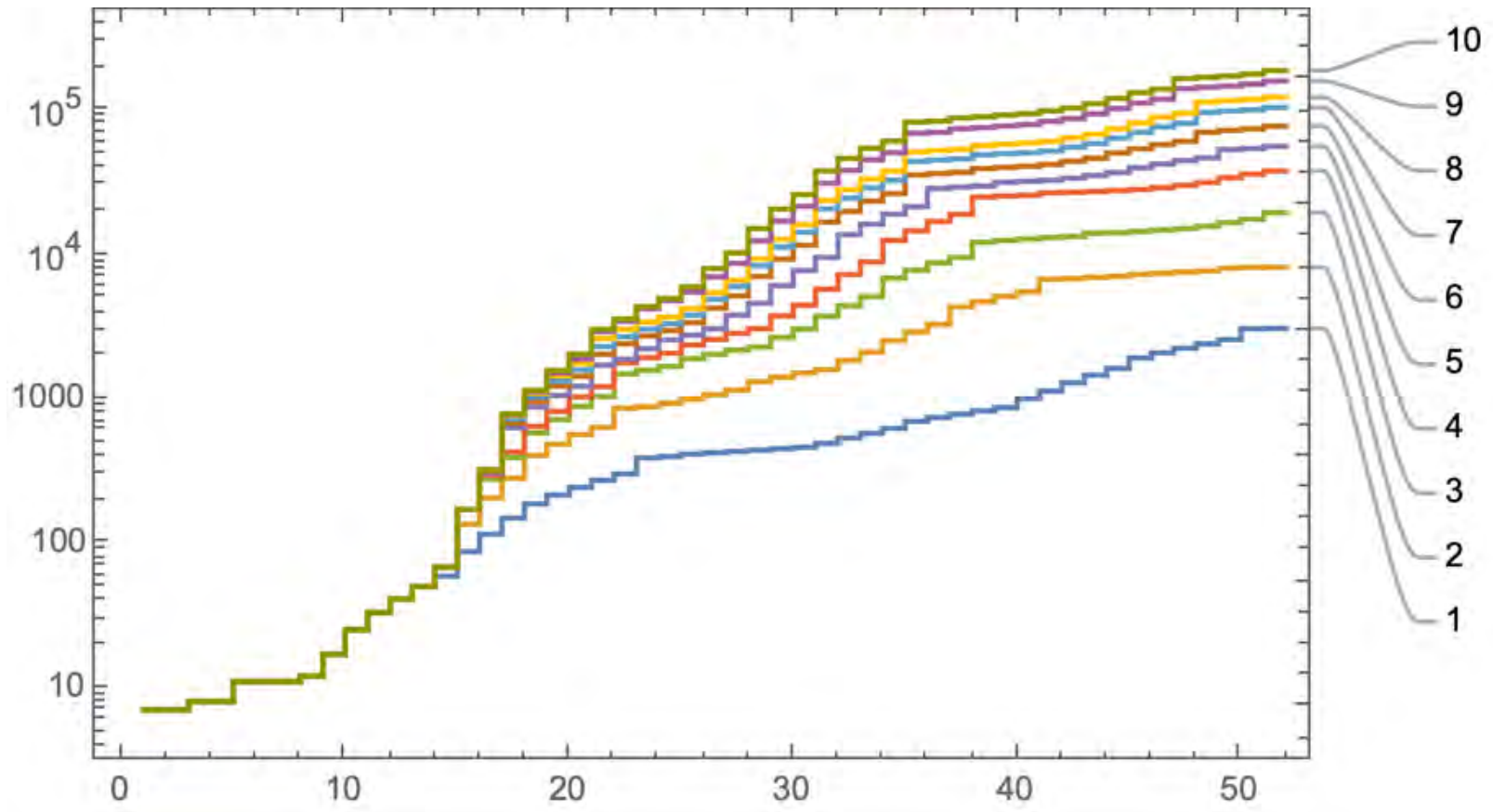

It's interesting to look at the ratios of sizes on successive steps for different updating schemes (still for s[s[s]][s][s][s][s]). Some schemes lead to much more "obviously simple" long-term behavior than others:

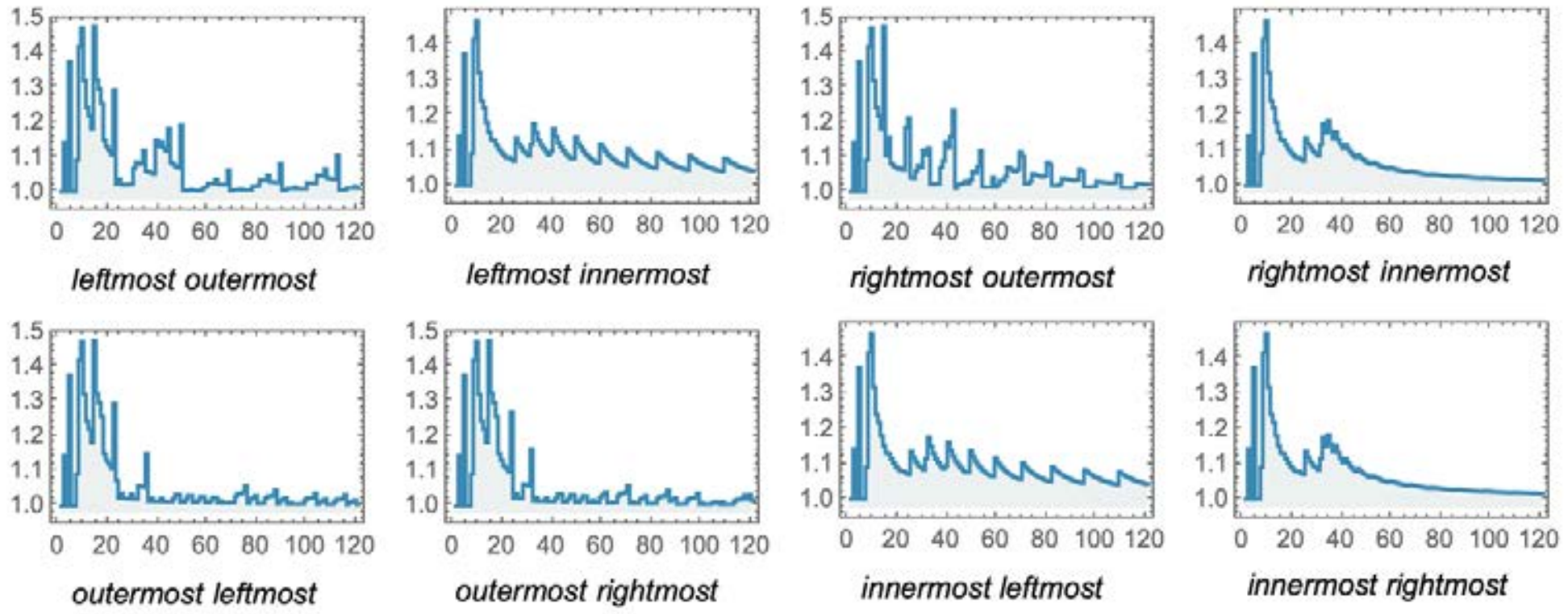

In fact, just changing the number of allowed matches (here for leftmost outermost) can have similar effects:

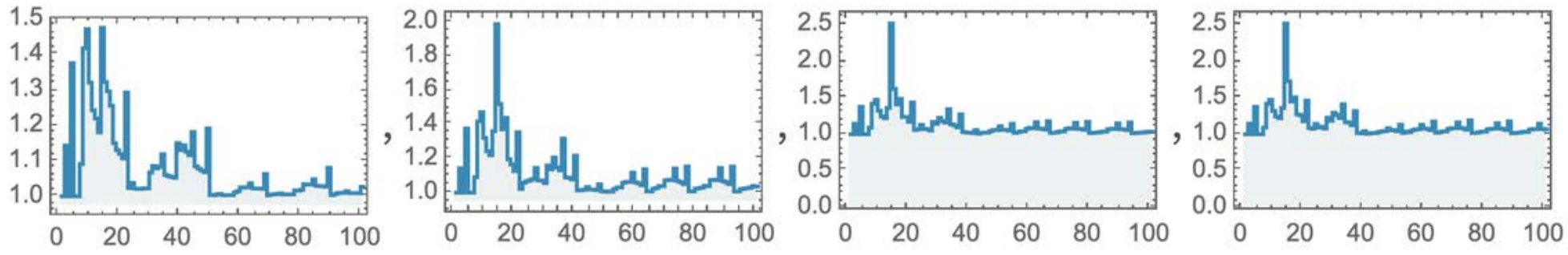

What about for other combinator expressions? Different updating schemes can lead to quite different behavior. Here's s[s[s]][s][s[s[s]]][k] (**S**(**SS**)**S**(**S**(**SS**))**K**):

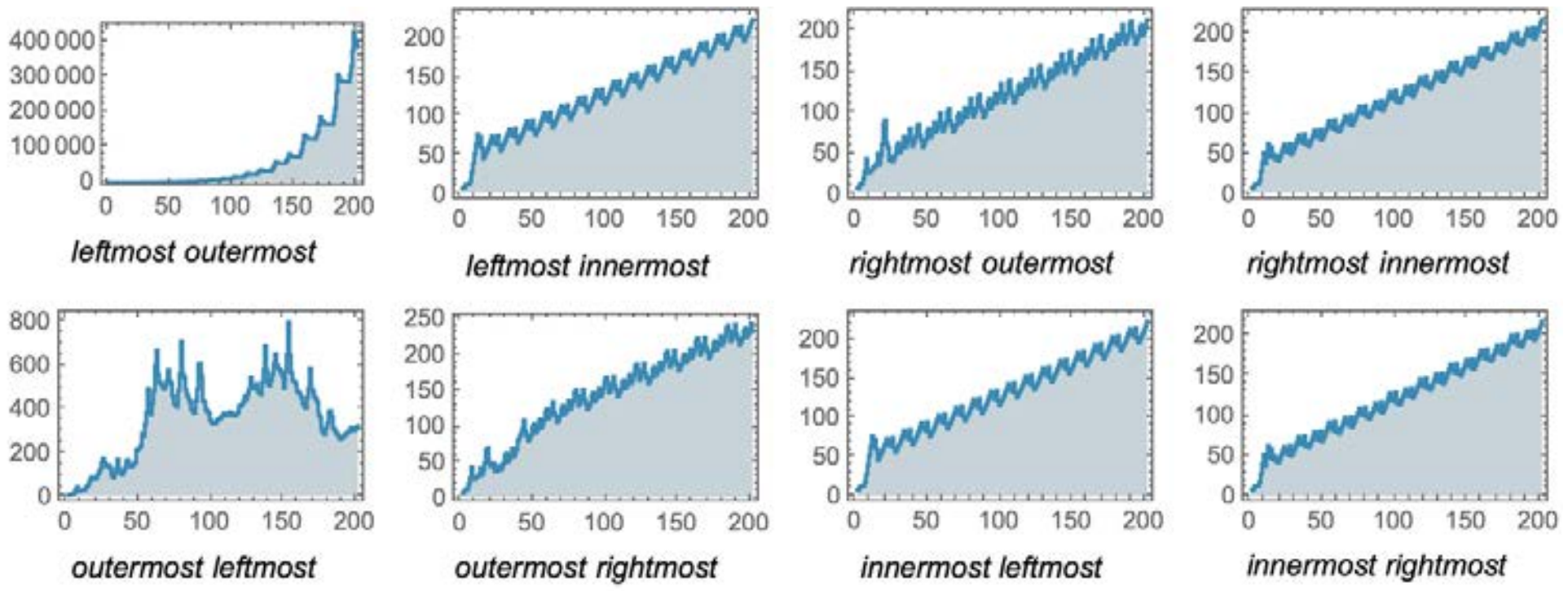

And here's s[s[s]][s][s][s][s[k]] (**S**(**SS**)**SSS**(**SK**))—which for some updating schemes gives purely periodic behavior (something which can't happen without a k in the original combinator expression):

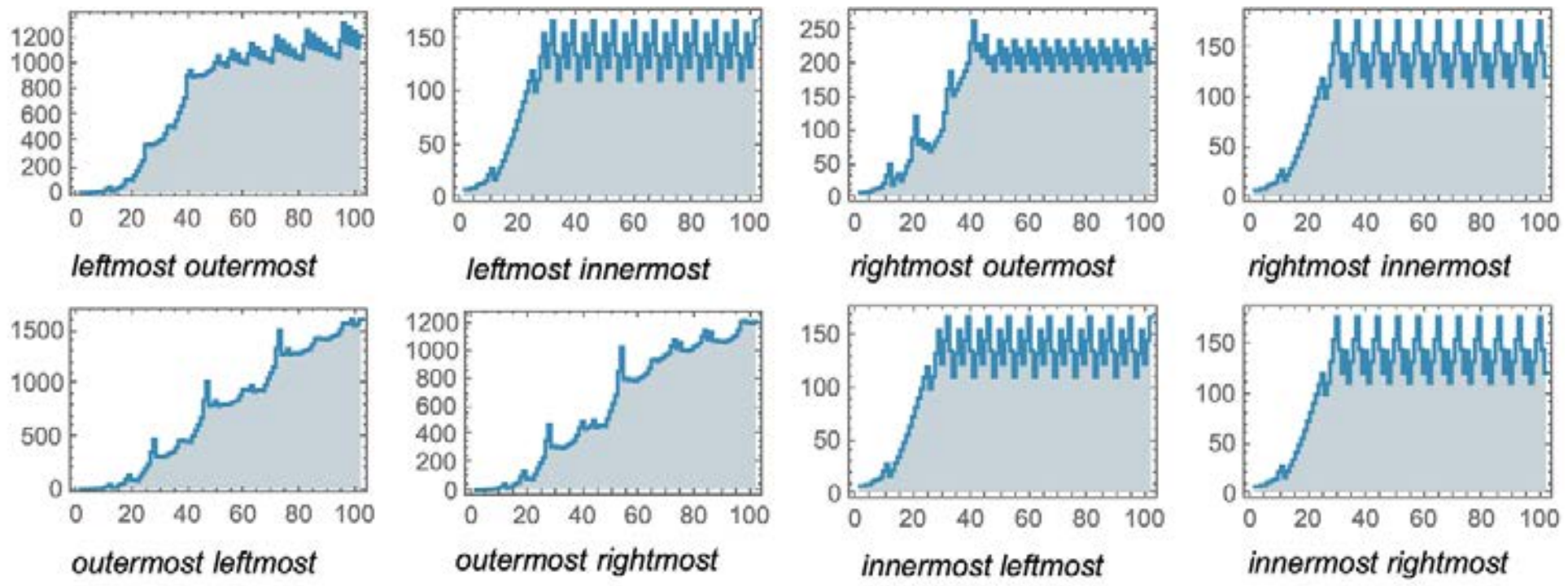

It's worth noting that—at least when there are k's involved—different updating schemes can even change whether the evaluation of a particular combinator expression ever terminates. This doesn't happen below size 8. But at size 8, here's what happens for example with s[s][s][s[s]][s][s][k] (**SSS**(**SS**)**SSK**):

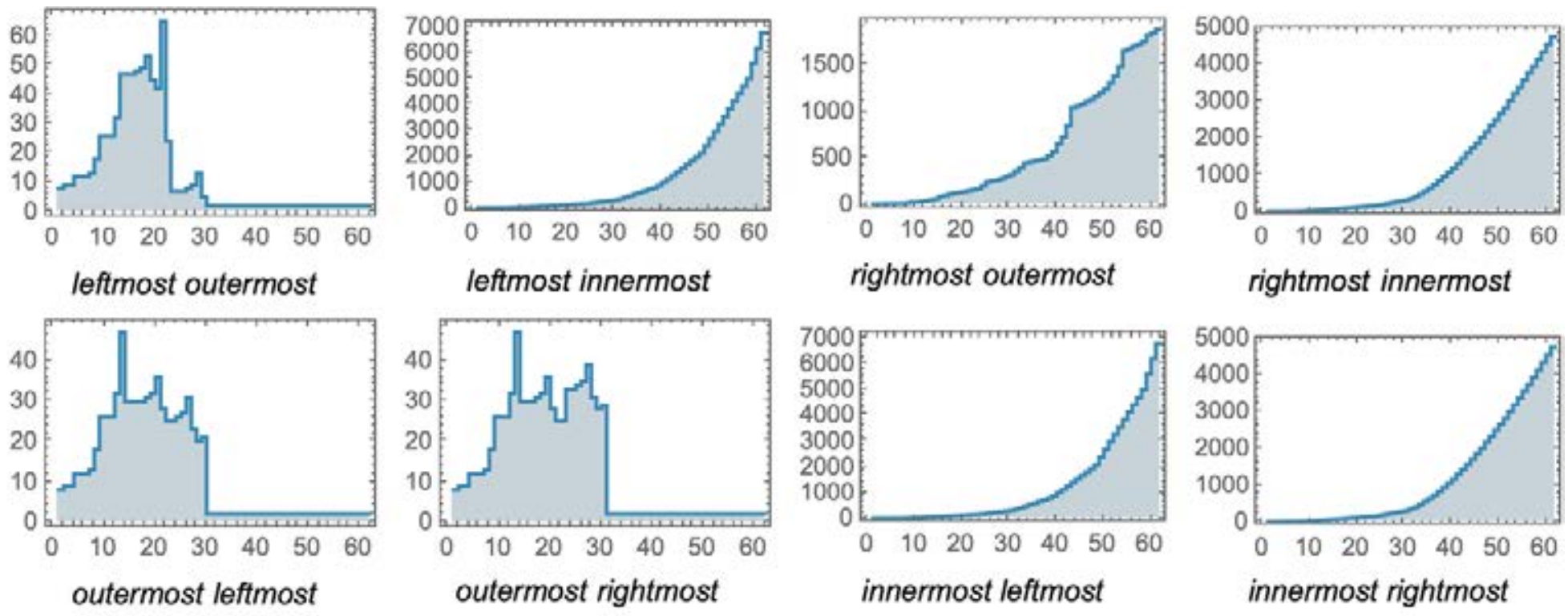

For some updating schemes it reaches a fixed point (always just s[k]) but for others it gives unbounded growth. The innermost schemes are the worst in terms of “missing fixed points”; they do it for 16 size-8 combinator expressions. But (as we mentioned earlier) leftmost outermost has the important feature that it’ll never miss a fixed point if one exists—though sometimes at the risk of taking an overly ponderous route to the fixed point.

But so if one’s applying combinator-like transformation rules in practice, what’s the best scheme to use? The Wolfram Language /. (ReplaceAll) operation in effect uses a leftmost-outermost scheme—but with an important wrinkle: instead of just using one match, it uses as many non-overlapping matches as possible.

Consider again the combinator expression:

s[s[s][s][s[s][k[k][s]][s]]][s][s][s[k[s][k]][k][s]]

In leftmost-outermost order the possible matches here are:

{{0}, {0, 0, 0, 1}, {0, 0, 0, 1, 1}, {0, 0, 0, 1, 1, 0, 1}, {1}, {1, 0, 0, 1}}

But the point is that the match at position {0} overlaps the match at position {0,0,0,1} (i.e. it is a tree ancestor of it). And in general the possible match positions form a partially ordered set, here:

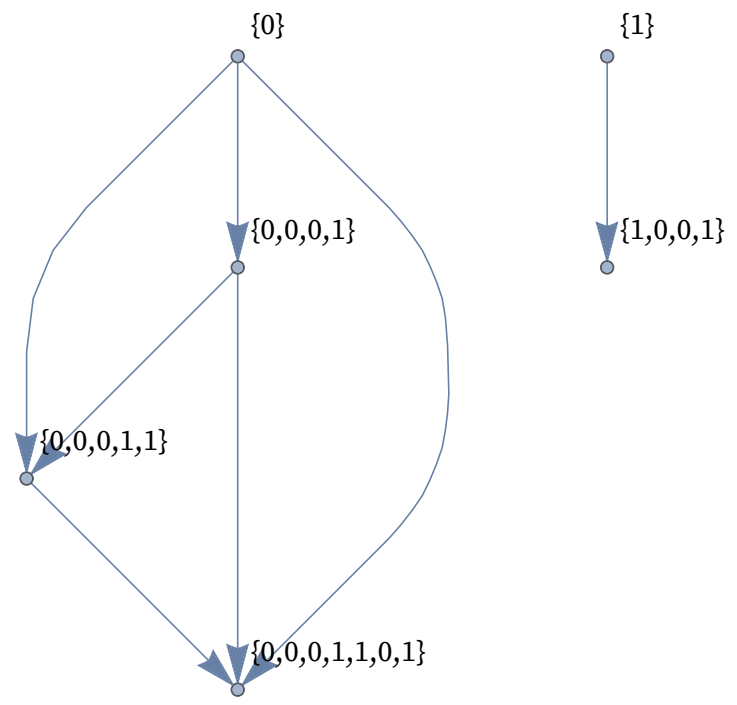

One possibility is always to use matches at the “bottom” of the partial order—or in other words, the very innermost matches. Inevitably these matches can’t overlap, so they can always be done in parallel, yielding a “parallel innermost” evaluation scheme that is potentially faster (though runs the risk of not finding a fixed point at all).

What /. does is effectively to use (in leftmost order) all the matches that appear at the “top” of the partial order. And the result is again typically faster overall updating. In the s[s][s][s[s]][s][s][k] example above, repeatedly applying /. (which is what //. does) finds the fixed point in 23 steps, while it takes ordinary one-replacement-at-a-time leftmost outermost updating 30 steps—and parallel innermost doesn’t terminate in this case:

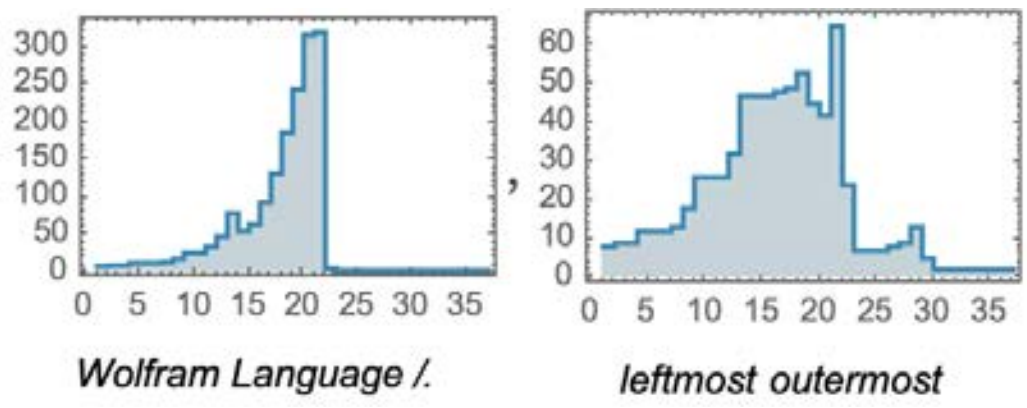

For s[s][s][s[s[s]]][k][s] (**SSS**(**S**(**SS**))**KS**) parallel innermost does terminate, getting a result in 27 steps compared to 26 for /.—but with somewhat smaller intermediate expressions:

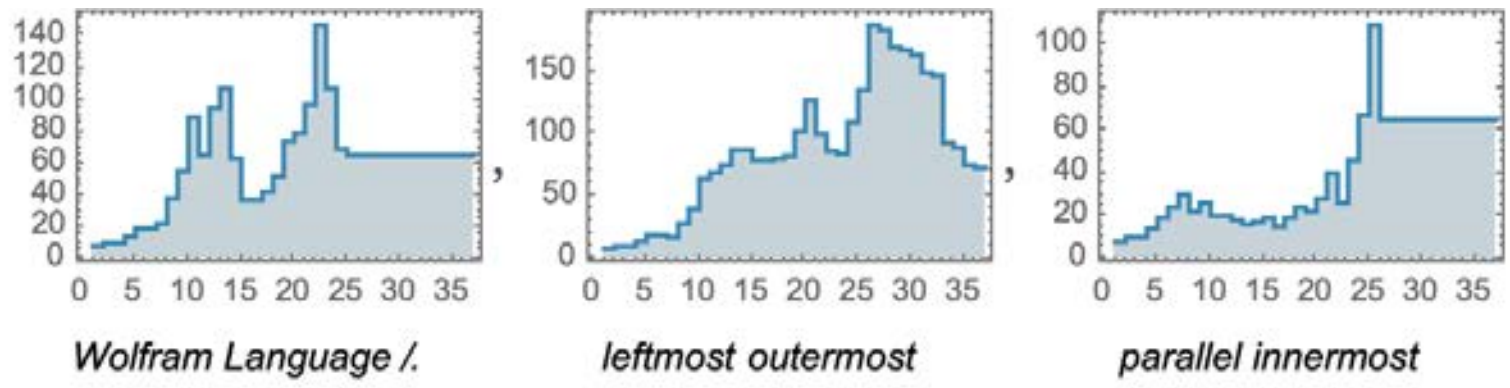

For a case in which there isn't a fixed point, however, /. will often lead to more rapid growth. For example, with s[s[s]][s][s][s][s] (**S**(**SS**)**SSSS**) it basically gives pure exponential $2^{t/2}$ growth (and eventually so does parallel innermost):

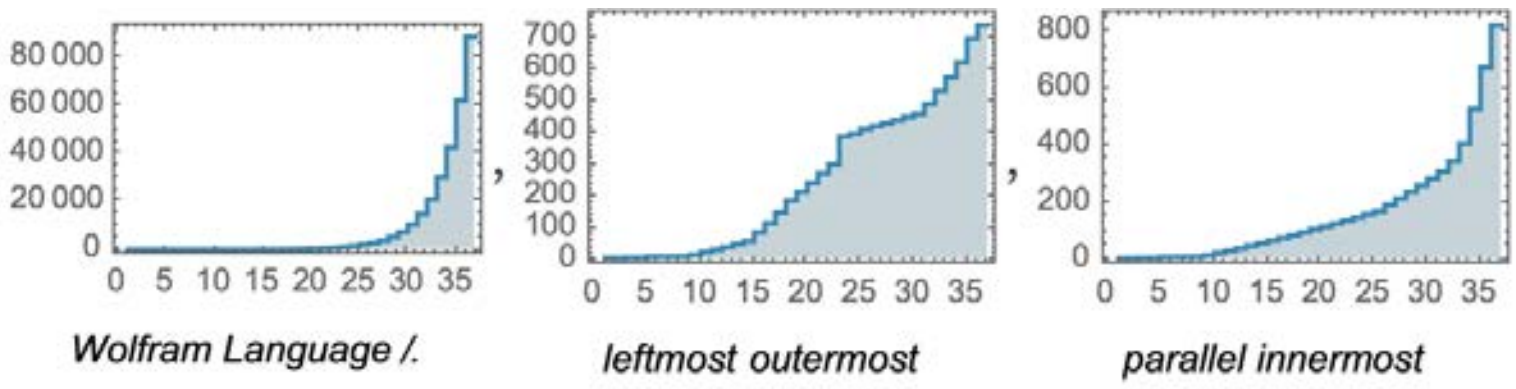

In A New Kind of Science I gave a bunch of results for combinators with /. updating, finding much of the same kind of behavior for "combinators in the wild" as we've seen here.

But, OK, so we've got the updating scheme of /. (and its repeated version //.), and we've got the updating scheme for automatic evaluation (with and without functions with "hold" attributes). But are there other updating schemes that might also be useful, and if so, how might we parametrize them?

I've wondered about this since I was first designing SMP—the forerunner to Mathematica and the Wolfram Language—more than 40 years ago. One place where the issue comes up is in automatic evaluation of recursively defined functions. Say one has a factorial function defined by:

```
f[1] = 1; f[n_] := n f[n – 1]
```

What will happen if one asks for f[0]? With the most obvious depth-first evaluation scheme, one will evaluate f[−1], f[−2], etc. forever, never noticing that everything is eventually going to be multiplied by 0, and so the result will be 0. If instead of automatic evaluation one was using //. all would be well—because it's using a different evaluation order:

```
f[0] //. f[n_] → n f[n – 1]
```

```
0
```

Let's consider instead the recursive definition of Fibonacci numbers (to make this more obviously "combinator like" we could for example use **Construct** instead of **Plus**):

```
f[1] = f[2] = 1; f[n_] := f[n – 1] + f[n – 2]
```

If you ask for f[7] you're essentially going to be evaluating this tree:

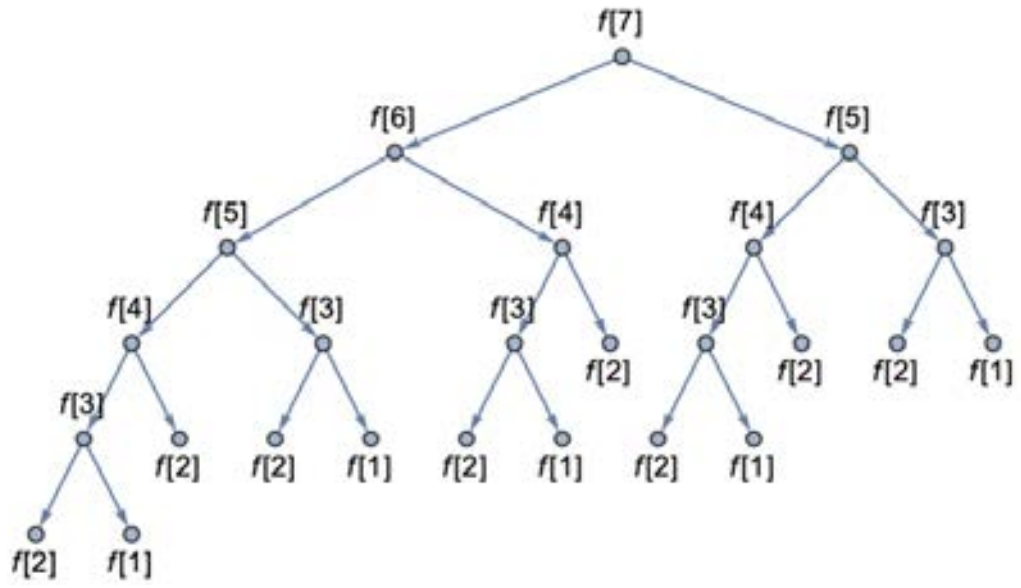

But the question is: how do you do it? The most obvious approach amounts to doing a depth-first scan of the tree—and doing about $\phi^n$ computations. But if you were to repeatedly use /. instead, you'd be doing more of a breadth-first scan, and it'd take more like $O(n^2)$ computations:

```
FixedPointList[# /. {f[1] → 1, f[2] → 1, f[n_] → f[n – 1] + f[n – 2]} &, f[7]]
```

```
{f[7], f[5] + f[6], f[3] + 2 f[4] + f[5], f[1] + f[2] + f[3] + 2 (f[2] + f[3]) + f[4],
  2 + f[1] + 2 f[2] + 2 (1 + f[1] + f[2]) + f[3], 11 + f[1] + f[2], 13, 13}
```

But how can one parametrize these different kinds of behavior? From our modern perspective in the Wolfram Physics Project, it's like picking different foliations—or different reference frames—in what amount to causal graphs that describe the dependence of one result on others. In relativity, there are some standard reference frames—like inertial frames parametrized by velocity. But in general it's not easy to "describe reasonable reference frames", and we're typically reduced to just talking about named metrics (Schwarzschild, Kerr, ...), much like here we're talking about "named updated orders" ("leftmost innermost", "outermost rightmost", ...).

But back in 1980 I did have an idea for at least a partial parametrization of evaluation orders. Here it is from section 3.1 of the SMP documentation:

The simplification of expressions proceeds as follows:

Numbers Ordinary numbers remain unchanged.

Symbols A symbol is replaced by the simplified form of any value assigned [3.2] to it.

Projections

1. Each filter is simplified in turn, unless the value ($k$) of any corresponding **Smp** property carried by the projector is **0**. (Future parallel-processing implementations may not respect this ordering.) The simplification of a filter is carried out until its value no longer changes, or until any projectors not carrying property **Rec** have appeared recursively at most $k$ times (see below). If any filter is found to be extended [4] with respect to the projector, then the projection is replaced or encased as specified in the relevant property list [4].

2. Projections with **Flat**, **Comm** or **Reor** properties [4, 7.7] are cast into canonical form.

What I called a "projection" then is what we'd call a function now; a "filter" is what we'd now call an argument. But basically what this is saying is that usually the arguments of a function are evaluated (or "simplified" in SMP parlance) before the function itself is evaluated. (Though note the ahead-of-its-time escape clause about "future parallel-processing implementations" which might evaluate arguments asynchronously.)

But here's the funky part: functions in SMP also had Smp and Rec properties (roughly, modern "attributes") that determined how recursive evaluation would be done. And in a first approximation, the concept was that Smp would choose between innermost and outermost, but then in the innermost case, Rec would say how many levels to go before "going outermost" again.

And, yes, nobody (including me) seems to have really understood how to use these things. Perhaps there's a natural and easy-to-understand way to parametrize evaluation order (beyond the /. vs. automatic evaluation vs. hold attributes mechanism in Wolfram Language), but I've never found it. And it's not encouraging here to see all the complexity associated with different updating schemes for combinators.

By the way, it's worth mentioning that there is always a way to completely specify evaluation order: just do something like procedural programming, where every "statement" is effectively numbered, and there can be explicit Goto's that say what statement to execute next. But in practice this quickly gets extremely fiddly and fragile—and one of the great values of functional programming is that it streamlines things by having "execution order" just implicitly determined by the order in which functions get evaluated (yes, with things like Throw and Catch also available).

And as soon as one's determining "execution order" by function evaluation order, things are immediately much more extensible: without having to specify anything else, there's automatically a definition of what to do, for example, when one gets a piece of input with more complex structure. If one thinks about it, there are lots of complex issues about when to recurse through different parts of an expression versus when to recurse through reevaluation. But the good news is that at least the way the Wolfram Language is designed, things in practice normally "just work" and one doesn't have to think about them.

Combinator evaluation is one exception, where, as we have seen, the details of evaluation order can have important effects. And presumably this dependence is in fact connected to why it's so hard to understand how combinators work. But studying combinator evaluation once again inspires one (or at least me) to try to find convenient parametrizations for evaluation order—perhaps now using ideas and intuition from physics.

## The World of the S Combinator

In the definitions of the combinators s and k

{s[x_][y_][z_] → x[z][y[z]], k[x_][y_] → x}

S is basically the one that "builds things up", while K is the one that "cuts things down". And historically, in creating and proving things with combinators, it was important to have the balance of both S and K. But what we've seen above makes it pretty clear that s alone can already do some pretty complicated things.

So it's interesting to consider the minimal case of combinators formed solely from S. For size n (i.e. LeafCount[n]), there are

$$\text{CatalanNumber}[n-1] = \frac{\text{Binomial}[2n+1, n+1]}{2n+1}$$

($\sim \frac{4^n}{n^{3/2}}$ for large n) possible such combinators, each of which can be characterized simply in terms of the sequence of bracket openings and closings it involves.

Some of these combinators terminate in a limited time, but above size 7 there are ones that do not:

| *size* | *total* | *nonterminating* | *fraction* |
|---|---|---|---|
| 2 | 1 | 0 | 0% |
| 3 | 2 | 0 | 0% |
| 4 | 5 | 0 | 0% |
| 5 | 14 | 0 | 0% |
| 6 | 42 | 0 | 0% |
| 7 | 132 | 2 | 1.5% |
| 8 | 429 | 41 | 9.6% |
| 9 | 1430 | 276 | 19% |
| 10 | 4862 | 1481 | 30% |
| 11 | 16 796 | 6829 | 41% |
| 12 | 58 786 | 29 288 | 50% |
| 13 | 742 900 | 119 946 | 16% |
| 14 | 2 674 440 | 477 885 | 18% |
| 15 | 9 694 845 | 1 870 502 | 19% |
| 16 | 35 357 670 | 7 238 607 | 20% |

And already there's something weird: the fraction of non-terminating combinator expressions steadily increases with size, then precipitously drops, then starts climbing again:

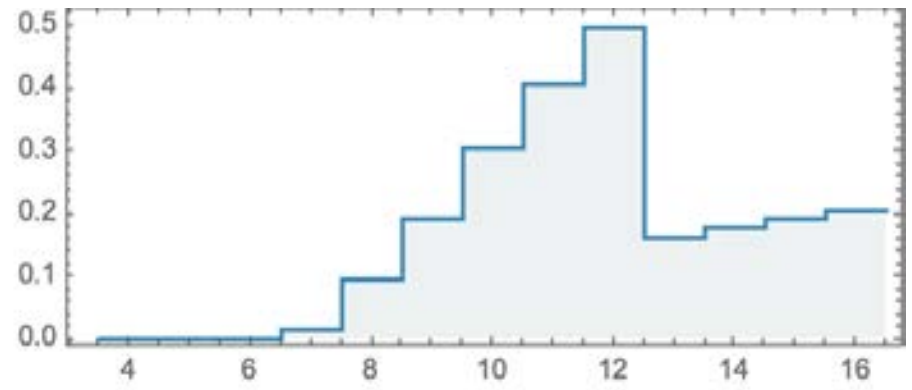

But let's look first at the combinator expressions whose evaluation does terminate. And, by the way, when we're dealing with S alone, there's no possibility of some evaluation schemes terminating and others not: they either all terminate, or none do. (This result was established in the 1940s from the fact that the S combinator—unlike K—in effect "conserves variables", making it an example of the so-called λI calculus.)

With leftmost-outermost evaluation, here are the halting time distributions, showing roughly exponential falloff with gradual broadening:

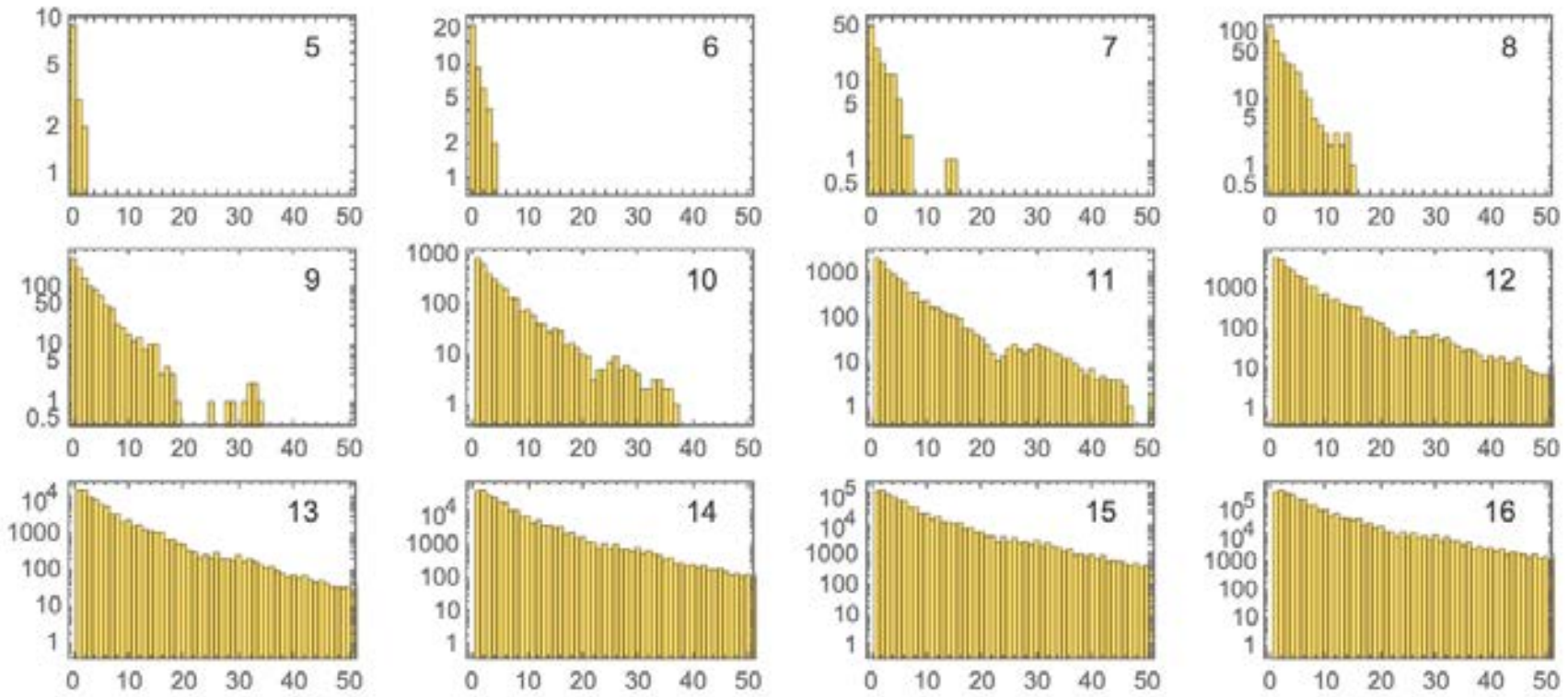

And here are the (leftmost-outermost) "champions"—the combinator expressions that survive longest (with leftmost-outermost evaluation) before terminating:

| *size* | *max steps* | *expression* | |
|---|---|---|---|
| 2 | 0 | s[s] | **ss** |
| 3 | 0 | s[s][s] | **sss** |
| 4 | 1 | s[s][s][s] | **ssss** |
| 5 | 2 | s[s][s][s][s] | **sssss** |
| 6 | 4 | s[s][s][s][s][s] | **ssssss** |
| 7 | 15 | s[s[s[s]]][s][s][s] | **s(s(ss))sss** |
| 8 | 15 | s[s[s[s[s]]][s][s][s]] | **s(s(s(ss))sss)** |
| 9 | 86 | s[s[s]][s[s]][s[s]][s][s] | **s(ss)(ss)(ss)ss** |
| 10 | 1109 | s[s[s][s]][s[s]][s][s][s][s] | **s(sss)(ss)ssss** |
| 11 | 1109 | s[s[s[s][s]][s[s]][s][s][s][s]] | **s(s(sss)(ss)ssss)** |
| 12 | 1444 | s[s[s]][s[s]][s[s][s][s][s][s]][s] | **s(ss)(ss)(ssssss)s** |
| 13 | 6317 | s[s[s]][s[s]][s[s][s][s][s][s][s]][s] | **s(ss)(ss)(sssssss)s** |
| 14 | 23679 | s[s[s]][s[s]][s[s][s][s][s][s][s][s]][s] | **s(ss)(ss)(ssssssss)s** |
| 15 | 131245 | s[s[s]][s[s]][s[s][s][s][s][s][s][s][s]][s] | **s(ss)(ss)(sssssssss)s** |
| 16 | 454708 | s[s[s]][s[s]][s[s][s][s][s][s][s][s][s][s]][s] | **s(ss)(ss)(ssssssssss)s** |

The survival (AKA halting) times grow roughly exponentially with size—and notably much slower than what we saw in the SK case above:

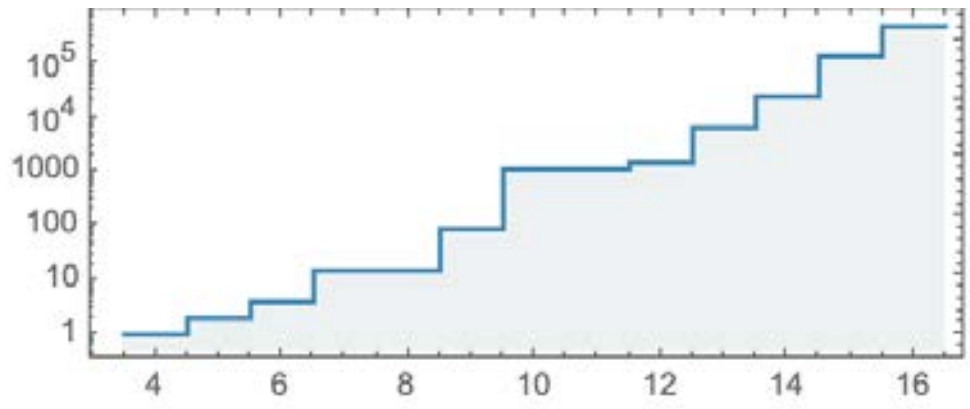

How do the champions actually behave? Here's what happens with for a sequence of sizes:

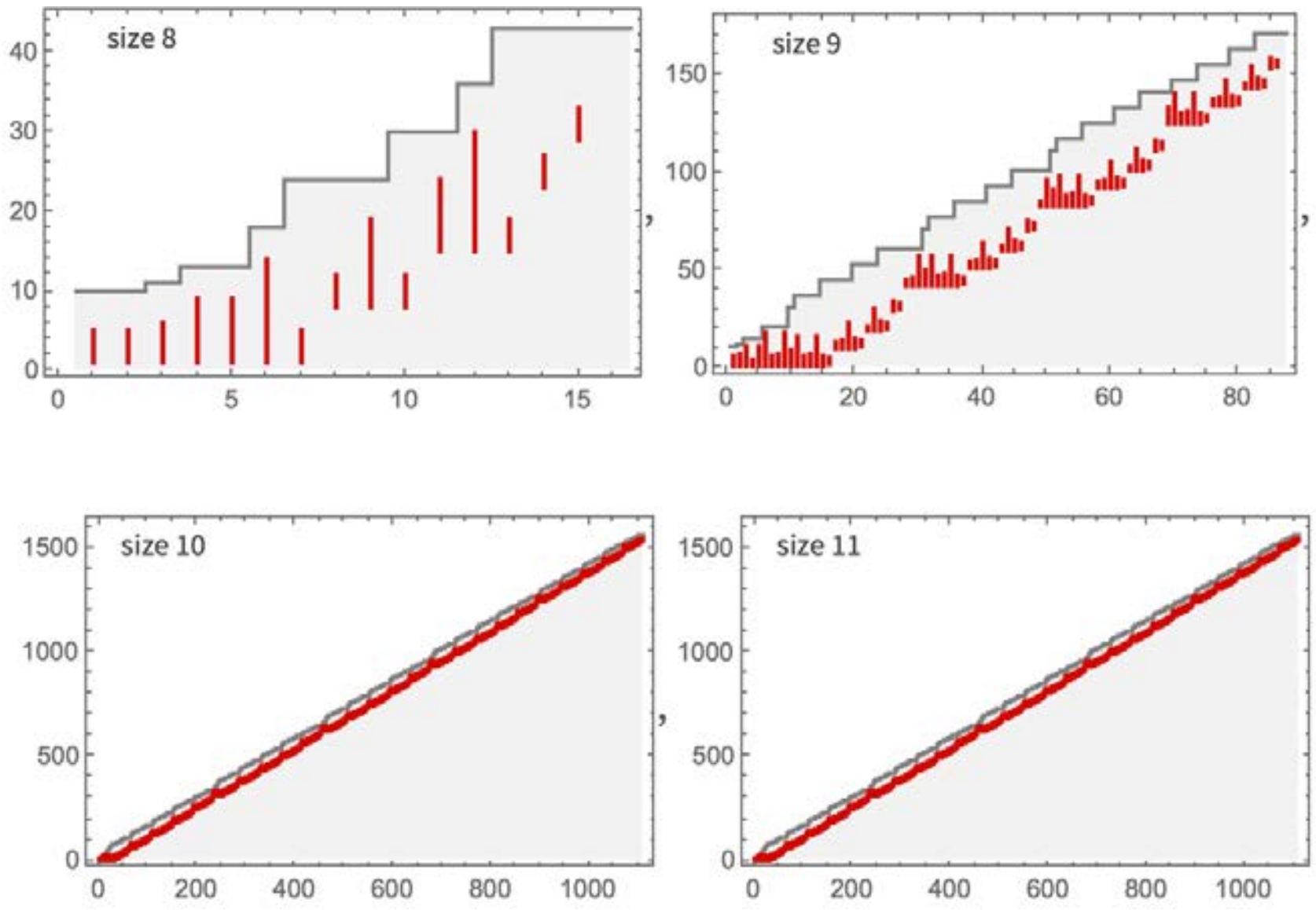

There's progressive increase in size, and then splat: the evolution terminates. Looking at the detailed behavior (here for size 9 with a "right-associative rendering") shows that what's going is quite systematic:

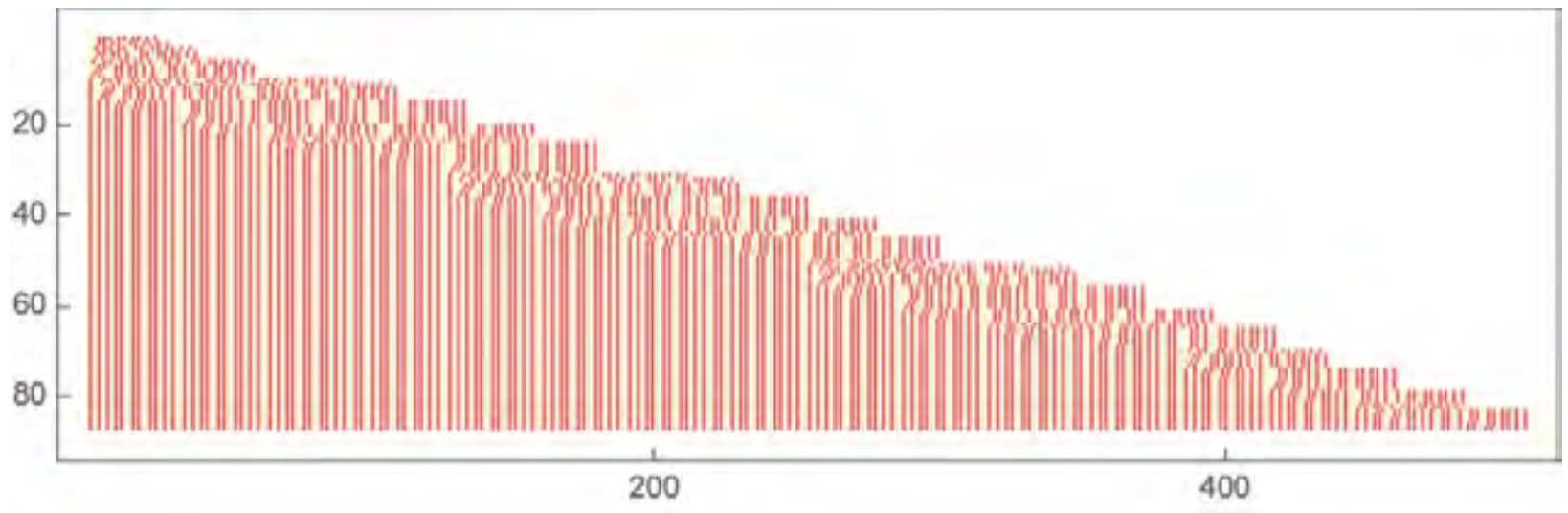

The differences again reflect the systematic character of the behavior:

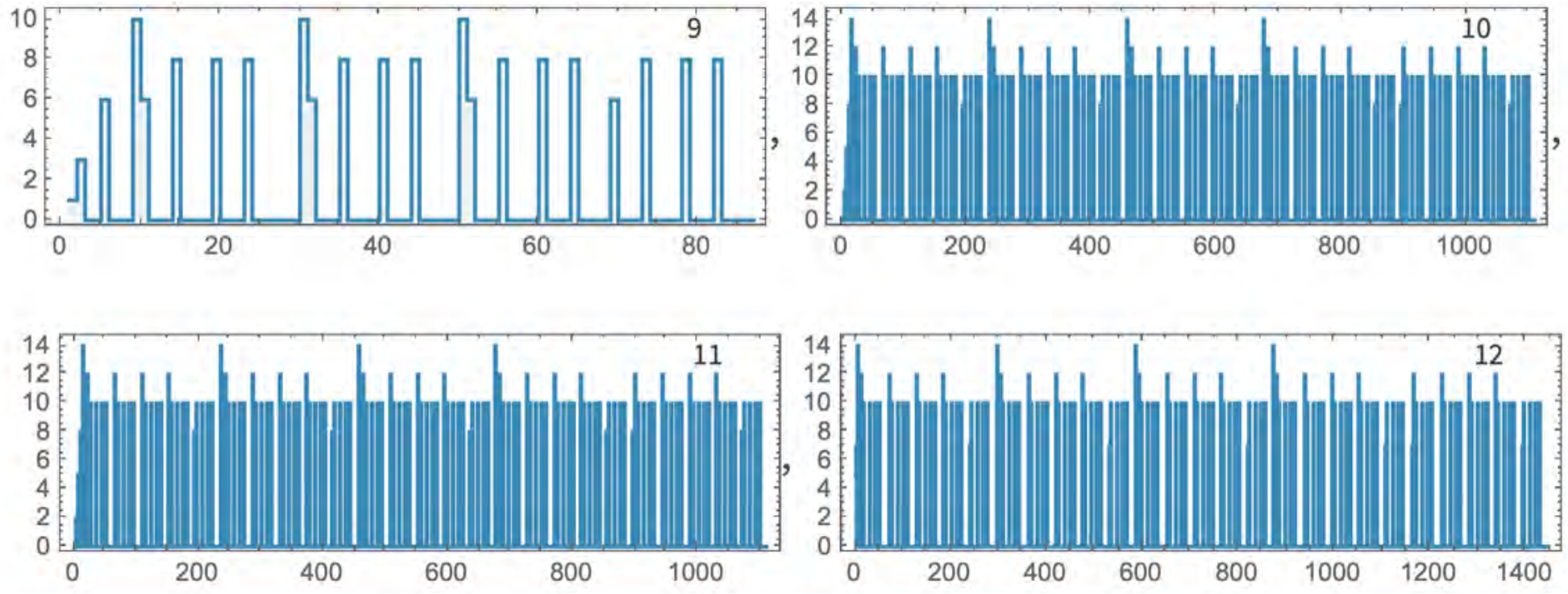

And it seems that what's basically happening is that the combinator is acting as a kind of digital counter that's going through an exponential number of steps—and ultimately building a very regular tree structure:

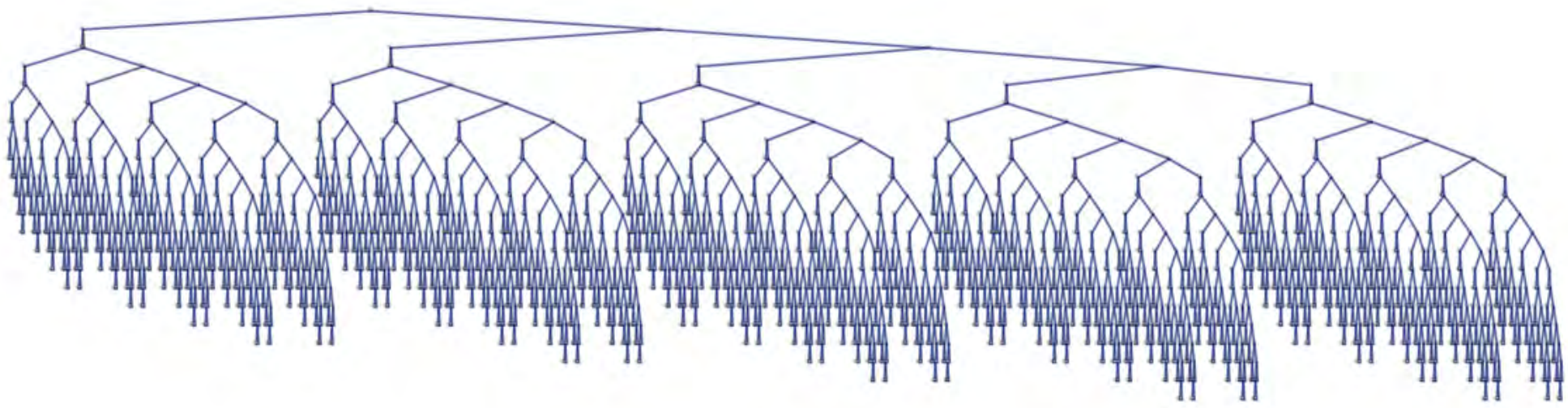

By the way, even though the final state is the same, the evolution is quite different with different evaluation schemes. And for example our "leftmost-outermost champions" actually terminate much faster with depth-first evaluation:

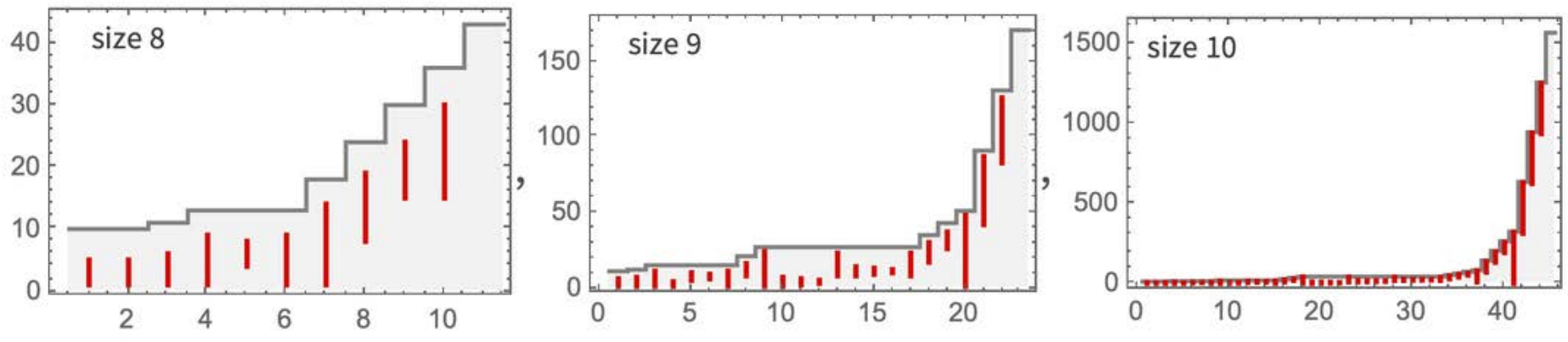

Needless to say, there can be different depth-first (AKA leftmost-innermost) champions, although—somewhat surprisingly—some turn out to be the same (but not sizes 8, 12, 13):

| size | max steps | expression | |
|---|---|---|---|
| 2 | 0 | s[s] | **SS** |
| 3 | 0 | s[s][s] | **SSS** |
| 4 | 1 | s[s][s][s] | **SSSS** |
| 5 | 2 | s[s][s][s][s] | **SSSSS** |
| 6 | 4 | s[s][s][s][s][s] | **SSSSSS** |
| 7 | 10 | s[s[s[s]]][s][s][s] | **S(S(SS))SSS** |
| 8 | 11 | s[s[s][s]][s][s][s][s] | **S(SSS)SSSS** |
| 9 | 22 | s[s[s]][s[s]][s[s]][s][s] | **S(SS)(SS)(SS)SS** |
| 10 | 44 | s[s[s][s]][s[s]][s][s][s][s] | **S(SSS)(SS)SSSS** |
| 11 | 44 | s[s[s[s][s]][s[s]][s][s][s][s]] | **S(S(SSS)(SS)SSSS)** |
| 12 | 48 | s[s][s[s[s][s]][s[s]]][s][s][s][s] | **SS(S(SSS)(SS))SSSS** |
| 13 | 55 | s[s[s][s]][s[s][s][s][s][s]][s][s][s] | **S(SSS)(SSSSSS)SSS** |

We can get a sense of what happens with all possible evaluation schemes if we look at the multiway graph. Here is the result for the size-8-leftmost-outermost champion s[s[s[s]]][s][s][s]:

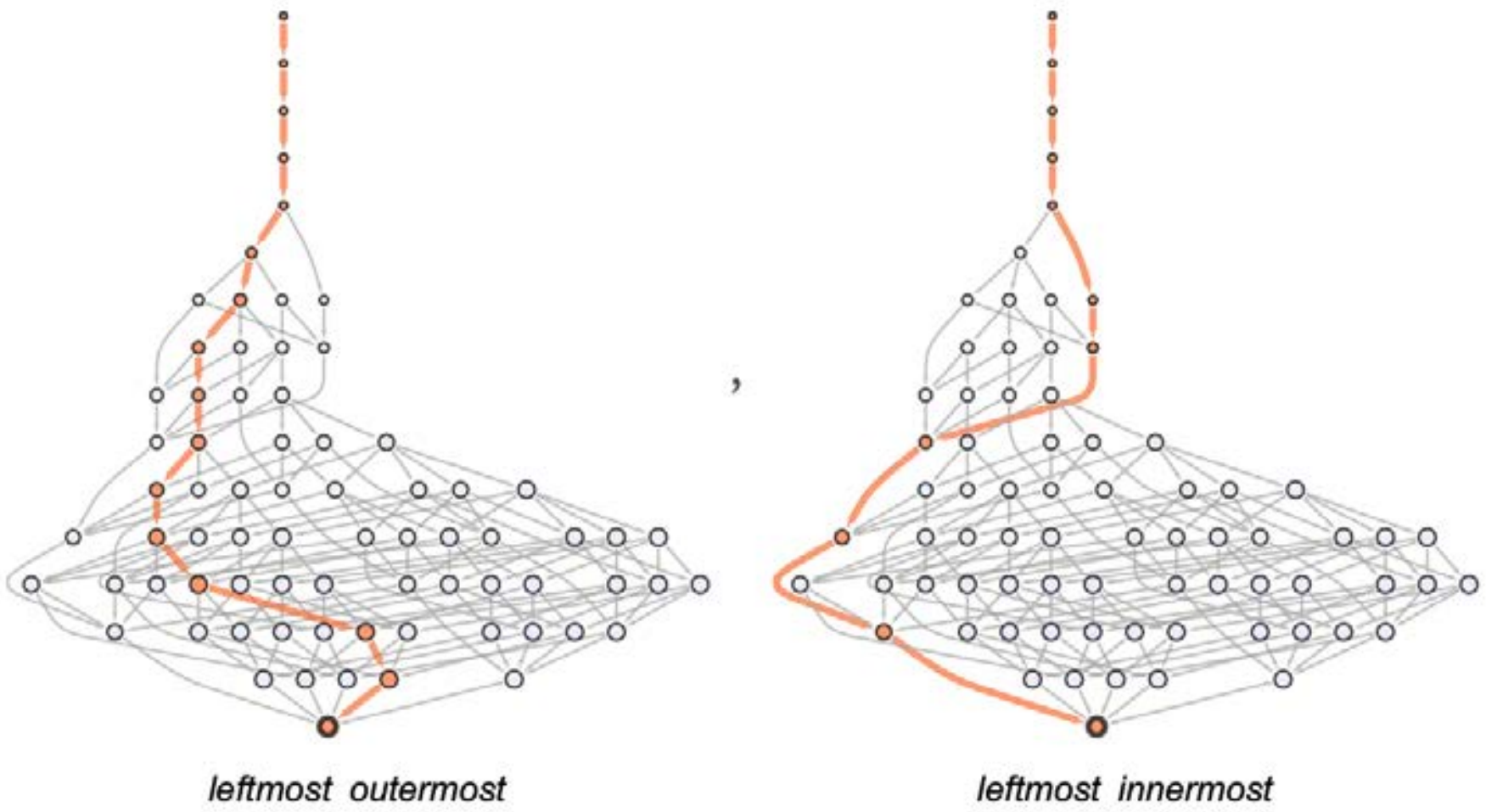

The number of expressions at successive levels in the multiway graph starts off growing quite exponentially, but after 12 steps it rapidly drops—eventually yielding a finite graph with 74 nodes (leftmost outermost is the "slowest" evaluation scheme—taking the maximum 15 steps possible):

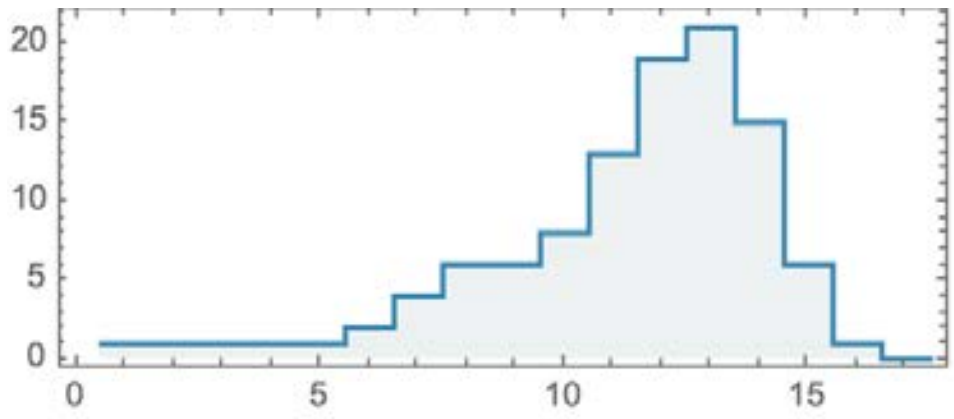

Even for the size-9 champion the full multiway graph is too large to construct explicitly. After 15 steps the number of nodes has reached 6598, and seems to be increasingly roughly like $2^t$—even though after at most 86 steps all “dangling ends” must have resolved, and the system must reach its fixed point:

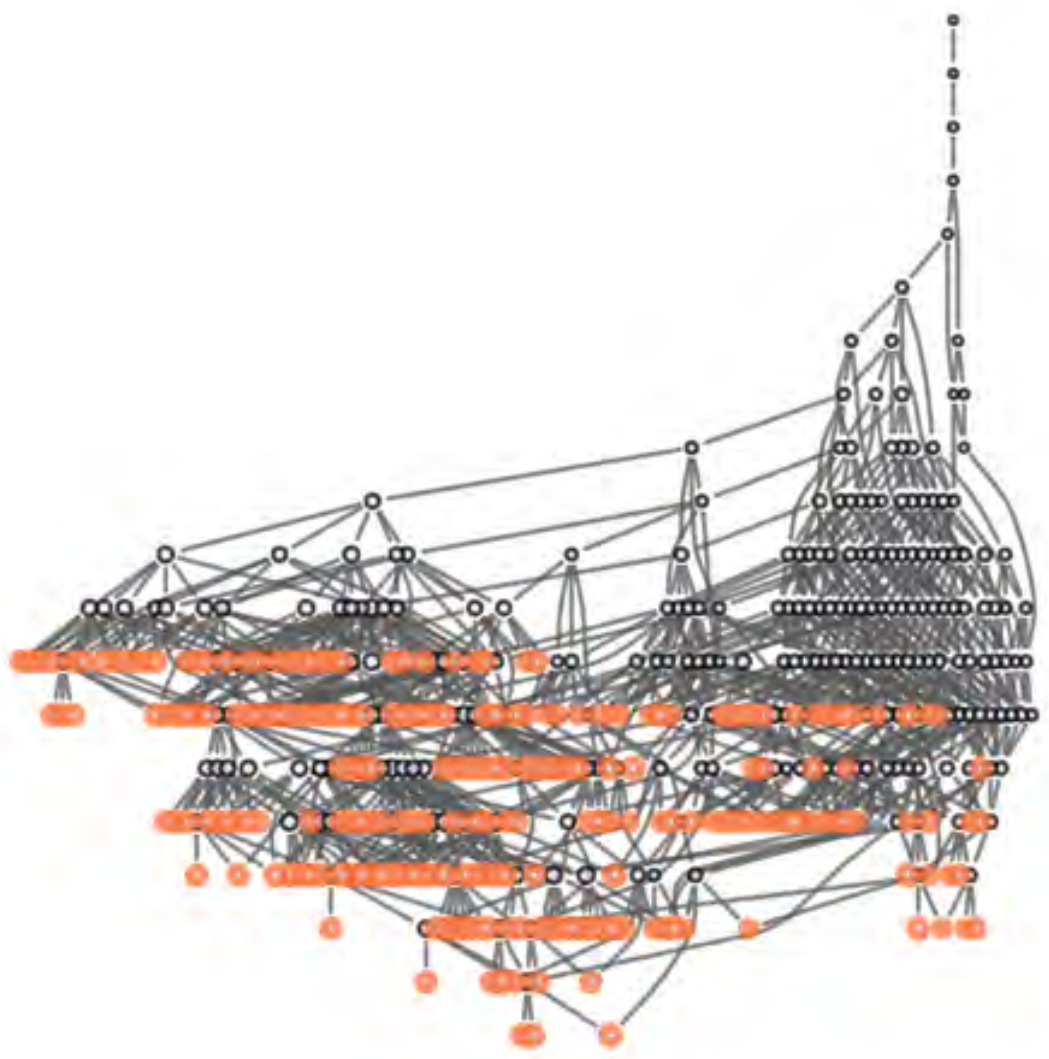

What happens with S combinator expressions that do not terminate? We already saw above some examples of the kind of growth in size one observes (say with leftmost-outermost evaluation). Here are examples with roughly exponential behavior, with differences between successive steps shown on a log scale:

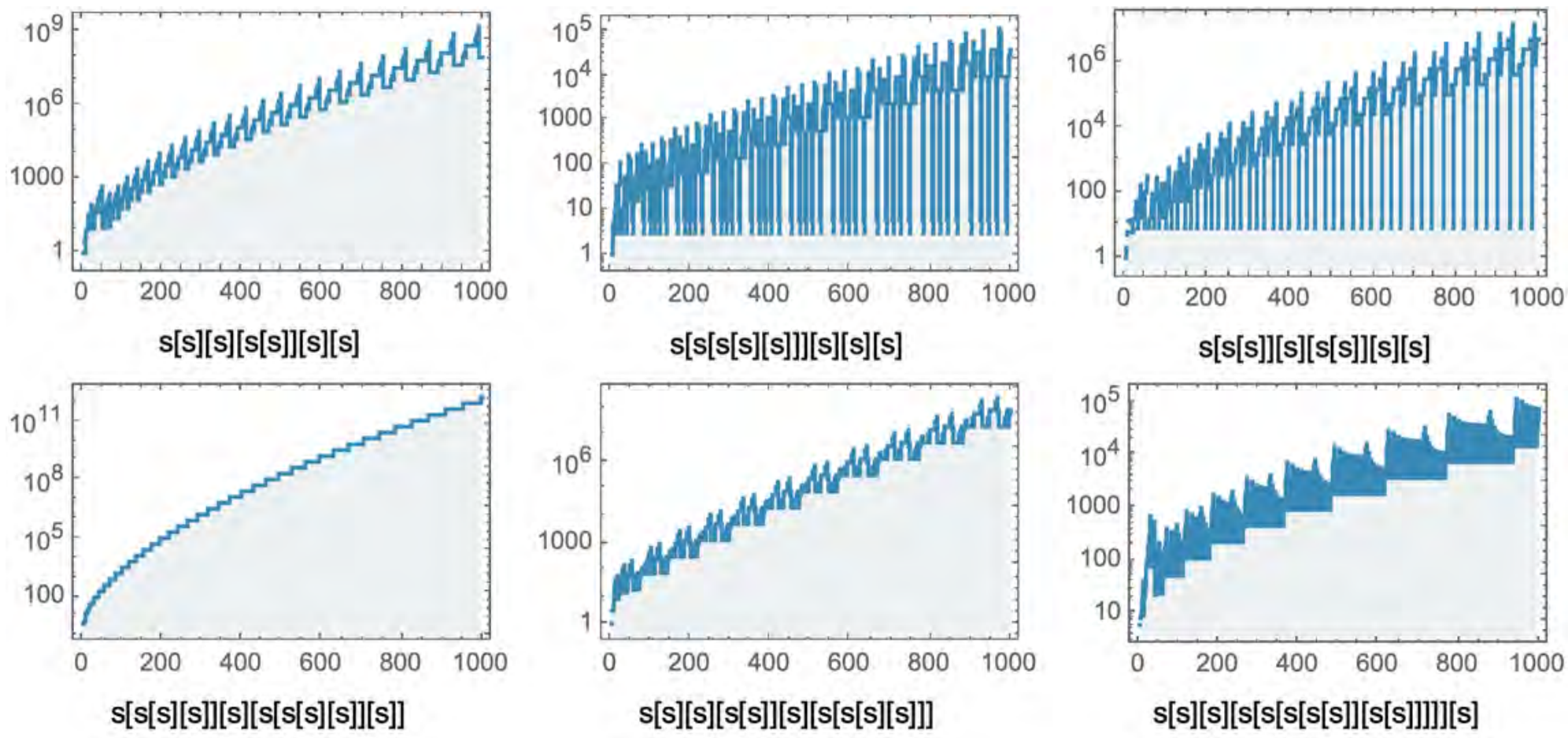

And here are examples of differences shown on a linear scale:

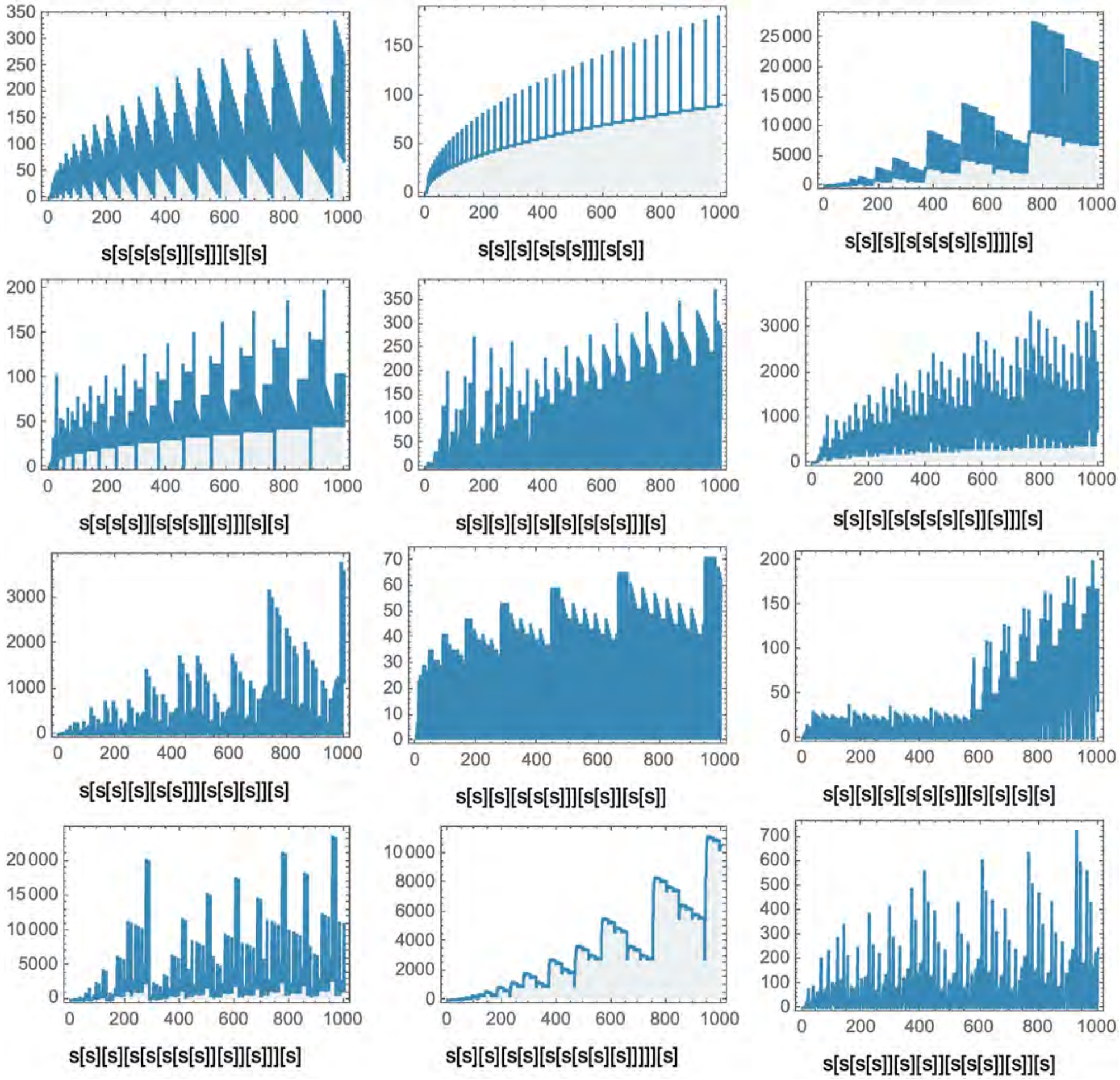

Sometimes there are fairly long transients, but what's notable is that among all the 8629 infinite-growth combinator expressions up to size 11 there are none whose evolution seems to show long-term irregularity in overall size. Of course, something like rule 30 also doesn't show irregularity in overall size; one has to look "inside" to see complex behavior—and difficulties of visualization make that hard to systematically do in the case of combinators.

But looking at the pictures above there seem to be a "limited number of ways" that combinator expressions grow without bound. Sometimes it's rather straightforward to see how the infinite growth happens. Here's a particularly "pure play" example: the size-9 case s[s[s[s]]][s[s[s]]][s[s]] (**S**(**S**(**SS**))(**S**(**SS**))(**SS**)) which evolves the same way with all evaluation schemes (in the pictures, the root of the match at each step is highlighted):

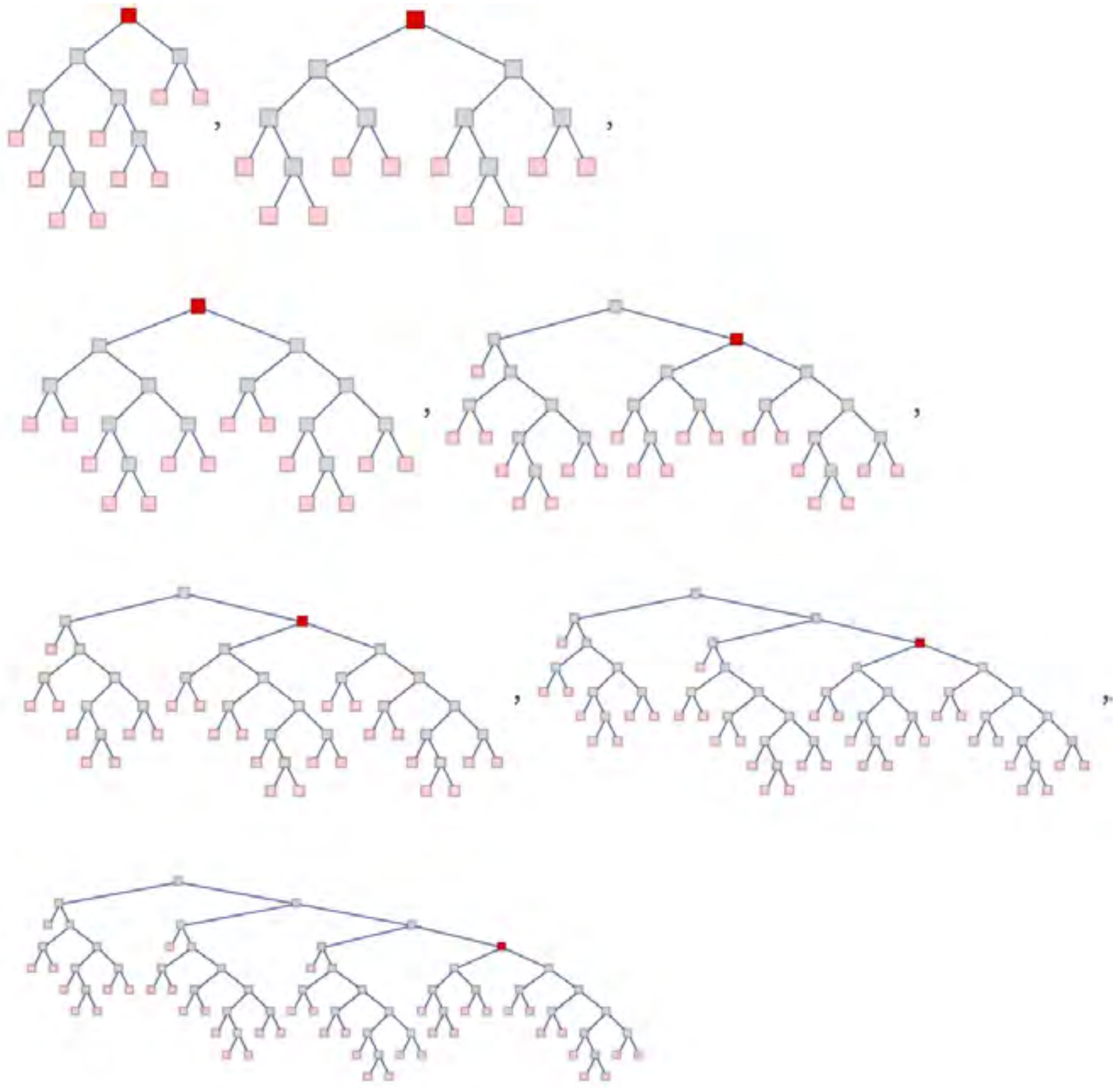

Looking at the subtree "below" each match we see and it is clear that there a definite progression which will keep going forever, leading to infinite growth.

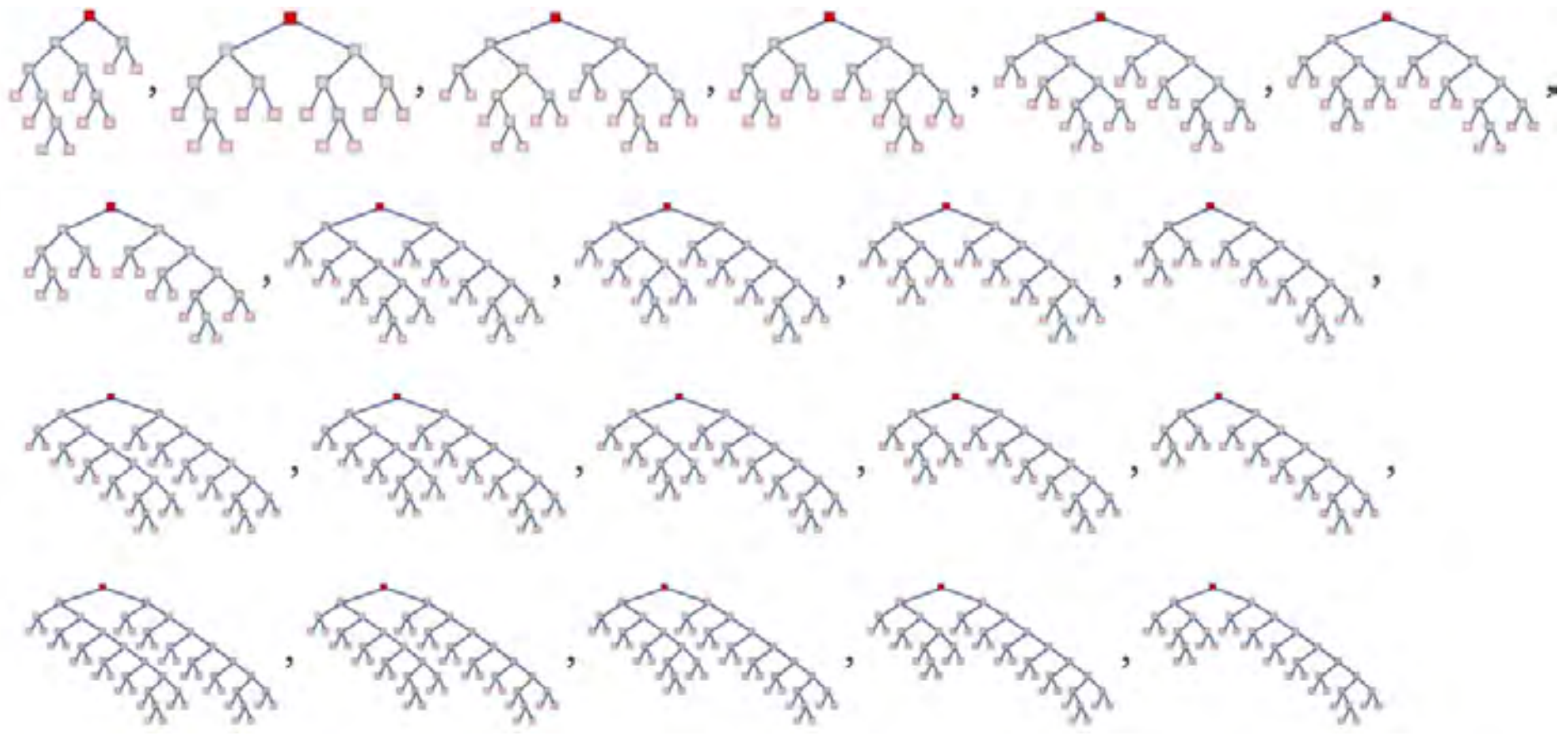

But if one looks at the corresponding sequence of subtrees for a case like the smallest infinite-growth combinator expression s[s][s][s[s]][s][s] (**SSS**(**SS**)**SS**), it's less immediately obvious what's going on:

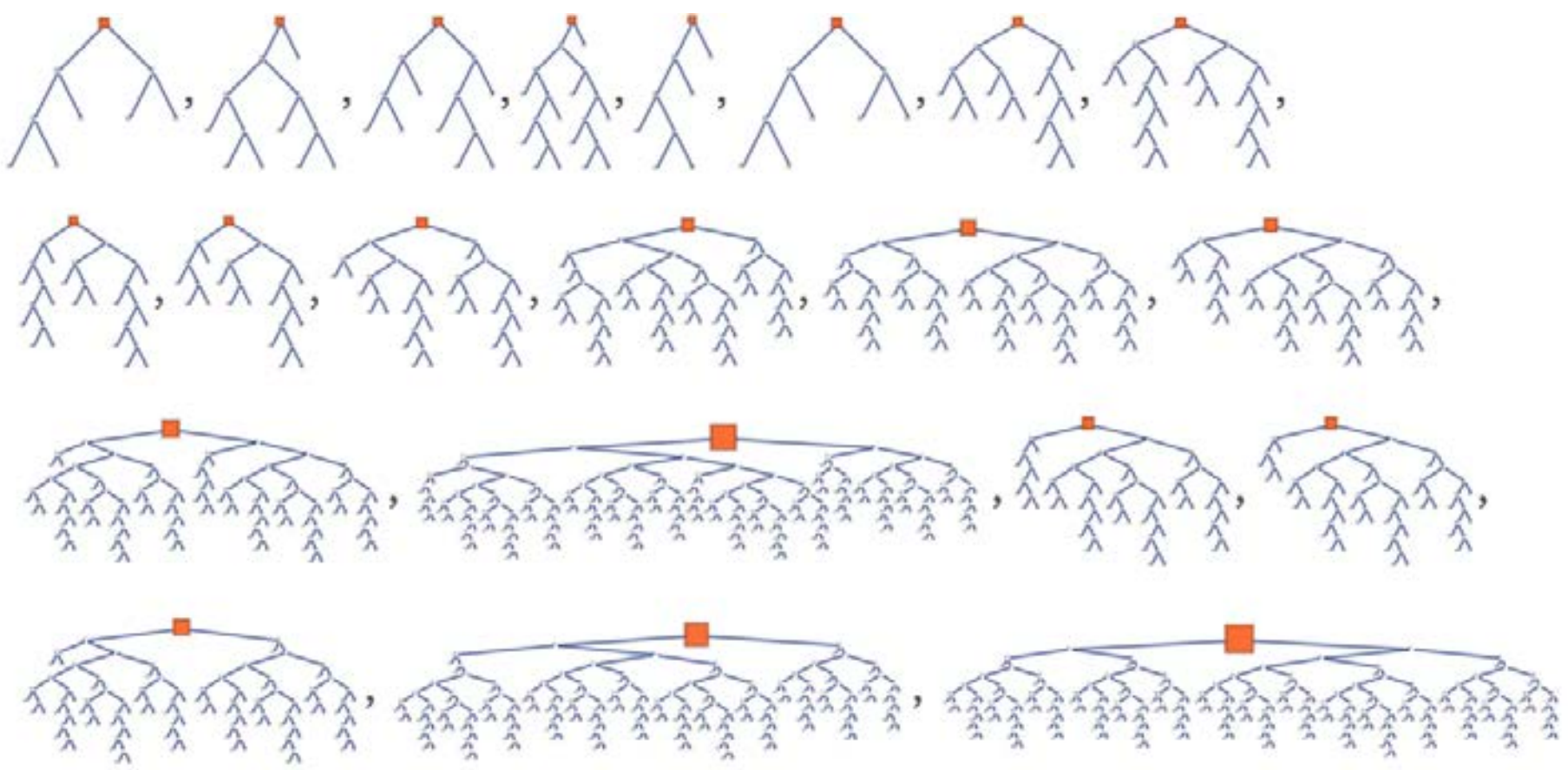

But there's a rather remarkable result from the end of the 1990s that gives one a way to "evaluate" combinator expressions, and tell whether they'll lead to infinite growth—and in particular to be able to say directly from an initial combinator expression whether it'll continue evolving forever, or will reach a fixed point.

One starts by writing a combinator expression like s[s[s[s]]][s[s[s]]][s[s]] (**S**(**S**(**SS**))(**S**(**SS**))(**SS**)) in an explicitly "functional" form:

f[f[f[s, f[s, f[s, s]]], f[s, f[s, s]]], f[s, s]]

Then one imagines f[x,y] as being a function with explicit (say, integer) values. One replaces s by some explicit value (say an integer), then defines values for f[1,1], f[1,2], etc.

As a first example, let's say that we take s = 1 and f[x_,y_]=x+y. Then we can "evaluate" the combinator expression above as and in this case the value at the root just counts the total size (i.e. LeafCount).

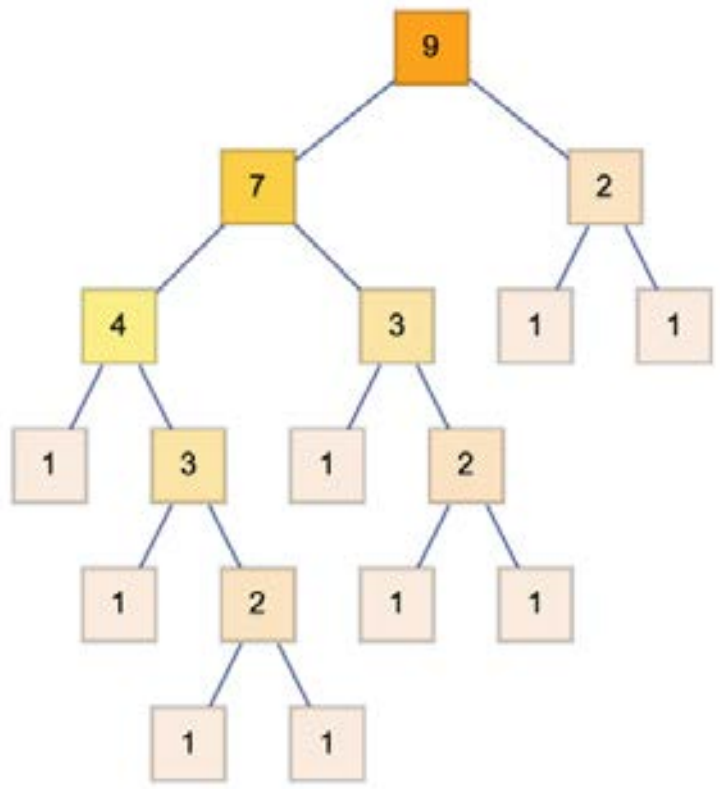

But by changing f one can probe other aspects of the combinator expression tree. And what was discovered in 2000 is that there's a complete way to test for infinite growth by setting up 39 possible values, and making f[x,y] be a particular ("tree automaton") "multiplication table" for these values:

| • | 0 | 1 | 2 | 3 | 4 | 5 | 6 | 7 | 8 | 9 | 10 | 11 | 12 | 13 | 14 | 15 | 16 | 17 | 18 | 19 | 20 | 21 | 22 | 23 | 24 | 25 | 26 | 27 | 28 | 29 | 30 | 31 | 32 | 33 | 34 | 35 | 36 | 37 | 38 |
|---|---|---|---|---|---|---|---|---|---|---|---|---|---|---|---|---|---|---|---|---|---|---|---|---|---|---|---|---|---|---|---|---|---|---|---|---|---|---|---|
| 0 | 3 | 7 | 7 | 5 | 6 | 11 | 18 | 18 | 18 | 13 | 18 | 15 | 15 | 18 | 18 | 18 | 18 | 28 | 28 | 23 | 26 | 26 | 28 | 28 | 28 | 28 | 28 | 28 | 28 | 28 | 28 | 28 | 28 | 28 | 28 | 28 | 28 | 28 | |
| 1 | 2 | 35 | 35 | 32 | 34 | 36 | 37 | 37 | 37 | 36 | 37 | 36 | 36 | 37 | 37 | 37 | 37 | 37 | 37 | 37 | 37 | 37 | 37 | 37 | 37 | 37 | 37 | 37 | 37 | 37 | 37 | 37 | 37 | 37 | 37 | 37 | 37 | 37 | |
| 2 | 2 | 36 | 36 | 33 | 36 | 37 | 37 | 37 | 37 | 37 | 37 | 37 | 37 | 37 | 37 | 37 | 37 | 37 | 37 | 37 | 37 | 37 | 37 | 37 | 37 | 37 | 37 | 37 | 37 | 37 | 37 | 37 | 37 | 37 | 37 | 37 | 37 | 37 | |
| 3 | 4 | 2 | 2 | 1 | 2 | 9 | 8 | 10 | 8 | 9 | 10 | 12 | 12 | 14 | 14 | 16 | 16 | 24 | 18 | 19 | 20 | 21 | 22 | 26 | 24 | 25 | 26 | 27 | 28 | 29 | 30 | 31 | 32 | 33 | 34 | 35 | 36 | 37 | |
| 4 | 1 | 36 | 36 | 27 | 36 | 36 | 37 | 37 | 37 | 37 | 37 | 37 | 37 | 37 | 37 | 37 | 37 | 37 | 37 | 37 | 37 | 37 | 37 | 37 | 37 | 37 | 37 | 37 | 37 | 37 | 37 | 37 | 37 | 37 | 37 | 37 | 37 | 37 | |
| 5 | 19 | 32 | 32 | 17 | 32 | 27 | 33 | 33 | 33 | 33 | 33 | 33 | 33 | 33 | 33 | 33 | 33 | 33 | 33 | 33 | 31 | 31 | 33 | 33 | 33 | 33 | 33 | 33 | 33 | 31 | 31 | 31 | 33 | 33 | 36 | 36 | 36 | 37 | |
| 6 | 20 | 34 | 34 | 29 | 34 | 36 | 36 | 36 | 36 | 36 | 36 | 36 | 36 | 36 | 36 | 36 | 36 | 36 | 36 | 36 | 36 | 36 | 36 | 36 | 36 | 36 | 36 | 36 | 36 | 36 | 36 | 36 | 36 | 36 | 36 | 36 | 36 | 37 | |
| 7 | 21 | 35 | 35 | 30 | 35 | 36 | 36 | 36 | 36 | 36 | 36 | 36 | 36 | 36 | 36 | 36 | 36 | 36 | 36 | 36 | 36 | 36 | 36 | 36 | 36 | 36 | 36 | 36 | 36 | 36 | 36 | 36 | 36 | 36 | 36 | 36 | 36 | 37 | |
| 8 | 20 | 36 | 36 | 31 | 36 | 37 | 37 | 37 | 37 | 37 | 37 | 37 | 37 | 37 | 37 | 37 | 37 | 37 | 37 | 37 | 37 | 37 | 37 | 37 | 37 | 37 | 37 | 37 | 37 | 37 | 37 | 37 | 37 | 37 | 37 | 37 | 37 | 37 | |
| 9 | 19 | 36 | 36 | 33 | 36 | 37 | 37 | 37 | 37 | 37 | 37 | 37 | 37 | 37 | 37 | 37 | 37 | 37 | 37 | 37 | 37 | 37 | 37 | 37 | 37 | 37 | 37 | 37 | 37 | 37 | 37 | 37 | 37 | 37 | 37 | 37 | 37 | 37 | |
| 10 | 21 | 36 | 36 | 31 | 36 | 37 | 37 | 37 | 37 | 37 | 37 | 37 | 37 | 37 | 37 | 37 | 37 | 37 | 37 | 37 | 37 | 37 | 37 | 37 | 37 | 37 | 37 | 37 | 37 | 37 | 37 | 37 | 37 | 37 | 37 | 37 | 37 | 37 | |
| 11 | 25 | 36 | 36 | 22 | 36 | 36 | 37 | 37 | 37 | 36 | 37 | 36 | 36 | 37 | 37 | 37 | 37 | 37 | 37 | 36 | 36 | 36 | 37 | 37 | 37 | 37 | 37 | 37 | 37 | 37 | 37 | 37 | 37 | 37 | 37 | 37 | 37 | 37 | |
| 12 | 25 | 36 | 36 | 33 | 36 | 37 | 37 | 37 | 37 | 37 | 37 | 37 | 37 | 37 | 37 | 37 | 37 | 37 | 37 | 37 | 37 | 37 | 37 | 37 | 37 | 37 | 37 | 37 | 37 | 37 | 37 | 37 | 37 | 37 | 37 | 37 | 37 | 37 | |
| 13 | 30 | 36 | 36 | 31 | 36 | 36 | 37 | 37 | 37 | 36 | 37 | 36 | 36 | 37 | 37 | 37 | 37 | 37 | 37 | 36 | 36 | 36 | 37 | 37 | 37 | 37 | 37 | 37 | 37 | 37 | 37 | 37 | 37 | 37 | 37 | 37 | 37 | 37 | |
| 14 | 30 | 36 | 36 | 31 | 36 | 37 | 37 | 37 | 37 | 37 | 37 | 37 | 37 | 37 | 37 | 37 | 37 | 37 | 37 | 37 | 37 | 37 | 37 | 37 | 37 | 37 | 37 | 37 | 37 | 37 | 37 | 37 | 37 | 37 | 37 | 37 | 37 | 37 | |
| 15 | 31 | 36 | 36 | 31 | 36 | 36 | 37 | 37 | 37 | 36 | 37 | 36 | 36 | 37 | 37 | 37 | 37 | 37 | 37 | 36 | 36 | 36 | 37 | 37 | 37 | 37 | 37 | 37 | 37 | 37 | 37 | 37 | 37 | 37 | 37 | 37 | 37 | 37 | |
| 16 | 31 | 36 | 36 | 31 | 36 | 37 | 37 | 37 | 37 | 37 | 37 | 37 | 37 | 37 | 37 | 37 | 37 | 37 | 37 | 37 | 37 | 37 | 37 | 37 | 37 | 37 | 37 | 37 | 37 | 37 | 37 | 37 | 37 | 37 | 37 | 37 | 37 | 37 | |
| 17 | 36 | 36 | 36 | 35 | 36 | 37 | 37 | 37 | 37 | 37 | 37 | 37 | 37 | 37 | 37 | 37 | 37 | | 37 | 37 | | | | 37 | | 37 | 37 | | 37 | | | | | | | | | | |
| 18 | 36 | 37 | 37 | 37 | 37 | 37 | 37 | 37 | 37 | 37 | 37 | 37 | 37 | 37 | 37 | 37 | 37 | 37 | 37 | 37 | 37 | 37 | 37 | 37 | 37 | 37 | 37 | 37 | 37 | 37 | 37 | 37 | 37 | 37 | 37 | 37 | 37 | 37 | |
| 19 | 27 | 37 | 37 | 36 | 37 | 37 | 37 | 37 | 37 | 37 | 37 | 37 | 37 | 37 | 37 | 37 | 37 | 37 | 37 | 37 | | | 37 | 37 | 37 | 37 | 37 | 37 | 37 | | | | 37 | 37 | | | | | |
| 20 | 32 | | | 37 | | | | | | | | | | | | | | | | | | | | | | | | | | | | | | | | | | | |
| 21 | 33 | | | 37 | | | | | | | | | | | | | | | | | | | | | | | | | | | | | | | | | | | |
| 22 | 37 | 37 | 37 | 36 | 37 | 37 | 37 | 37 | 37 | 37 | 37 | 37 | 37 | 37 | 37 | 37 | 37 | | 37 | | | | | 37 | | | 37 | | 37 | | | | | | | | | | |
| 23 | 36 | 36 | 36 | 36 | 36 | 36 | 37 | 37 | 37 | 36 | 37 | 36 | 36 | 37 | 37 | 37 | 37 | 37 | 37 | 37 | 37 | 37 | 37 | 37 | 37 | 37 | 37 | 37 | 37 | 37 | 37 | 37 | 37 | 37 | 37 | 37 | 37 | 37 | |
| 24 | 36 | 36 | 36 | 36 | 36 | 37 | 37 | 37 | 37 | 37 | 37 | 37 | 37 | 37 | 37 | 37 | 37 | | 37 | 37 | | | | 37 | | 37 | 37 | | 37 | | | | | | | | | | |
| 25 | 36 | 37 | 37 | 37 | 37 | 37 | 37 | 37 | 37 | 37 | 37 | 37 | 37 | 37 | 37 | 37 | 37 | 37 | 37 | 37 | | | 37 | 37 | 37 | 37 | 37 | 37 | 37 | | | | 37 | 37 | | | | | |
| 26 | 36 | 36 | 36 | 36 | 36 | 37 | 37 | 37 | 37 | 37 | 37 | 37 | 37 | 37 | 37 | 37 | 37 | 37 | 37 | 37 | 37 | 37 | 37 | 37 | 37 | 37 | 37 | 37 | 37 | 37 | 37 | 37 | 37 | 37 | 37 | 37 | 37 | 37 | |
| 27 | 37 | 37 | 37 | 37 | 37 | 37 | 37 | 37 | 37 | 37 | 37 | 37 | 37 | 37 | 37 | 37 | 37 | | 37 | 37 | | | | 37 | | 37 | 37 | | 37 | | | | | | | | | | |
| 28 | 37 | 37 | 37 | 37 | 37 | 37 | 37 | 37 | 37 | 37 | 37 | 37 | 37 | 37 | 37 | 37 | 37 | 37 | 37 | 37 | 37 | 37 | 37 | 37 | 37 | 37 | 37 | 37 | 37 | 37 | 37 | 37 | 37 | 37 | 37 | 37 | 37 | 37 | |
| 29 | 34 | | | 37 | | | | | | | | | | | | | | | | | | | | | | | | | | | | | | | | | | | |
| 30 | 36 | | | 37 | | | | | | | | | | | | | | | | | | | | | | | | | | | | | | | | | | | |
| 31 | 37 | | | 37 | | | | | | | | | | | | | | | | | | | | | | | | | | | | | | | | | | | |
| 32 | 36 | 37 | 37 | 37 | 37 | 37 | 37 | 37 | 37 | 37 | 37 | 37 | 37 | 37 | 37 | 37 | 37 | | 37 | | | | | 37 | | | 37 | | 37 | | | | | | | | | | |
| 33 | 37 | 37 | 37 | 37 | 37 | 37 | 37 | 37 | 37 | 37 | 37 | 37 | 37 | 37 | 37 | 37 | 37 | | 37 | | | | | 37 | | | 37 | | 37 | | | | | | | | | | |
| 34 | 35 | | | | | | | | | | | | | | | | | | | | | | | | | | | | | | | | | | | | | | |
| 35 | 36 | | | | | | | | | | | | | | | | | | | | | | | | | | | | | | | | | | | | | | |
| 36 | 37 | | | | | | | | | | | | | | | | | | | | | | | | | | | | | | | | | | | | | | |
| 37 | | | | | | | | | | | | | | | | | | | | | | | | | | | | | | | | | | | | | | | |
| 38 | | | | | | | | | | | | | | | | | | | | | | | | | | | | | | | | | | | | | | | |

Bright red (value 38) represents the presence of an infinite growth seed—and once one exists, f makes it propagate up to the root of the tree. And with this setup, if we replace s by the value 0, the combinator expression above can be "evaluated" as:

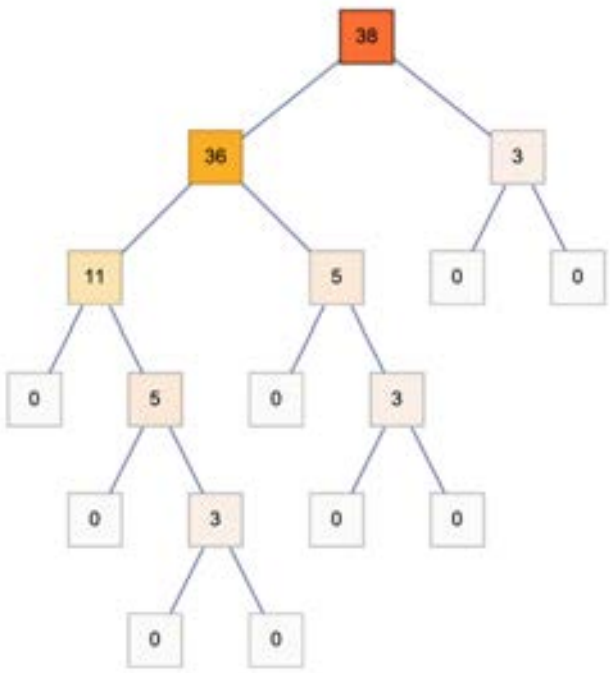

At successive steps in the evolution we get:

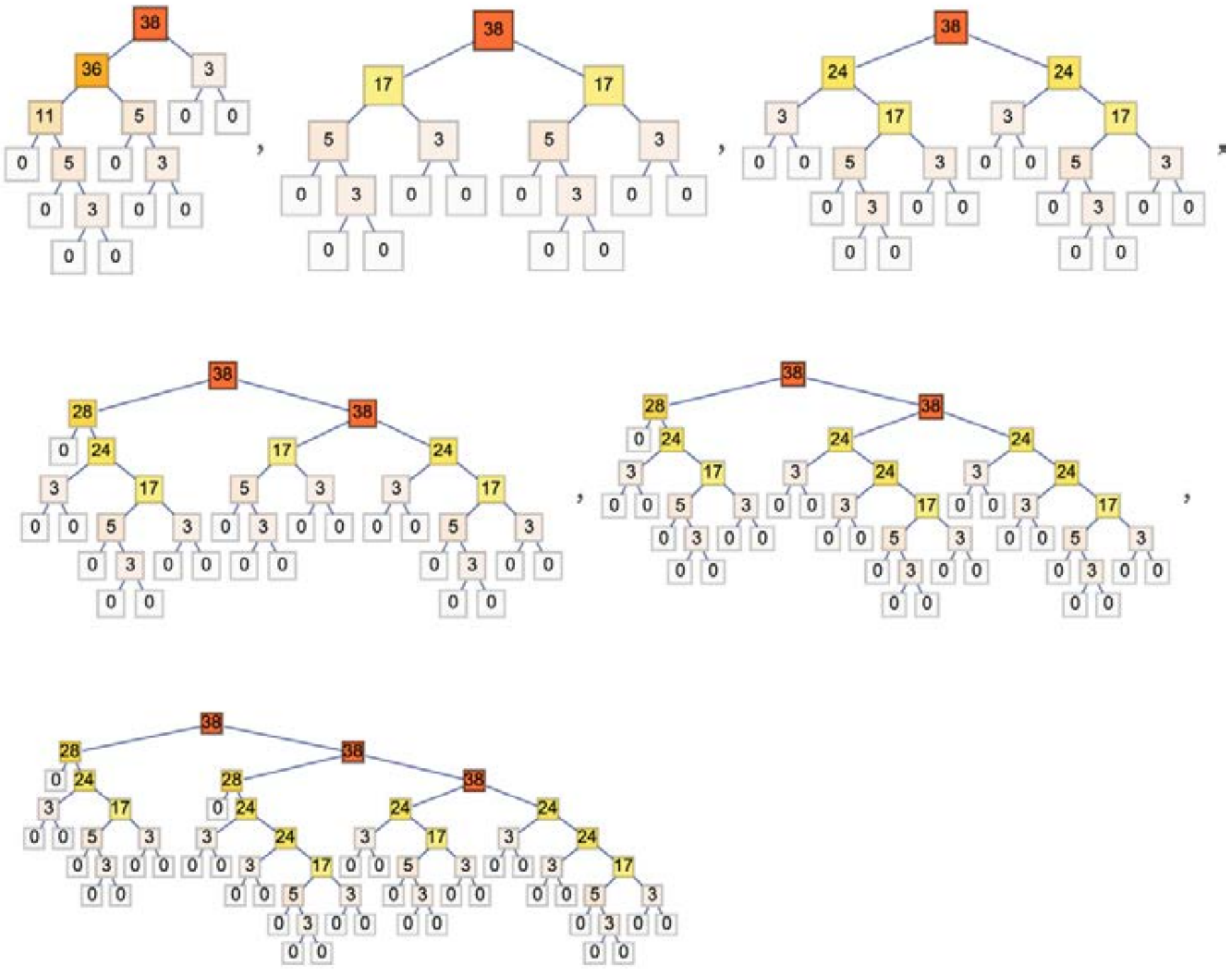

Or after 8 steps:

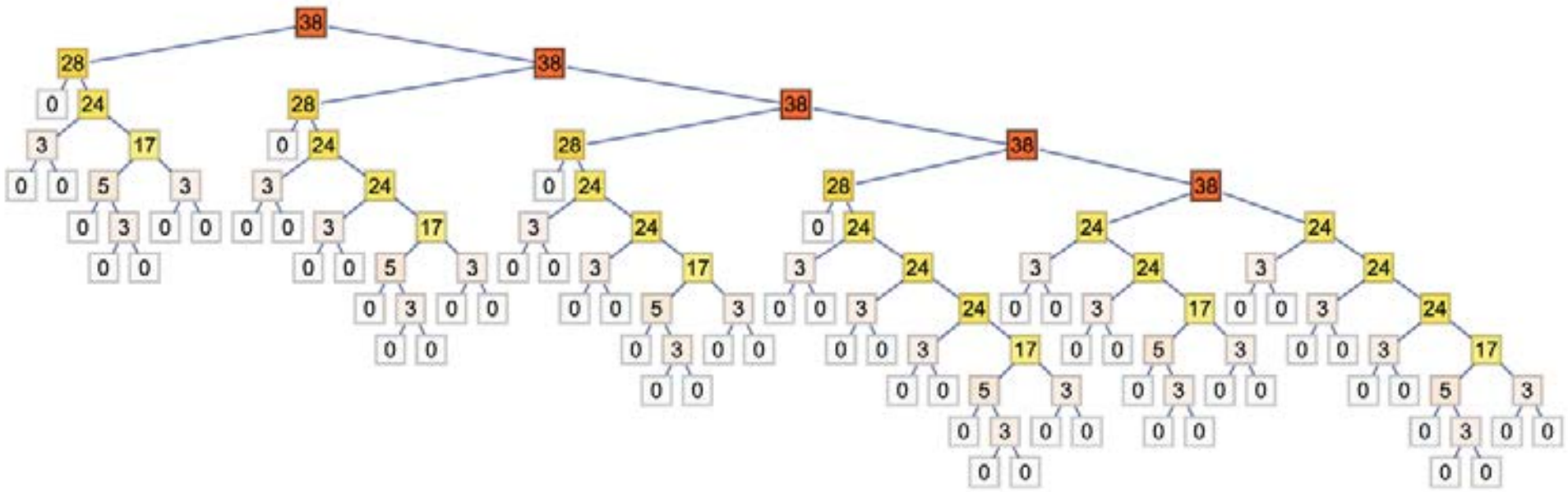

The “lowest 38” is always at the top of the subtree where the match occurs, serving as a “witness” of the fact that this subtree is an infinite growth seed.

Here are some sample size-7 combinator expressions, showing how the two that lead to infinite growth are identified:

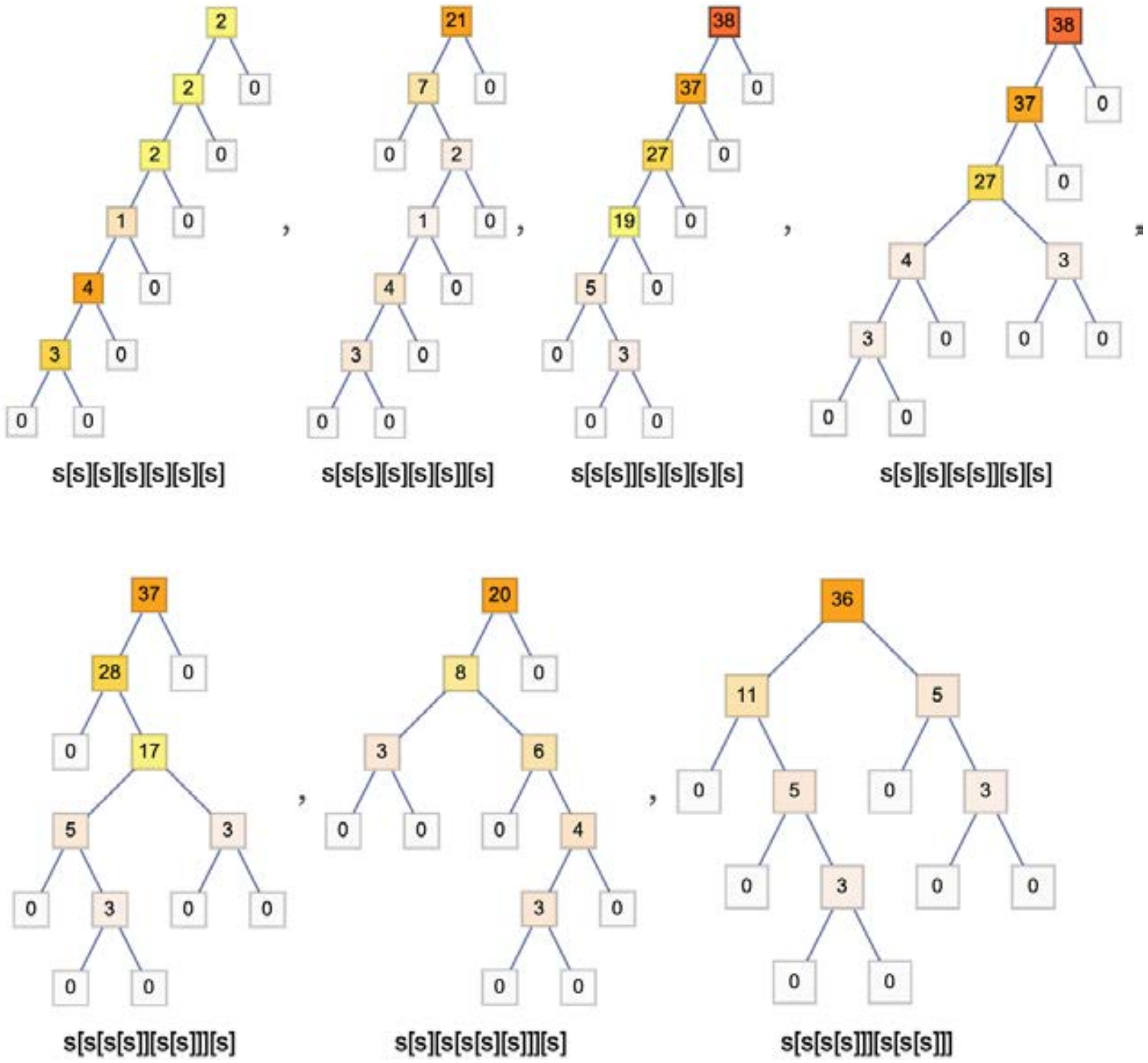

If we were dealing with combinator expressions involving both S and K we know that it's in general undecidable whether a particular expression will halt. So what does it mean that there's a decidable way to determine whether an expression involving only S halts?

One might assume it's a sign that S alone is somehow computationally trivial. But there's more to this issue. In the past, it has often been thought that a "computation" must involve starting with some initial ("input") state, then ending up at a fixed point corresponding to a final result. But that's certainly not how modern computing in practice works. The computer and its operating system do not completely stop when a particular computation is finished. Instead, the computer keeps running, but the user is given a signal to come and look at something that provides the output for the computation.

There's nothing fundamentally different about how computation universality works in a setup like this; it's just a "deployment" issue. And indeed the simplest possible examples of universality in cellular automata and Turing machines have been proved this way.

So how might this work for S combinator expressions? Basically any sophisticated computation has to live on top of an infinite combinator growth process. Or, put another way, the computation has to exist as some kind of "transient" of potentially unbounded length, that in effect "modulates" the infinite growth "carrier".

One would set up a program by picking an appropriate combinator expression from the infinite collection that lead to infinite growth. Then the evolution of the combinator expression would "run" the program. And one would use some computationally bounded process (perhaps a bounded version of a tree automaton) to identify when the result of the computation is ready—and one would "read it out" by using some computationally bounded "decoder".

My experience in the computational universe—as captured in the Principle of Computational Equivalence—is that once the behavior of a system is not "obviously simple", the system will be capable of sophisticated computation, and in particular will be computation universal. The S combinator is a strange and marginal case. At least in the ways we have looked at it here, its behavior is not "obviously simple". But we have not quite managed to identify things like the kind of seemingly random behavior that occurs in a system like rule 30, that are a hallmark of sophisticated computation, and probably computation universality.

There are really two basic possibilities. Either the S combinator alone is capable of sophisticated computation, and there is, for example, computational irreducibility in determining the outcome of a long S combinator evolution. Or the S combinator is fundamentally computationally reducible—and there is some approach (and maybe some new direction in mathematics) that "cracks it open", and allows one to readily predict everything that an S combinator expression will do.

I'm not sure which way it's going to go—although my almost-uniform experience over the last four decades has been that when I think some system is "too simple" to "do anything interesting" or show sophisticated computation, it eventually proves me wrong, often in bizarre and unexpected ways. (In the case of the S combinator, a possibility—like I found for example in register machines—is that sophisticated computation might first reveal itself in very subtle effects, like seemingly random off-by-one patterns.)

But whatever happens, it's amazing that 100 years after the invention of the S combinator there are still such big mysteries about it. In his original paper, Moses Schönfinkel expressed his surprise that something as simple as S and K were sufficient to achieve what we would now call universal computation. And it will be truly remarkable if in fact one can go even further, and S alone is sufficient: a minimal example of universal computation hiding in plain sight for a hundred years.

(By the way, in addition to ordinary "deterministic" combinator evolution with a particular evaluation scheme, one can also consider the "nondeterministic" case corresponding to all possible paths in the multiway graph. And in that case there's a question of categorizing infinite graphs obtained by non-terminating S combinator expressions—perhaps in terms of transfinite numbers.)

# Causal Graphs and the Physicalization of Combinators

Not long ago one wouldn't have had any reason to think that ideas from physics would relate to combinators. But our Wolfram Physics Project has changed that. And in fact it looks as if methods and intuition from our physics project—and the connections they make to things like relativity—may give some interesting new insights into combinators, and may in fact make their operation a little less mysterious.

In our physics project we imagine that the universe consists of a very large number of abstract elements ("atoms of space") connected by relations—as represented by a hypergraph. The behavior of the universe—and the progression of time—is then associated with repeated rewriting of this hypergraph according to a certain set of (presumably local) rules.

It's certainly not the same as the way combinators work, but there are definite similarities. In combinators, the basic "data structure" is not a hypergraph, but a binary tree. But combinator expressions evolve by repeated rewriting of this tree according to rules that are local on the tree.

There's a kind of intermediate case that we've often used as a toy model for aspects of physics (particularly quantum mechanics): string substitution systems. A combinator expression can be written out "linearly" (say as s[s][s][s[s[s]]][k][s]), but really it's tree-structured and hierarchical. In a string substitution system, however, one just has plain strings, consisting of sequences of characters, without any hierarchy. The system then evolves by repeatedly rewriting the string by applying some local string substitution rule.

For example, one could have a rule like {"A"→"BBB","BB"→"A"}. And just like with combinators, given a particular string—like "BBA"—there are different possible choices about where to apply the rule. And—again like with combinators—we can construct a multiway graph to represent all possible sequences of rewritings:

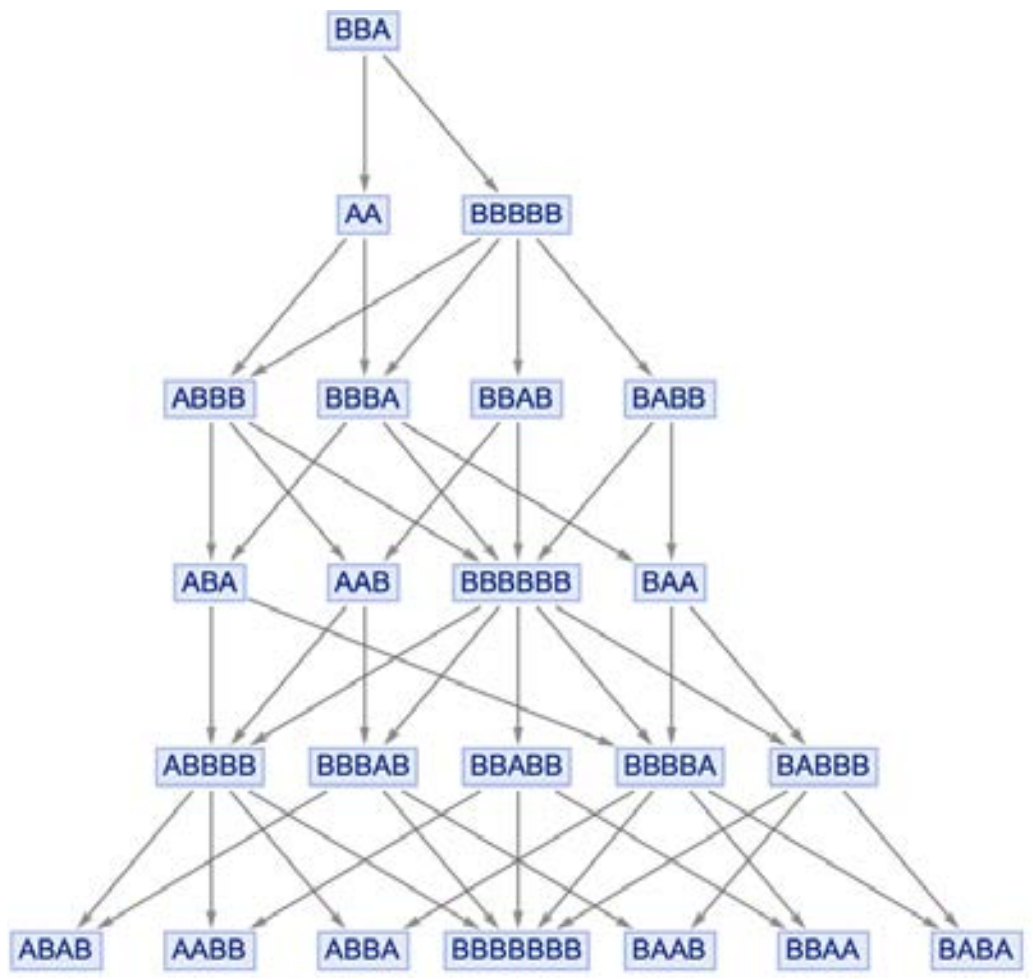

And again as with combinators we can define a particular "evaluation order" that determines which of the possible updates to the string to apply at each step—and that defines a path through the multiway graph.

For strings there aren't really the same notions of "innermost" and "outermost", but there are "leftmost" and "rightmost". Leftmost updating in this case would give the evolution history

{BBA, AA, BBBA, ABA, BBBBA, ABBA, BBBBBA, ABBBA, BBBBBBA, ABBBBA, BBBBBBBA}

which corresponds to the path:

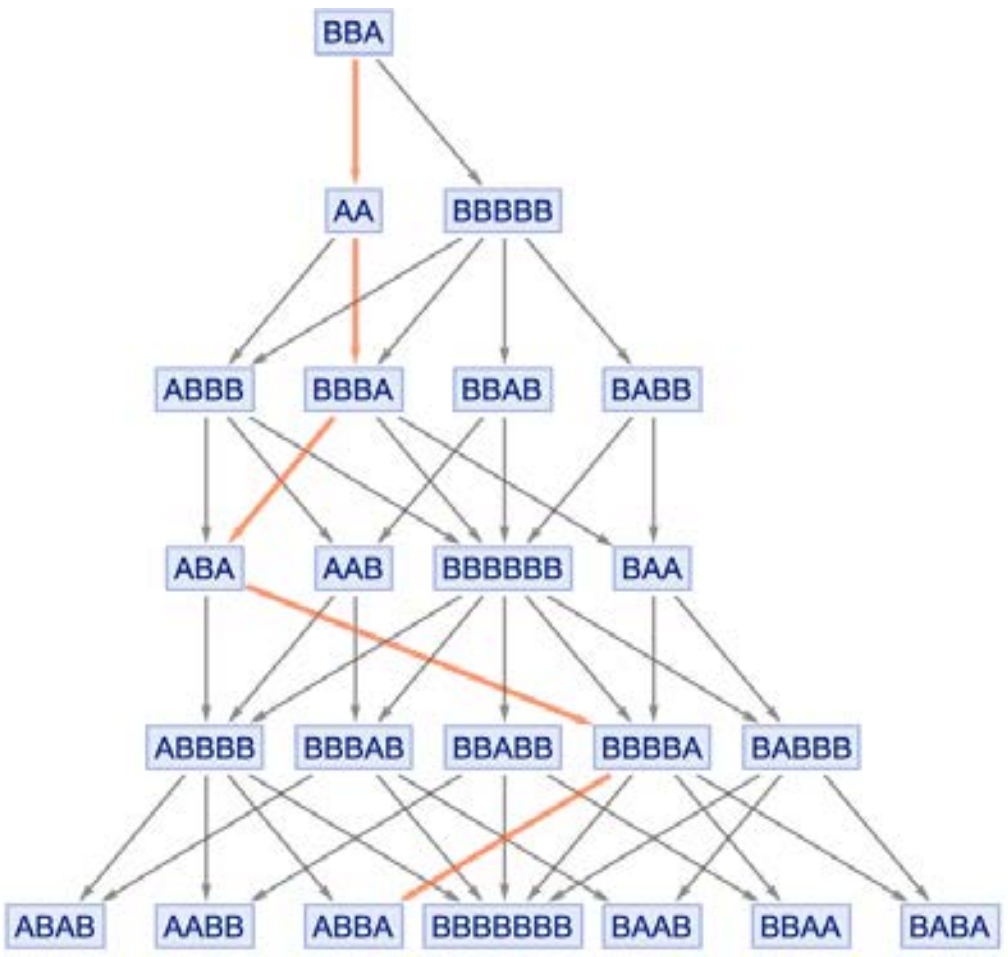

Here's the underlying evolution corresponding to that path, with the updating events indicated in yellow:

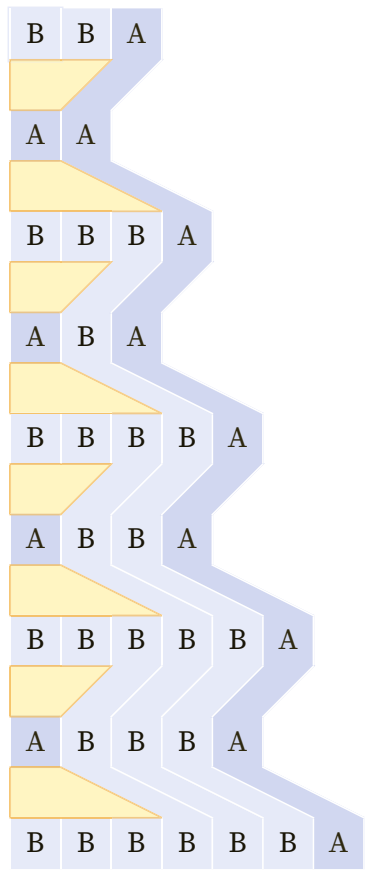

But now we can start tracing the “causal dependence” of one event on another. What characters need to have been produced as “output” from a preceding event in order to provide “input” to a new event? Let’s look at a case where we have a few more events going on:

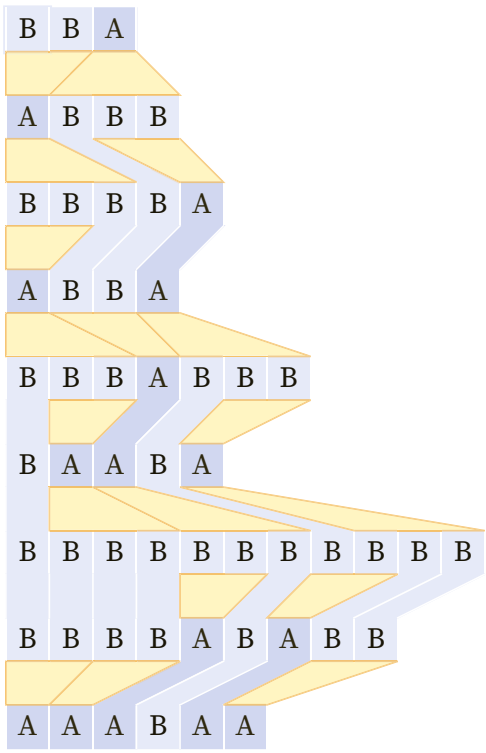

But now we can draw a causal graph that shows causal relationships between events, i.e. which events have to have happened in order to enable subsequent events:

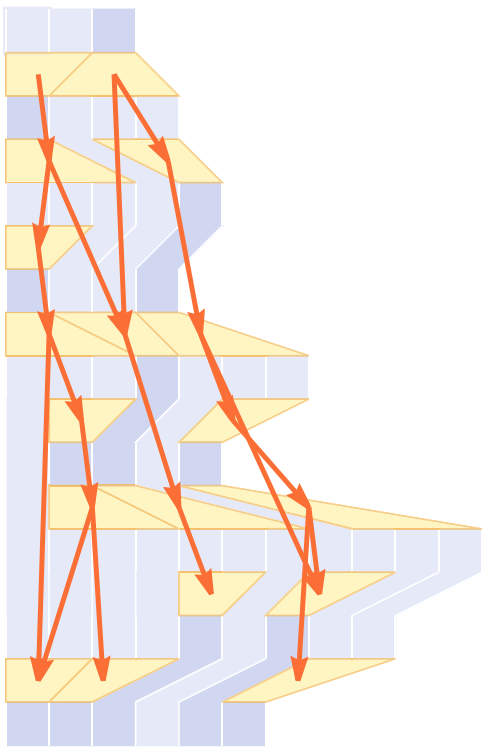

And at a physics level, if we’re an observer embedded in the system, operating according to the rules of the system, all we can ultimately “observe” is the “disembodied” causal graph, where the nodes are events, and the edges represent the causal relationships between these events:

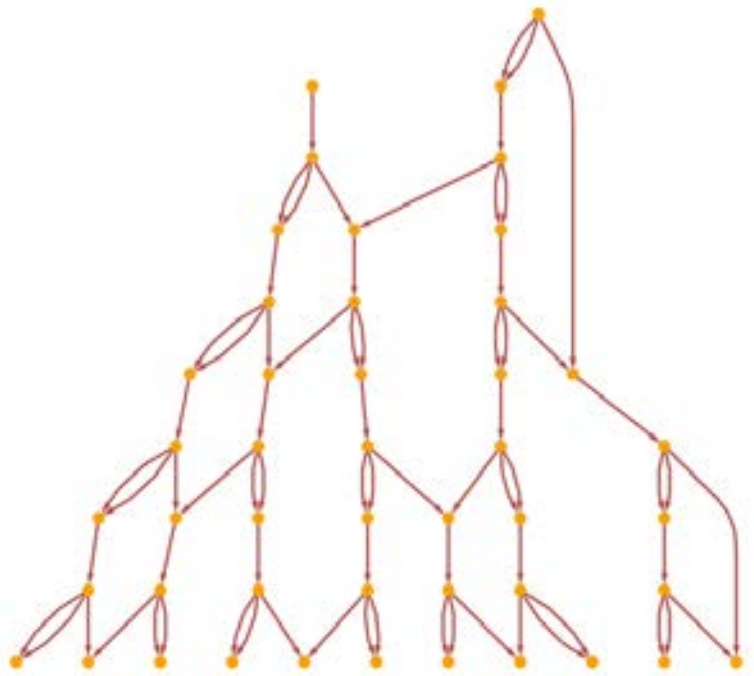

So how does this relate to combinators? Well, we can also create causal graphs for those—to get a different view of "what's going on" during combinator evolution.

There is significant subtlety in exactly how "causal dependence" should be defined for combinator systems (when is a copied subtree "different"?, etc.). Here I'll use a straightforward definition that'll give us an indication of how causal relationships in combinators work, but that's going to require further refinement to fit in with other definitions we want.

Imagine we just write out combinator expressions in a linear way. Then here's a combinator evolution:

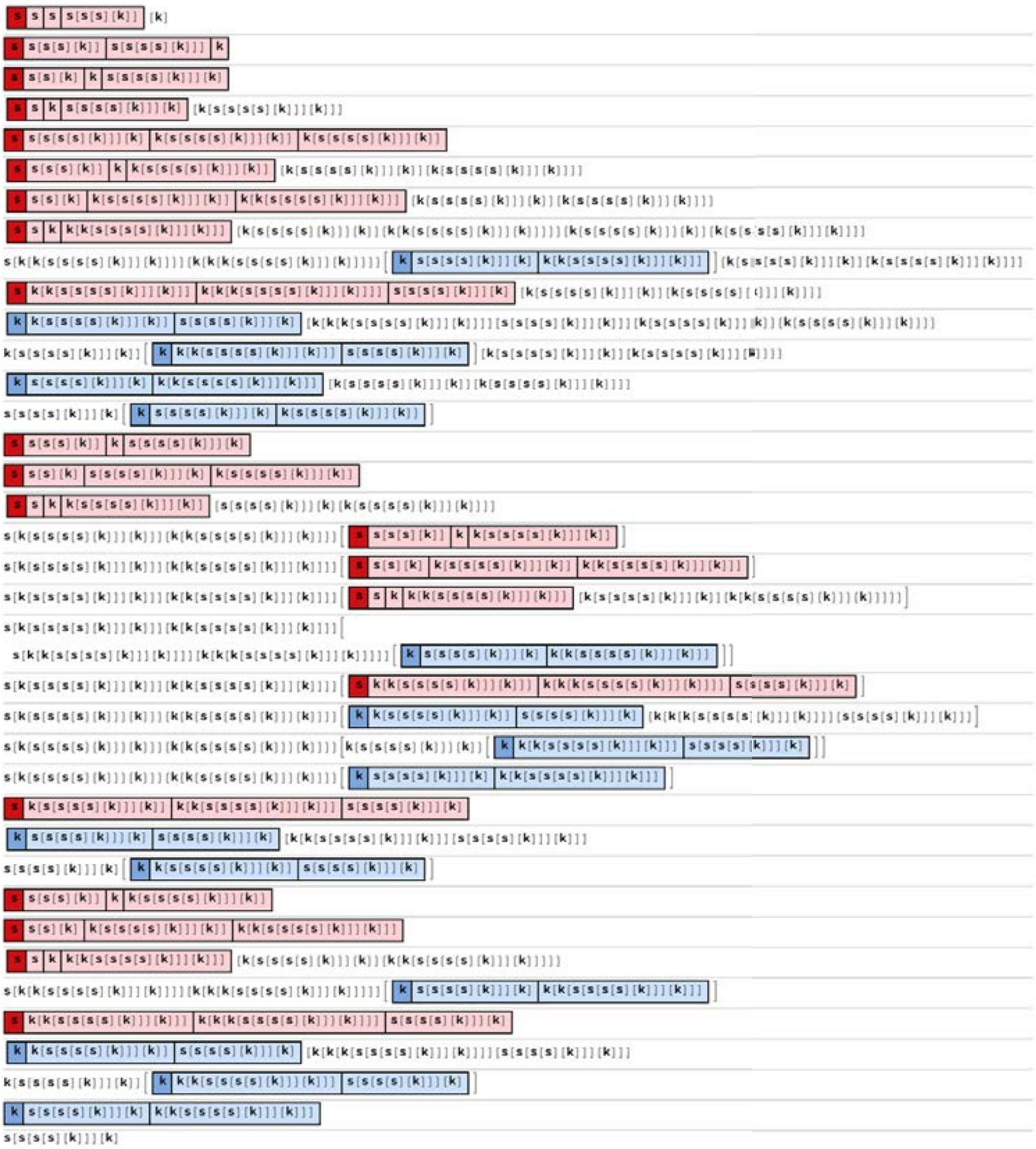

To understand causal relationships we need to trace "what gets rewritten to what"—and which previous rewriting events a given rewriting events "takes its input from". It's helpful to look at the rewriting process above in terms of trees:

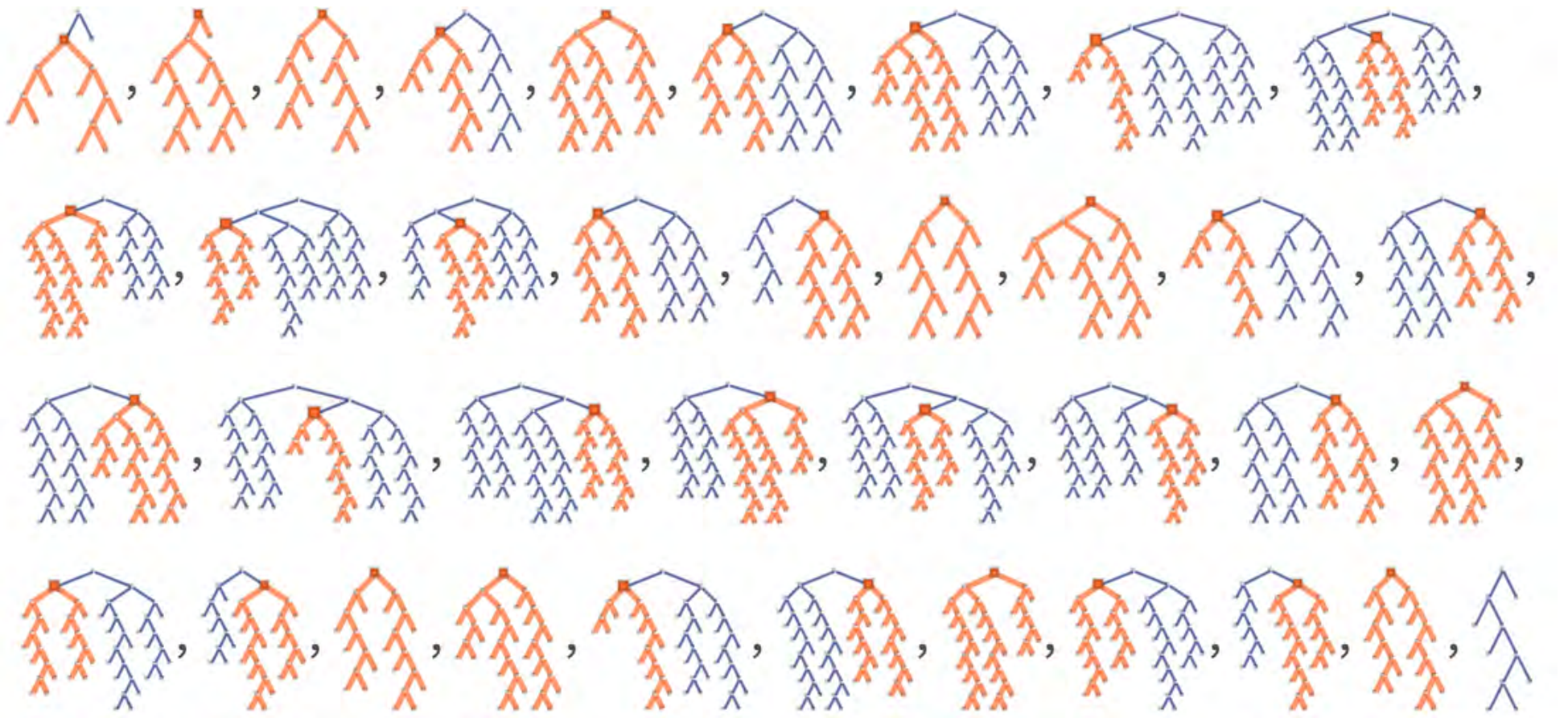

Going back to a textual representation, we can show the evolution in terms of "states", and the "events" that connect them. Then we can trace (in orange) what the causal relationships between the events are:

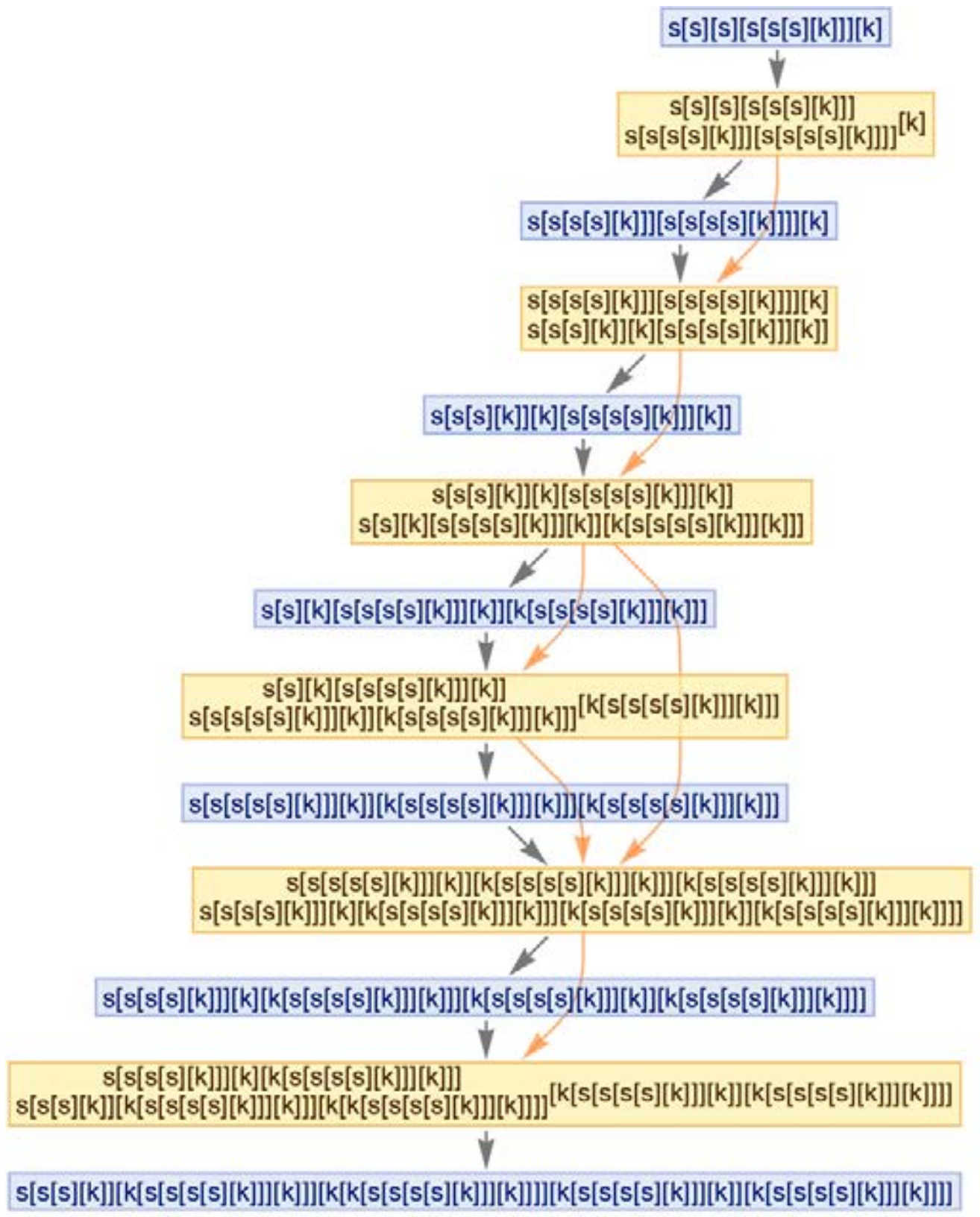

Continuing this for a few more steps we get:

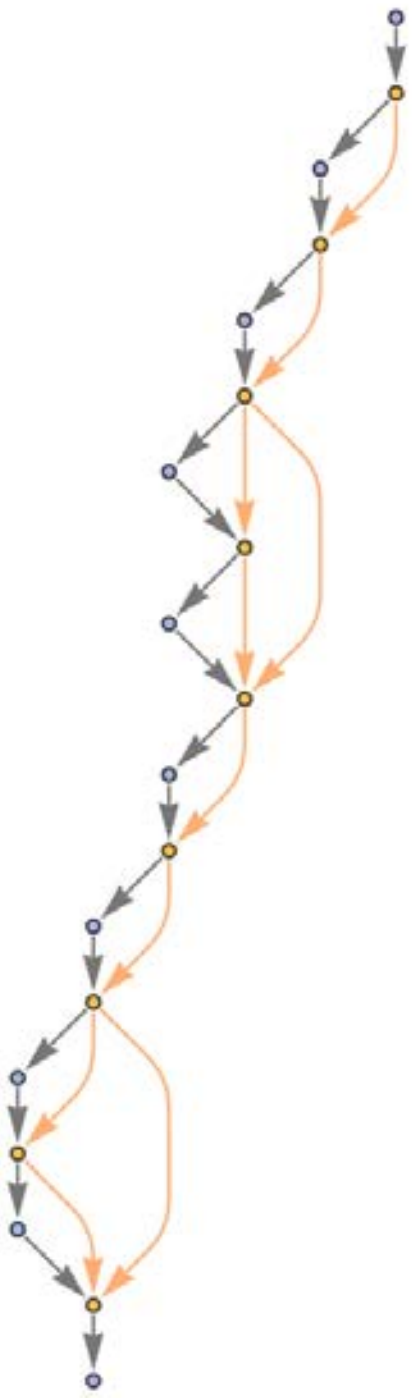

Now keeping only the causal graph, and continuing until the combinator evolution terminates, we get:

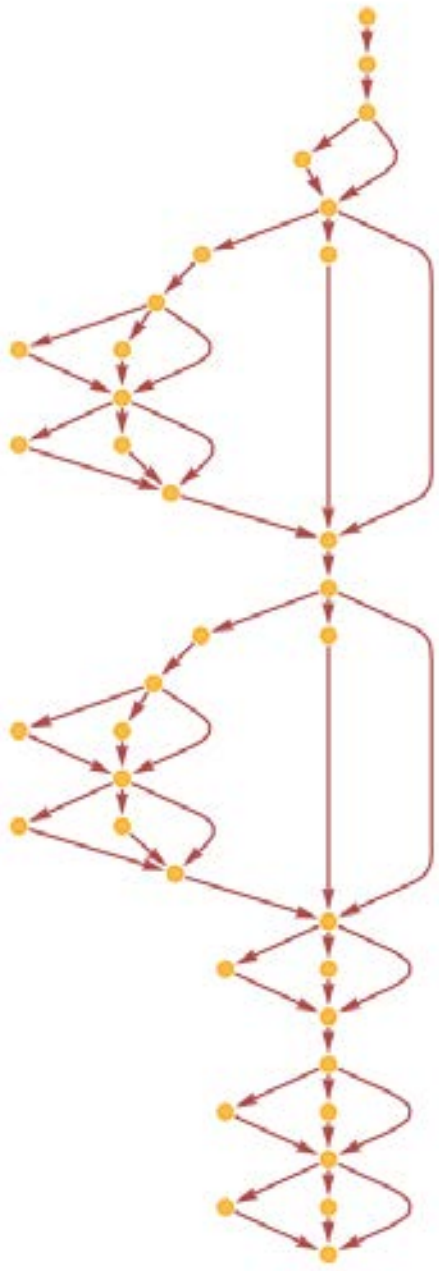

It's interesting to compare this with a plot that summarizes the succession of rewriting events:

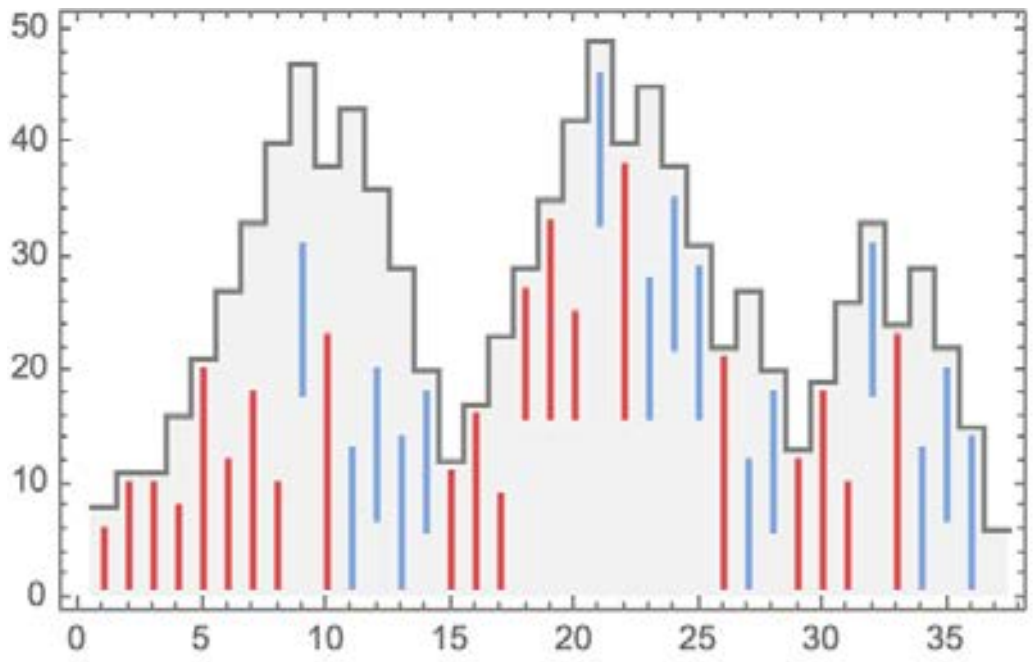

So what are we actually seeing in the causal graph? Basically it's showing us what "threads of evaluation" occur in the system. When there are different parts of the combinator expression that are in effect getting updated independently, we see multiple causal edges running in parallel. But when there's a synchronized evaluation that affects the whole system, we just see a single thread—a single causal edge.

The causal graph is in a sense giving us a summary of the structure of the combinator evolution, with many details stripped out. And even when the size of the combinator expression grows rapidly, the causal graph can still stay quite simple. So, for example, the growing combinator s[s][s][s[s]][s][s] has a causal graph that forms a linear chain with simple "side loops" that get systematically further apart:

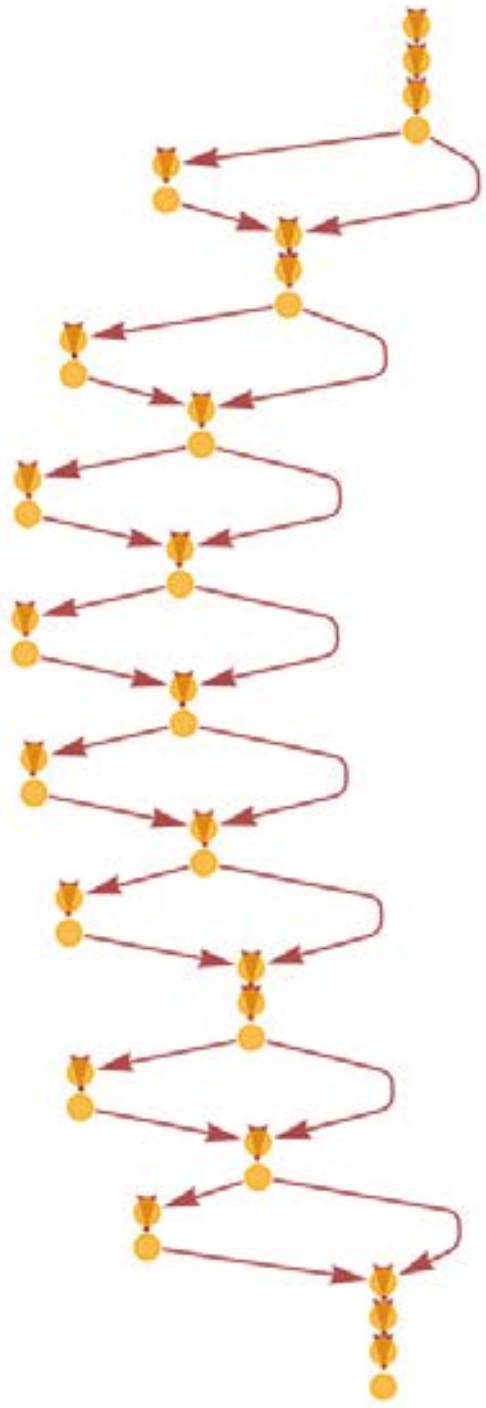

Sometimes it seems that the growth dies out because different parts of the combinator system become causally disconnected from each other:

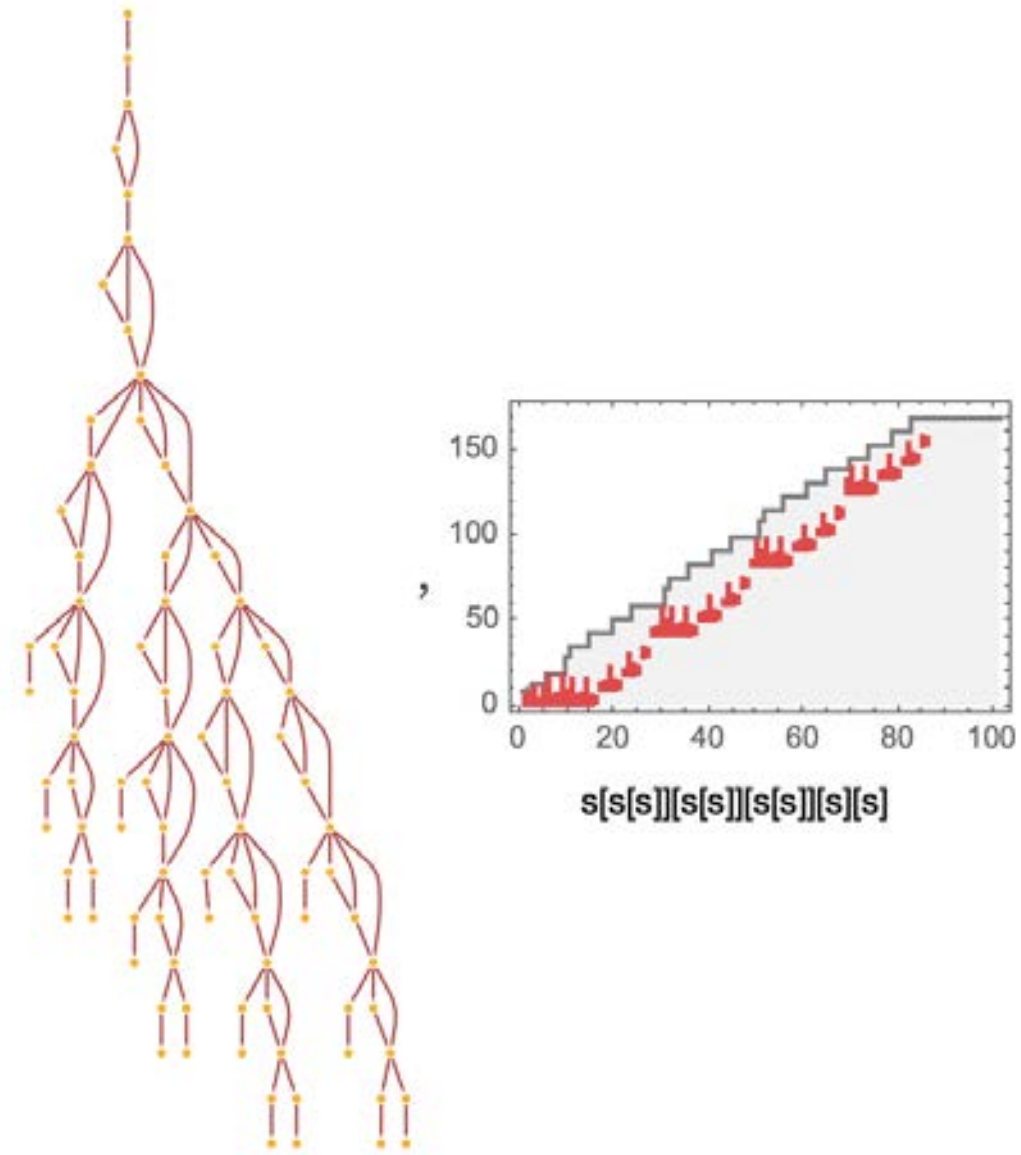

Here are a few other examples:

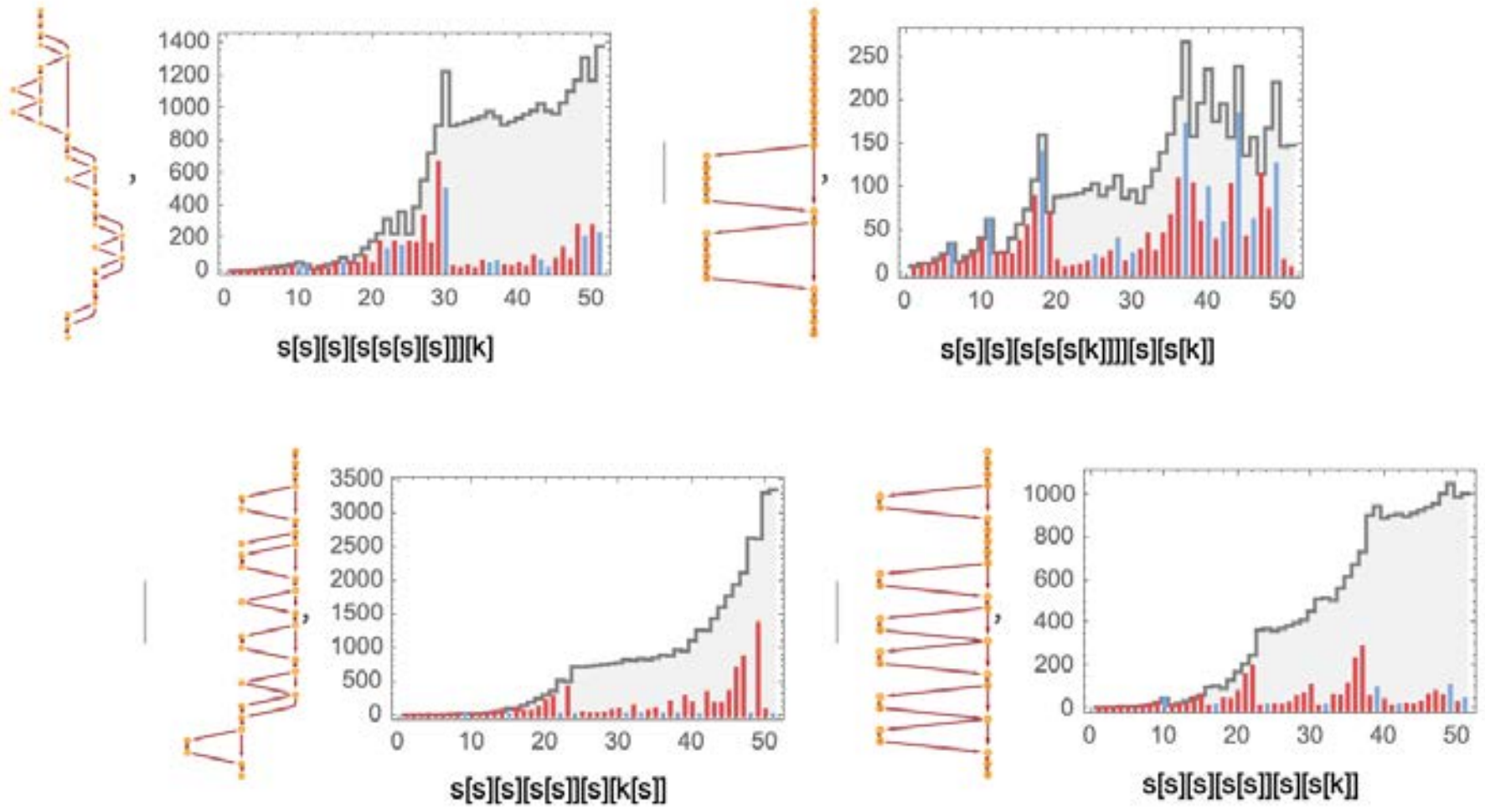

But do such causal graphs depend on the evaluation scheme used? This turns out to be a subtle question that depends sensitively on definitions of identity for abstract expressions and their subexpressions.

The first thing to say is that combinators are confluent, in the sense that different evaluation schemes—even if they take different paths—must always give the same final result whenever the evolution of a combinator expression terminates. And closely related to this is the fact that in the multiway graph for a combinator system, any branching must be accompanied by subsequent merging.

For both string and hypergraph rewriting rules, the presence of these properties is associated with another important property that we call causal invariance. And causal invariance is precisely the property that causal graphs produced by different updating orders must always be isomorphic. (And in our model of physics, this is what leads to relativistic invariance, general covariance, objective measurement in quantum mechanics, etc.)

So is the same thing true for combinators? It's complicated. Both string and hypergraph rewriting systems have an important simplifying feature: when you update something in them, it's reasonable to think of the thing you update as being "fully consumed" by the updating event, with a "completely new thing" being created as a result of the event.

But with combinators that's not such a reasonable picture. Because when there's an updating event, say for s[x][y][z], x can be a giant subtree that you end up "just copying", without, in a sense, "consuming" and "reconstituting". In the case of strings and hypergraphs, there's a clear distinction between elements of the system that are "involved in an update", and ones that aren't. But in a combinator system, it's not so obvious whether nodes buried deep in a subtree that's "just copied" should be considered "involved" or not.

There's a complicated interplay with definitions used in constructing multiway graphs. Consider a string rewriting system. Start from a particular state and then apply rewriting rules in all possible ways:

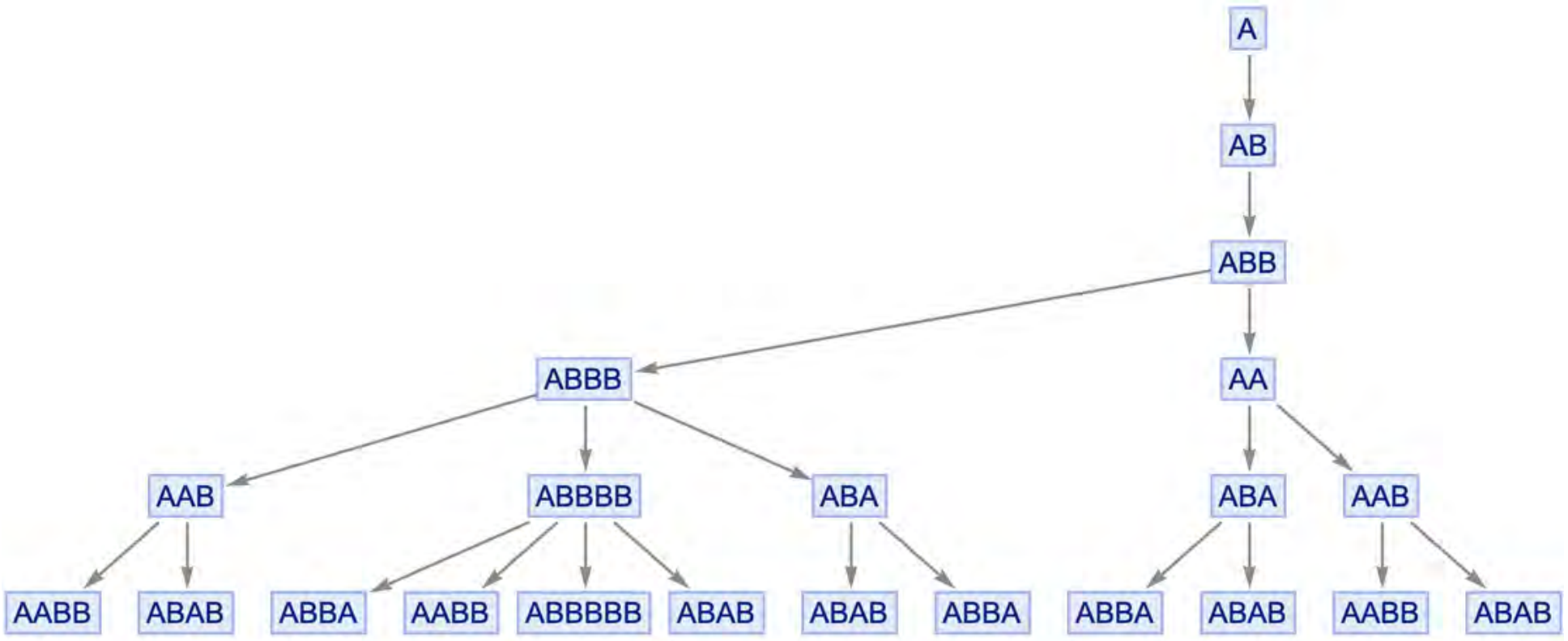

Absent anything else, this will just generate a tree of results. But the crucial idea behind multiway graphs is that when states are identical, they should be merged, in this case giving:

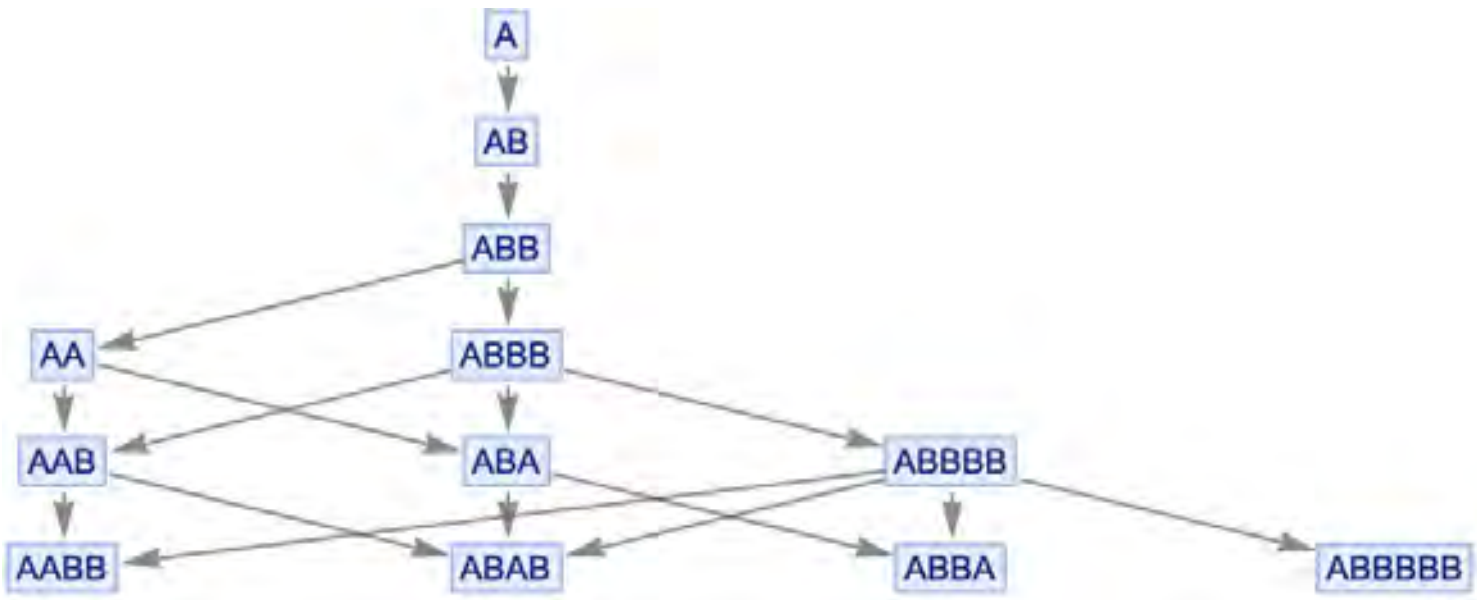

For strings it's very obvious what "being identical" means. For hypergraphs, the natural definition is hypergraph isomorphism. What about for combinators? Is it pure tree isomorphism, or should one take into account the "provenance" of subtrees?

(There are also questions like whether one should define the nodes in the multiway graph in terms of "instantaneous states" at all, or whether instead they should be based on "causal graphs so far", as obtained with particular event histories.)

These are subtle issues, but it seems pretty clear that with appropriate definitions combinators will show causal invariance, so that (appropriately defined) causal graphs will be independent of evaluation scheme.

By the way, in addition to constructing causal graphs for particular evolution histories, one can also construct multiway causal graphs representing all possible causal relationships both within and between different branches of history. This shows the multiway graph for the (terminating) evolution of s[s][s][s[s[k]]][k], annotated with casual edges:

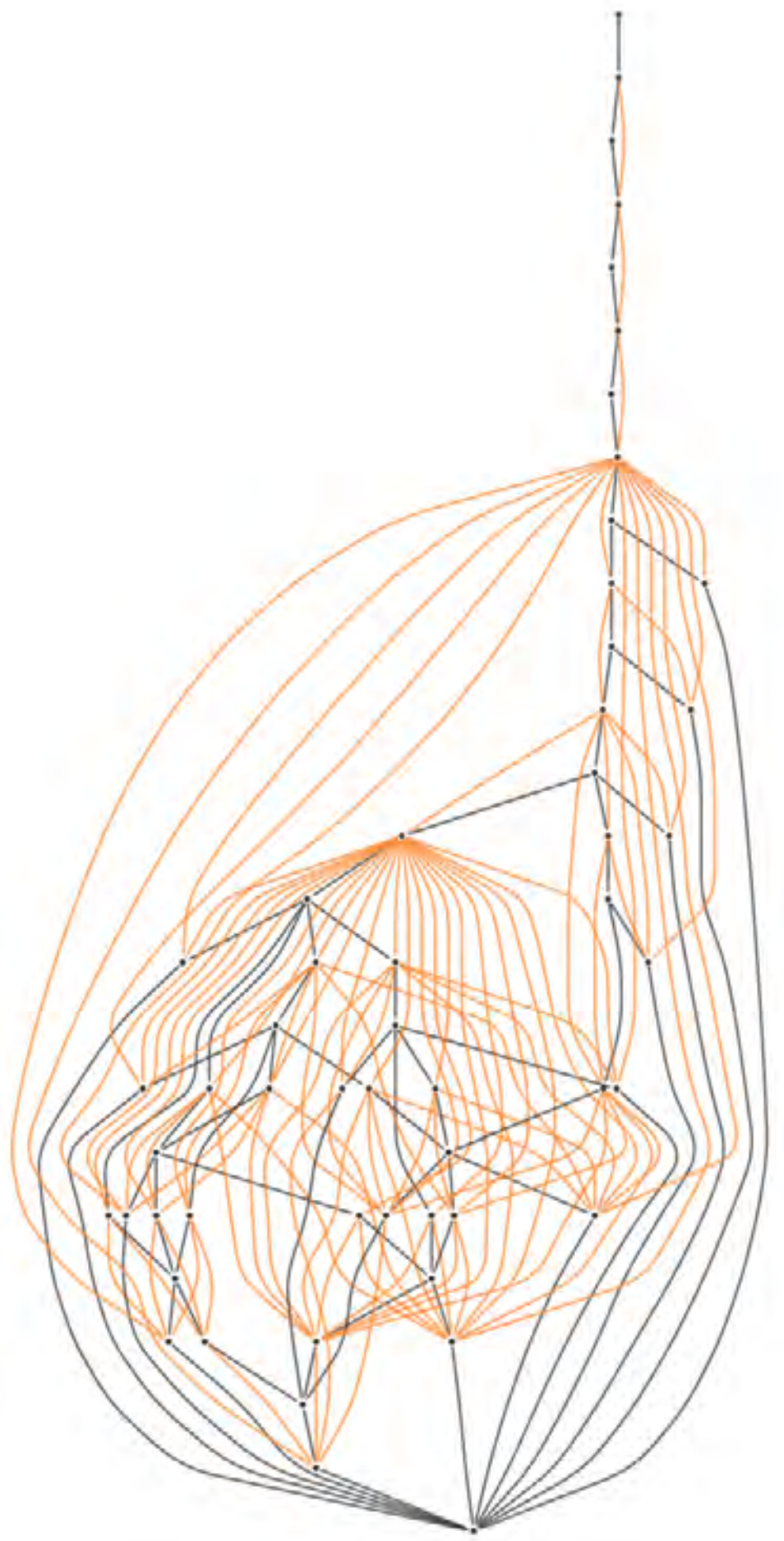

And here's the multiway causal graph alone in this case:

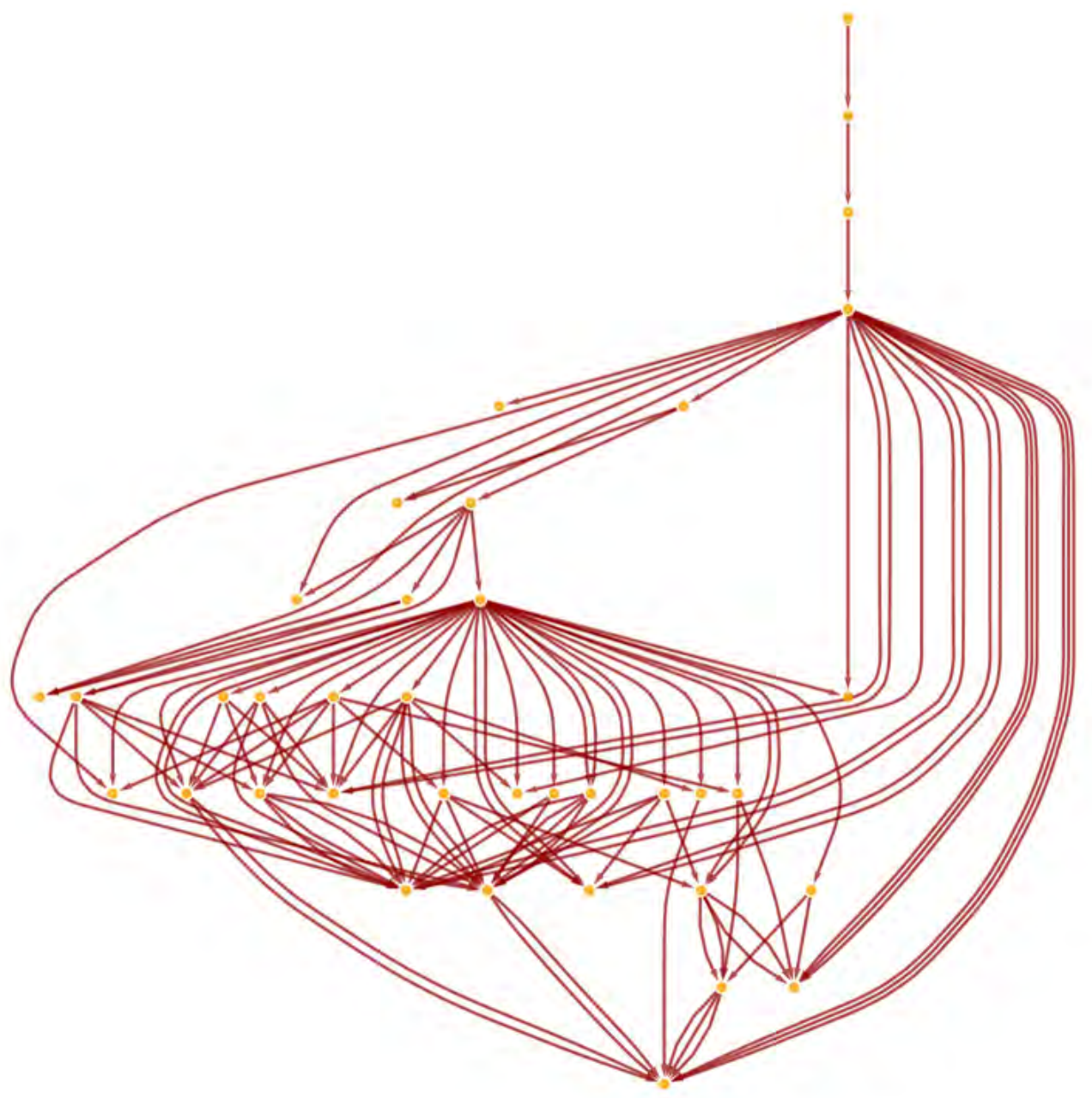

(And, yes, the definitions don't all quite line up here, so the individual instances of causal graphs that can be extracted here aren't all the same, as causal invariance would imply.)

The multiway causal graph for s[s[s]][s][s][s][s] shows a veritable explosion of causal edges:

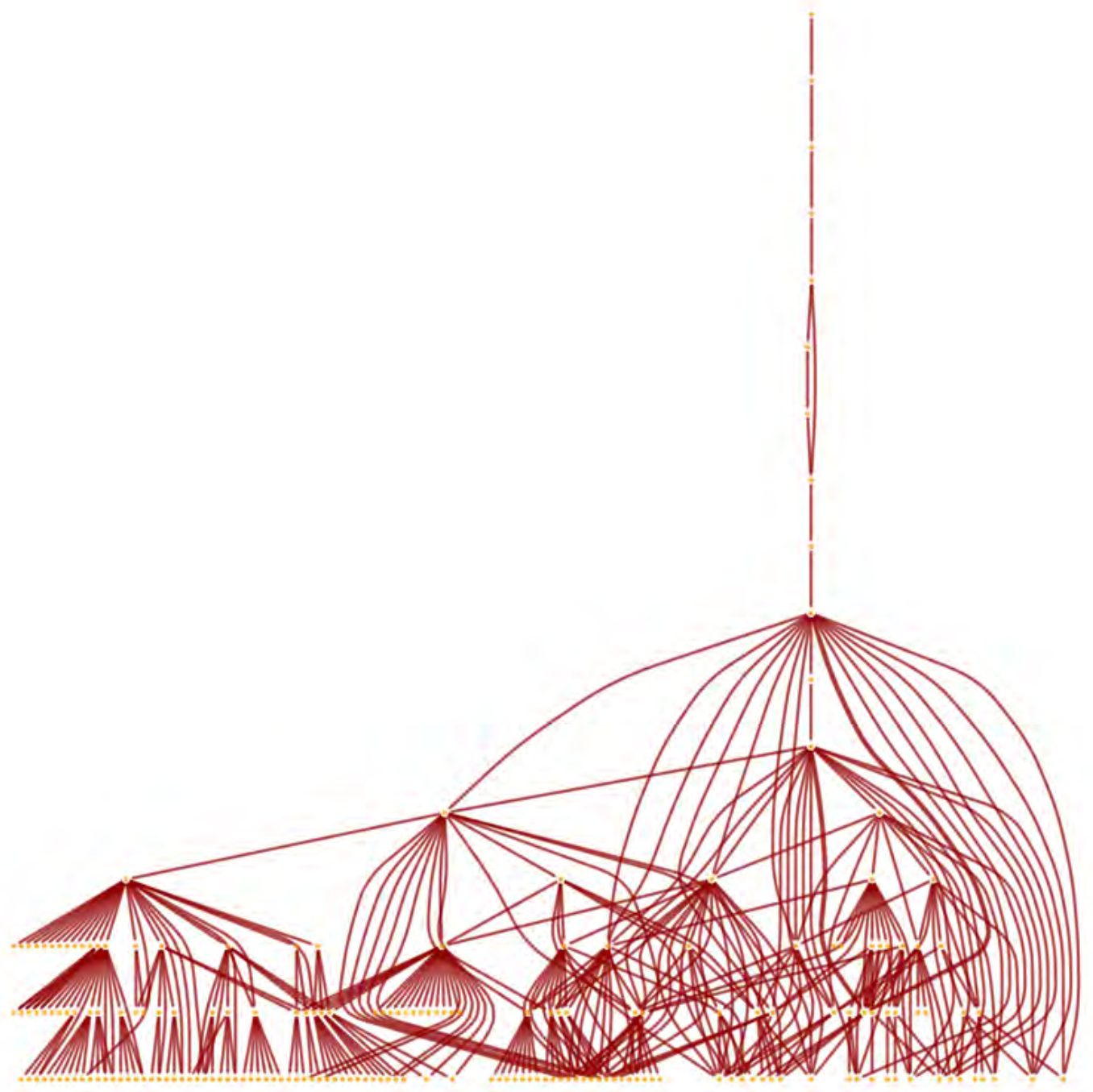

In our model of physics, the causal graph can be thought of as a representation of the structure of spacetime. Events that follow from each other are "timelike separated". Events that can be arranged so that none are timelike separated can be considered to form a "spacelike slice" (or a "surface of simultaneity"), and to be spacelike separated. (Different foliations of the causal graph correspond to different "reference frames" and identify different sets of events as being in the same spacelike slice.)

When we're dealing with multiway systems it's also possible for events to be associated with different "threads of history"—and so to be branchlike separated. But in combinator systems, there's yet another form of separation between events that's possible—that we can call "treelike separation".

Consider these two pairs of updating events:

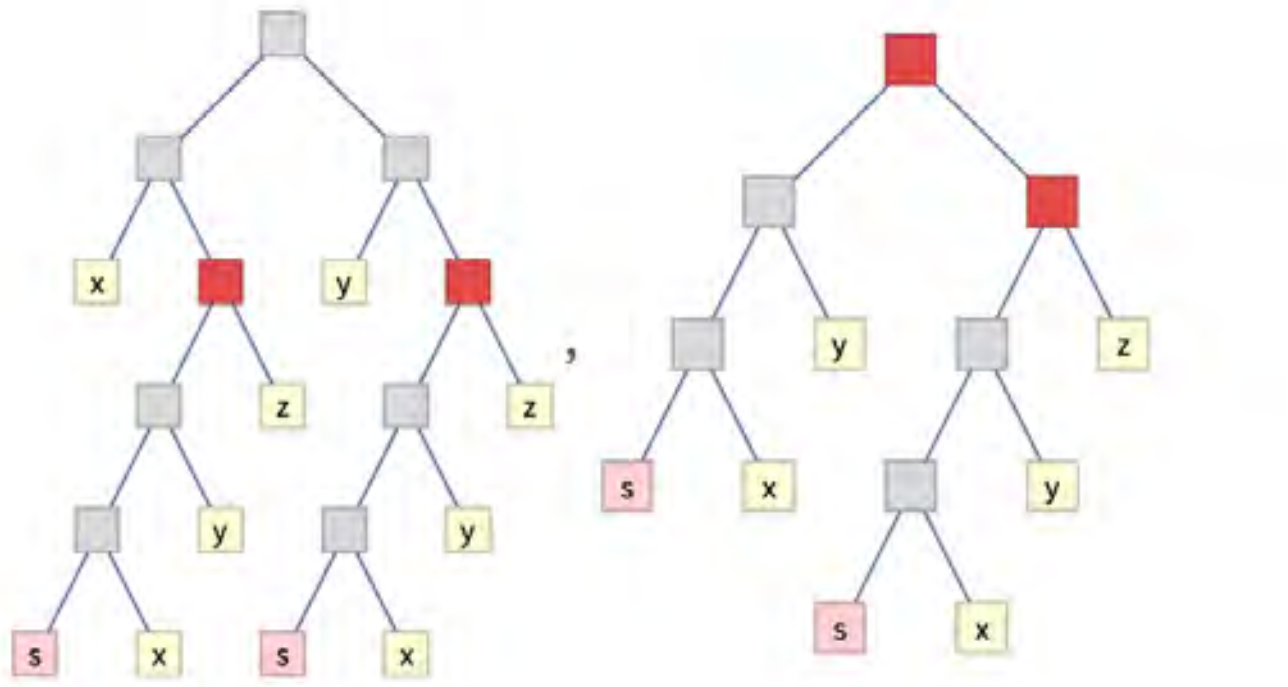

In the first case, the events are effectively "spacelike separated". They are connected by being in the same combinator expression, but they somehow appear at "distinct places". But what about the second case? Again the two events are connected by being in the same combinator expression. But now they're not really "at distinct places"; they're just "at distinct scales" in the tree.

One feature of hypergraph rewriting systems is that in large-scale limits the hypergraphs they produce can behave like continuous manifolds that potentially represent physical space, with hypergraph distances approximating geometric distances. In combinator systems there is almost inevitably a kind of nested structure that may perhaps be reminiscent of scale-invariant critical phenomena and ideas like scale relativity. But I haven't yet seen combinator systems whose limiting behavior produces something like finite-dimensional "manifold-like" space.

It's common to see "event horizons" in combinator causal graphs, in which different parts of the combinator system effectively become causally disconnected. When combinators reach fixed points, it's as if "time is ending"—much as it does in spacelike singularities in spacetime. But there are no doubt new "treelike" limiting phenomena in combinator systems, that may perhaps be reflected in properties of hyperbolic spaces.

One important feature of both string and hypergraph rewriting systems is that their rules are generally assumed to be somehow local, so that the future effect of any given element must lie within a certain “cone of influence”. Or, in other words, there’s a light cone which defines the maximum spacelike separation of events that can be causally connected when they have a certain timelike separation. In our model of physics, there’s also an “entanglement cone” that defines maximum branchlike separation between events.

But what about in combinator systems? The rules aren’t really “spatially local”, but they are “tree local”. And so they have a limited “tree cone” of influence, associated with a “maximum treelike speed”—or, in a sense, a maximum speed of scale change.

Rewriting systems based strings, hypergraphs and combinator expressions all have different simplifying and complexifying features. The relation between underlying elements (“characters arranged in sequence”) is simplest for strings. The notion of what counts as the same element is simplest for hypergraphs. But the relation between the “identities of elements” is probably simplest for combinator expressions.

Recall that we can always represent a combinator expression by a DAG in which we “build up from atoms”, sharing common subexpressions all the way up:

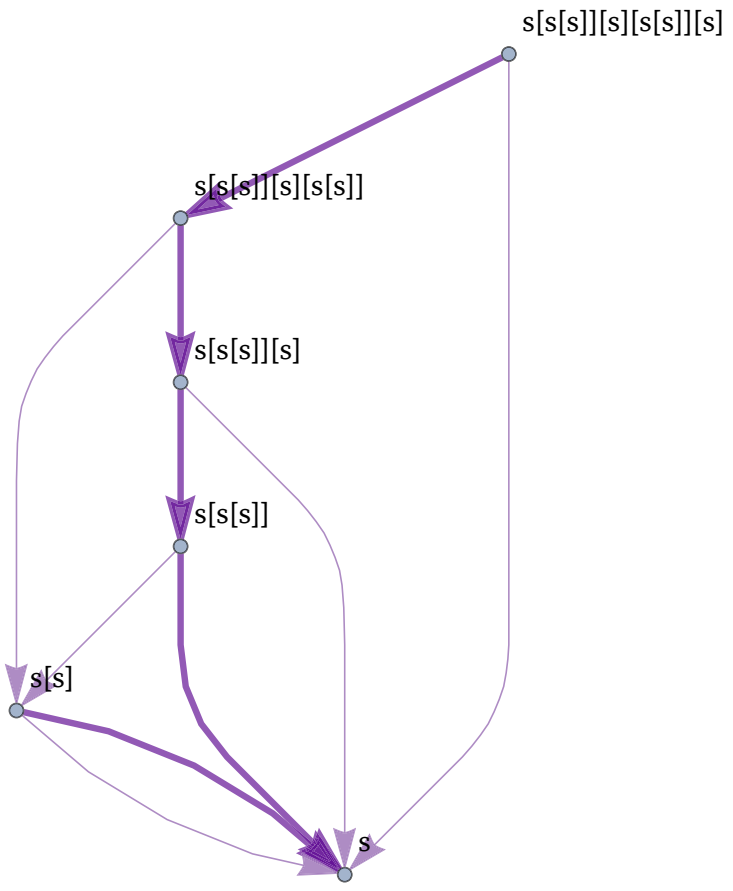

But what does combinator evolution look like in this representation? Let’s start from the extremely simple case of k[x][y], which in one step becomes just x. Here’s how we can represent this evolution process in DAGs:

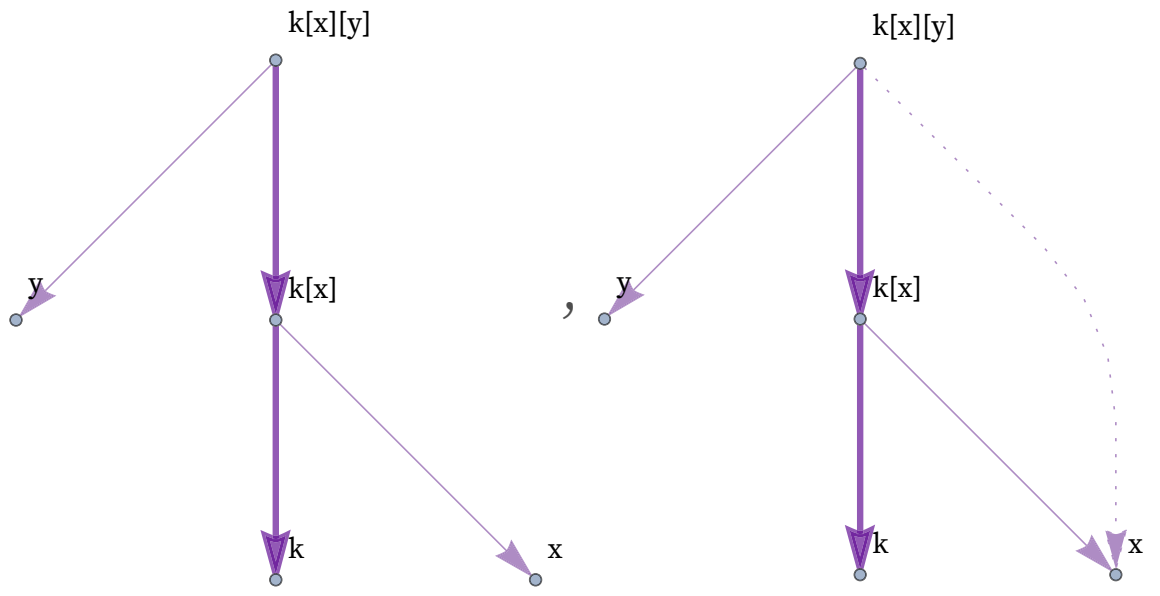

The dotted line in the second DAG indicates an update event, which in this case transforms k[x][y] to the “atom” x.

Now let’s consider s[x][y][z]. Once again there’s a dotted line that signifies the evolution:

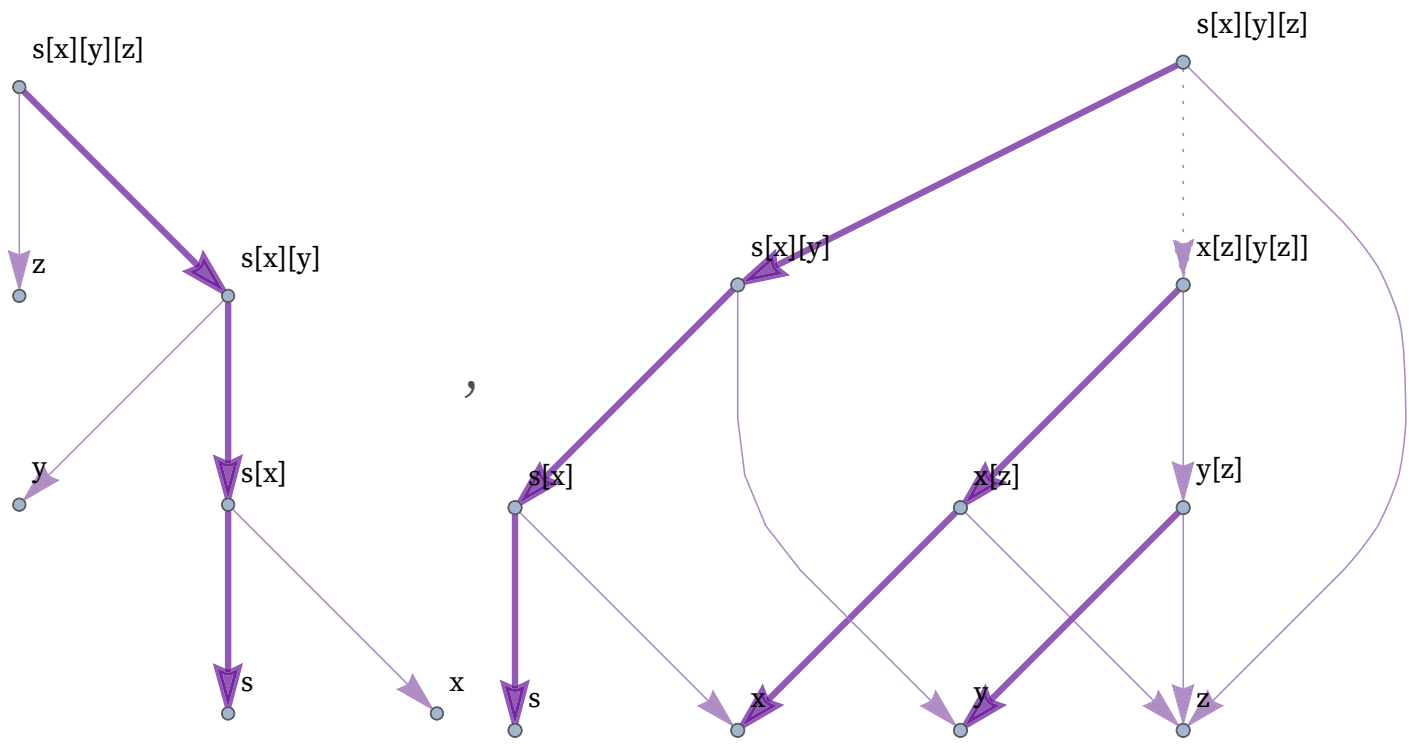

Now let’s add an extra wrinkle: consider not k[x][y] but s[k[x][y]]. The outer s doesn’t really do anything here. But it still has to be accounted for, in the sense that it has to be “wrapped back around” the x that comes from k[x][y]→x. We can represent that “rewrapping” process, by a “tree pullback pseudoevent” indicated by the dotted line:

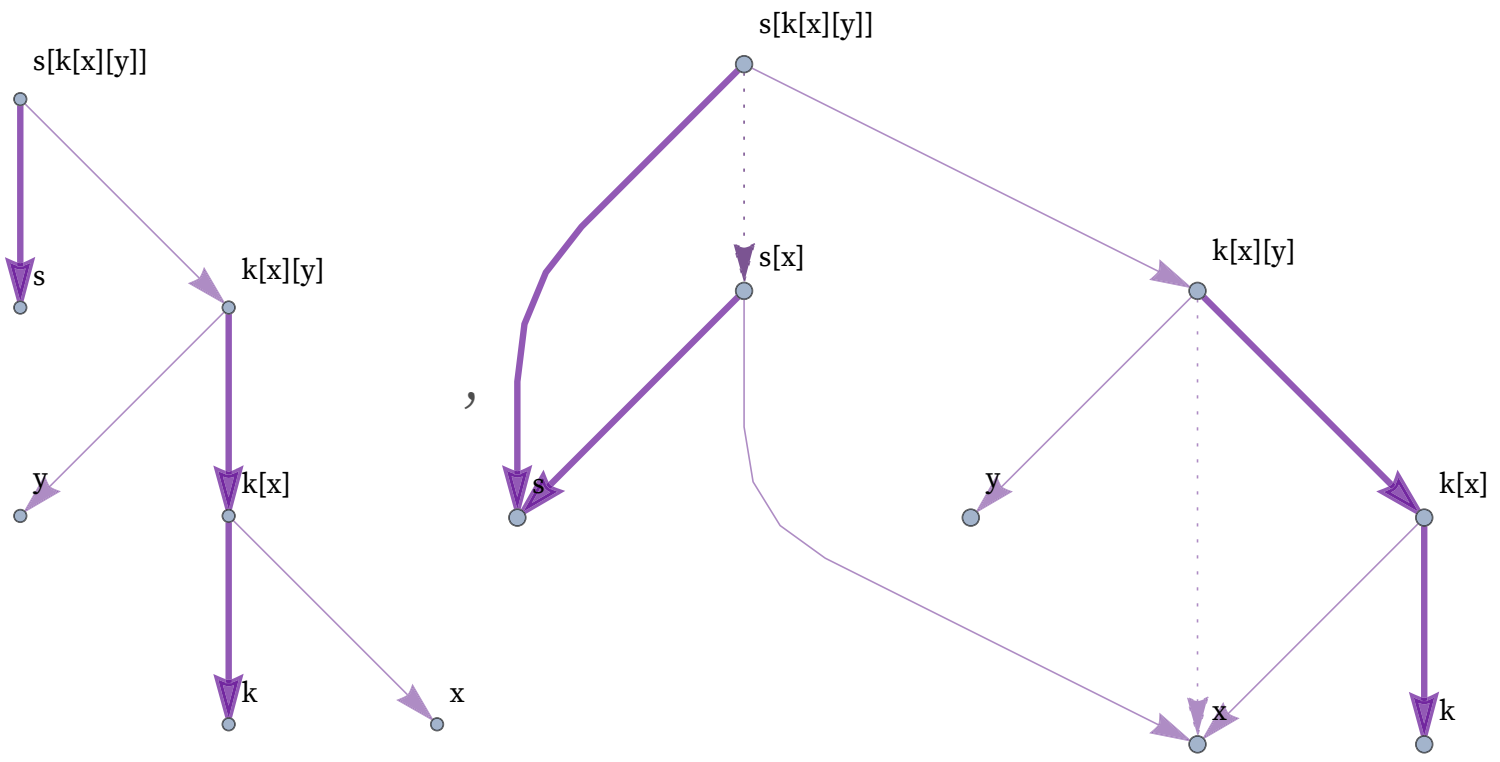

If a given event happens deep inside a tree, there’ll be a whole sequence of “pullback pseudoevents” that “reconstitute the tree”.

Things get quite complicated pretty quickly. Here’s the (leftmost-outermost) evolution of s[s[s]][s][k][s] to its fixed point in terms of DAGs:

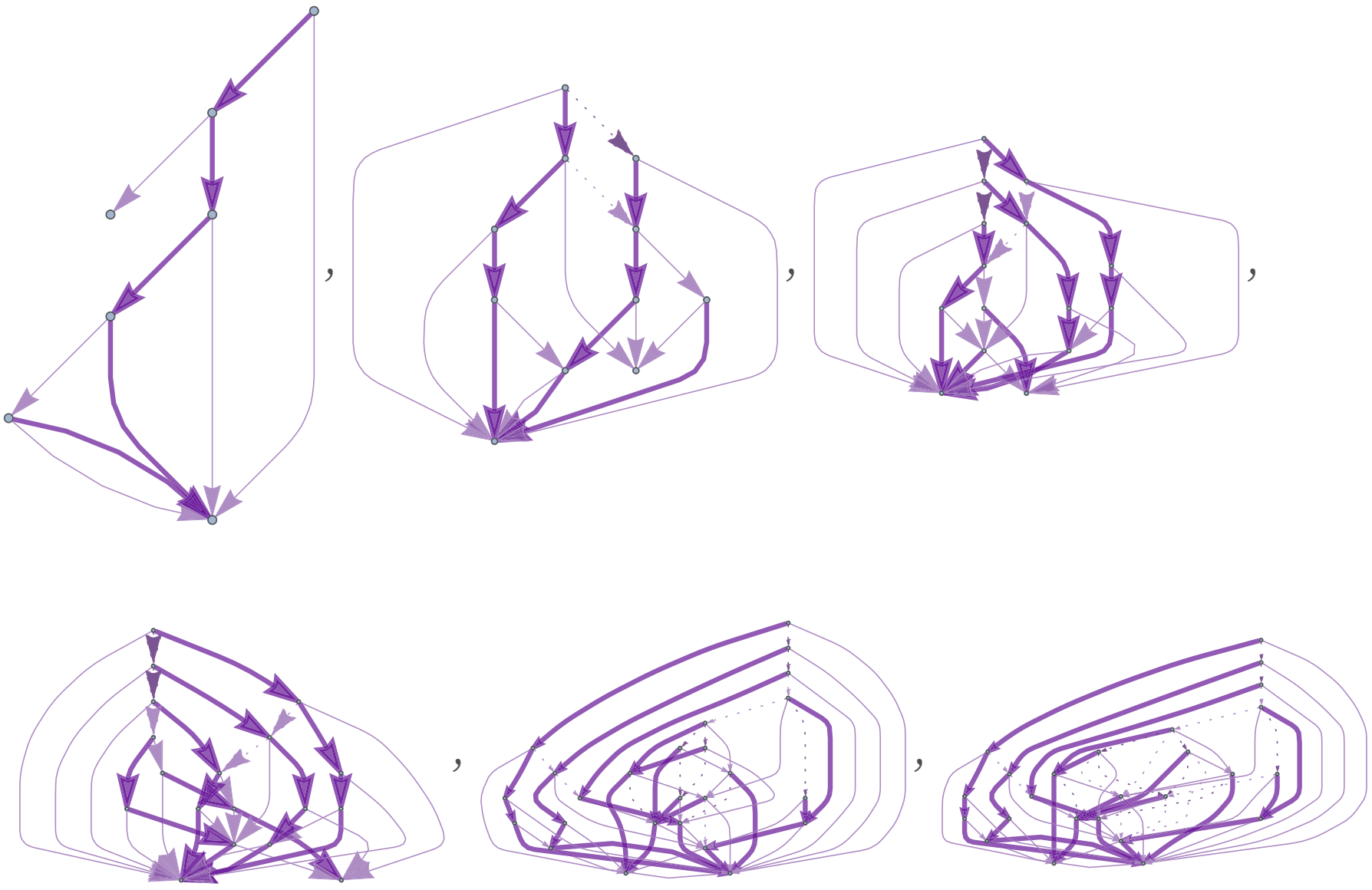

Or with labels:

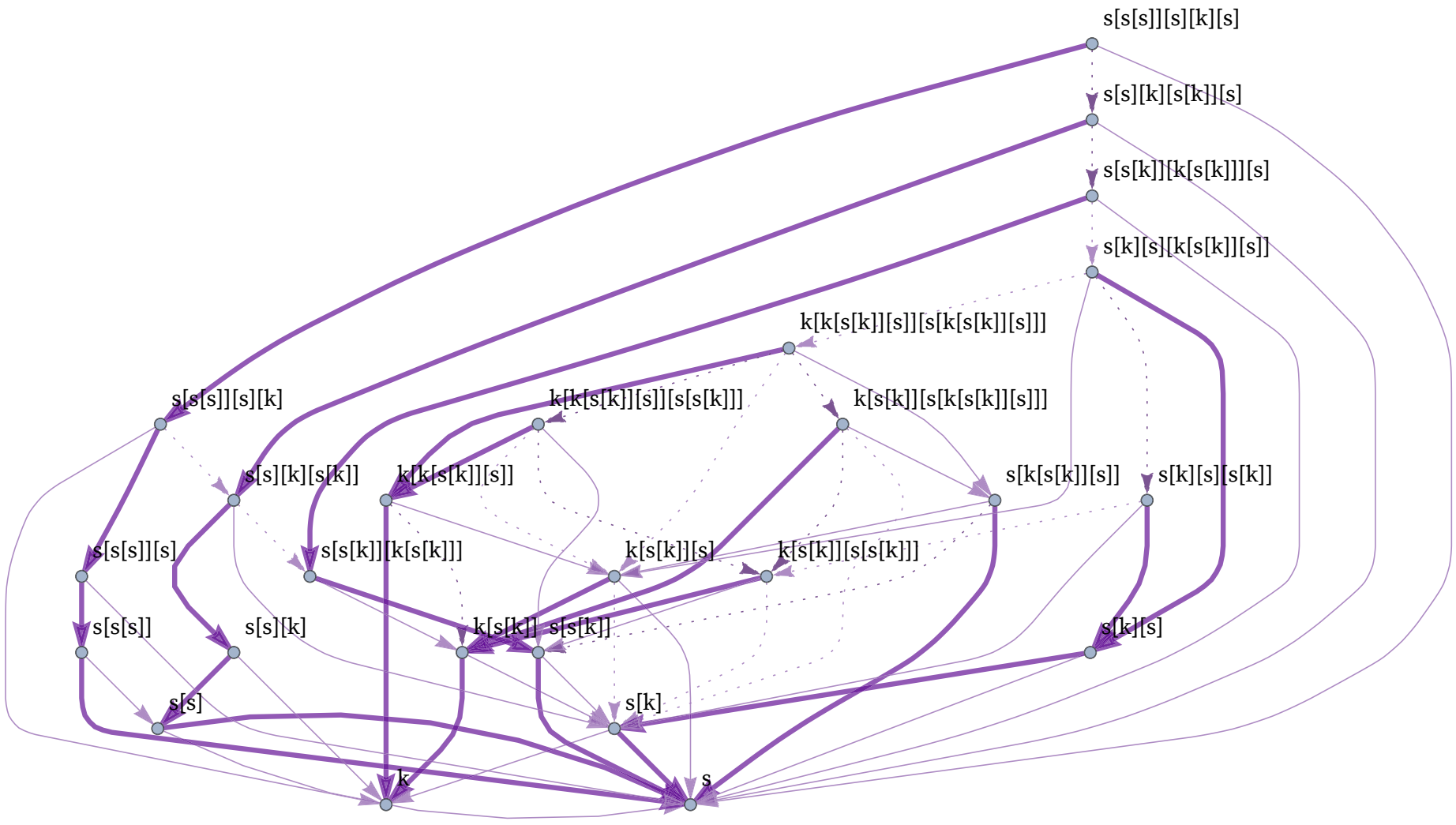

One notable feature is that this final DAG in a sense encodes the complete history of the evolution—in a "maximally shared" way. And from this DAG we can construct a causal graph—whose nodes are derived from the edges in the DAG representing update events and pseudoevents. It's not clear how to do this in the most consistent way—particularly when it comes to handling pseudoevents. But here's one possible version of a causal graph for the evolution of s[s[s]][s][k][s] to its fixed point—with the yellow nodes representing events, and the gray ones pseudoevents:

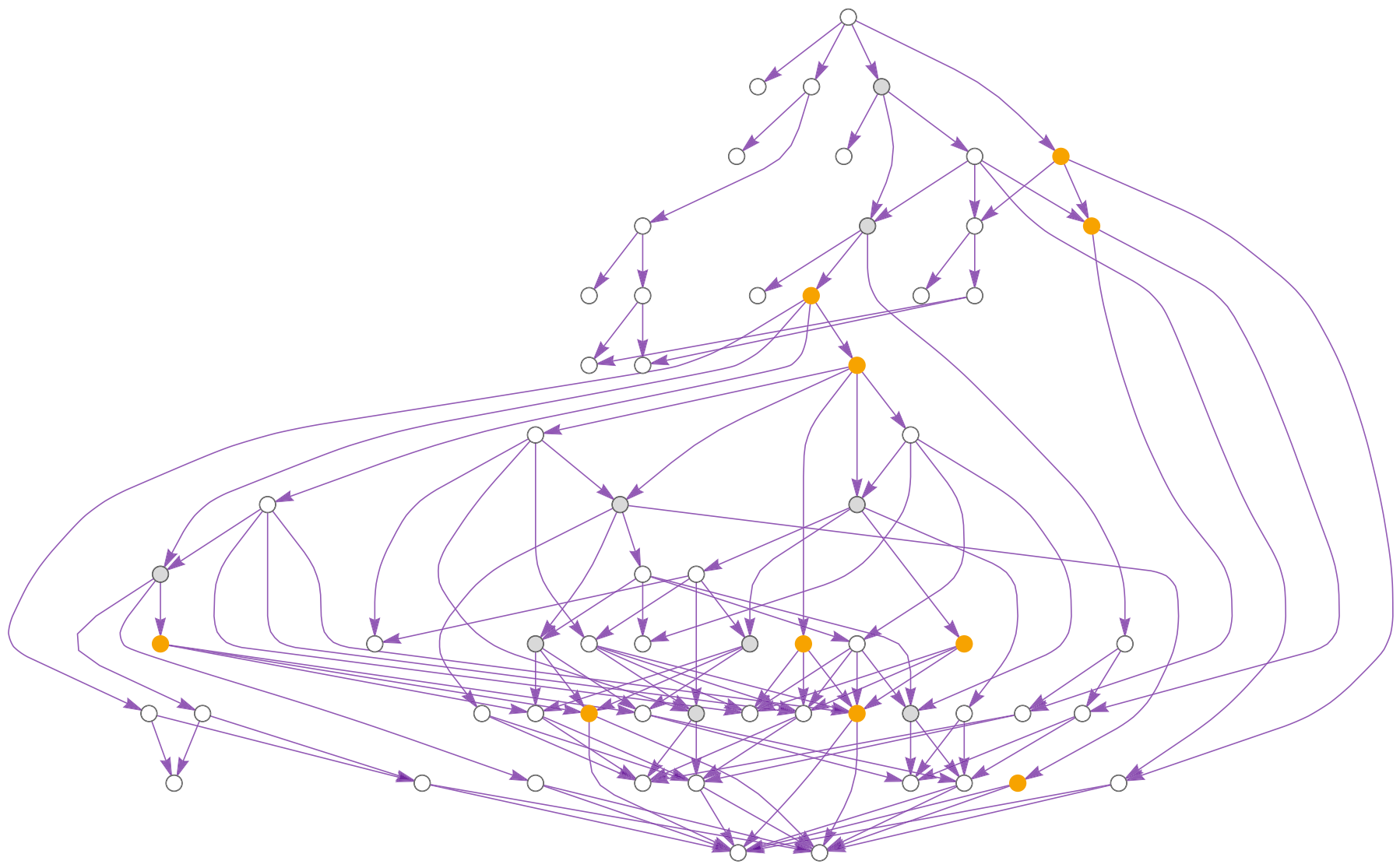

## Combinator Expressions as Dynamical Systems

Start with all possible combinator expressions of a certain size, say involving only S. Some are immediately fixed points. But some only evolve to fixed points. So how are the possible fixed points distributed in the set of all possible combinator expressions?

For size 6 there are 42 possible combinator expressions, and all evolve to fixed points—but only 27 distinct ones. Here are results for several combinator sizes:

| *n* | 1 | 2 | 3 | 4 | 5 | 6 | 7 | 8 | 9 | 10 | 11 | 12 |
|---|---|---|---|---|---|---|---|---|---|---|---|---|
| all | 1 | 1 | 2 | 5 | 14 | 42 | 132 | 429 | 1430 | 4862 | 16796 | 58786 |
| terminating | 1 | 1 | 2 | 5 | 14 | 42 | 130 | 388 | 1154 | 3381 | 9967 | 29498 |
| distinct | 1 | 1 | 2 | 4 | 10 | 27 | 77 | 213 | 592 | 1637 | 4574 | 12899 |
| fraction distinct | 1 | 1 | 1 | 0.8 | 0.71 | 0.64 | 0.59 | 0.55 | 0.51 | 0.48 | 0.46 | 0.44 |

As the size of the combinator expression goes up, the fraction of distinct fixed points seems to systematically go down:

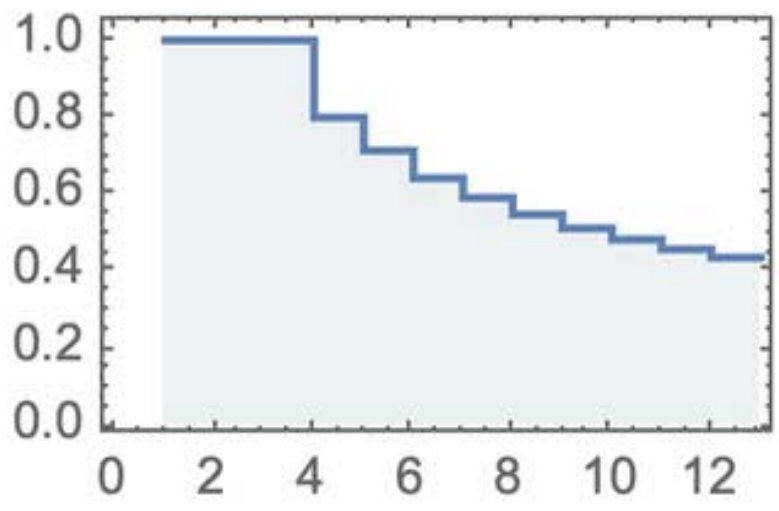

And what this shows is that combinator evolution is in a sense a "contractive" process: starting from all possible expressions, there's only a certain "attractor" of expressions that survives. Here's a "state transition graph" for initial expressions of size 9 computed with leftmost-outermost evaluation (we'll see a more general version in the next section):

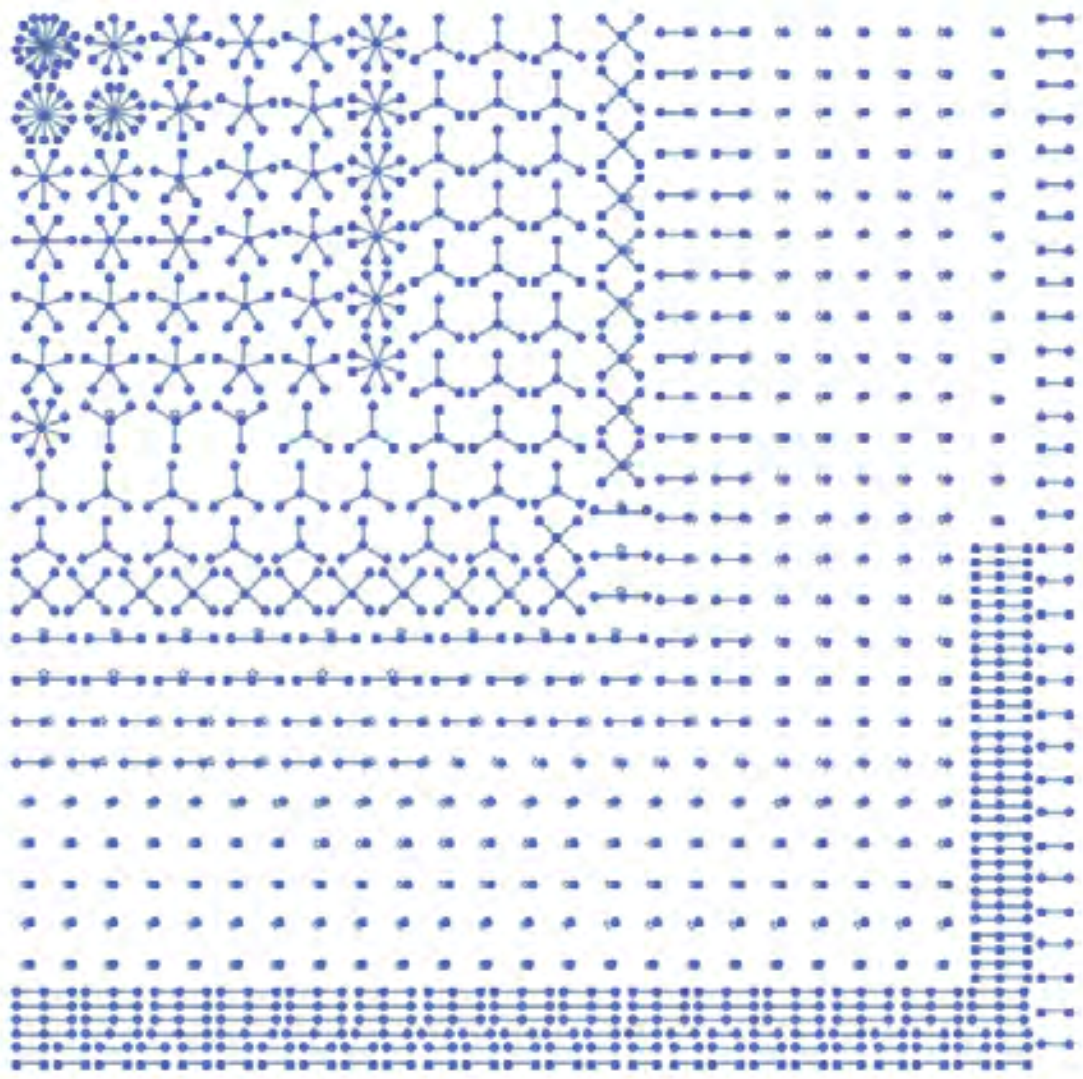

This shows the prevalence of different fixed-point sizes as a function of the size of the initial expression:

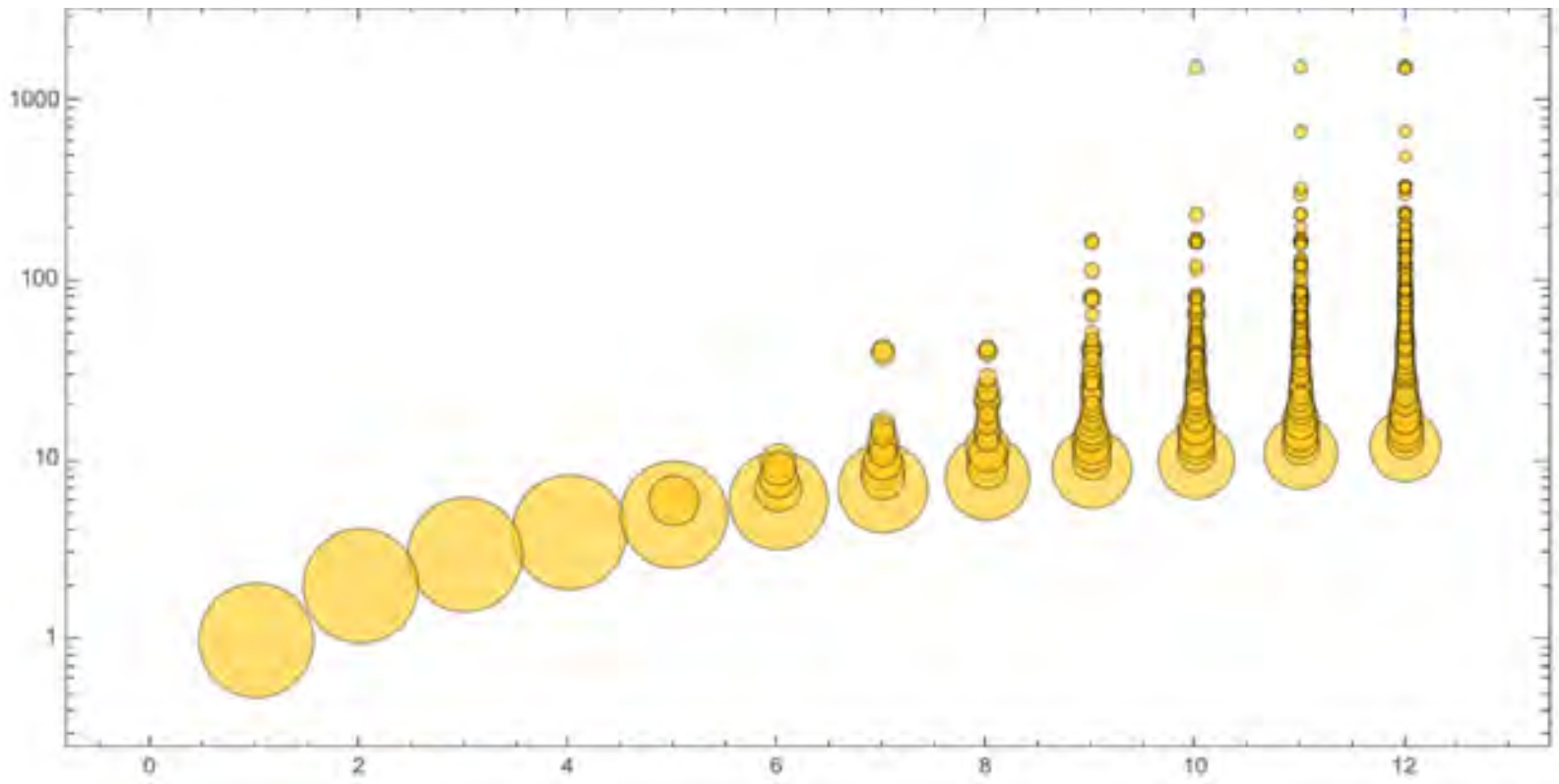

What about the cases that don't reach fixed points? Can we somehow identify different equivalent classes of infinite combinator evolutions (perhaps analogously to the way we can identify different transfinite numbers)? In general we can look at similarities between the multiway systems that are generated, since these are always independent of updating scheme (see the next section).

But something else we can do for both finite and infinite evolutions is to consider the set of subexpressions common to different steps in the evolution—or across different evolutions. Here's a plot of the number of copies of the ultimately most frequent subexpressions at successive steps in the (leftmost-outermost) evolution of s[s][s][s[s]][s][s] (**SSS**(**SS**)**SS**):

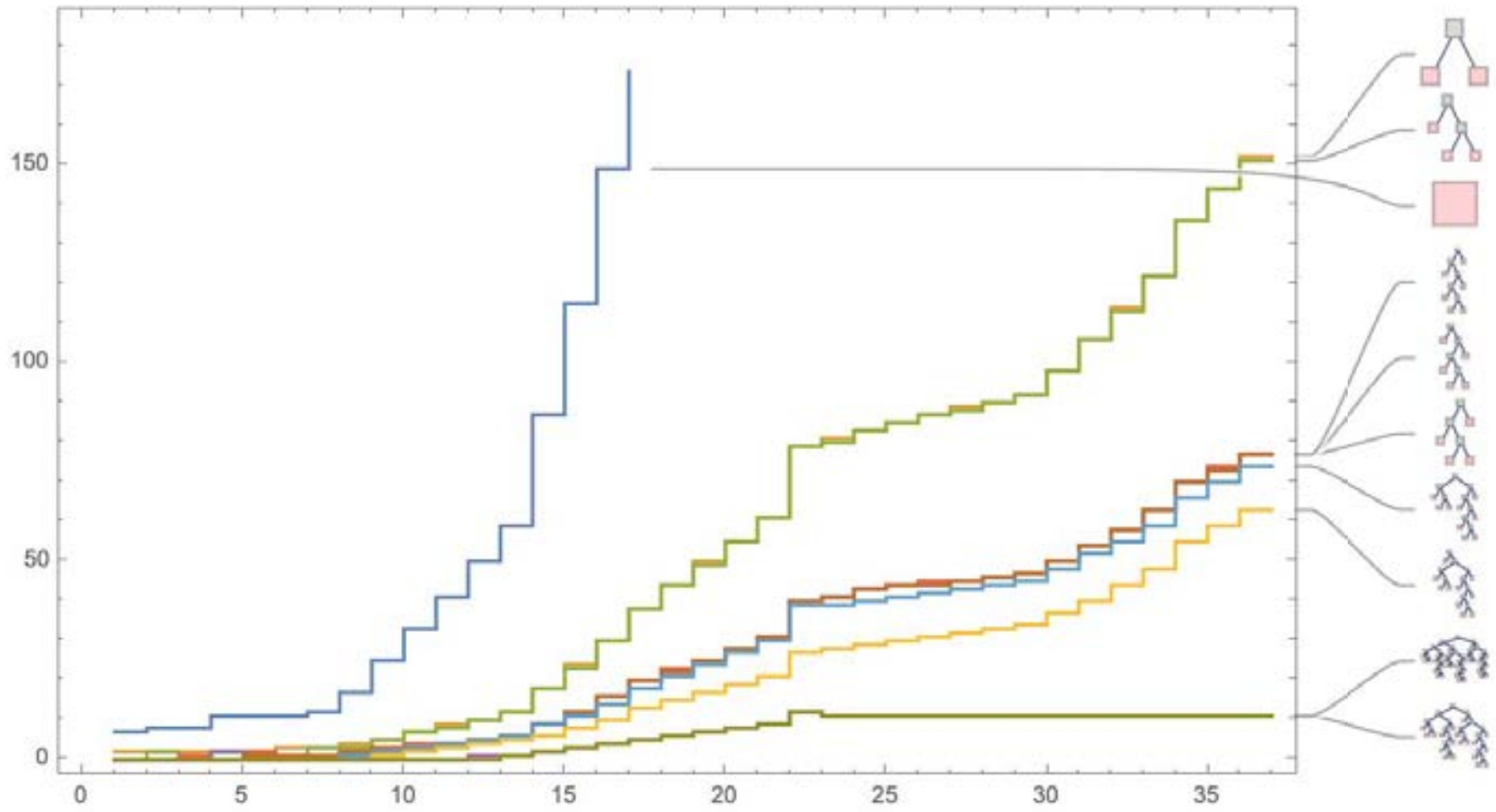

The largest subexpression shown here has size 29. And as the picture makes clear, most subexpressions do not appear with substantial frequency; it's only a thin set that do.

Looking at the evolution of all possible combinator expressions up to size 8, one sees gradual "freezing out" of certain subexpressions (basically as a result of their involvement in halting), and continued growth of others:

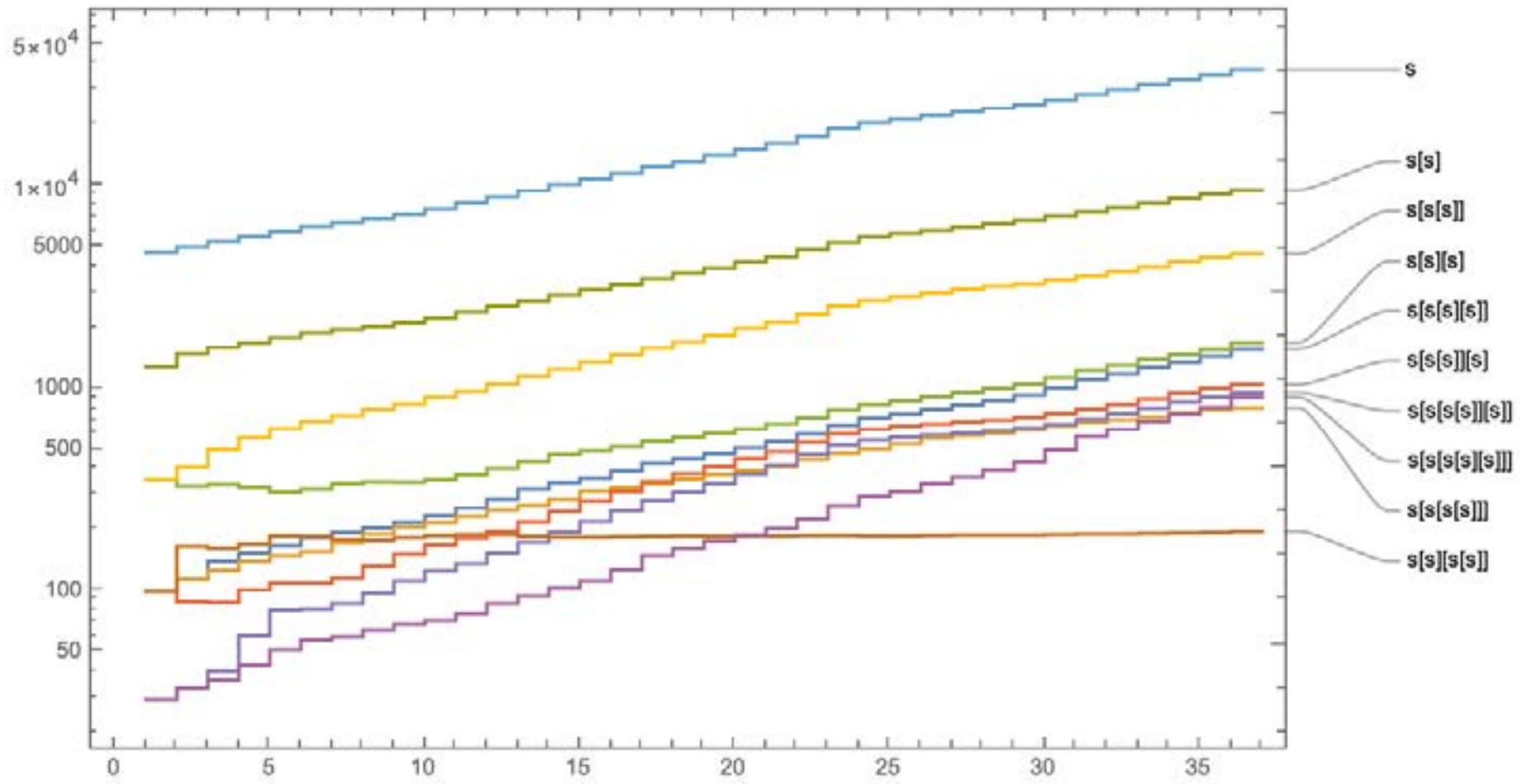

In an attempt to make contact with traditional dynamical systems theory it's interesting to try to map combinator expressions to numbers. A straightforward way to do this (particularly when one's only dealing with expressions involving S) is to use Polish notation, which represents

s[s[s]][s[s[s[s]][s]][s]][s[s[s[s]][s[s[s[s]][s]][s]]]][s[s[s[s]][s[s[s[s]][s]][s]]]]

as

••••s•ss••s••s•ssss•s••s•ss••s••s•ssss•s••s•ss••s••s•ssss

or the binary number

111101001101101000010110100110110100001011010011011010000

i.e., in decimal:

137 839 369 892 767 440

Represented in terms of numbers like this, we can plot all subexpressions which arise in the evolution of s[s][s][s[s]][s][s] (**SSS**(**SS**)**SS**):

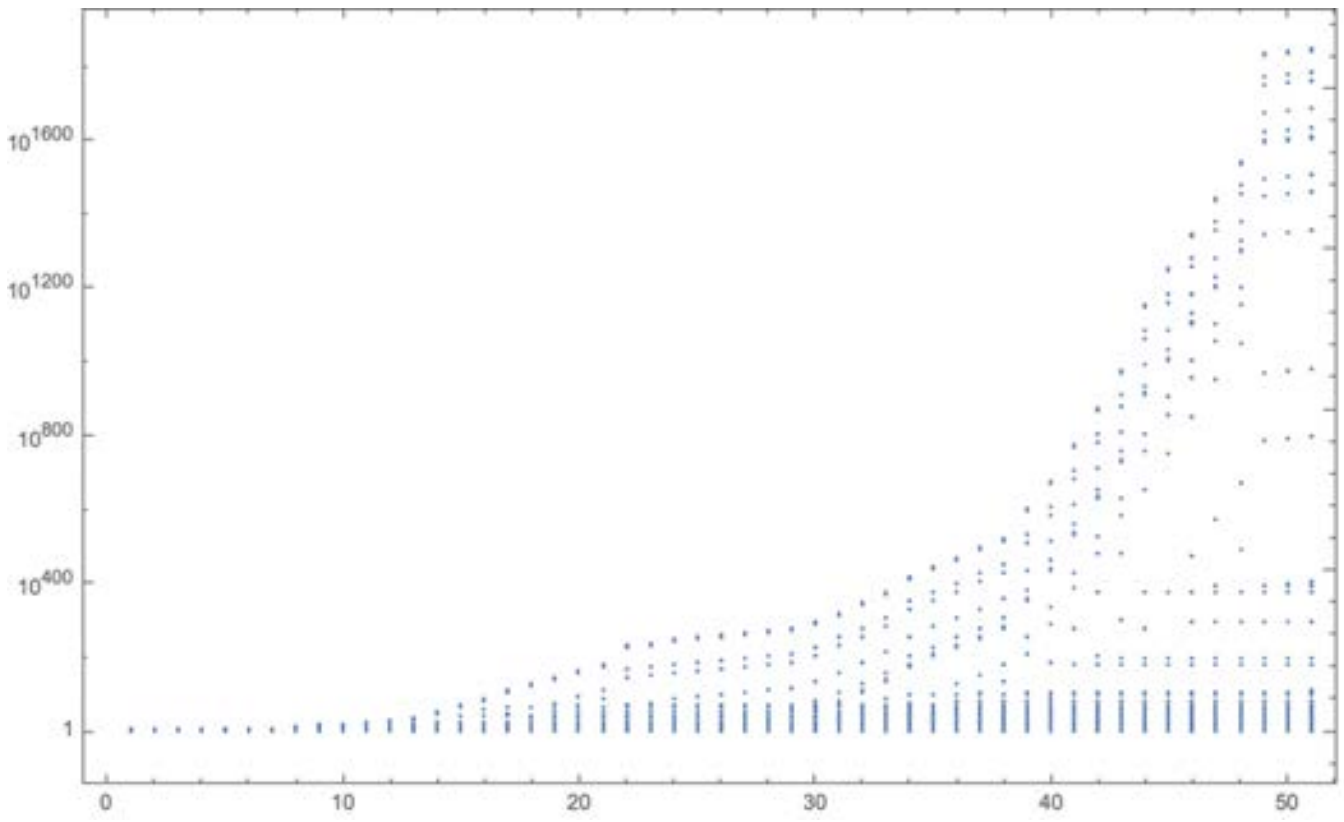

Making a combined picture for all combinator expressions up to size 8, one gets:

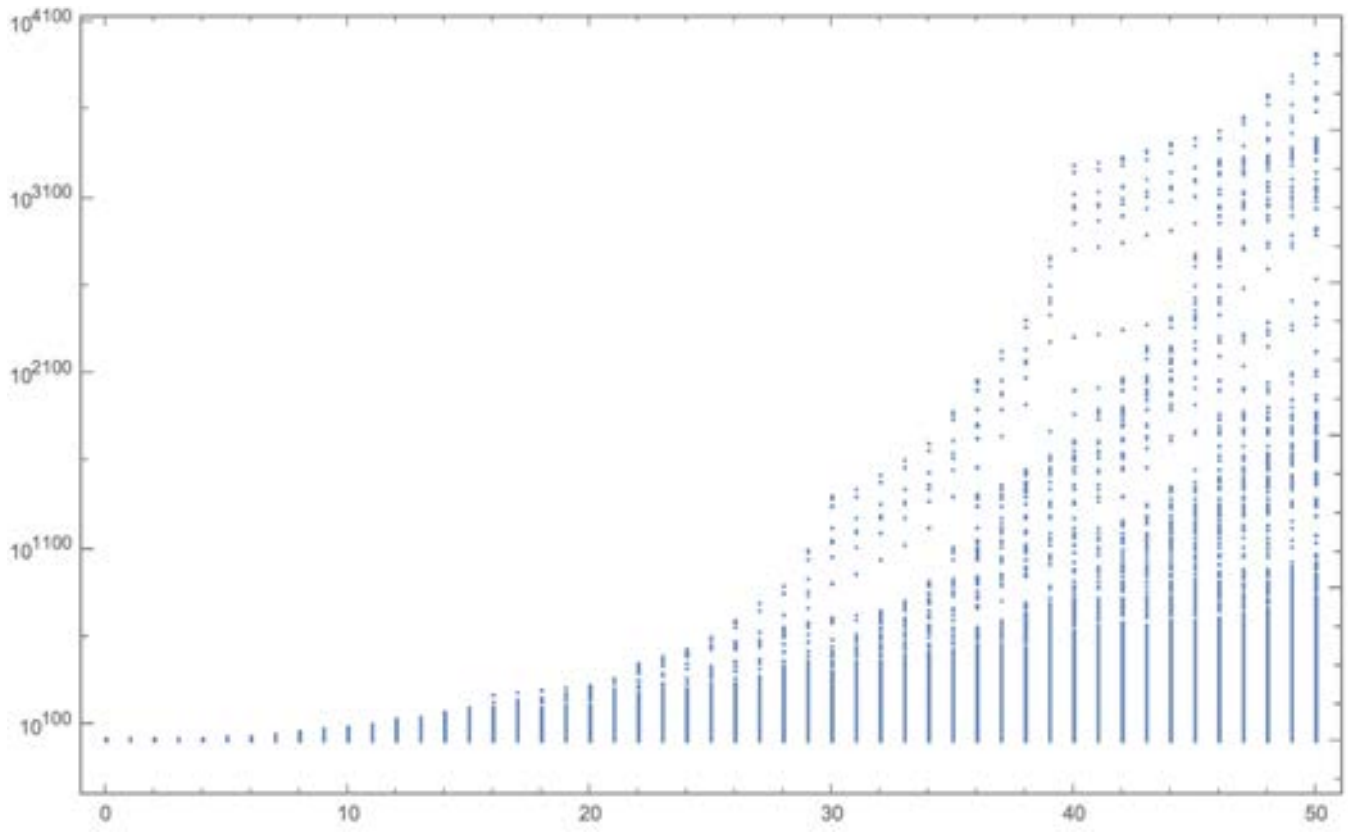

There's definitely some structure: one's not just visiting every possible subexpression. But quite what the limiting form of this might be is not clear.

Another type of question to ask is what the effect of a small change in a combinator expression is on its evolution. The result will inevitably be somewhat subtle—because there is both spacelike and treelike propagation of effects in the evolution.

As one example, though, consider evolving s[s][s][s[s]][s][s] (**SSS**(**SS**)**SS**) for 20 steps (to get an expression of size 301). Now look at the effect of changing a single s in this expression to s[s], and then evolving the result. Here are the sizes of the expressions that are generated:

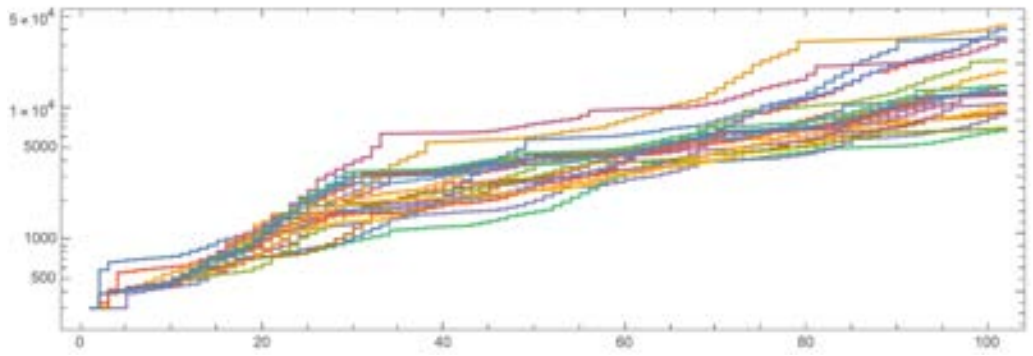

## Equality and Theorem Proving for Combinators

How do you tell if two combinator expressions are equal? It depends what you mean by "equal". The simplest definition—that we've implicitly used in constructing multiway graphs—is that expressions are equal only if they're syntactically exactly the same (say they're both s[k][s[s]]).

But what about a more semantic definition, that takes into account the fact that one combinator expression can be transformed to another by the combinator rules? The obvious thing to say is that combinator expressions should be considered equal if they can somehow be transformed by the rules into expressions that are syntactically the same.

And so long as the combinators evolve to fixed points this is in principle straightforward to tell. Like here are four syntactically different combinator expressions that all evolve to the same fixed point, and so in a semantic sense can be considered equal:

| | | | |
|---|---|---|---|
| s[s][k][k][s[k]] | s[s[s]][s][k][k] | s[s][s][k][s][k] | s[k[s]][k[k]][k] |
| s[k][k[k]][s[k]] | s[s][k][s[k]][k] | s[k][s[k]][s][k] | k[s][k][k[k][k]] |
| k[s[k]][k[k][s[k]]] | s[s[k]][k[s[k]]][k] | k[s][s[k][s]][k] | s[k[k][k]] |
| s[k] | s[k][k][k[s[k]][k]] | s[k] | s[k] |
| | k[k[s[k]][k]][k[k[s[k]][k]]] | | |
| | k[s[k]][k] | | |
| | s[k] | | |

One can think of the fixed point as representing a canonical form to which combinator expressions that are equal can be transformed. One can also think of the steps in the evolution as corresponding to steps in a proof of equality.

But there's already an issue—that's associated with the fundamental fact that combinators are computation universal. Because in general there's no upper bound on how many steps it can take for the evolution of a combinator expression to halt (and no general a priori way to even tell if it'll halt at all). So that means that there's also no upper bound on the "length of proof" needed to show by explicit computation that two combinators are equal. Yes, it might only take 12 steps to show that this is yet another combinator equal to s[k]:

```
s[s[s]][s][s][s][k]
s[s][s][s[s]][s][k]
s[s[s]][s[s[s]]][s][k]
s[s][s][s[s[s]][s]][k]
s[s[s[s]][s]][s[s[s[s]][s]]][k]
s[s[s]][s][k][s[s[s[s]][s]][k]]
s[s][k][s[k]][s[s[s[s]][s]][k]]
s[s[k]][k[s[k]]][s[s[s[s]][s]][k]]
s[k][s[s[s[s]][s]][k]][k[s[k]][s[s[s[s]][s]][k]]]
k[k[s[k]][s[s[s[s]][s]][k]]][s[s[s[s]][s]][k][k[s[k]][s[s[s[s]][s]][k]]]]
k[s[k]][s[s[s[s]][s]][k]]
s[k]
```

But it could also take 31 steps (and involve an intermediate expression of size 65):

```
s[s[s]][s][s][s][s][k]
s[s][s][s[s]][s][s][k]
s[s[s]][s[s[s]]][s][s][k]
s[s][s][s[s[s]][s]][s][k]
s[s[s[s]][s]][s[s[s[s]][s]]][s][k]
s[s[s]][s][s][s[s[s[s]][s]][s]][k]
s[s][s][s[s]][s[s[s[s]][s]][s]][k]
s[s[s]][s[s[s]]][s[s[s[s]][s]][s]][k]
s[s][s[s[s[s]][s]][s]][s[s[s]][s[s[s[s]][s]][s]]][k]
s[s[s[s]][s[s[s[s]][s]][s]]][s[s[s[s]][s]][s][s[s[s]][s[s[s[s]][s]][s]]]][k]
s[s[s]][s[s[s[s]][s]][s]][k][s[s[s[s]][s]][s][s[s[s]][s[s[s[s]][s]][s]]][k]]
s[s][k][s[s[s[s]][s]][s][k]][s[s[s[s]][s]][s][s[s[s]][s[s[s[s]][s]][s]]][k]]
s[s[s[s[s]][s]][s][k]][k[s[s[s[s]][s]][s][k]]][s[s[s[s]][s]][s][s[s[s]][s[s[s[s]][s]][s]]][k]]
s[s[s[s]][s]][s][k][s[s[s[s]][s]][s][s[s[s]][s[s[s[s]][s]][s]]][k]][k[s[s[s[s]][s]][s][k]][s[s[s[s]][s]][s][s[s[s]][s[s[s[s]][s]][s]]][k]]
s[s[s]][s][k][s[k]][s[s[s[s]][s]][s][s[s[s]][s[s[s[s]][s]][s]]][k]][k[s[s[s[s]][s]][s][k]][s[s[s[s]][s]][s][s[s[s]][s[s[s[s]][s]][s]]][k]]
s[s][k][s[k]][s[k]][s[s[s[s]][s]][s][s[s[s]][s[s[s[s]][s]][s]]][k]][k[s[s[s[s]][s]][s][k]][s[s[s[s]][s]][s][s[s[s]][s[s[s[s]][s]][s]]][k]]
s[s[k]][k[s[k]]][s[k]][s[s[s[s]][s]][s][s[s[s]][s[s[s[s]][s]][s]]][k]][k[s[s[s[s]][s]][s][k]][s[s[s[s]][s]][s][s[s[s]][s[s[s[s]][s]][s]]][k]]
s[k][s[k]][k[s[k]][s[k]]][s[s[s[s]][s]][s][s[s[s]][s[s[s[s]][s]][s]]][k]][k[s[s[s[s]][s]][s][k]][s[s[s[s]][s]][s][s[s[s]][s[s[s[s]][s]][s]]][k]]
k[k[s[k]][s[k]]][s[k][k[s[k]][s[k]]]][s[s[s[s]][s]][s][s[s[s]][s[s[s[s]][s]][s]]][k]][k[s[s[s[s]][s]][s][k]][s[s[s[s]][s]][s][s[s[s]][s[s[s[s]][s]][s]]][k]]
k[s[k]][s[k]][s[s[s[s]][s]][s][s[s[s]][s[s[s[s]][s]][s]]][k]][k[s[s[s[s]][s]][s][k]][s[s[s[s]][s]][s][s[s[s]][s[s[s[s]][s]][s]]][k]]
s[k][s[s[s[s]][s]][s][s[s[s]][s[s[s[s]][s]][s]]][k]][k[s[s[s[s]][s]][s][k]][s[s[s[s]][s]][s][s[s[s]][s[s[s[s]][s]][s]]][k]]
k[k[s[s[s[s]][s]][s][k]][s[s[s[s]][s]][s][s[s[s]][s[s[s[s]][s]][s]]][k]][
 s[s[s[s]][s]][s][s[s[s]][s[s[s[s]][s]][s]]][k][k[s[s[s[s]][s]][s][k]][s[s[s[s]][s]][s][s[s[s]][s[s[s[s]][s]][s]]][k]]]
k[s[s[s[s]][s]][s][k]][s[s[s[s]][s]][s][s[s[s]][s[s[s[s]][s]][s]]][k]]
s[s[s[s]][s]][s][k]
s[s[s]][s][k][s[k]]
s[s][k][s[k]][s[k]]
s[s[k]][k[s[k]]][s[k]]
s[k][s[k]][k[s[k]][s[k]]]
k[k[s[k]][s[k]]][s[k][k[s[k]][s[k]]]]
k[s[k]][s[k]]
s[k]
```

We know that if we use leftmost-outermost evaluation, then any combinator expression that has a fixed point will eventually evolve to it (even though we can't in general know how long it will take). But what about combinator expressions that don't have fixed points? How can we tell if they're "equal" according to our definition?

Basically we have to be able to tell if there are sequences of transformations under the combinator rules that cause the expressions to wind up syntactically the same. We can think of these sequences of transformations as being like possible paths of evolution. So then in effect what we're asking is whether there are paths of evolution for different combinators that intersect.

But how can we characterize what possible paths of evolution might exist for all possible evaluation schemes? Well, that's what the multiway graph does. And in terms of multiway graphs there's then a concrete way to ask about equality (or, really, equivalence) between combinator expressions. We basically just need to ask whether there is some appropriate path between the expressions in the multiway graph.

There are lots of details, some of which we'll discuss later. But what we're basically dealing with is a quintessential example of the problem of theorem proving in a formal system. There are different ways to set things up. But as one example, we could take our system to define certain axioms that transform expressions. Applying these axioms in all possible ways generates a multiway graph with expressions as nodes. But then the statement that there's a theorem that expression A is equal to expression B (in the sense that it can be transformed to it) becomes the statement that there's a way to get from A to B in the graph—and giving a path can then be thought of as giving a proof of the theorem.

As an example, consider the combinator expressions:

s[s][s[s][s[s[s]]][k]][k[s[s][s[s[s]]][k]]]

s[k[s[k][s[s[s]][k]]]][k[s[s][s[s[s]]][k]]]

Constructing a multiway graph one can then find a path

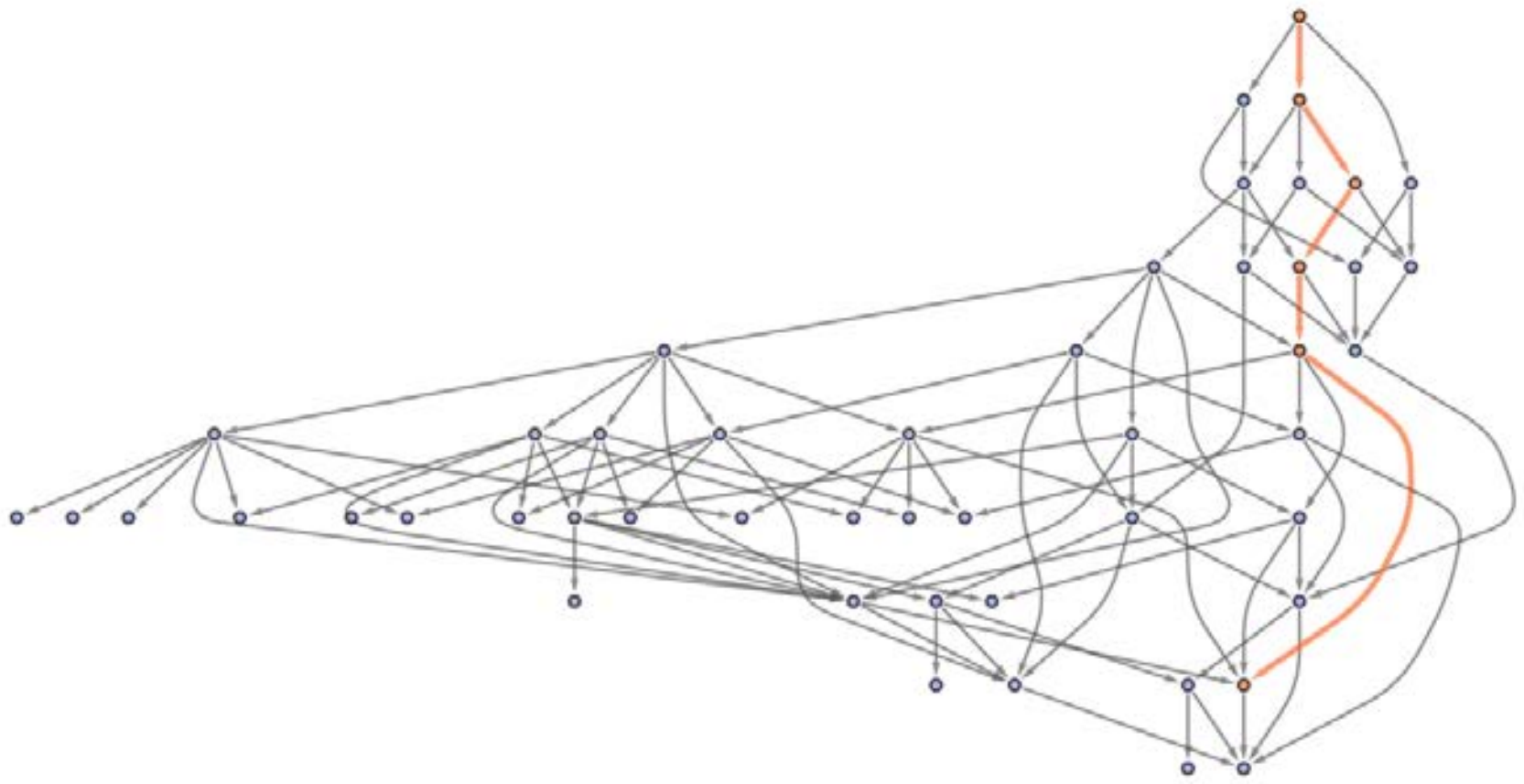

which corresponds to the proof that one can get from one of these expressions to the other:

s[s][s[s][s[s[s]]][k]][k[s[s][s[s[s]]][k]]]
s[k[s[s][s[s[s]]][k]]][s[s][s[s[s]]][k][k[s[s][s[s[s]]][k]]]]
s[k[s[k][s[s[s]][k]]]][s[s][s[s[s]]][k][k[s[s][s[s[s]]][k]]]]
s[k[s[k][s[s[s]][k]]]][s[k][s[s[s]][k]][k[s[s][s[s[s]]][k]]]]
s[k[s[k][s[s[s]][k]]]][k[k[s[s][s[s[s]]][k]]][s[s[s]][k][k[s[s][s[s[s]]][k]]]]]
s[k[s[k][s[s[s]][k]]]][k[s[s][s[s[s]]][k]]]

In this particular case, both expressions eventually reach a fixed point. But consider the expressions:

s[s[s[s][s]]][k]][s[s[s[s[s][s]]][k]]][k[s[s[s[s][s]]][k]]]

s[s[s[s][s]]][k][s[s[s[s[s][s]]][k]][k[s[s[s[s][s]]][k]]]]

Neither of these expressions evolves to a fixed point. But there's still a path in the (ultimately infinite) multiway graph between them

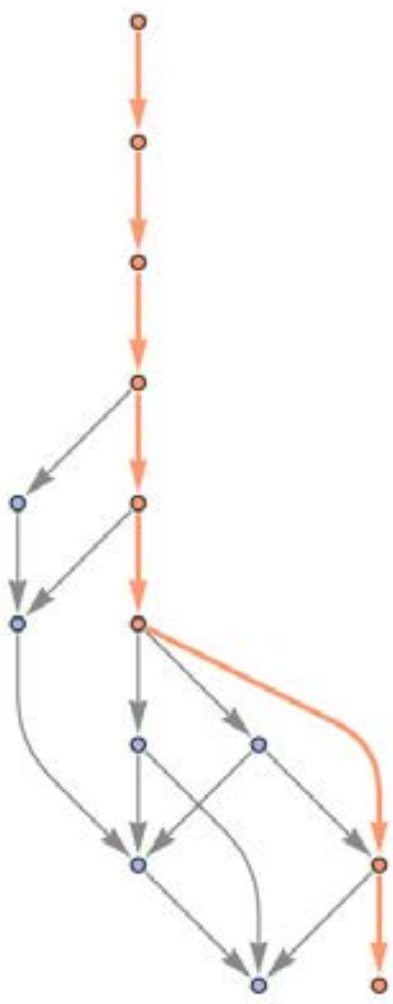

corresponding to the equivalence proof:

```
s[s[s[s[s][s]]][k]][s[s[s[s[s][s]]][k]]][k[s[s[s[s][s]]][k]]]
s[s[s[s][s]]][k][k[s[s[s[s][s]]][k]]][s[s[s[s[s][s]]][k]][k[s[s[s[s][s]]][k]]]]
s[s[s][s]][k[s[s[s[s][s]]][k]]][k[k[s[s[s[s][s]]][k]]]][s[s[s[s[s][s]]][k]][k[s[s[s[s][s]]][k]]]]
s[s][s][k[k[s[s[s[s][s]]][k]]]][k[s[s[s[s][s]]][k]][k[k[s[s[s[s][s]]][k]]]]][s[s[s[s[s][s]]][k]][k[s[s[s[s][s]]][k]]]]
s[k[k[s[s[s[s][s]]][k]]]][s[k[k[s[s[s[s][s]]][k]]]]][k[s[s[s[s][s]]][k]][k[k[s[s[s[s][s]]][k]]]]][s[s[s[s[s][s]]][k]][k[s[s[s[s][s]]][k]]]]
k[k[s[s[s[s][s]]][k]]][k[s[s[s[s][s]]][k]][k[k[s[s[s[s][s]]][k]]]]][
  s[k[k[s[s[s[s][s]]][k]]]][k[s[s[s[s][s]]][k]][k[k[s[s[s[s][s]]][k]]]]]][s[s[s[s[s][s]]][k]][k[s[s[s[s][s]]][k]]]]
k[s[s[s[s][s]]][k]][s[k[k[s[s[s[s][s]]][k]]]][k[s[s[s[s][s]]][k]][k[k[s[s[s[s][s]]][k]]]]]][s[s[s[s[s][s]]][k]][k[s[s[s[s][s]]][k]]]]
s[s[s[s][s]]][k][s[s[s[s[s][s]]][k]][k[s[s[s[s][s]]][k]]]]
```

But with our definition, two combinator expressions can still be considered equal even if one of them can't evolve into the other: it can just be that among the possible ancestors (or, equivalently for combinators, successors) of the expressions there's somewhere an expression in common. (In physics terms, that their light cones somewhere overlap.)

Consider the expressions:

```
{s[s[s][s]][s][s[s][k]], s[s][k][s[s[s][k]]][k]}
```

Neither terminates, but it still turns out that there are paths of evolution for each of them that lead to the same expression:

| | |
|---|---|
| s[s[s][s]][s][s[s][k]] | s[s][k][s[s[s][k]]][k] |
| s[s][s][s[s][k]][s[s[s][k]]] | s[s[s[s][k]]][k[s[s[s][k]]]][k] |
| s[s[s][k]][s[s[s][k]]][s[s[s][k]]] | s[s[s][k]][k][k[s[s[s][k]]][k]] |
| | s[s[s][k]][k][s[s[s][k]]] |
| | s[s][k][s[s[s][k]]][k[s[s[s][k]]]] |
| | s[s[s[s][k]]][k[s[s[s][k]]]][k[s[s[s][k]]]] |
| | s[s[s][k]][k[s[s[s][k]]]][k[s[s[s][k]]][k[s[s[s][k]]]]] |
| | s[s][k][k[s[s[s][k]]][k[s[s[s][k]]]]][k[s[s[s][k]]][k[s[s[s][k]]][k[s[s[s][k]]]]]] |
| | s[s][k][k[s[s[s][k]]][k[s[s[s][k]]]]][s[s[s][k]]] |
| | s[s][k][s[s[s][k]]][s[s[s][k]]] |
| | s[s[s[s][k]]][k[s[s[s][k]]]][s[s[s][k]]] |
| | s[s[s][k]][s[s[s][k]]][k[s[s[s][k]]][s[s[s][k]]]] |
| | s[s[s][k]][s[s[s][k]]][s[s[s][k]]] |

If we draw a combined multiway graph starting from the two initial expressions, we can see the converging paths:

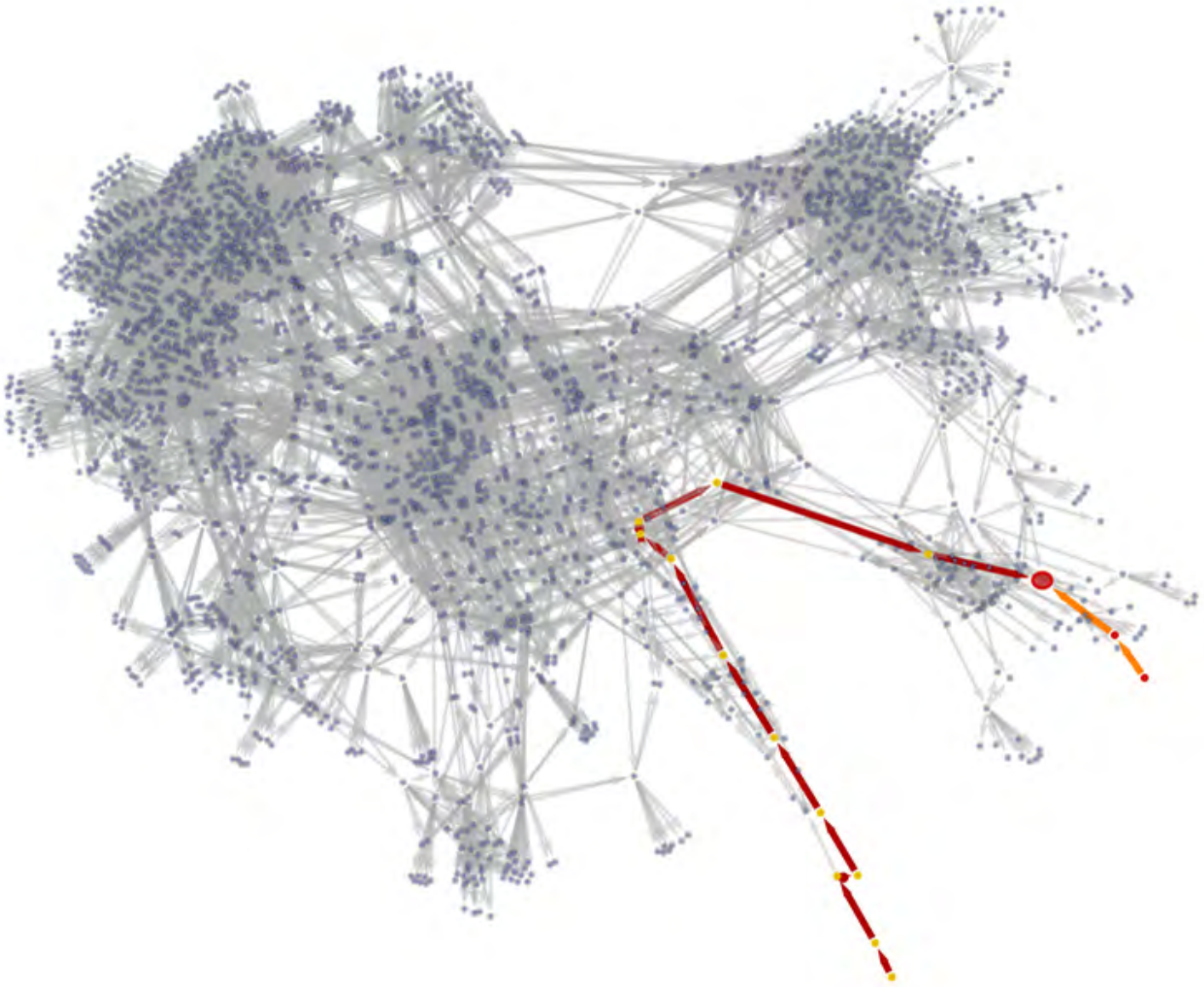

But is there a more systematic way to think about relations between combinator expressions? Combinators are in a sense fundamentally computational constructs. But one can still try to connect them with traditional mathematics, and in particular with abstract algebra.

And so, for example, it's common in the literature of combinators to talk about "combinatory algebra", and to write an expression like

s[k][s[s[k[s[s]][s]][s]][k]][k[s[k]][s[s[k[s[s]][s]][s]][k]]]

as

**S•K•(S•(S•(K•(S•S)•S)•S)•K)•(K•(S•K)•(S•(S•(K•(S•S)•S)•S)•K))**

where now one imagines that • ("application") is like an algebraic operator that "satisfies the relations"

$$\{\mathbf{S}\bullet x\bullet y\bullet z = x\bullet z\bullet(y\bullet z),\ \mathbf{K}\bullet x\bullet y = x\}$$

with "constants" **S** and **K**. To determine whether two combinator expressions are equal one then has to see if there's a sequence of "algebraic" transformations that can go from one to the other. The setup is very similar to what we've discussed above, but the "two-way" character of the rules allows one to directly use standard equational logic theorem-proving methods (although because combinator evolution is confluent one never strictly has to use reversed rules).

So, for example, to prove s[k[s]][k[k]][k]==s[s][s][k][s][k] or

**S(KS)(KK)K = SSSKSK**

one applies a series of transformations based on the S and K "axioms" to parts of the left- and right-hand sides to eventually reduce the original equation to a tautology:

| | |
|---|---|
| s•s•s•k •s•k == s•(k•s)•(k•k)•k | s•x•y•z → x•z•(y•z) |
| s•k•(s•k)•s•k == s•(k•s)•(k•k)•k | s•x•y•z → x•z•(y•z) |
| s•k•(s•k)•s•k == k•s•k• k•k•k | k•x•y → x |
| s•k•(s•k)•s •k == k•s•k•k | s•x•y•z → x•z•(y•z) |
| k•s•(s•k•s)•k == k•s•k •k | k•x•y → x |
| k•s•(s•k•s)•k == s•k | k•x•y → x |
| ■ | True |

One can give the outline of this proof as a standard FindEquationalProof proof graph:

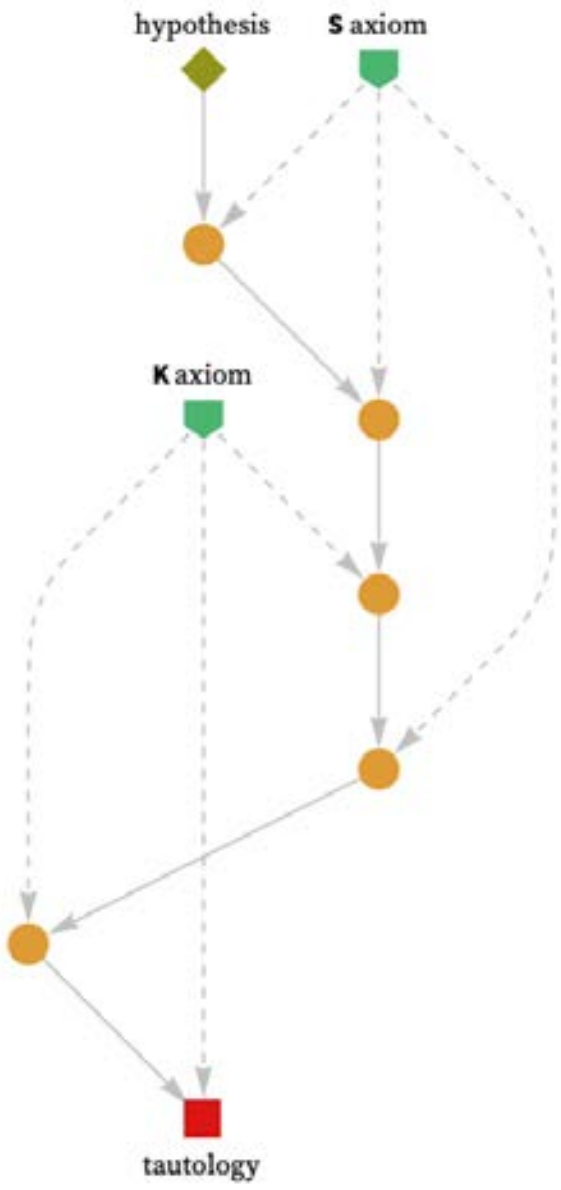

The yellowish dots correspond to the "intermediate lemmas" listed above, and the dotted lines indicate which lemmas use which axioms.

One can establish a theorem like

**S(KS)(KK)K = S(SS)SSSK**

with a slightly more complex proof:

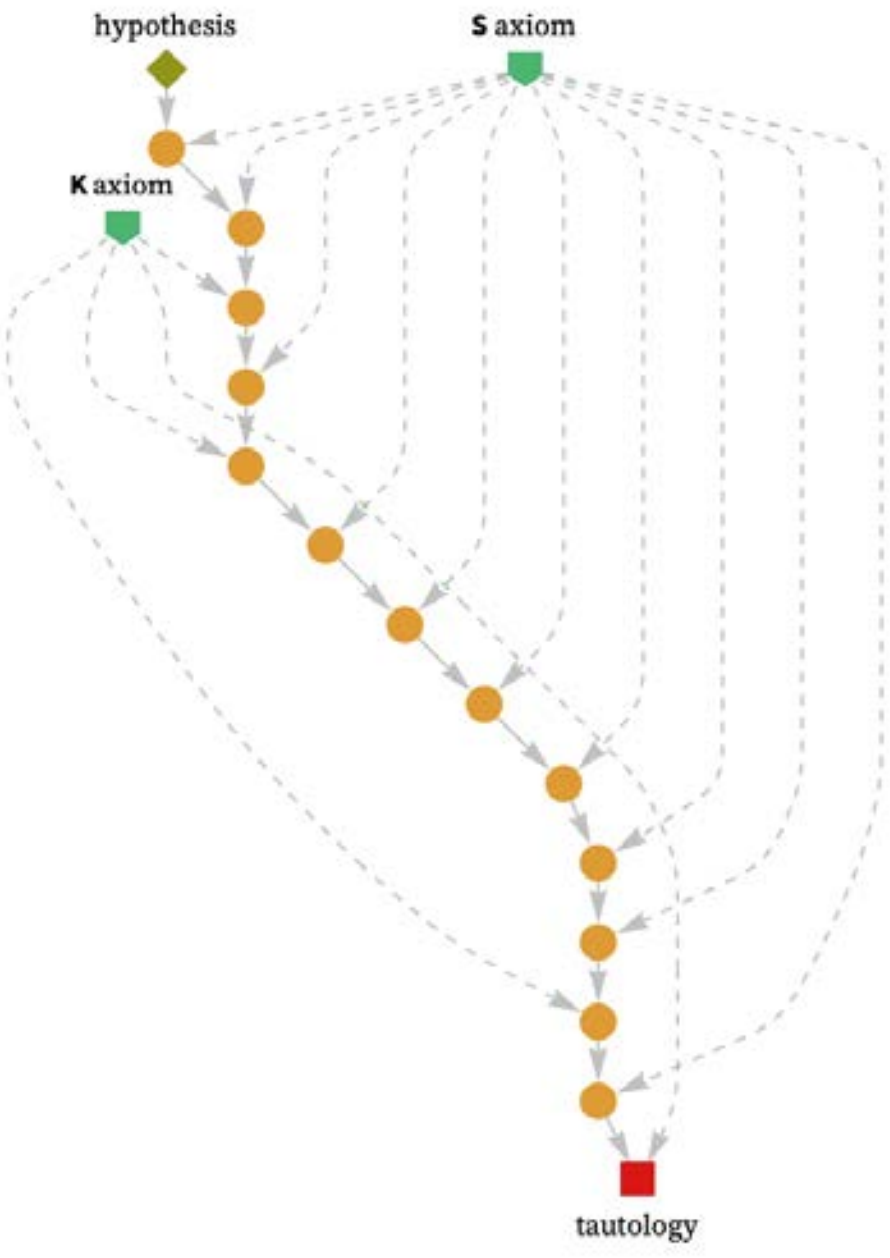

One feature of this proof is that because the combinator rules are confluent—so that different branches in the multiway system always merge—the proof never has to involve critical pair lemmas representing equivalences between branches in the multiway system, and so can consist purely of a sequence of "substitution lemmas".

There's another tricky issue, though. And it has to do with taking "everyday" mathematical notions and connecting them with the precise symbolic structure that defines combinators and their evolution. As an example, let's say you have combinators a and b. It might seem obvious that if a is to be considered equal to b, then it must follow that a[x]==b[x] for all x.

But actually saying this is true is telling us something about what we mean by "equal", and to specify this precisely we have to add the statement as a new axiom.

In our basic setup for proving anything to do with equality (or, for that matter, any equivalence relation), we're already assuming the basic features of equivalence relations (reflexivity, symmetry, transitivity):

$$x = x$$
$$x = y \Rightarrow y = x$$
$$(x = y) \wedge (y = z) \Rightarrow x = z$$

In order to allow us to maintain equality while doing substitutions we also need the axiom:

$$(x = y) \wedge (z = u) \Rightarrow x \bullet z = y \bullet u$$

And now to specify that combinator expressions that are considered equal also "do the same thing" when applied to equal expressions, we need the "extensionality" axiom:

$$x = y \Rightarrow x \bullet z = y \bullet z$$

The previous axioms all work in pure "equational logic". But when we add the extensionality axiom we have to explicitly use full first-order logic—with the result that we get more complicated proofs, though the same basic methods apply.

## Lemmas and the Structure of Combinator Space

One feature of the proofs we've shown above is that each intermediate lemmas just involves direct use of one or other of the axioms. But in general, lemmas can use lemmas, and one can "recursively" build up a proof much more efficiently than just by always directly using the axioms.

But which lemmas are best to use? If one's doing ordinary human mathematics—and trying to make proofs intended for human consumption—one typically wants to use "famous lemmas" that help create a human-relatable narrative. But realistically there isn't likely to be a "human-relatable narrative" for most combinator equivalence theorems (or, at least there won't be until or unless thinking in terms of combinators somehow becomes commonplace).

So then there's a more "mechanical" criterion: what lemmas do best at reducing the lengths of as many proofs as much as possible? There's some trickiness associated with translations between proofs of equalities and proofs that one expression can evolve into another. But roughly the question boils downs to this. When we construct a multiway graph of combinator evolution, each event—and thus each edge—is just the application of a single combinator "axiom".

But if instead we do transformations based on more sophisticated lemmas we can potentially get from one expression to another in fewer steps. In other words, if we "cache" certain combinator transformations, can we make finding paths in combinator multiway graphs systematically more efficient?

To find all possible "combinator theorems" from a multiway system, we should start from all possible combinator expressions, then trace all possible paths to other expressions. It's a little like what we did in the previous section—except now we want to consider multiway evolution with all possible evaluation orders.

Here's the complete multiway graph starting from all size-4 combinator expressions:

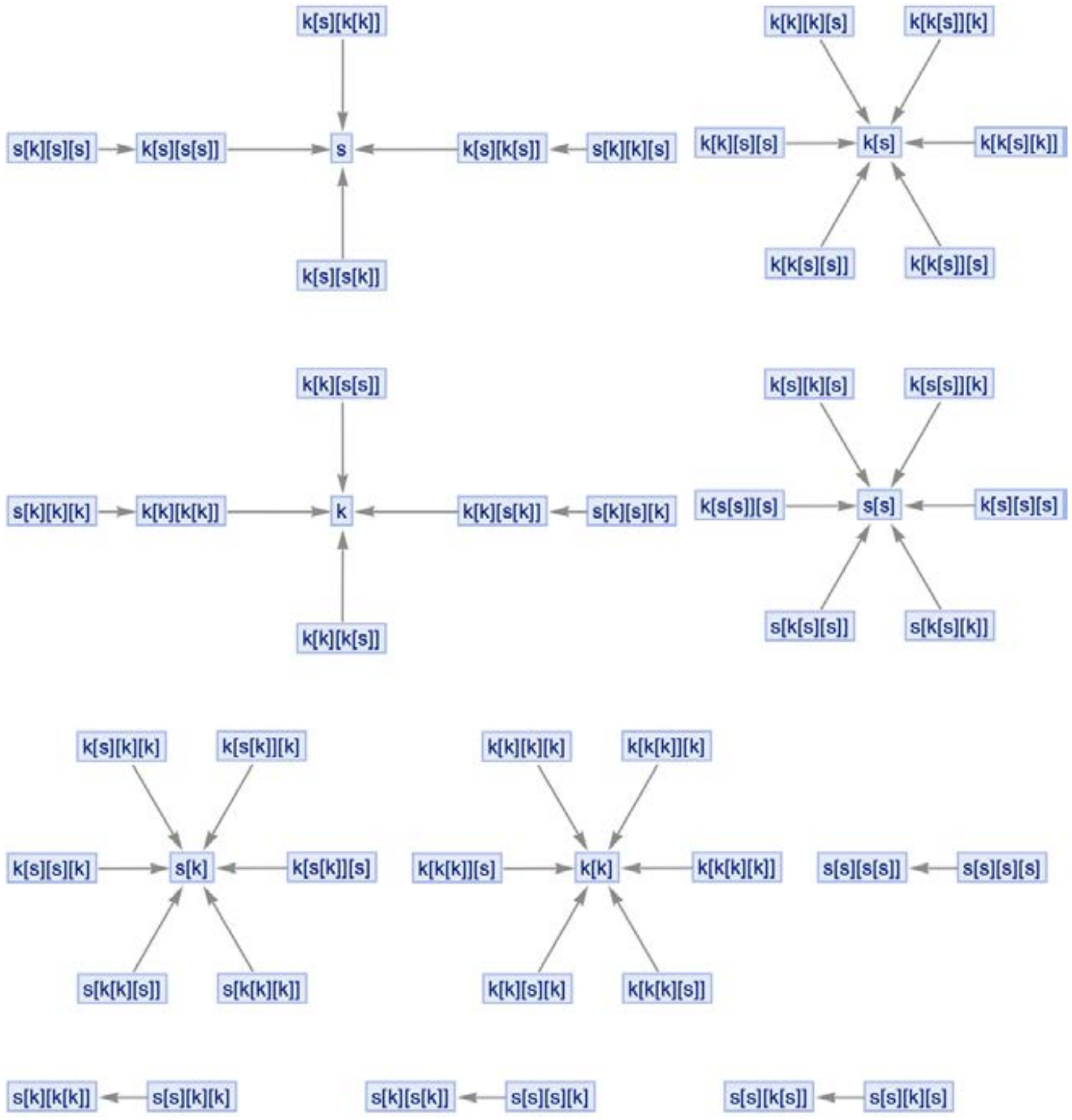

Up to size 6, the graph is still finite (with each disconnected component in effect corresponding a separate "fixed point attractor"):

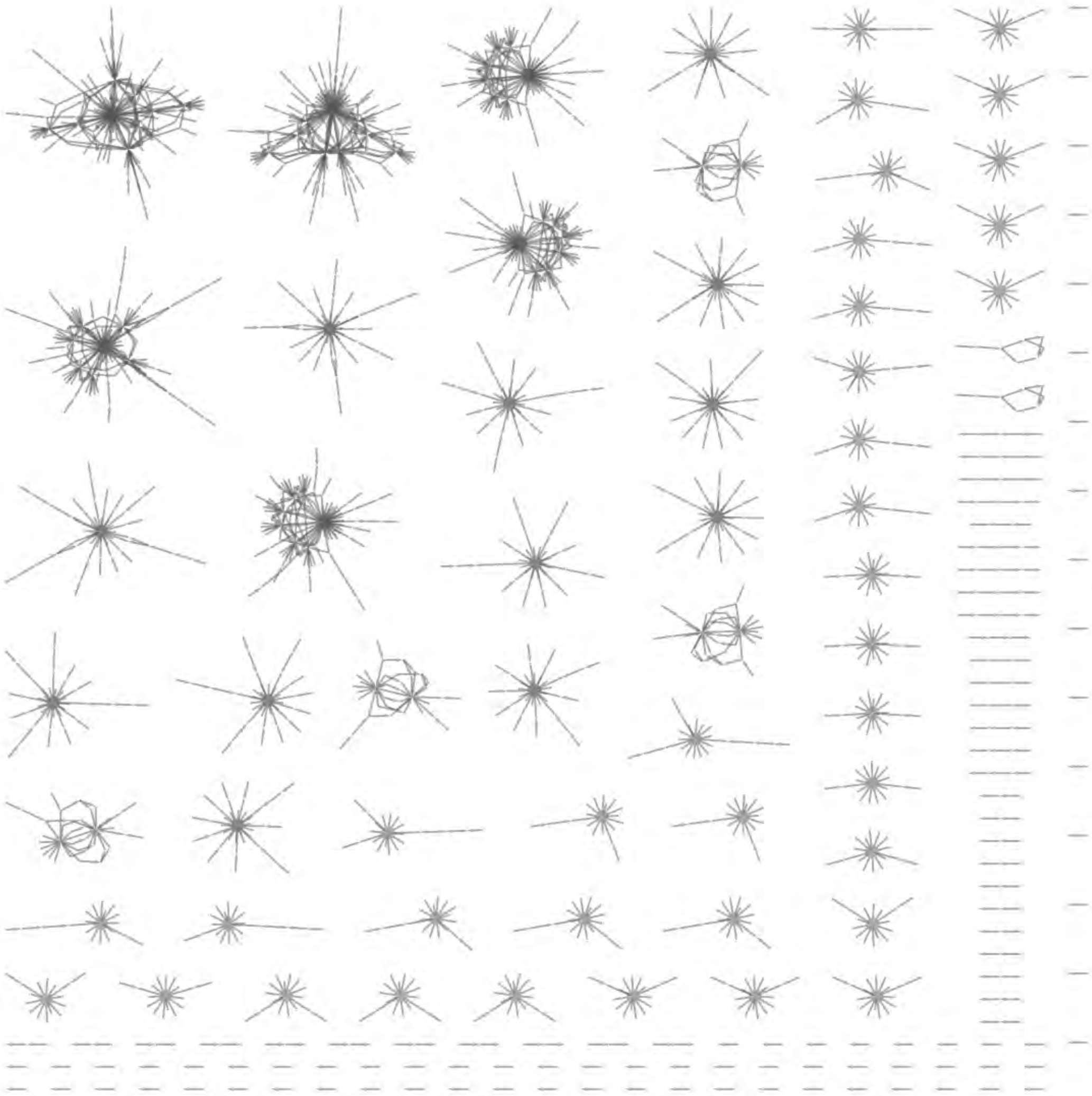

For size 7 and above, it becomes infinite. Here’s the beginning of the graph for size-8 expressions involving only S:

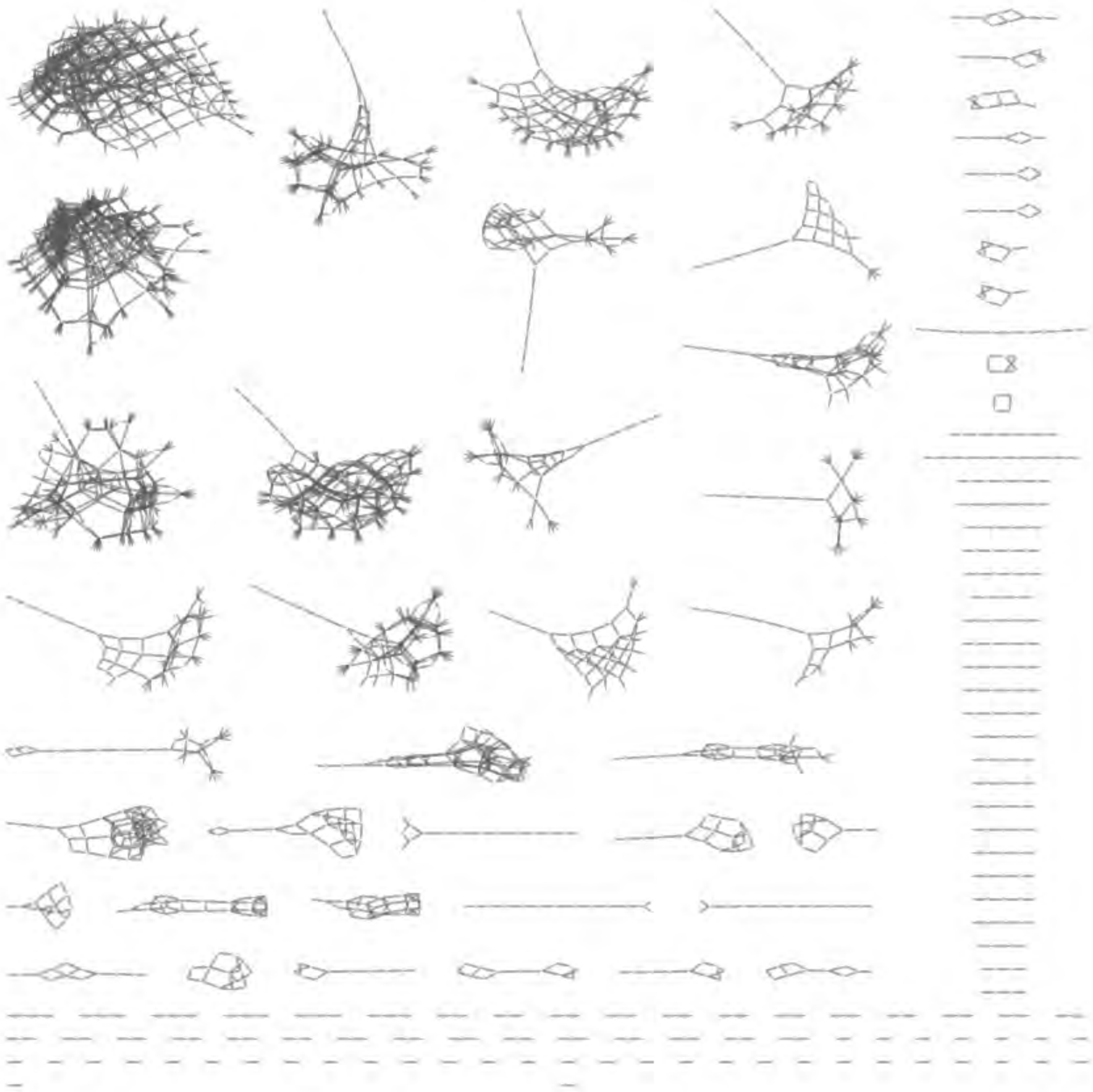

If one keeps only terminating cases, one gets for size 8:

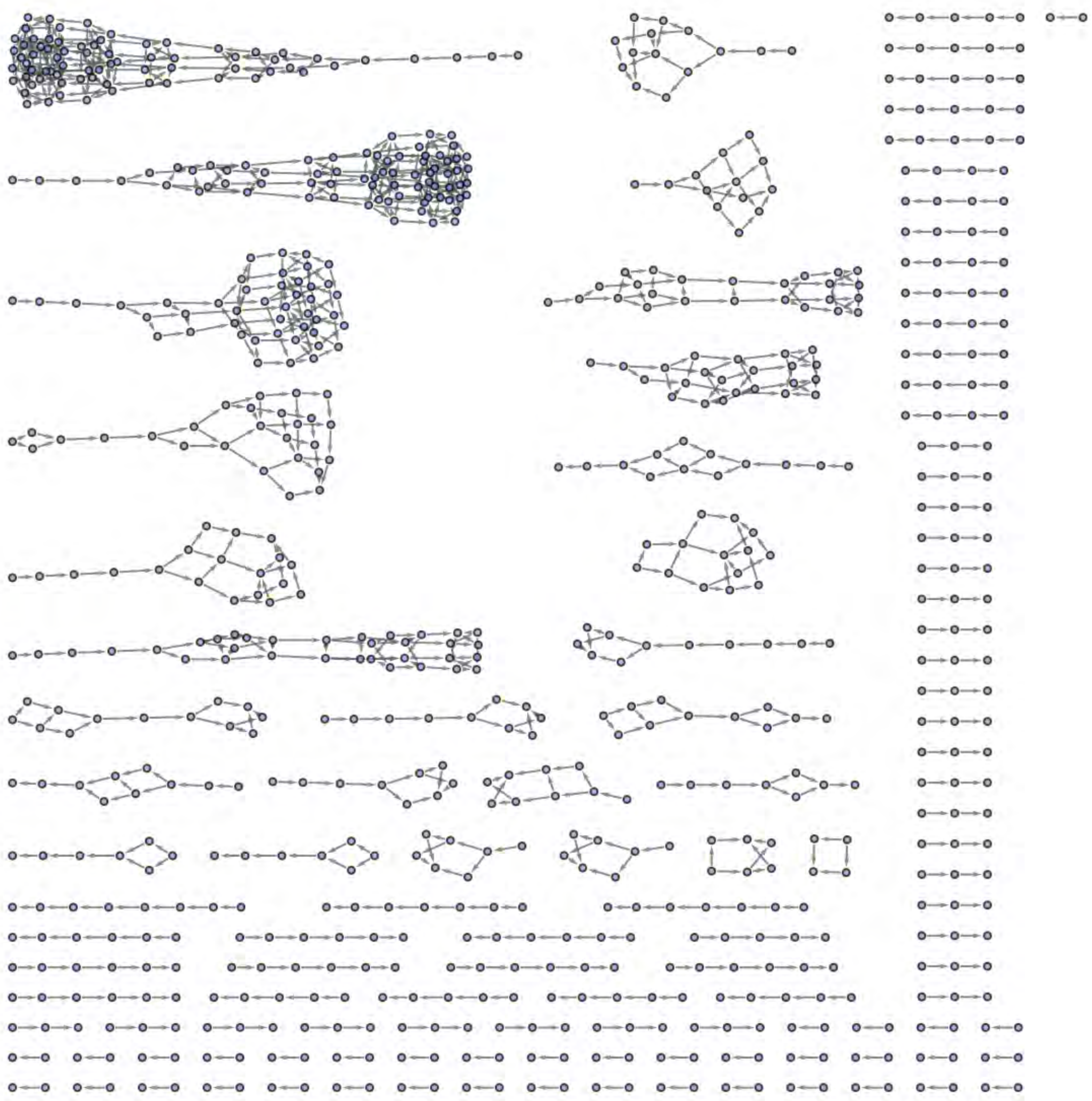

For size 9:

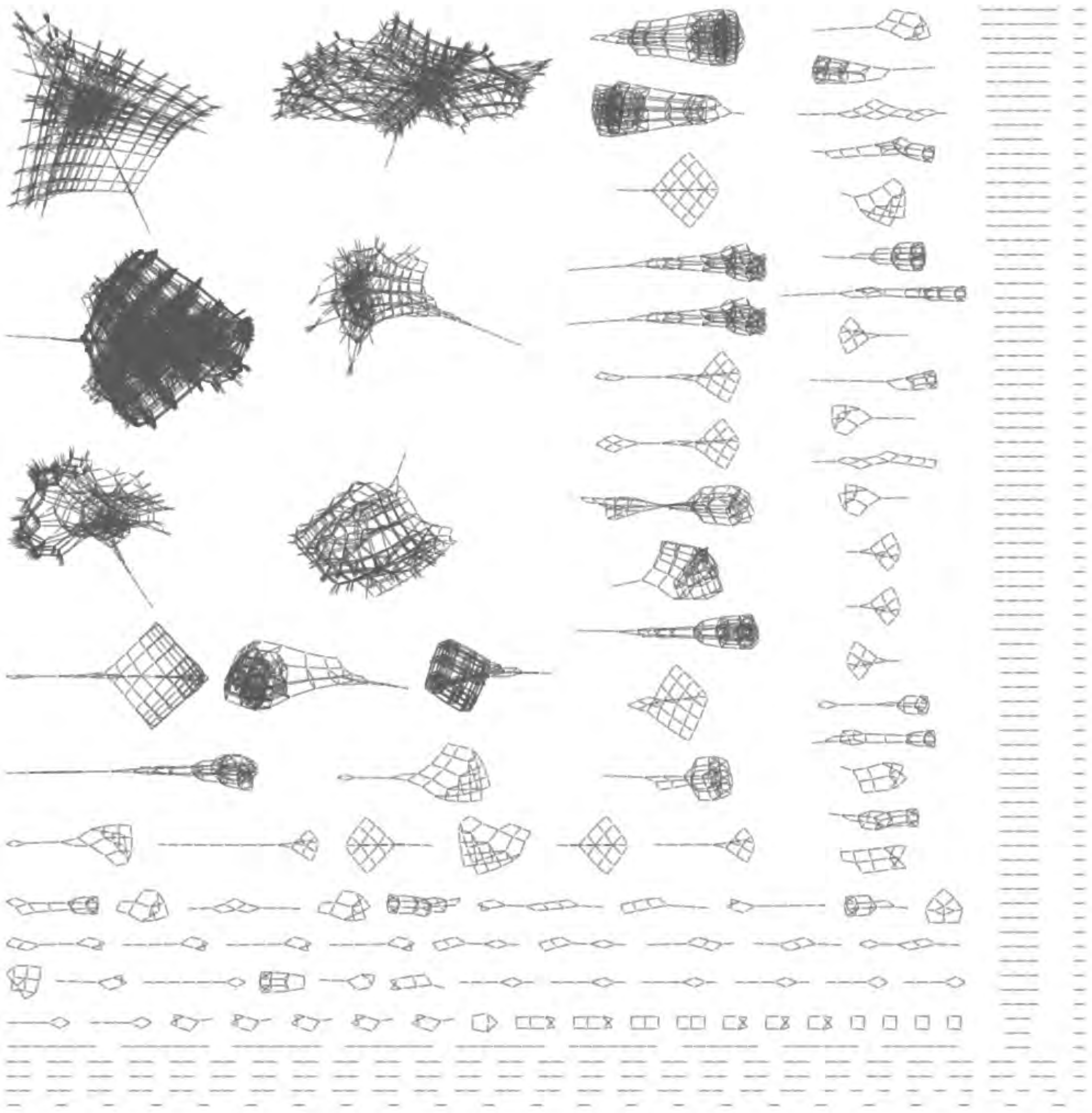

And for size 10:

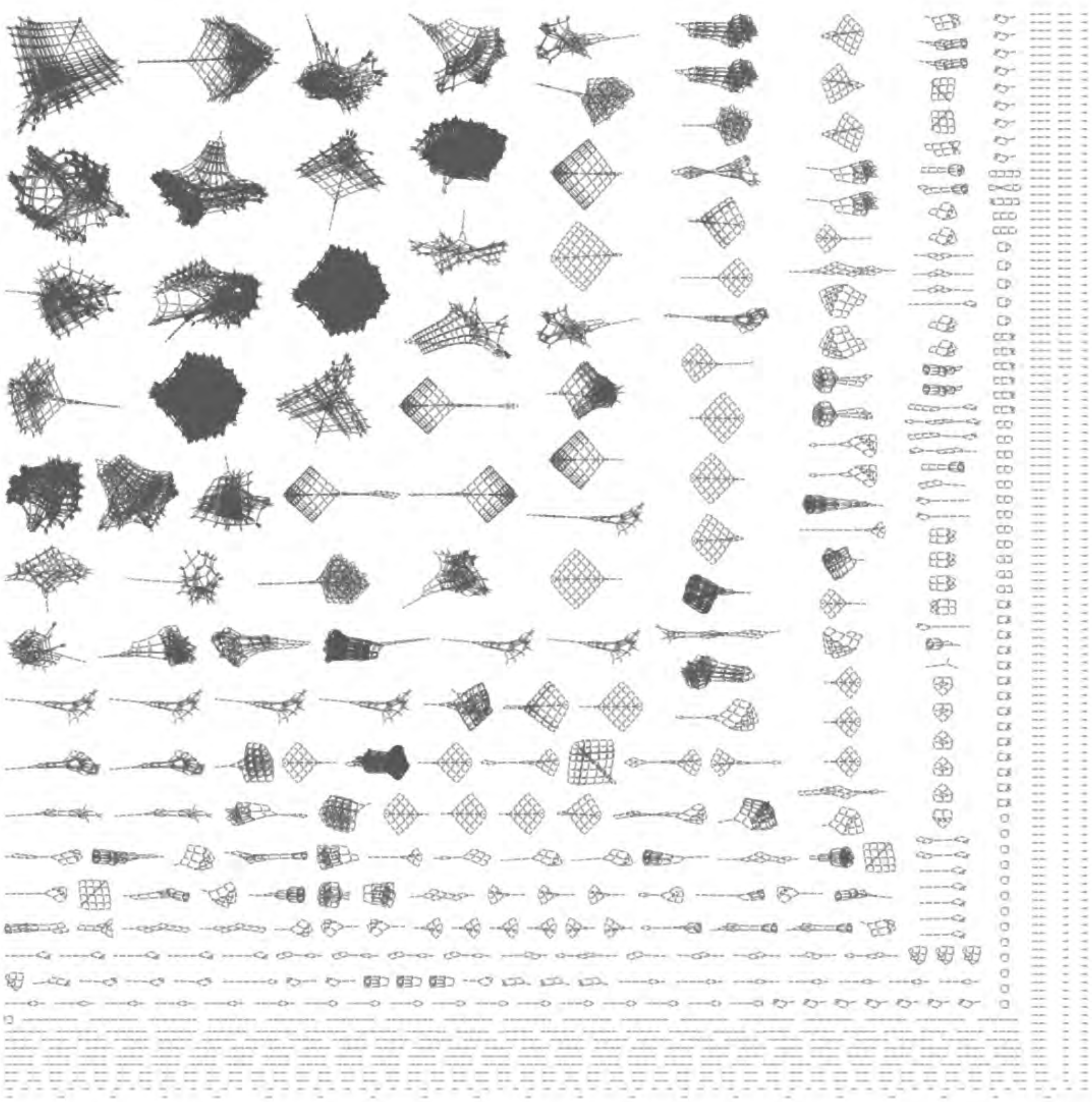

To assess the “most useful” transformations for “finding equations” there’s more to do: not only do we need to track what leads to what, but we also need to track causal relationships. And this leads to ideas like using lemmas that have the largest number of causal edges associated with them.

But are there perhaps other ways to find relations between combinator expressions, and combinator theorems? Can we for example figure out what combinator expressions are “close to” what others? In a sense what we need is to define a “space of combinator expression” with some appropriate notion of nearness.

One approach would just be to look at "raw distances" between trees—say based on asking how many edits have to be made to one tree to get to another. But an approach that more closely reflects actual features of combinators is to think about the concept of branchial graphs and branchial space that comes from our Physics Project.

Consider for example the multiway graph generated from s[s[s]][s][s[s]][s] (**S**(**SS**)**S**(**SS**)**S**):

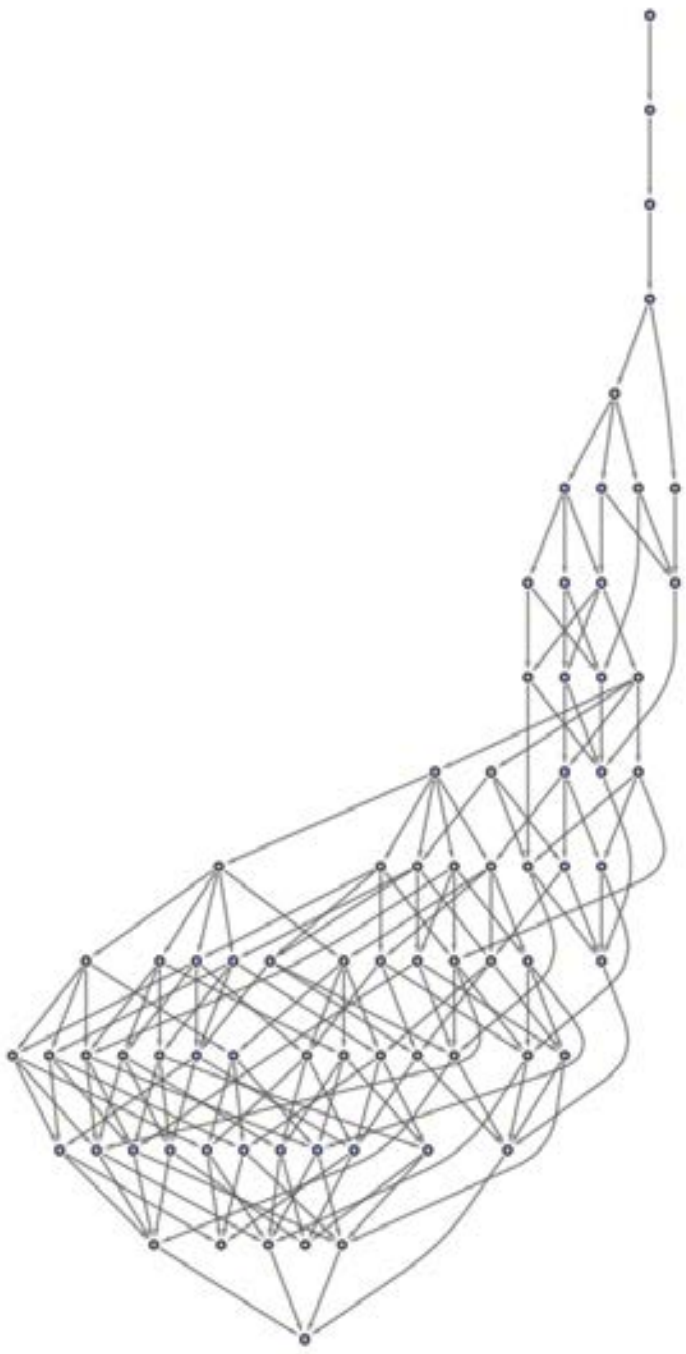

Now consider a foliation of this graph (and in general there will be many possible foliations that respect the partial order defined by the multiway graph):

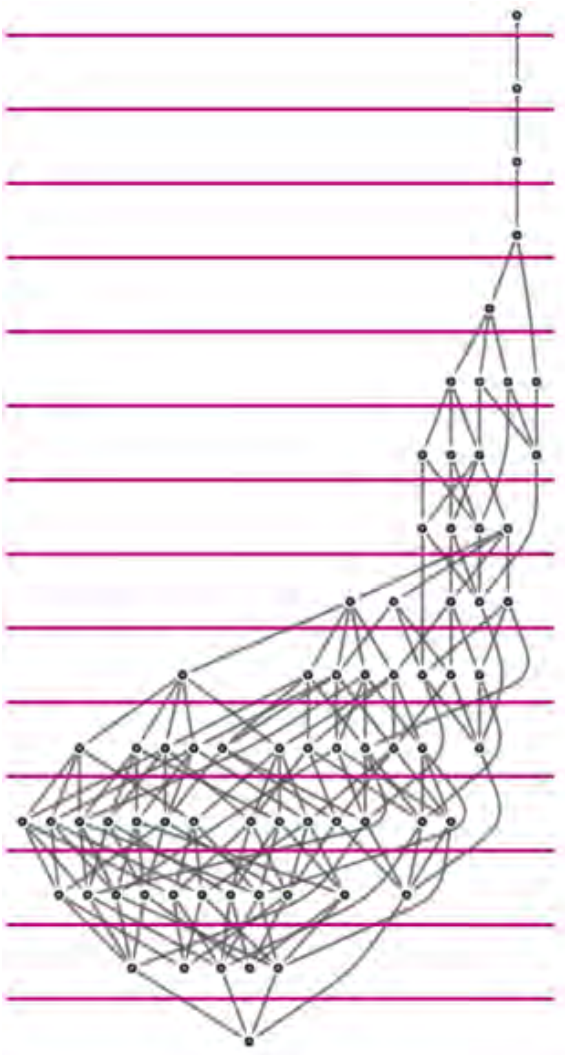

In each slice, we can then define—as in our Physics Project—a branchial graph in which nodes are joined when they have an immediate common ancestor in the multiway graph. In the case shown here, the branchial graphs in successive slices are:

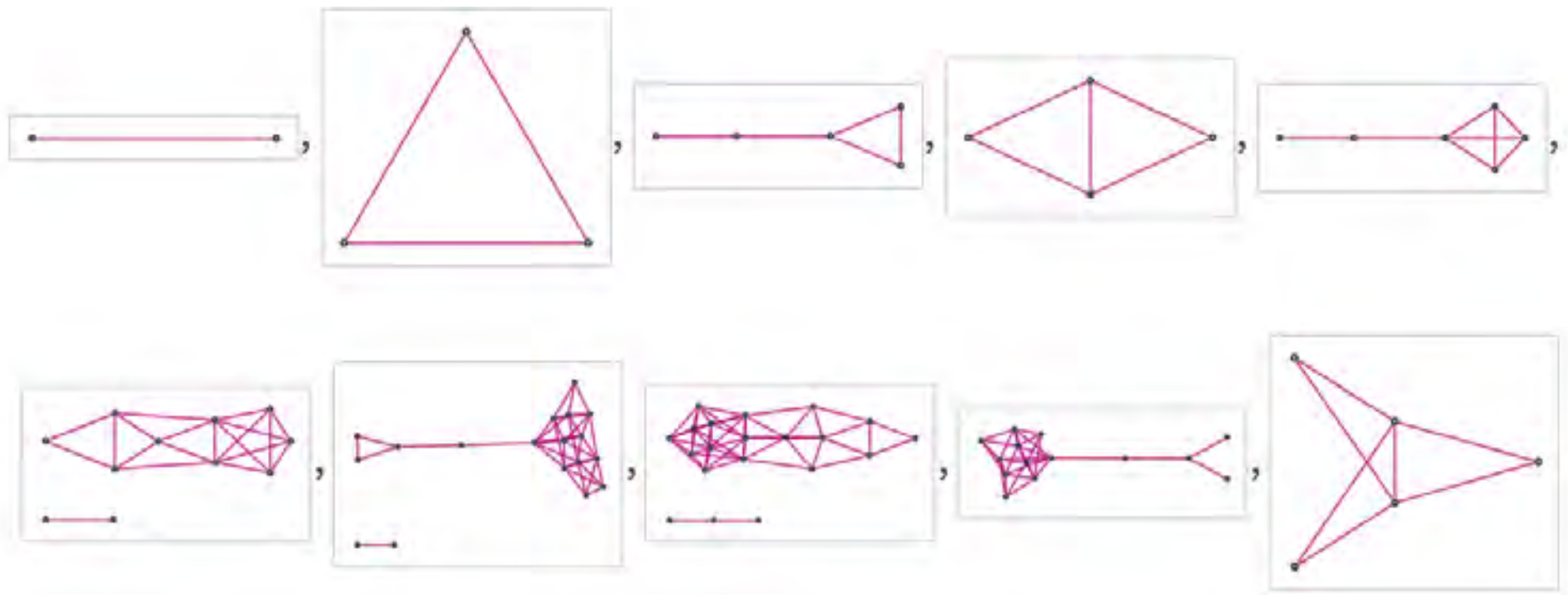

If we consider a combinator expression like s[s][s][s[s]][s][s] (**SSS**(**SS**)**SS**) that leads to infinite growth, we can ask what the "long-term" structure of the branchial graph will be. Here are the results after 18 and 19 steps:

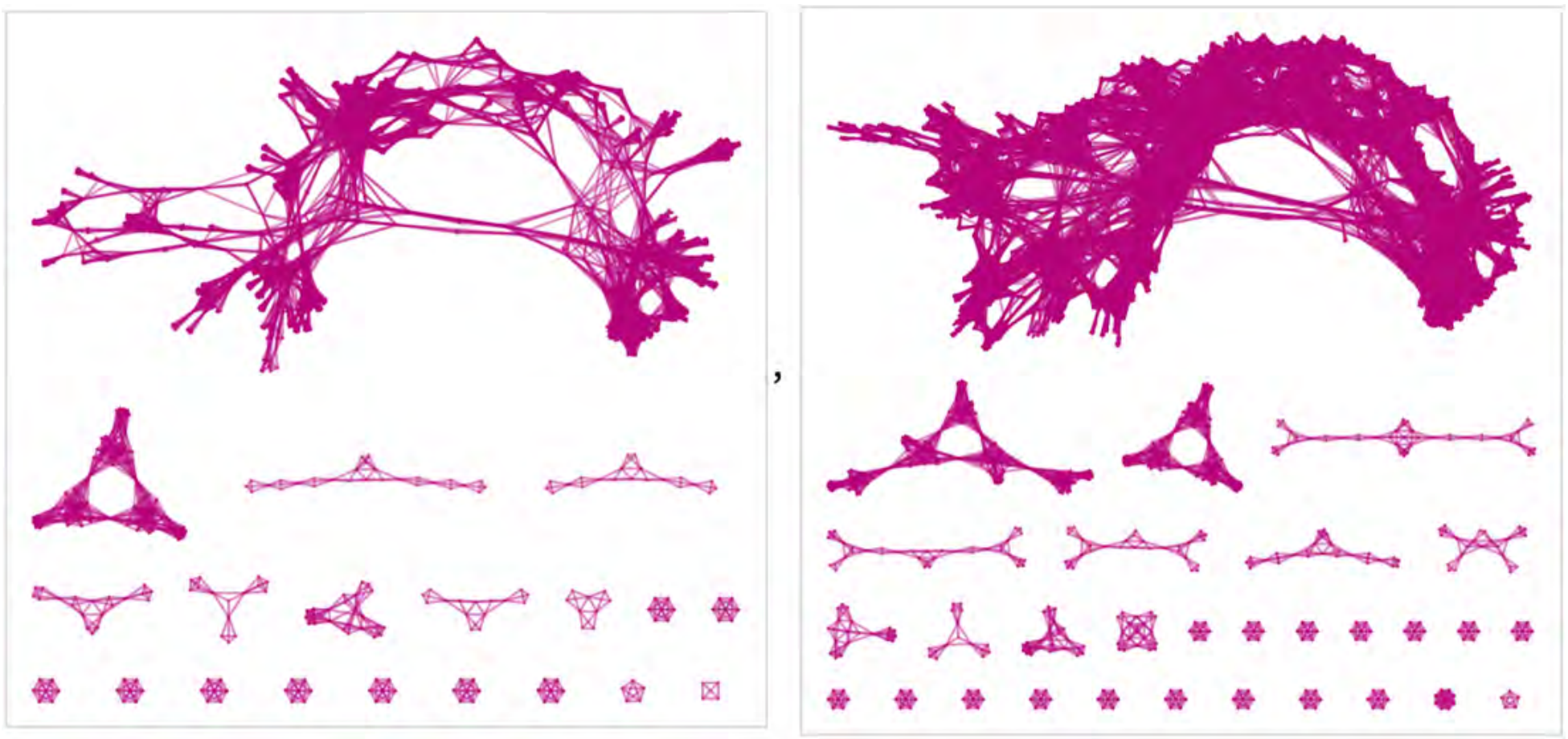

The largest connected components here contain respectively 1879 and 10,693 combinator expressions. But what can we say about their structure? One thing suggested by our Physics Project is to try to "fit them to continuous spaces". And a first step in doing that is to estimate their effective dimension—which one can do by looking at the growth in the volume of a "geodesic ball" in the graph as a function of its radius:

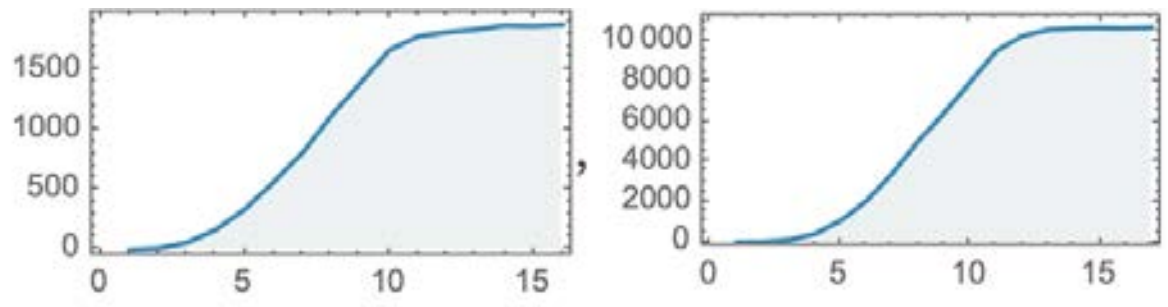

The result for distances small compared to the diameter of the graph is close to quadratic growth—suggesting that there is some sense in which the space of combinator expressions generated in this way may have a limiting 2D manifold structure.

It’s worth pointing out that different foliations of the multiway graph (i.e. using different “reference frames”) will lead to different branchial graphs—but presumably the (suitably defined) causal invariance of combinator evolution will lead to relativistic-like invariance properties of the branchial graphs.

Somewhat complementary to looking at foliations of the multiway graph is the idea of trying to find quantities that can be computed for combinator expressions to determine whether the combinator expressions can be equal. Can we in essence find hash codes for combinator expressions that are equal whenever the combinator expressions are equal?

In general we’ve been looking at “purely symbolic” combinator expressions—like:

**K•(K•(S•K)•(S•K))•(S•K•(K•(S•K)•(S•K)))**

But what if we consider S, K to have definite, say numerical, values, and • to be some kind of generalized multiplication operator that combines these values? We used this kind of approach above in finding a procedure for determining whether S combinator expressions will evolve to fixed points. And in general each possible choice of “multiplication functions” (and S, K “constant values”) can be viewed in mathematical terms as setting up a "model" (in the model-theoretic sense) for the “combinatory algebra”.

As a simple example, let’s consider a finite model in which there are just 2 possible values, and the “multiplication table” for the • operator is:

| • | 1 | 2 |
|---|---|---|
| 1 | 2 | 1 |
| 2 | 2 | 2 |

If we consider S combinator expressions of size 5, there are a total of 14 such expressions, in 10 equivalence classes, that evolve to different fixed points. If we now “evaluate the trees” according to our “model for •” we can see that within each equivalence class the value accumulated at the root of the tree is always the same, but differs between at least some of the equivalence classes:

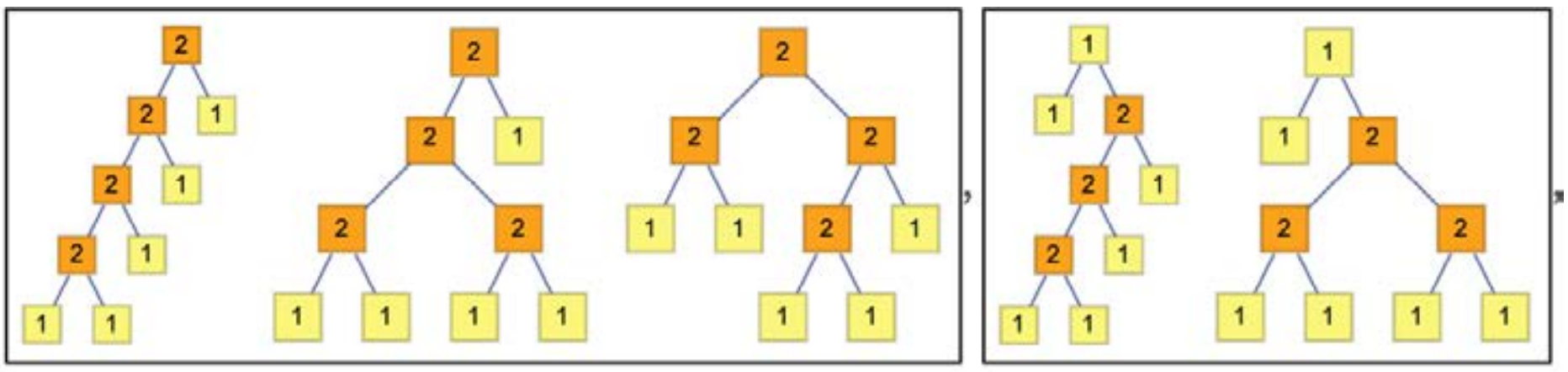

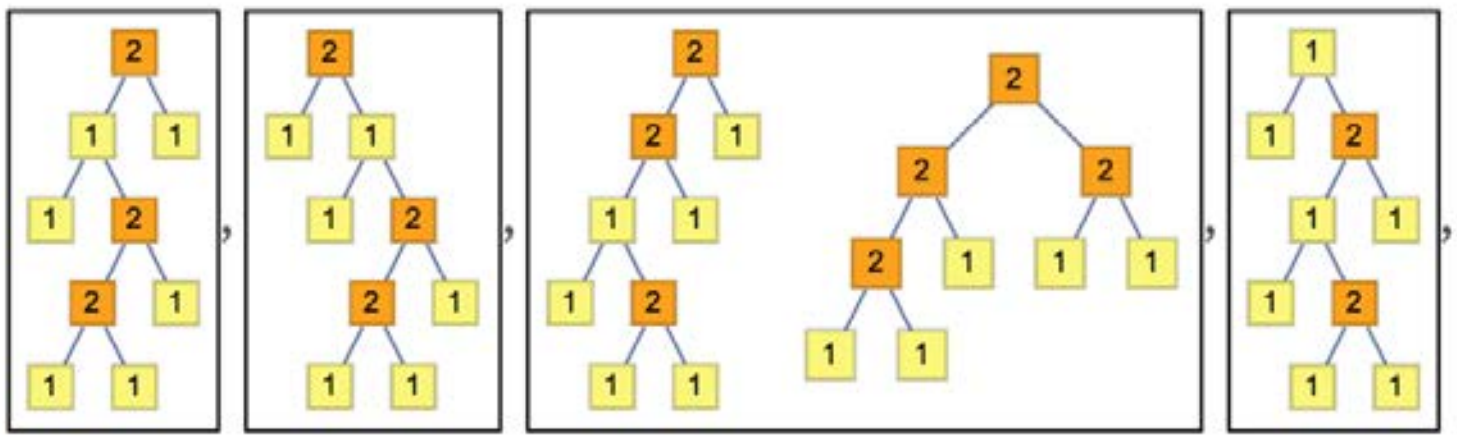

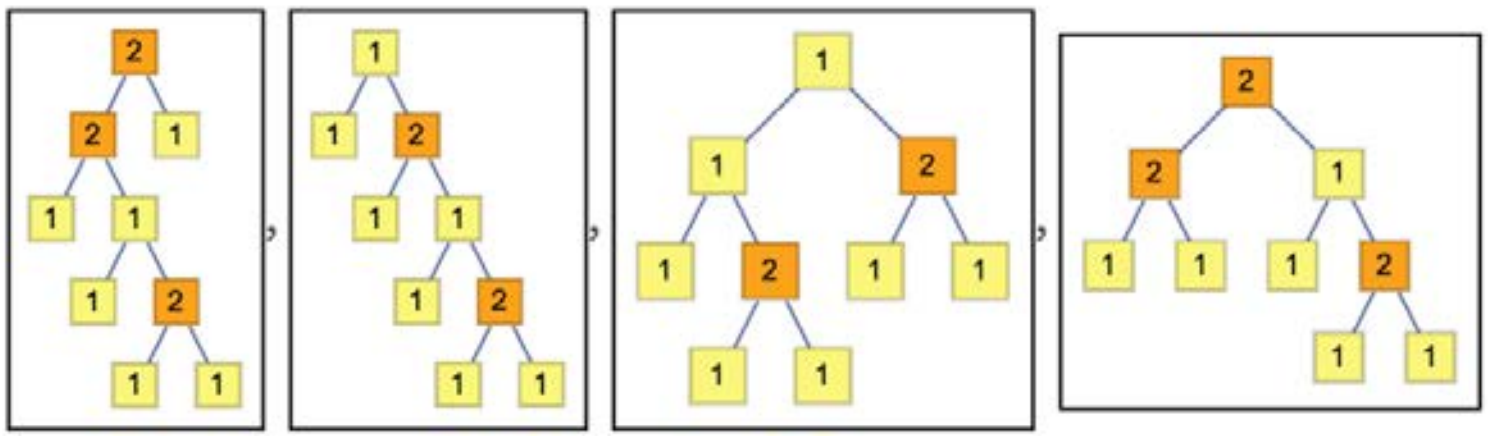

If we look at larger combinator expressions this all keeps working—until we get to two particular size-10 expressions, which have the same fixed point, but different “values”:

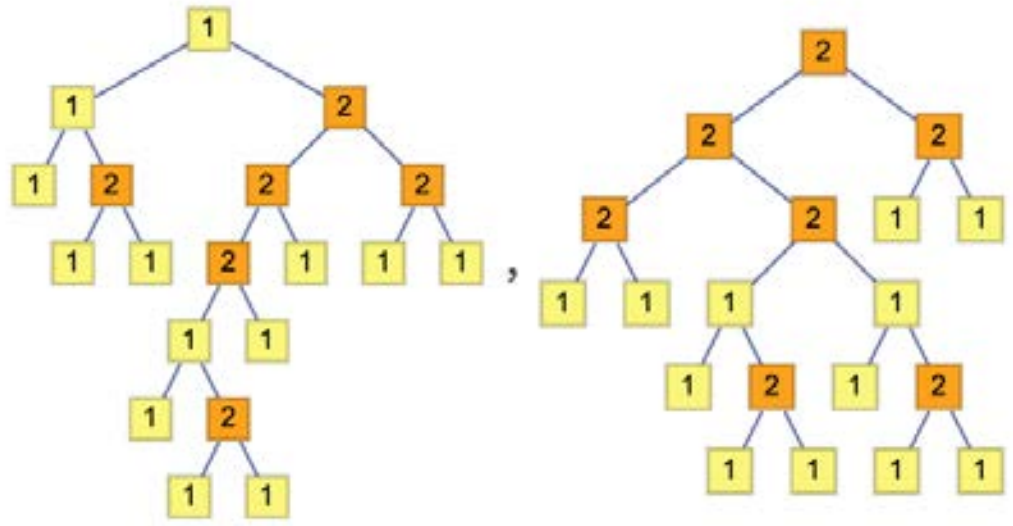

Allowing 3 possible values, the longest-surviving models are

| • | 1 | 2 | 3 |
|---|---|---|---|
| 1 | 2 | 3 | 2 |
| 2 | 2 | 2 | 2 |
| 3 | 2 | 2 | 1 |

,

| • | 1 | 2 | 3 |
|---|---|---|---|
| 1 | 3 | 3 | 2 |
| 2 | 3 | 1 | 3 |
| 3 | 3 | 3 | 3 |

but these both fail at size 13 (e.g. for s[s][s[s]][s[s[s[s[s]][s][s]]][s[s]]], s[s[s]][s[s[s[s[s]][s[s[s]]]]]][s[s]]).

The fact that combinator equivalence is in general undecidable means we can't expect to find a computationally finite "valuation procedure" that will distinguish all inequivalent combinator expressions. But it's still conceivable that we could have a scheme to distinguish some classes of combinator expressions from others—in essence through the values of a kind of "conserved quantity for combinators".

Another approach is to consider directly "combinator axioms" like

{**S**•x•y•z=x•z•(y•z), **K**•x•y=x}

and simply ask if there are models of •, S and K that satisfy them. Assuming a finite "multiplication table", there's no way to do this for K, and thus for S and K together. For S alone, however, there are already 8 2-valued models, and 285 3-valued ones.

The full story is more complicated, and has been the subject of a fair amount of academic work on combinators over the past half century. The main result is that there are models that are in principle known to exist, though they're infinite and probably can't be explicitly constructed.

In the case of something like arithmetic, there are formal axioms (the Peano axioms). But we know that (even though Gödel's theorem shows that there are inevitably also other, exotic, non-standard models) there's a model of these axioms that is the ordinary integers. And our familiarity with these and their properties makes us feel that the Peano axioms aren't just formal axioms; they're axioms "about" something, namely integers.

What are the combinator axioms "about"? There's a perfectly good interpretation of them in terms of computational processes. But there doesn't seem to be some "static" set of constructs—like the integers—that give one more insight about what combinators "really are". Instead, it seems, combinators are in the end just through and through computational.

# Empirical Computation Theory with Combinators

We've talked a lot here about what combinators "naturally do". But what about getting combinators to do something specific—for example to perform a particular computation we want?

As we saw by example at the beginning of this piece, it's not difficult to take any symbolic structure and "compile it" to combinators. Let's say we're given:

f[y[x]][y][x]

There's then a recursive procedure that effectively builds "function invocations" out of S's and "stops computations" with K's. And using this we can "compile" our symbolic expression to the (slightly complicated) combinator expression:

s[s[k[s]][s[k[s[k[s]]]][s[s[k[s]][s[k[s[k[s]]]][s[s[k[s]][s[k[k]][s[k[s]][k]]]][k[s[k[s[s[k][k]]]][k]]]]]][k[k[s[k][k]]]]]][k[k]]

To "compute our original expression" we just have to take this combinator expression ("■"), form ■[f][x][y], then apply the combinator rules and find the fixed point:

```
s[s[k[s]][s[k[s[k[s]]]][s[s[k[s]][s[k[s[k[s]]]][s[s[k[s]][s[k[k]][s[k[s]][k]]]][k[s[k[s[s[k][k]]][k]]]]][k[k[s[k][k]]]]]][k[k]][f][x][y]
s[k[s]][s[k[s[k[s]]]][s[s[k[s]][s[k[s[k[s]]]][s[s[k[s]][s[k[k]][s[k[s]][k]]]][k[s[k[s[s[k][k]]][k]]]]][k[k[s[k][k]]]]]][f][k[k][f]][x][y]
k[s][f][s[k[s[k[s]]]][s[s[k[s]][s[k[s[k[s]]]][s[s[k[s]][s[k[k]][s[k[s]][k]]]][k[s[k[s[s[k][k]]][k]]]]][k[k[s[k][k]]]]][f][k[k][f]][x][y]
s[s[k[s[k[s]]]][s[s[k[s]][s[k[s[k[s]]]][s[s[k[s]][s[k[k]][s[k[s]][k]]]][k[s[k[s[s[k][k]]][k]]]]][k[k[s[k][k]]]][f]][k[k][f]][x][y]
s[k[s[k[s]]]][s[s[k[s]][s[k[s[k[s]]]][s[s[k[s]][s[k[k]][s[k[s]][k]]]][k[s[k[s[s[k][k]]][k]]]]][k[k[s[k][k]]]][f][x][k[k][f][x]][y]
k[s[k[s]]][f][s[s[k[s]][s[k[s[k[s]]]][s[s[k[s]][s[k[k]][s[k[s]][k]]]][k[s[k[s[s[k][k]]][k]]]]][k[k[s[k][k]]]][f]][x][k[k][f][x]][y]
s[k[s]][s[s[k[s]][s[k[s[k[s]]]][s[s[k[s]][s[k[k]][s[k[s]][k]]]][k[s[k[s[s[k][k]]][k]]]]][k[k[s[k][k]]]][f]][x][k[k][f][x]][y]
k[s][x][s[s[k[s]][s[k[s[k[s]]]][s[s[k[s]][s[k[k]][s[k[s]][k]]]][k[s[k[s[s[k][k]]][k]]]]][k[k[s[k][k]]]][f][x]][k[k][f][x]][y]
s[s[s[k[s]][s[k[s[k[s]]]][s[s[k[s]][s[k[k]][s[k[s]][k]]]][k[s[k[s[s[k][k]]][k]]]]][k[k[s[k][k]]]][f][x]][k[k][f][x]][y]
s[s[k[s]][s[k[s[k[s]]]][s[s[k[s]][s[k[k]][s[k[s]][k]]]][k[s[k[s[s[k][k]]][k]]]]][k[k[s[k][k]]]][f][x][y][k[k][f][x][y]]
s[k[s]][s[k[s[k[s]]]][s[s[k[s]][s[k[k]][s[k[s]][k]]]][k[s[k[s[s[k][k]]][k]]]][f][k[k[s[k][k]]][f]][x][y][k[k][f][x][y]]
k[s][f][s[k[s[k[s]]]][s[s[k[s]][s[k[k]][s[k[s]][k]]]][k[s[k[s[s[k][k]]][k]]]][f]][k[k[s[k][k]]][f]][x][y][k[k][f][x][y]]
s[s[k[s[k[s]]]][s[s[k[s]][s[k[k]][s[k[s]][k]]]][k[s[k[s[s[k][k]]][k]]]][f]][k[k[s[k][k]]][f]][x][y][k[k][f][x][y]]
s[k[s[k[s]]]][s[s[k[s]][s[k[k]][s[k[s]][k]]]][k[s[k[s[s[k][k]]][k]]]][f][x][k[k[s[k][k]]][f][x]][y][k[k][f][x][y]]
k[s[k[s]]][f][s[s[k[s]][s[k[k]][s[k[s]][k]]]][k[s[k[s[s[k][k]]][k]]][f]][x][k[k[s[k][k]]][f][x]][y][k[k][f][x][y]]
s[k[s]][s[s[k[s]][s[k[k]][s[k[s]][k]]]][k[s[k[s[s[k][k]]][k]]][f]][x][k[k[s[k][k]]][f][x]][y][k[k][f][x][y]]
k[s][x][s[s[k[s]][s[k[k]][s[k[s]][k]]]][k[s[k[s[s[k][k]]][k]]][f][x]][k[k[s[k][k]]][f][x]][y][k[k][f][x][y]]
s[s[s[k[s]][s[k[k]][s[k[s]][k]]]][k[s[k[s[s[k][k]]][k]]][f][x]][k[k[s[k][k]]][f][x]][y][k[k][f][x][y]]
s[s[k[s]][s[k[k]][s[k[s]][k]]]][k[s[k[s[s[k][k]]][k]]][f][x][y][k[k[s[k][k]]][f][x][y]][k[k][f][x][y]]
s[k[s]][s[k[k]][s[k[s]][k]]][f][k[s[k[s[s[k][k]]][k]]][f]][x][y][k[k[s[k][k]]][f][x][y]][k[k][f][x][y]]
k[s][f][s[k[k]][s[k[s]][k]][f]][k[s[k[s[s[k][k]]][k]]][f]][x][y][k[k[s[k][k]]][f][x][y]][k[k][f][x][y]]
s[s[k[k]][s[k[s]][k]][f]][k[s[k[s[s[k][k]]][k]]][f]][x][y][k[k[s[k][k]]][f][x][y]][k[k][f][x][y]]
s[k[k]][s[k[s]][k]][f][x][k[s[k[s[s[k][k]]][k]]][f][x]][y][k[k[s[k][k]]][f][x][y]][k[k][f][x][y]]
k[k][f][s[k[s]][k][f]][x][k[s[k[s[s[k][k]]][k]]][f][x]][y][k[k[s[k][k]]][f][x][y]][k[k][f][x][y]]
k[s[k[s]][k][f]][x][k[s[k[s[s[k][k]]][k]]][f][x]][y][k[k[s[k][k]]][f][x][y]][k[k][f][x][y]]
s[k[s]][k][f][k[s[k[s[s[k][k]]][k]]][f][x]][y][k[k[s[k][k]]][f][x][y]][k[k][f][x][y]]
k[s][f][k[f]][k[s[k[s[s[k][k]]][k]]][f][x]][y][k[k[s[k][k]]][f][x][y]][k[k][f][x][y]]
s[k[f]][k[s[k[s[s[k][k]]][k]]][f][x]][y][k[k[s[k][k]]][f][x][y]][k[k][f][x][y]]
k[f][y][k[s[k[s[s[k][k]]][k]]][f][x][y]][k[k[s[k][k]]][f][x][y]][k[k][f][x][y]]
f[k[s[k[s[s[k][k]]][k]]][f][x][y]][k[k[s[k][k]]][f][x][y]][k[k][f][x][y]]
f[s[k[s[s[k][k]]][k][x][y]][k[k[s[k][k]]][f][x][y]][k[k][f][x][y]]
f[k[s[s[k][k]]][x][k[x]][y]][k[k[s[k][k]]][f][x][y]][k[k][f][x][y]]
f[s[s[k][k]][k[x]][y]][k[k[s[k][k]]][f][x][y]][k[k][f][x][y]]
f[s[k][k][y][k[x][y]]][k[k[s[k][k]]][f][x][y]][k[k][f][x][y]]
f[k[y][k[y]][k[x][y]]][k[k[s[k][k]]][f][x][y]][k[k][f][x][y]]
f[y[k[x][y]]][k[k[s[k][k]]][f][x][y]][k[k][f][x][y]]
f[y[x]][k[k[s[k][k]]][f][x][y]][k[k][f][x][y]]
f[y[x]][k[s[k][k]][x][y]][k[k][f][x][y]]
f[y[x]][s[k][k][y]][k[k][f][x][y]]
f[y[x]][k[y][k[y]]][k[k][f][x][y]]
f[y[x]][y][k[k][f][x][y]]
f[y[x]][y][k[x][y]]
f[y[x]][y][x]
```

But is this the "best combinator way" to compute this result?

There are various different things we could mean by "best". Smallest program? Fastest program? Most memory-efficient program? Or said in terms of combinators: Smallest combinator expression? Smallest number of rule applications? Smallest intermediate expression growth?

In computation theory one often talks theoretically about optimal programs and their characteristics. But when one's used to studying programs "in the wild" one can start to do empirical studies of computation-theoretic questions—as I did, for example, with simple Turing machines in *A New Kind of Science*.

Traditional computation theory tends to focus on asymptotic results about "all possible programs". But in empirical computation theory one's dealing with specific programs—and in practice there's a limit to how many one can look at. But the crucial and surprising fact that comes from studying the computational universe of "programs in the wild" is that actually even very small programs can show highly complex behavior that's in some sense typical of all possible programs. And that means that it's realistic to get intuition—and results—about computation-theoretic questions just by doing empirical investigations of actual, small programs.

So how does this work with combinators? An immediate question to ask is: if one wants a particular expression, what are all the possible combinator expressions that will generate it?

Let's start with a seemingly trivial case: x[x]. With the compilation procedure we used above we get the size-7 combinator expression

s[s[k][k]][s[k][k]]

which (with leftmost outermost evaluation) generates x[x] in 6 steps:

| |
|---|
| **s**[**s**[**k**][**k**]][**s**[**k**][**k**]][x] |
| **s**[**k**][**k**][x][**s**[**k**][**k**][x]] |
| **k**[x][**k**[x]][**s**[**k**][**k**][x]] |
| x[**s**[**k**][**k**][x]] |
| x[**k**[x][**k**[x]]] |
| x[x] |

But what happens if we just start enumerating possible combinator expressions? Up to size 5, none compute x[x]. But at size 6, we have:

| |
|---|
| **s**[**s**[**s**]][**s**][**s**[**k**]][x] |
| **s**[**s**][**s**[**k**]][**s**[**s**[**k**]]][x] |
| **s**[**s**[**s**[**k**]]][**s**[**k**][**s**[**s**[**k**]]]][x] |
| **s**[**s**[**k**]][x][**s**[**k**][**s**[**s**[**k**]]][x]] |
| **s**[**k**][**s**[**k**][**s**[**s**[**k**]]][x]][x[**s**[**k**][**s**[**s**[**k**]]][x]]] |
| **k**[x[**s**[**k**][**s**[**s**[**k**]]][x]]][**s**[**k**][**s**[**s**[**k**]]][x][x[**s**[**k**][**s**[**s**[**k**]]][x]]]] |
| x[**s**[**k**][**s**[**s**[**k**]]][x]] |
| x[**k**[x][**s**[**s**[**k**]][x]]] |
| x[x] |

So we can "save" one unit of program size, but at the "cost" of taking 9 steps, and having an intermediate expression of size 21.

What if we look at size 7 programs? There are a total of 11 that work (including the one from our "compiler"):

{s[s[s[s]]][s][s[k]], s[s[s]][s[k]][s[k]], s[s][s[k]][s[s[k]]], s[s[s[k]]][s[k][s]], s[s][s[k]][s[k][s]],
s[s[s[k]]][s[k][k]], s[s][s[k]][s[k][k]], s[s[k][s]][s[k][s]], s[s[k][s]][s[k][k]], s[s[k][k]][s[k][s]], s[s[k][k]][s[k][k]]}

How do these compare in terms of "time" (i.e. number of steps) and "memory" (i.e. maximum intermediate expression size)? There are 4 distinct programs that all take the same time and memory, there are none that are faster, but there are others that are slowest (the slowest taking 12 steps):

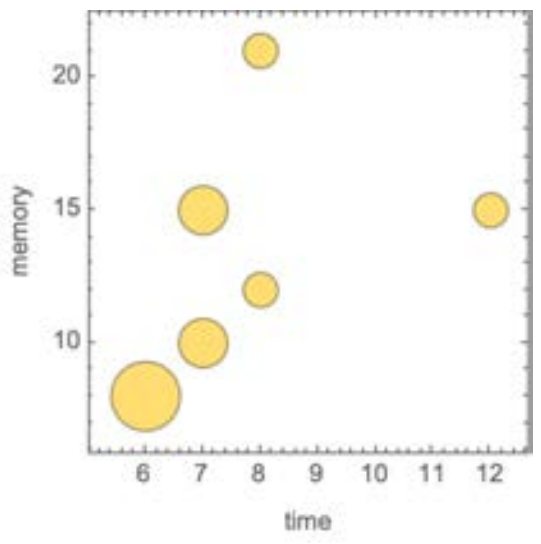

What happens with larger programs? Here's a summary:

| size | programs | min time | min memory | max time | max memory | median time | median memory |
|---|---|---|---|---|---|---|---|
| 6 | 1 | 9 | 21 | 9 | 21 | 9 | 21 |
| 7 | 11 | 6 | 8 | 12 | 21 | 7 | 10 |
| 8 | 95 | 6 | 9 | 18 | 39 | 10 | 15 |
| 9 | 730 | 6 | 10 | 67 | 207 | 8 | 14 |
| 10 | 5754 | 6 | 11 | 375 | 937 | 10 | 15 |

Here are the distributions of times (dropping outliers)—implying (as the medians above suggest) that even a randomly picked program is likely to be fairly fast:

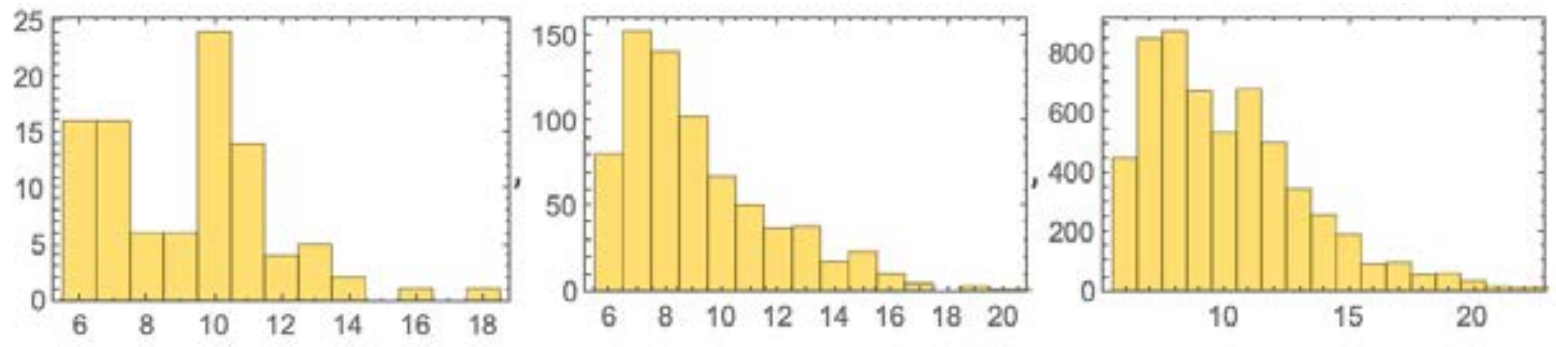

And here's the distribution of time vs. memory on a log-log scale:

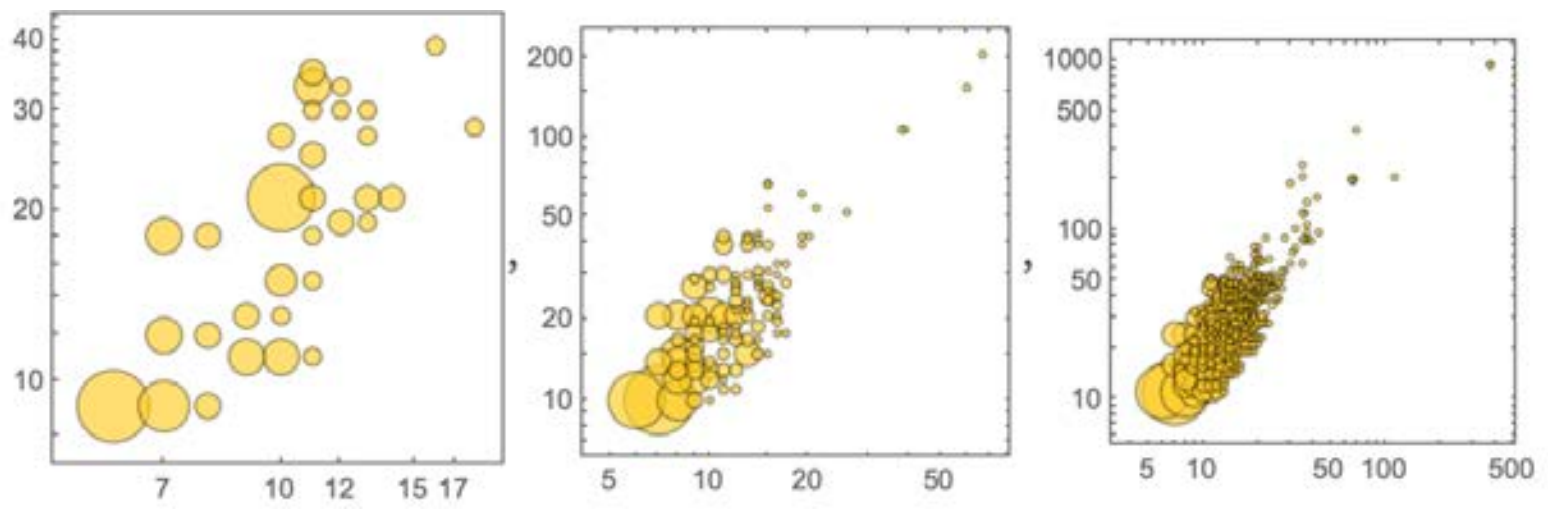

At size 10, the slowest and most memory-intensive program is s[s[s][k][s[s[s[s]]]]][s][k] (**S**(**SSK**(**S**(**S**(**SS**))))**SK**):

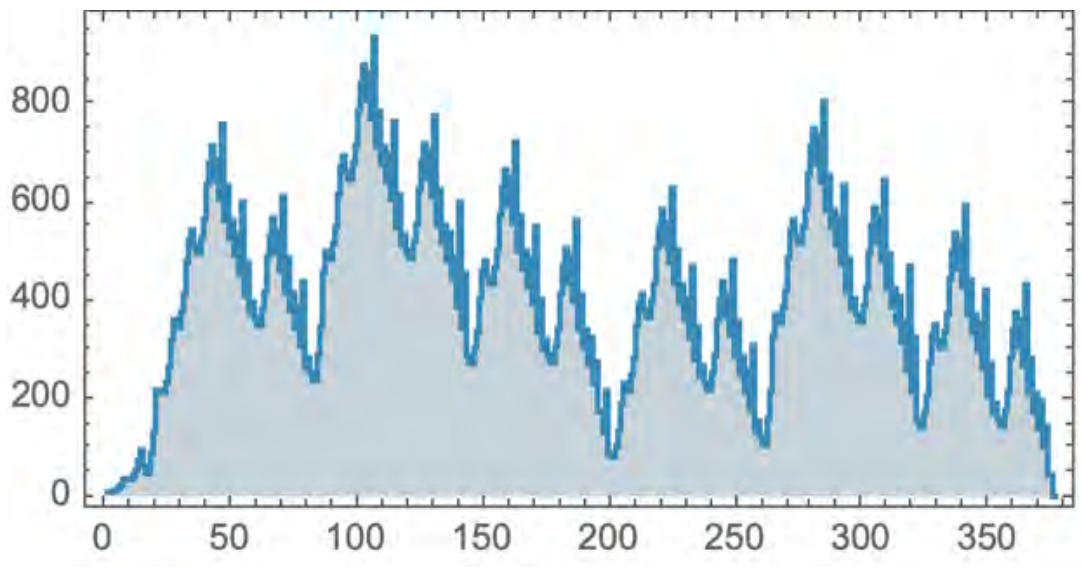

There are so many other questions one can ask. For example: how similar are the various fastest programs? Do they all "work the same way"? At size 7 they pretty much seem to:

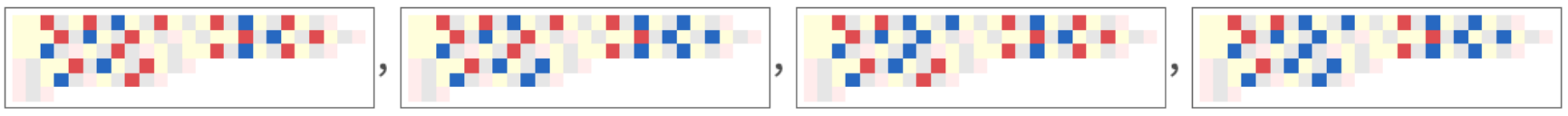

At size 8 there are a few "different schemes" that start to appear:

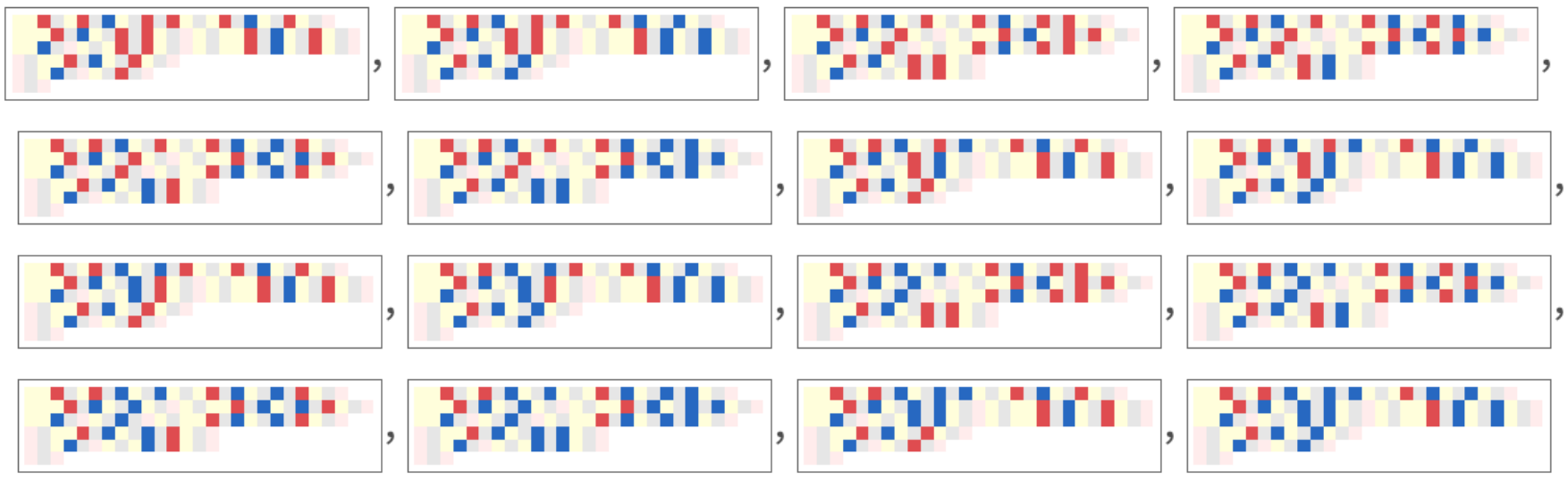

Then one can start to ask questions about how these fastest programs are laid out in the kind of "combinator space" we discussed in the last section—and whether there are good incremental ("evolutionary") ways to find these fastest programs.

Another type of question has to do with the running of our programs. In everything we've done so far in this section, we've used a definite evaluation scheme: leftmost outermost. And in using this definite scheme, we can think of ourselves as doing "deterministic combinator computation". Bu we can also consider the complete multiway system of all possible updating sequences—which amounts to doing non-deterministic computation.

Here's the multiway graph for the size-6 case we considered above, highlighting the leftmost-outermost evaluation path:

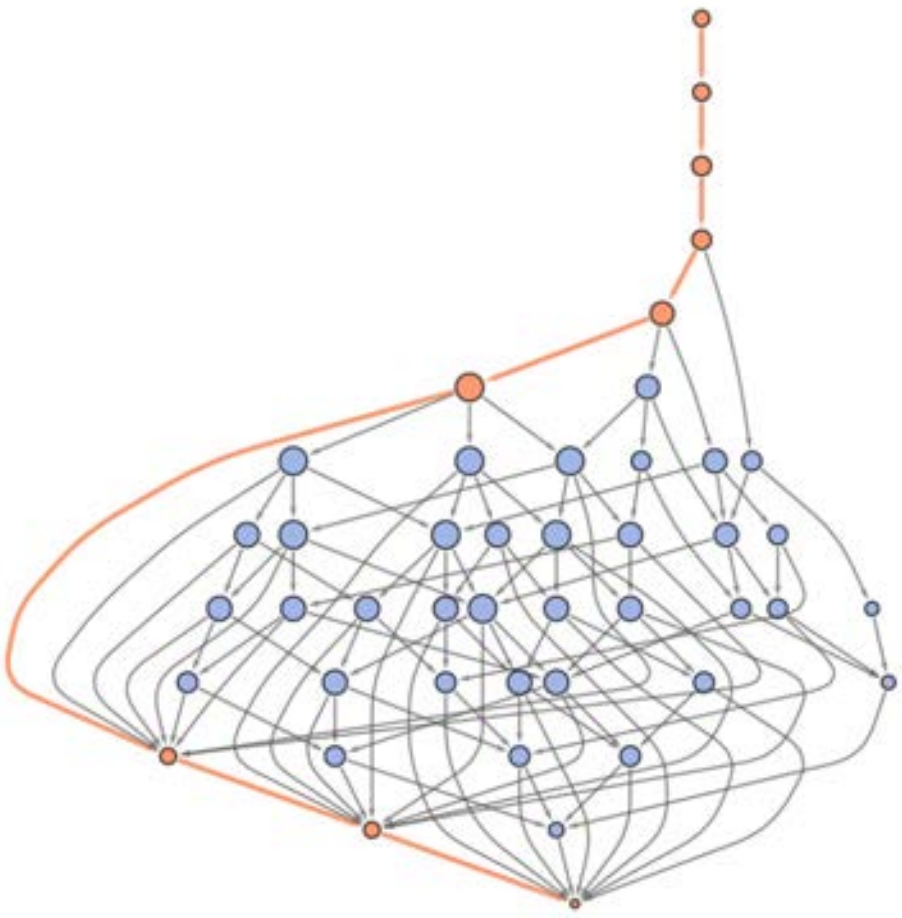

And, yes, in this case leftmost outermost happens to follow a fastest path here. Some other possible schemes are somewhat slow in comparison—with the maximum time being 13 and the maximum intermediate expression size being 21.

At size 7 the multiway graphs for all the leftmost-outermost-fastest programs are the same—and are very simple—among other things making it seem that in retrospect the size-6 case "only just makes it":

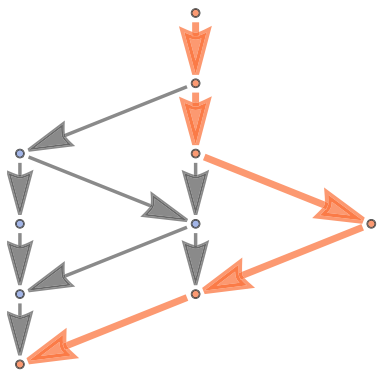

At size 8 there "two ideas" among the 16 cases:

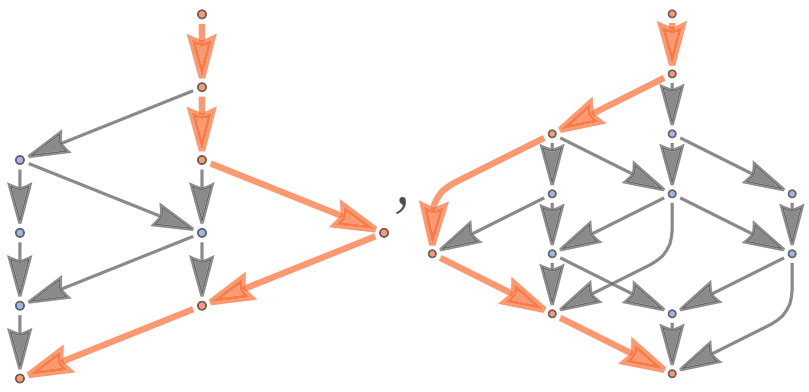

At size 9 there are "5 ideas" among 80 cases:

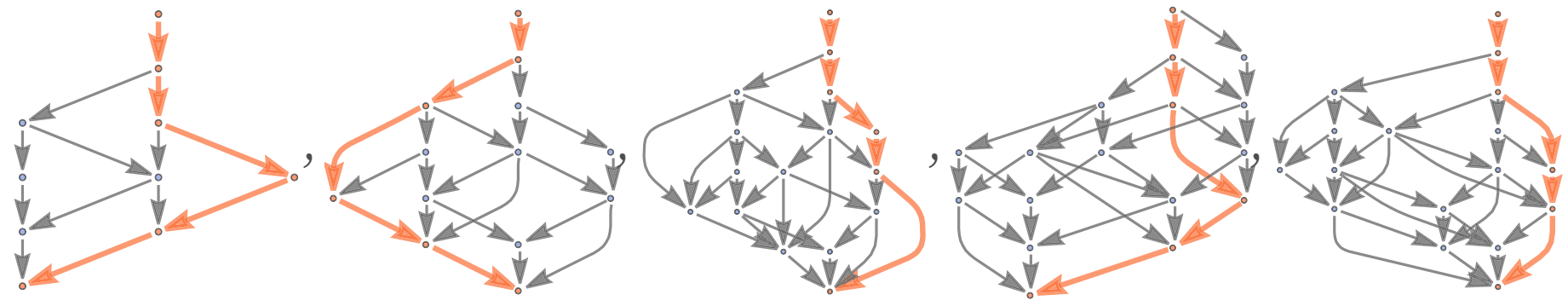

And at size 10 things are starting to get more complicated:

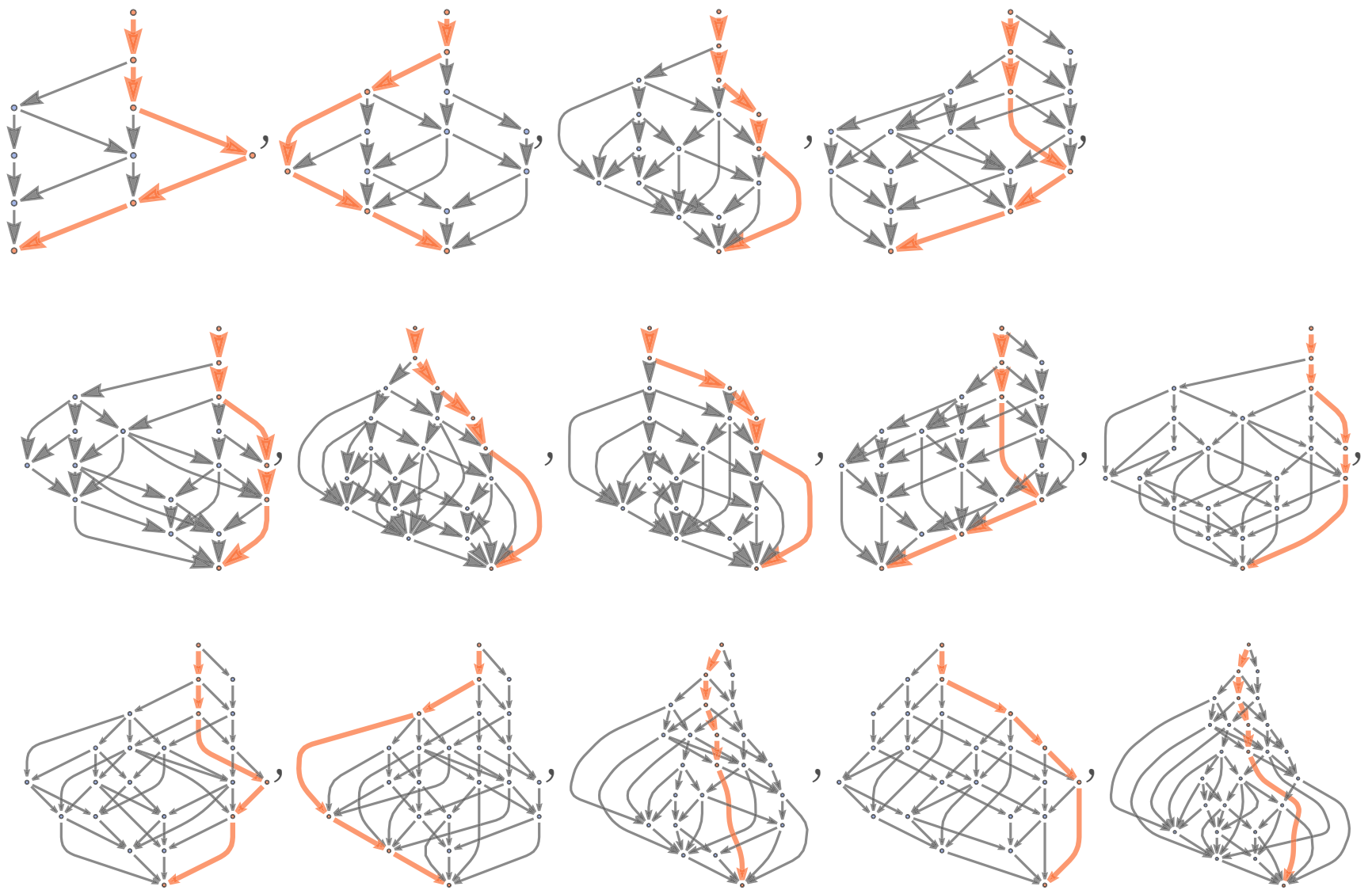

But if we don't look at only at leftmost-outermost-fastest programs? At size 7 here are the multiway graphs for all combinator expressions that compute x[x]:

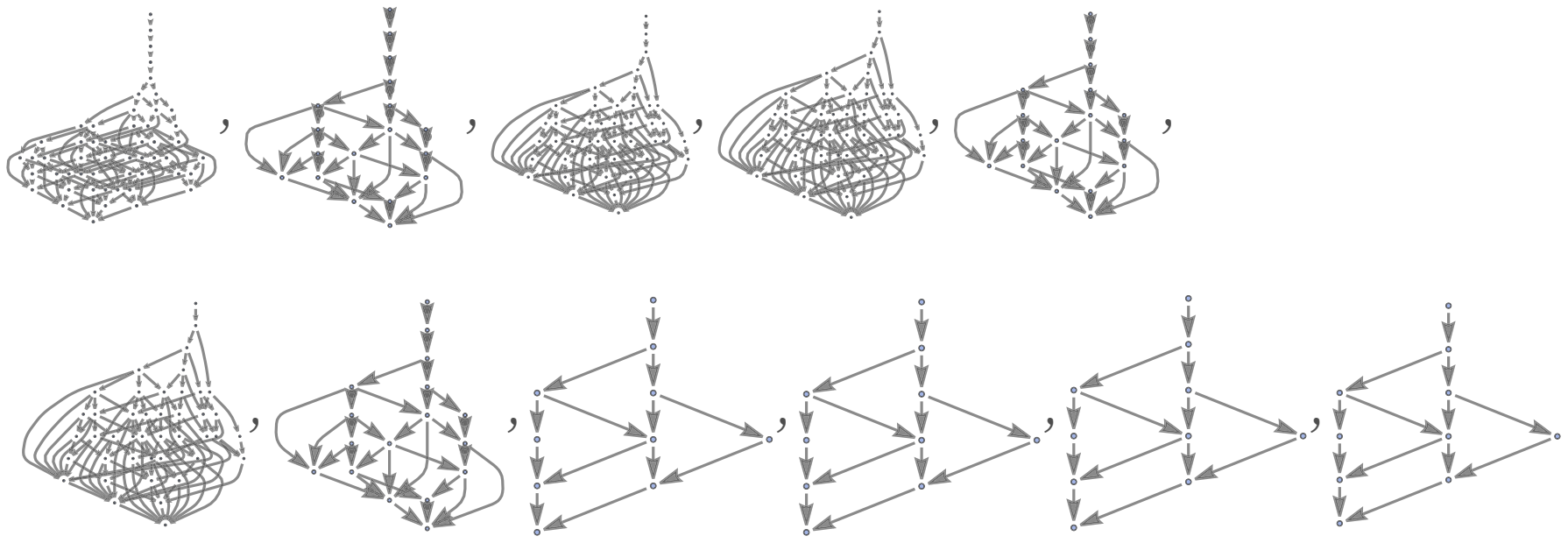

So if one operates "non-deterministically"—i.e. one can follow any path in the multiway graph, not just the leftmost outermost evaluation scheme one—can one compute the answer faster? The answer in this particular case is no.

But what about at size 8? Of the 95 programs that compute x[x], in most cases the situation is like for size 7 and leftmost outermost gives the fastest result. But there are some wilder things that can happen.

Consider for example

s[s[s[s]]][k[s[k]]][s]

Here's the complete multiway graph in this case (with 477 nodes altogether):

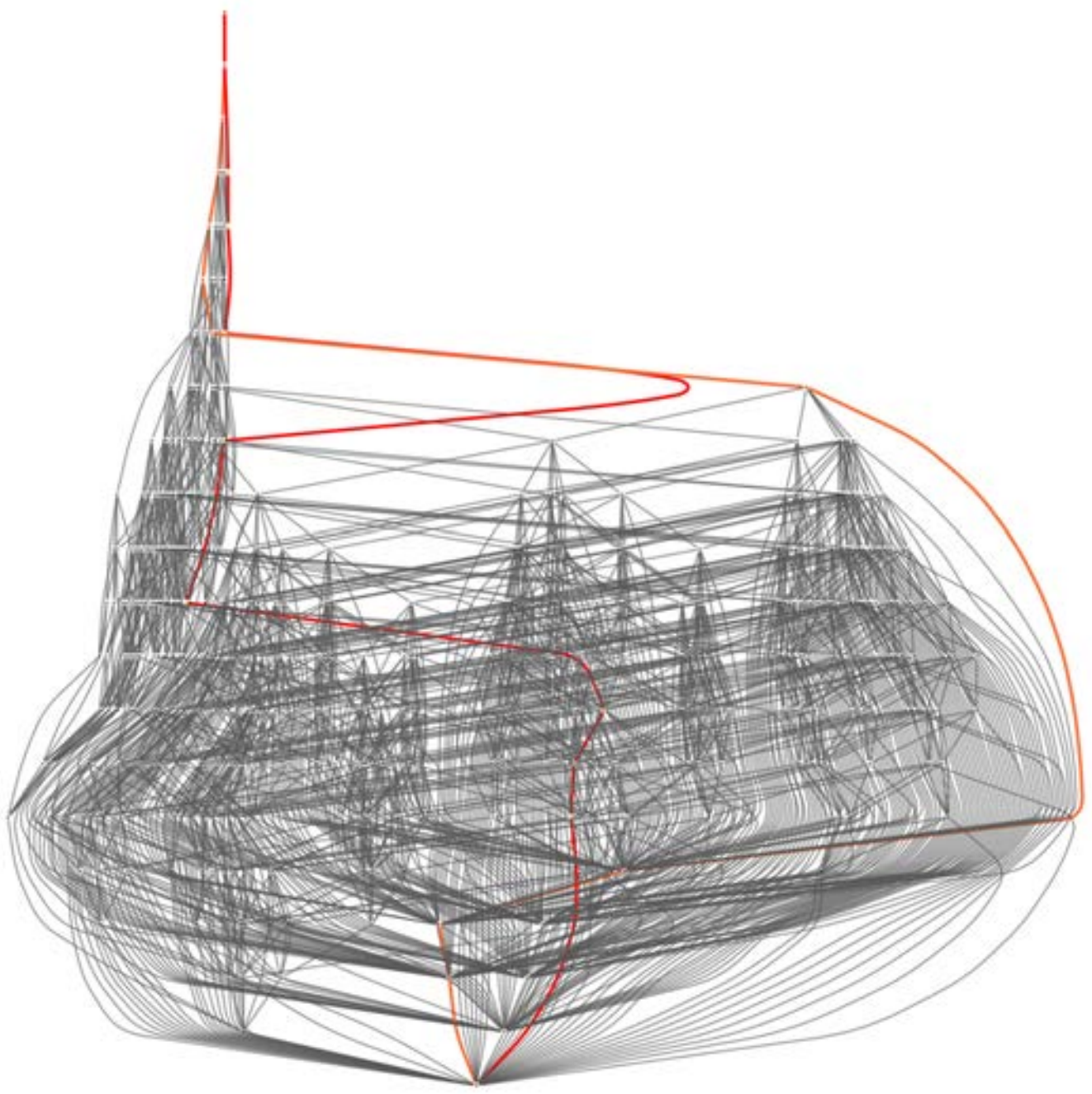

Two paths are indicated: the one in orange is the leftmost outermost evaluation—which takes 12 steps in this case. But there's also another path, shown in red—which has length 11. Here's a comparison:

| | |
|---|---|
| 1 | s[s[s[s]]][k[s[k]]][s][x] |
| 2 | s[s[s]][s][k[s[k]][s]][x] |
| 3 | s[s][k[s[k]][s]][s[k[s[k]][s]]][x] |
| 4 | s[s[k[s[k]][s]]][k[s[k]][s][s[k[s[k]][s]]]][x] |
| 5 | s[k[s[k]][s]][x][k[s[k]][s][s[k[s[k]][s]]][x]] |
| 6 | k[s[k]][s][k[s[k]][s][s[k[s[k]][s]]][x]][x[k[s[k]][s][s[k[s[k]][s]]][x]]] |
| 7 | s[k][k[s[k]][s][s[k[s[k]][s]]][x]][x[k[s[k]][s][s[k[s[k]][s]]][x]]] |
| 8 | k[x[k[s[k]][s][s[k[s[k]][s]]][x]]][k[s[k]][s][s[k[s[k]][s]]][x][x[k[s[k]][s][s[k[s[k]][s]]][x]]]] |
| 9 | x[k[s[k]][s][s[k[s[k]][s]]][x]] |
| 10 | x[s[k][s[k[s[k]][s]]][x]] |
| 11 | x[k[x][s[k[s[k]][s]][x]]] |
| 12 | x[x] |
| 1 | s[s[s[s]]][k[s[k]]][s][x] |
| 2 | s[s[s]][s][k[s[k]][s]][x] |
| 3 | s[s[s]][s][s[k]][x] |
| 4 | s[s][s[k]][s[s[k]]][x] |
| 5 | s[s[s[k]]][s[k][s[s[k]]]][x] |
| 6 | s[s[k]][x][s[k][s[s[k]]][x]] |
| 7 | s[k][s[k][s[s[k]]][x]][x[s[k][s[s[k]]][x]]] |
| 8 | k[x[s[k][s[s[k]]][x]]][s[k][s[s[k]]][x][x[s[k][s[s[k]]][x]]]] |
| 9 | k[x[k[x][s[s[k]][x]]]][s[k][s[s[k]]][x][x[s[k][s[s[k]]][x]]]] |
| 10 | k[x[x]][s[k][s[s[k]]][x][x[s[k][s[s[k]]][x]]]] |
| 11 | x[x] |

To get a sense of the "amount of non-determinism" that can occur, we can look at the number of nodes in successive layers of the multiway graph—essentially the number of "parallel threads" present at each "non-deterministic step":

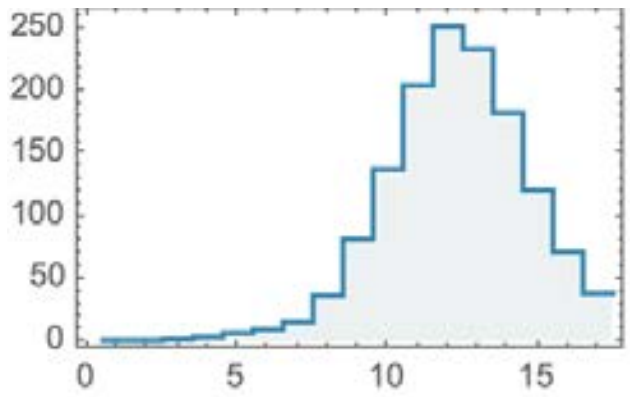

What about size-8 programs for x[x]? There are 9 more—similar to this one—where the non-deterministic computation is one step shorter. (Sometimes—as for s[s[s]][s][s[k[k][s]]]—the multiway graph is more complicated, in this case having 1661 nodes.)

But there are some other things that happen. And a dramatic one is that there can paths that just don't terminate at all. s[s[s[s][s]]][s][s[k]] gives an example. Leftmost-outermost evaluation reaches a fixed point after 14 steps. But overall the multiway graph grows exponentially (already having size 24,705 after 14 steps)—yielding eventually an infinite number of infinite paths: non-deterministic threads that in a sense get "lost forever".

So far all we've talked about here is the computation of the one—seemingly trivial—object x[x]. But what about computing other things? Imagine we have a combinator expression ■ that we apply to x to form ■[x]. If when we "evaluate" this with the combinator rules it reaches a fixed point we can say this is the result of the computation. But a key point is that most of the time this "result" won't just contain x; it'll still have "innards of the computation"—in the form of S's and K's—in it.

Out of all 2688 combinator expressions of size 6, 224 compute x. Only one (that we saw above) computes something more complicated: x[x]. At size 7, there are 11 programs that compute x[x], and 4 that compute x[x][x]. At size 8 the things that can be computed are:

| | |
|---|---|
| x | 8926 |
| x[x] | 95 |
| x[x[x]] | 13 |
| x[x][x] | 13 |
| x[x[x]][x] | 3 |
| x[x[x]][x[x]][x[x[x[x]][x[x]]]][x[x[x[x]][x[x]]]] | 1 |

At size 9 the result is:

| | |
|---|---|
| x | 58678 |
| x[x] | 730 |
| x[x[x]] | 55 |
| x[x][x] | 174 |
| x[x[x][x]] | 5 |
| x[x][x[x]] | 16 |
| x[x[x]][x] | 6 |
| x[x][x][x] | 3 |
| x[x][x[x[x]]] | 9 |
| x[x][x[x][x]] | 4 |
| x[x[x]][x[x]] | 7 |
| x[x[x][x]][x] | 1 |
| x[x][x[x][x[x][x]]] | 1 |
| x[x[x][x]][x[x][x]] | 2 |
| x[x][x[x[x]]][x[x]] | 3 |
| x[x[x]][x[x]][x[x[x[x]][x[x]]][x[x[x]][x[x]]]] | 1 |
| x[x[x][x[x][x]]][x[x[x][x[x][x]]][x[x][x[x][x]]]] | 1 |

In a sense what we’re seeing here are the expressions (or objects) of “low algorithmic information content” with respect to combinator computation: those for which the shortest combinator program that generates them is just of length 9. In addition to shortest program length, we can also ask about expressions generated within certain time or intermediate-expression-size constraints.

What about the other way around? How large a program does one need to generate a certain object? We know that x[x] can be generated with a program of size 6. It turns out x[x[x]] needs a program of size 8:

| | | |
|---|---|---|
| x[x][x] | s[s[s]][s[s]][s[k]] | 7 |
| x[x[x]] | s[s[s[s]]][s][s][s[k]] | 8 |

Here are the shortest programs for objects of size 4:

| | | |
|---|---|---|
| x[x][x][x] | s[s][s[s[s]]][s[s]][s[k]] | 9 |
| x[x[x][x]] | s[s][s[s[s][k]]][s[k][s]] | 9 |
| x[x[x]][x] | s[s[s[s][k]]][s[k][s]] | 8 |
| x[x[x[x]]] | s[s][s[s[s[s[k]]]]][s[k][s]] | 10 |
| x[x][x[x]] | s[s[s[s[s]]]][s[s]][s[k]] | 9 |

Our original "straightforward compiler" generates considerably longer programs: to get an object involving only x's of size n it produces a program of length 4n – 1 (i.e. 15 in this case).

It's interesting to compare the different situations here. x[x[x]][x[x]][x[x[x[x]][x[x]]]][x[x[x[x]][x[x]]]] (of size 17) can be generated by the program s[s[s]][s][s[s][s[k]]] (of size 8). But the shortest program that can generate x[x[x[x]]] (size 4) is of length 10. And what we're seeing is that different object can have very different levels of "algorithmic redundancy" under combinator computation.

Clearly we could on to investigate objects that involve not just x, but also y, etc. And in general there's lots of empirical computation theory that one can expect to do with combinators.

As one last example, one can ask how large a combinator expression is needed to "build to a certain size", in the sense that the combinator expression evolves to a fixed point with that size. Here is the result for all sizes up to 100, both for S,K expressions, and for expressions with S alone (the dotted line is $\log_{\frac{3}{2}}(n)$):

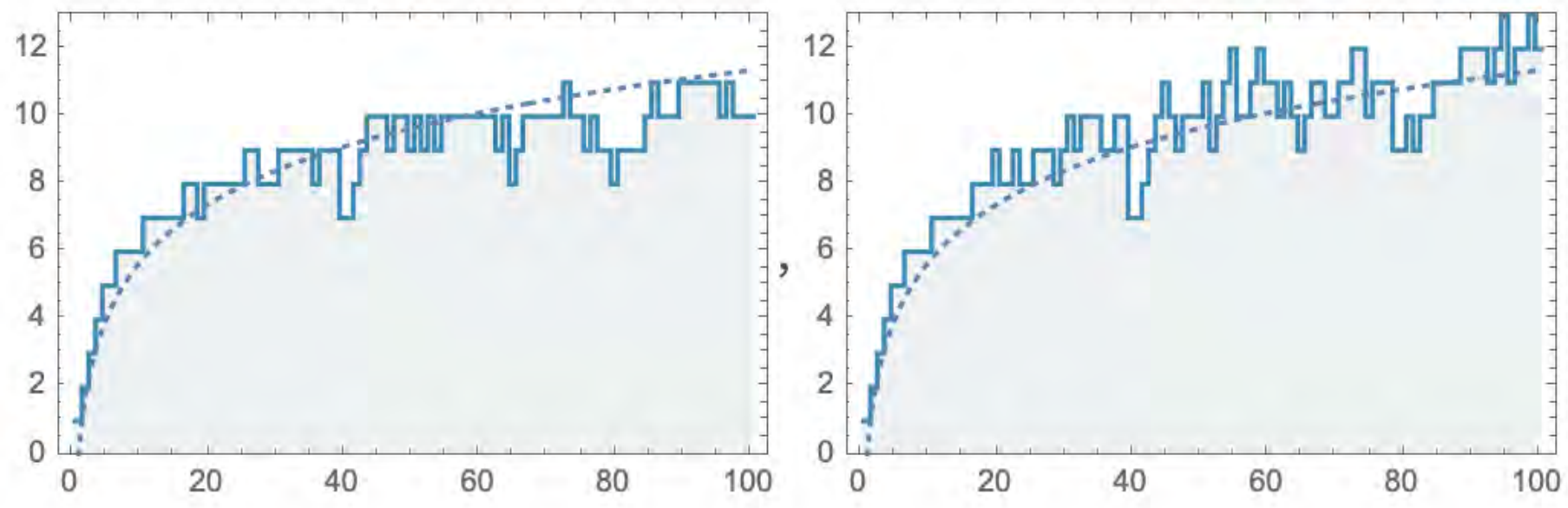

By the way, we can also ask in general about programs that involve only S, without K. If one wants ■[x] to evaluate to an expression involving only x this isn't possible if one only uses S. But as we discussed above, it's still perfectly possible to imagine "doing a computation" only using S: one just can't expect to have the result delivered directly on its own. Instead, one must run some kind of procedure to extract the result from a "wrapper" that contains S's.

What about practical computations? The most obvious implementation of combinators on standard modern computer systems isn't very efficient because it tends to involve extensive copying of expressions. But by using things like the DAG approach discussed above it's perfectly possible to make it efficient.

What about physical systems? Is there a way to do “intrinsically combinator” computation? As I discussed above, our model of fundamental physics doesn’t quite align with combinators. But closer would be computations that can be done with molecules. Imagine a molecule with a certain structure. Now imagine that another molecule reacts with it to produce a molecule with a new structure. If the molecules were tree-like dendrimers, it’s at least conceivable that one can get something like a combinator transformation process.

I’ve been interested for decades in using ideas gleaned from exploring the computational universe to do molecular-scale computation. Combinators as such probably aren’t the best “raw material", but understanding how computation works with combinators is likely to be helpful.

And just for fun we can imagine taking actual expressions—say from the evolution of s[s][s][s[s]][s][s]—and converting them to “molecules” just using standard chemical SMILES strings (with C in place of S):

**{SSS(SS)SS, S(SS)(S(SS))SS, SSS(S(SS)S)S, S(S(SS)S)(S(S(SS)S))S,
S(SS)SS(S(S(SS)S)S), SSS(SS)(S(S(SS)S)S), S(SS)(S(SS))(S(S(SS)S)S),
SS(S(S(SS)S)S)(S(SS)(S(S(SS)S)S)), S(S(SS)(S(S(SS)S)S))(S(S(SS)S)S(S(SS)(S(S(SS)S)S))),
S(S(SS)(S(S(SS)S)S))(S(SS)S(S(SS)(S(S(SS)S)S))(S(S(SS)(S(S(SS)S)S)))),
S(S(SS)(S(S(SS)S)S))(SS(S(SS)(S(S(SS)S)S))(S(S(SS)(S(S(SS)S)S)))(S(S(SS)(S(S(SS)S)S))))}**

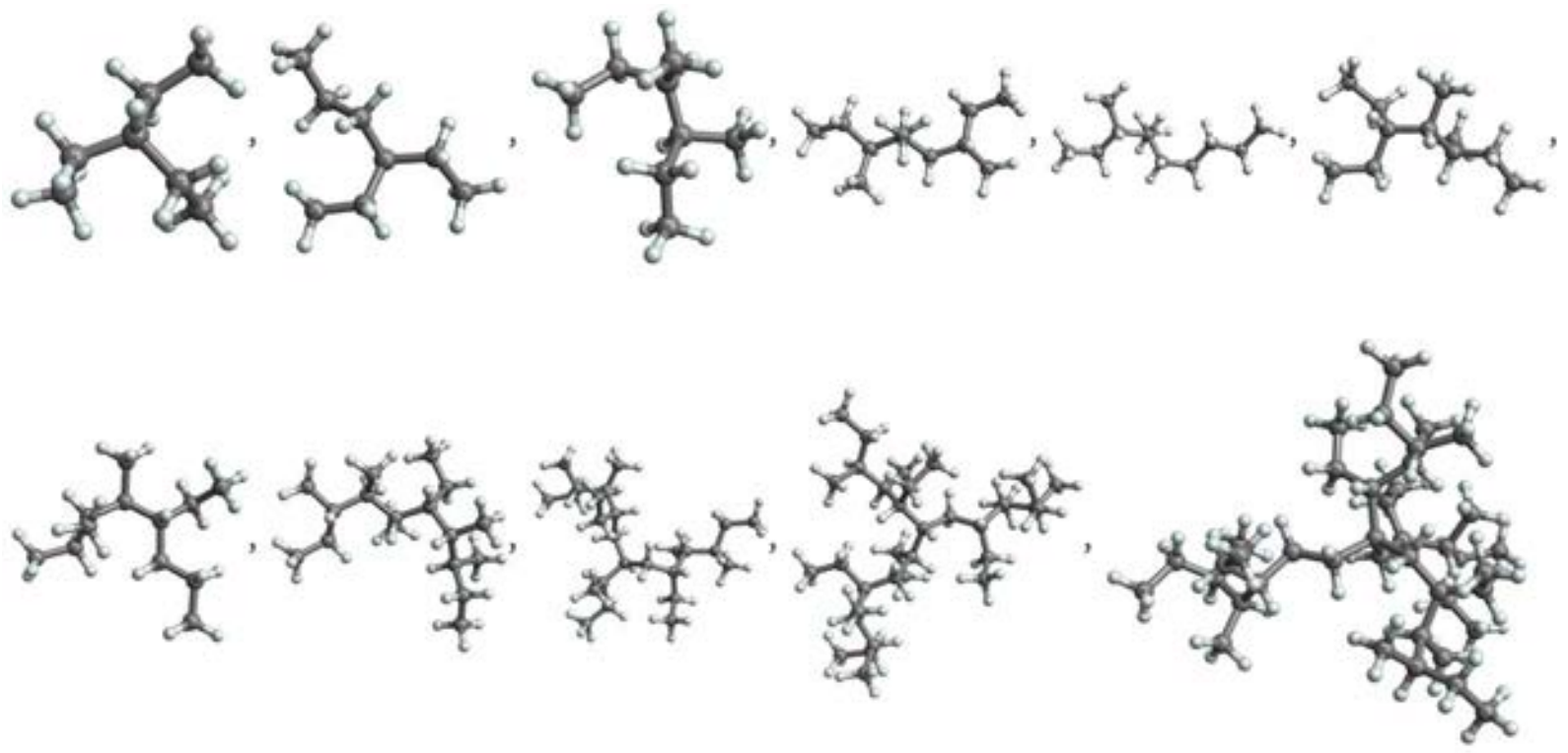

## The Future of Combinators

S and K at first seem so simple, so basic. But as we’ve seen here, there’s an immense richness to what they can do. It’s a story I’ve seen played out many times across the computational universe. But in a sense it’s particularly remarkable for combinators because they were invented so early, and they seem so very simple.

There’s little question that even a century after they were invented, combinators are still hard to get one’s head around. Perhaps if computation and computer technology had developed differently, we’d now find combinators easier to understand. Or perhaps the way our brains are made, they’re just intrinsically difficult.

In a sense what makes combinators particularly difficult is the extent to which they're both featureless and fundamentally dynamic in their structure. When we apply the ideas of combinators in practical "human-oriented" computing—for example in the Wolfram Language—we annotate what's going on in a variety of ways. But with the Wolfram Physics Project we now have the idea that what happens at the lowest level in the physics of our universe is something much more like "raw combinators".

The details are different—we're dealing with hypergraphs, not trees—but many of the concepts are remarkably similar. Yes, a universe made with combinators probably won't have anything like space in the way we experience it. But a lot of ideas about updating processes and multiway systems are all there in combinators.

For most of their history, combinators have been treated mainly as a kind of backstop for proofs. Yes, it is possible to avoid variables, construct everything symbolically, etc. But a century after they were invented, we can now see that combinators in their own right have much to contribute.

What happens if we don't just think about combinators in general, but actually look at what specific combinators do? What happens if we do experiments on combinators? In the past some elaborate behavior of a particular combinator expression might have just seemed like a curiosity. But now that we have the whole paradigm that I've developed from studying the computational universe we can see how such things fit in, and help build up a coherent story about the ways of computation.

In *A New Kind of Science* I looked a bit at the behavior of combinators; here I've done more. But there's still vastly more to explore in the combinator universe—and many surprises yet to uncover. Doing it will both advance the general science of the computational universe, and will give us a new palette of phenomena and intuition with which to think about other computational systems.

There are things to learn for physics. There are things to learn for language design. There are things to learn about the theoretical foundations of computer science. There may also be things to learn for models of concrete systems in the natural and artificial world—and for the construction of useful technology.

As we look at different kinds of computational systems, several stand out for their minimalism. Particularly notable in the past have been cellular automata, Turing machines and string substitution systems. And now there are also the systems from our Wolfram Physics Project—that seem destined to have all sorts of implications even far beyond physics. And there are also combinators.

One can think of cellular automata, for example, as minimal systems that are intrinsically organized in space and time. The systems from our Wolfram Physics Project are minimal systems that purely capture relations between things. And combinators are in a sense minimal systems that are intrinsically about programs—and whose fundamental structure and operation revolve around the symbolic representation of programs.

What can be done with such things? How should we think about them?

Despite the passage of a century—and a substantial body of academic work—we're still just at the beginning of understanding what can be done with combinators. There's a rich and fertile future ahead, as we begin the second combinator century, now equipped with the ideas of symbolic computational language, the phenomena of the computational universe, and the computational character of fundamental physics.

## Historical & Other Notes

I'm writing elsewhere about the origin of combinators, and about their interaction with the history of computation. But here let me make some remarks more specific to this piece.

Combinators were invented in 1920 by Moses Schönfinkel (hence the centenary), and since the late 1920s there's been continuous academic work on them—notably over more than half a century by Haskell Curry.

A classic summary of combinators from a mathematical point of view is the book: Haskell B. Curry and Robert Feys, Combinatory Logic (1958). More recent treatments (also of lambda calculus) include: H. P. Barendregt, The Lambda Calculus (1981) and J. Roger Hindley and Jonathan P. Seldin, Lambda-Calculus and Combinators (1986).

In the combinator literature, what I call "combinator expressions" are often called "terms" (as in "term rewriting systems"). The part of the expression that gets rewritten is often called the "redex"; the parts that get left over are sometimes called the "residuals". The fixed point to which a combinator expression evolves is often called its "normal form", and expressions that reach fixed points are called "normalizing".

Forms like a[b[a][c]] that I "immediately apply to arguments" are basically lambda expressions, written in Wolfram Language using **Function**. The procedure of "compiling" from lambda expressions to combinators is sometimes called bracket abstraction. As indicated by examples at the end of this piece, there are many possible methods for doing this.

The scheme for doing arithmetic with combinators at the beginning of this piece is based on work by Alonzo Church in the 1930s, and uses so-called "Church numerals". The idea of encoding logic by combinators was discussed by Schönfinkel in his original paper, though the specific minimal encoding I give was something I found by explicit computational search in just the past few weeks. Note that if one uses s[k] for **True** and k for **False** (as in the rule 110 cellular automaton encoding) the minimal forms for the Boolean operators are:

| | | | | |
|---|---|---|---|---|
| 0 | □□□□ | True | k[k[k]] | K(KK) |
| 1 | □□□■ | Nand | s[s[k[s[s[s][k[k[k]]]]]]][s] | S(S(K(S(SS(K(KK))))))S |
| 2 | □□■□ | Implies | s[s][k[k[k]]] | SS(K(KK)) |
| 3 | □□■■ | Not | s[s][s[s[s[s[k]]][s]]][k[k]] | SS(S(S(S(SK))S))(KK) |
| 4 | □■□□ | | s[k[s[s[s][k[k[k]]]]]][k] | S(K(S(SS(K(KK)))))K |
| 5 | □■□■ | Not | k[s[s[s][k[k[k]]]][s]] | K(S(SS(K(KK)))S) |
| 6 | □■■□ | Equal | s[s][k[s[s[s][k[k[k]]]][s]]] | SS(K(S(SS(K(KK)))S)) |
| 7 | □■■■ | Nor | s[s[s[s[s][k[k[k[k]]]]]][k[s]]] | S(S(S(SS(K(K(KK)))))(KS)) |
| 8 | ■□□□ | Or | s[s[s]][s][s[k]] | S(SS)S(SK) |
| 9 | ■□□■ | Xor | s[s[s[s[s]][s[s[s[k]]]][s]]][k] | S(S(S(SS)(S(S(SK)))S))K |
| 10 | ■□■□ | Last | s[k] | SK |
| 11 | ■□■■ | | s[s[s[k]]][s] | S(S(SK))S |
| 12 | ■■□□ | First | k[k][s] | KKS |
| 13 | ■■□■ | | s[k[s[s[s[s[k]]][s]]]][k] | S(K(S(S(S(SK))S)))K |
| 14 | ■■■□ | And | s[s][k] | SSK |
| 15 | ■■■■ | False | k[k[s[k]]] | K(K(SK)) |

The uniqueness of the fixed point for combinators is a consequence of the Church–Rosser property for combinators from 1941. It is closely related to the causal invariance property that appears in our model of physics.

There's been a steady stream of specific combinators defined for particular mathematical purposes. An example is the Y combinator s[s][k][s[k[s[s][s[s[s][k]]]]][k]], which has the property that for any x, Y[x] can be proved to be equivalent to x[Y[x]], and "recurses forever". Here's how Y[x] grows if one just runs it with leftmost outermost evaluation (and it produces expressions of the form Nest[x, _, n] at step $n^2 + 7n$):

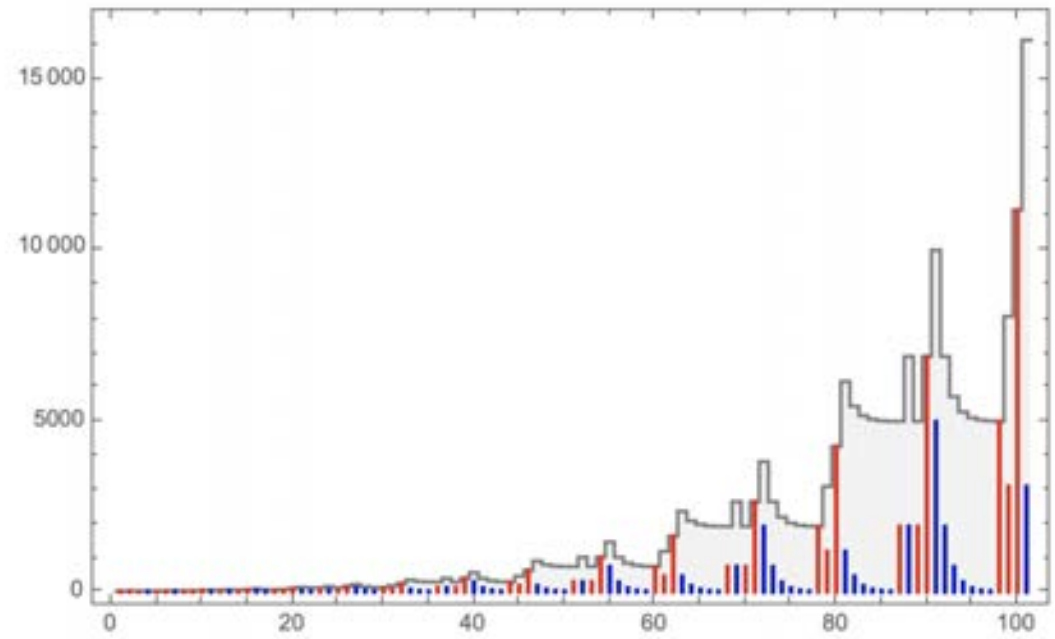

The Y combinator was notably used by Paul Graham in 2005 to name his Y Combinator startup accelerator. And perhaps channeling the aspirations of startups the "actual" Y combinator goes through many ups and downs but (with leftmost outermost evaluation) reaches size 1 billion ("unicorn") after 494 steps—and after 1284 steps reaches more-dollars-than-in-the-world size: 508,107,499,710,983.

Empirical studies of the actual behavior of combinators “in the wild” have been pretty sparse. The vast majority of academic work on combinators has been done by hand, and without the overall framework of A New Kind of Science the detailed behavior of actual combinators mostly just seemed like a curiosity.

I did fairly extensive computational exploration of combinators (and in general what I called “symbolic systems”) in the 1990s for A New Kind of Science. Page 712 summarized some combinator behavior I found (with /. evaluation):

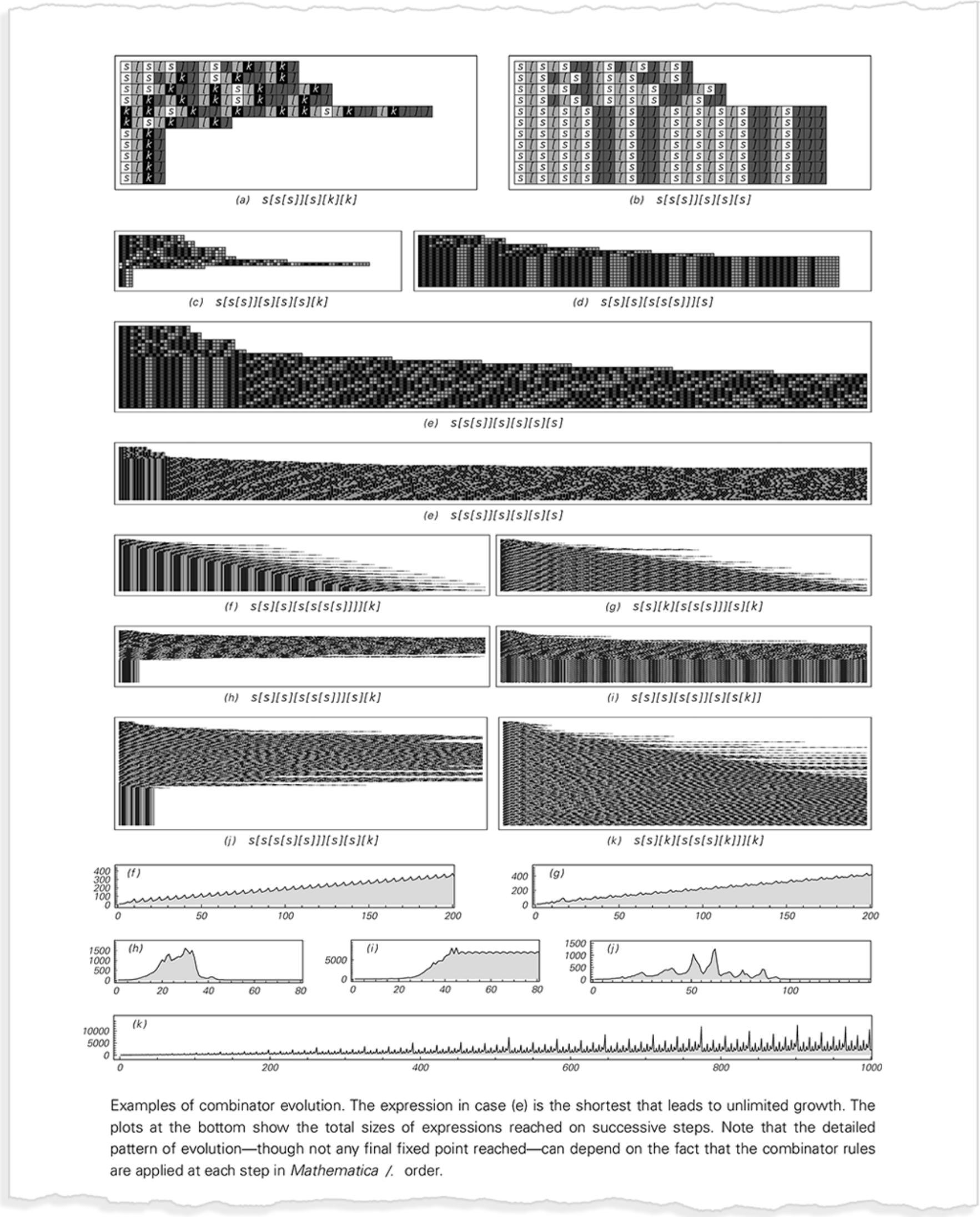

Examples of combinator evolution. The expression in case (e) is the shortest that leads to unlimited growth. The plots at the bottom show the total sizes of expressions reached on successive steps. Note that the detailed pattern of evolution—though not any final fixed point reached—can depend on the fact that the combinator rules are applied at each step in *Mathematica* /. order.

I don’t know to what extent the combinator results in A New Kind of Science were anticipated elsewhere. Long-time combinator enthusiast Henk Barendregt for example recently pointed me to a paper of his from 1976 mentioning non-termination in S combinator expressions:

The length of an S-term is the number of its S's. If $a_n$ is the number of S-terms with length n, then by the formula of Catalan (cf. [3] p. 64)

$$a_n = \frac{1}{2n-1}\binom{2n-1}{n}.$$

The first values of $a_n$ are indicated in fig. 1.
Let $b_n$ be the number of S-terms of length n without a nf.
Mr. Duboué has calculated by computer upper bounds for $b_n$, for $n < 10$, see fig. 1.

| n | 1 | 2 | 3 | 4 | 5 | 6 | 7 | 8 | 9 | 10 |
|---|---|---|---|---|---|---|---|---|---|---|
| $a_n$ | 1 | 1 | 2 | 5 | 14 | 42 | 132 | 429 | 1430 | 4862 |
| $b_n$ | 0 | 0 | 0 | 0 | 0 | 0 | 2 | ⩽39 | ⩽231 | |

(fig. 1)

The bounds are not exact, since the computer only reduced a term a (large) finite number of times in order to conclude that it might be non-normal. For n = 7, theorem 6.4 proves that the bound is exact.

7.1. Notations. C[ ] is a context containing one or more holes.
$F^0X = X$; $F^{n+1} = F(F^nX)$.
$M \xrightarrow{\circledast} N \iff CL \vdash M \longrightarrow C[N]$ for some context C[ ], and $M \not\equiv C\,N$.

The procedure I describe for determining the termination of S combinator expression was invented by Johannes Waldmann at the end of the 1990s. (The detailed version that I used here came from Jörg Endrullis.)

What we call multiway systems have been studied in different ways in different fields, under different names. In the case of combinators, they are basically Böhm trees (named after Corrado Böhm).

I've concentrated here on the original S, K combinators; in recent livestreams, as in A New Kind of Science, I've also been exploring other combinator rules.

## Thanks, etc.

Matthew Szudzik has helped me with combinator matters since 1998 (and has given a lecture on combinators almost every year for the past 18 years at our Wolfram Summer School). Roman Maeder did a demo implementation of combinators in Mathematica in 1988, and has now added `CombinatorS` etc. to Version 12.2 of Wolfram Language.

I've had specific help on this piece from Jonathan Gorard, Jose Martin-Garcia, Eric Paul, Ed Pegg, Max Piskunov, and particularly Mano Namuduri, as well as Jeremy Davis, Sushma Kini, Amy Simpson and Jessica Wong. We've had recent interactions about combinators with a four-academic-generation sequence of combinator researchers: Henk Barendregt, Jan Willem Klop, Jörg Endrullis and Roy Overbeek.

## References

*Links to references are included within the body of this document.*